\documentclass[twocolumn]{aa}
\pdfoutput=1 
\usepackage[english]{babel}
\usepackage{graphicx}
\usepackage{epstopdf}
\usepackage{epsfig}
\usepackage{natbib}
\usepackage{aalongtable}
\usepackage{booktabs}
\usepackage{multirow}
\usepackage{rotating}
\usepackage{pdfpages}
\usepackage{amsmath}
\usepackage{amssymb,color,latexsym,longtable,array}
\usepackage{txfonts}
\usepackage{lscape}
\usepackage{morefloats}
\usepackage{floatrow}
\DeclareFloatFont{tinyFont}{\small}
\bibpunct{(}{)}{;}{a}{}{,}
\graphicspath{{Images/}{Images/Specs/}{Images/newSED/}{Images/Profiles/}}

\begin{document}

\title{Properties of optically selected BL Lac candidates from the SDSS
\thanks{Based on observations collected with the NTT on La Silla (Chile)
operated by the European Southern Observatory under proposal 082.B-0133.}
\thanks{Based on observations collected at the Centro Astron\'{o}mico
Hispano Alem\'{a}n (CAHA), operated jointly by the Max-Planck-Institut
f\"ur Astronomie and the Instituto de Astrofisica de Canarias.}
\thanks{Based on observations made with the Nordic Optical Telescope,
operated on the island of La Palma jointly by Denmark, Finland, Iceland,
Norway, and Sweden, in the Spanish Observatorio del Roque de los
Muchachos of the Instituto de Astrofisica de Canarias.}}

\author{
S. D. K\"ugler\inst{1,3} \and
K. Nilsson\inst{2} \and
J. Heidt\inst{1} \and
J. Esser\inst{1} \and
T. Schultz\inst{1,4}
}

\institute{
ZAH, Landessternwarte Heidelberg, K\"onigstuhl 12, 69117 Heidelberg, Germany \\
email: dkuegler@lsw.uni-heidelberg.de
\and
Finnish Centre for Astronomy with ESO (FINCA), University of Turku,
V\"ais\"al\"antie 20, FI-21500 Piikki\"o, Finland
\and
Heidelberger Institut f\"ur Theoretische Studien (HITS), Schloss-Wolfsbrunnenweg 35, D-69118 Heidelberg, Germany
\and
Haus der Astronomie Heidelberg, K\"onigstuhl 17, 69117 Heidelberg, Germany
}
\date{Received/Accepted}

\abstract{Deep optical surveys open the avenue for 
find large numbers of BL Lac objects that are hard to identify 
because they lack the unique properties classifying them as such. 
While radio or X-ray surveys 
typically reveal dozens of sources, recent compilations based on 
optical criteria alone have increased the number of BL Lac candidates 
considerably.
However, these compilations are subject to biases and 
may contain a substantial number of contaminating sources.} 
  {In this paper we extend our analysis of 182 optically 
selected BL Lac object candidates from the SDSS with respect to an earlier study.
The main goal is to determine the number of bona fide BL Lac objects in 
this sample. }
  { We examine their variability characteristics, determine their
broad-band radio-UV SEDs, and search for the presence of a host 
  galaxy. 
In addition we present 
  new optical spectra for 27 targets with improved S/N with
  respect to the SDSS spectra.} 
{At least 59\% of our targets have shown variability between SDSS DR2 
  and our observations by more than 0.1-0.27 mag depending on the 
  telescope used. A host galaxy was detected in 36\% 
  of our targets. The host galaxy type and luminosities are consistent 
  with earlier studies of BL Lac host galaxies. Simple fits to broad-band
SEDS for 104 targets of our sample derived synchrotron peak frequencies
between  $13.5 \leq \mathrm{log}_{10}\left(\nu_{\mathrm{peak}}\right) \leq 16$ with 
a peak at $\mathrm{log}_{10} \sim 14.5$.  Our new optical spectra do 
not reveal any new redshift for any of our objects.
Thus the sample contains a large number of bona fide BL Lac objects and 
seems to contain a substantial fraction of intermediate-frequency peaked BL Lacs.}
{}

\maketitle

\section{Introduction}

Active galactic nuclei (AGN) are compact and extremely luminous 
objects located in the center of galaxies. They are dominated by
a super massive black hole that is continuously fed by 
infalling gas from the surrounding environment. 
Their emission is  strongly variable and consists of a continuum emission 
from an accretion disk that dominates the optical to X-ray regime and
strong infrared emission that is attributed to a surrounding dusty torus. 
In addition high-velocity gas is present that manifests itself by 
emission lines in the optical spectrum. In about 10\% of the AGN,
synchrotron radiation from a jet dominates, which led to a 
distinction between radio-loud and radio-quiet AGN. 
They come in different flavors all of which can well be explained within the 
so-called ``unified scheme'' (e.g., \citealp{1995PASP..107..803U}),
where the major difference between each of them is the inclination between
the geometry of the system and the line of sight to the observer.

BL Lacertae objects (BL Lacs) are a subclass of AGN characterized by 
strong variability across the entire electromagnetic spectrum
 on timescales down to minutes, as well as high and variable polarization. 
With some exceptions, where narrow emission
lines from the AGN itself or absorption lines from 
their host galaxy are present, their optical spectra are mostly featureless. 
This is due to strong Doppler-boosting of dominating jet emission, which 
leads to an outshining of the host galaxy or accretion disk in many cases
and to the well-known effect of superluminal motion.
Within the unified scheme, BL Lacs are thought to be Fanaroff-Riley class I 
radio galaxies \citep{1974MNRAS.167P..31F}, whose jets are aligned within a few 
degrees to the line of sight. 
Not surprisingly, they are rare. In the latest AGN catalog of  
\citet{2010A&A...518A..10V}, fewer than $1\%$ objects are listed as BL Lacs. 

Although apparently simple, there has been an intense discussion of
the defining characteristics of a BL Lac since their detection.  For
example, \citet{1991ApJ...374..431S} classified objects as a BL Lac
when their optical spectrum did not contain lines with a rest-frame
equivalent width (RFEW) exceeding 5\AA, while
\citet{1991ApJS...76..813S}) used the criteria of synchrotron emission
dominating the optical spectrum, as well as optical (variable)
polarization.  The former criterion was violated at least
once by the prototype BL Lac itself \citep{1995ApJ...452L...5V}.  
For a long time it is known that the properties of a given BL Lac sample 
are depending on the selection frequency. 
Early BL Lac 
samples were formed by identifying the optical counterparts of radio
\citep{1991ApJ...374..431S} or X-ray \citep{1996ApJS..104..251P}
sources and finding targets with featureless optical spectra. The
former method favored BL Lacs with synchrotron peaks in the infrared
range, while the latter favors targets with their synchrotron emission
peaking in the UV to X-ray regime, leading to an apparent bimodality in
source properties when viewed in the $\alpha_{ox}$ - $\alpha_{ro}$
plane (e.g., \citealp{1995ApJ...444..567P}). Later surveys employing
cross-correlations between radio and x-ray surveys
\citep{1999ApJ...525..127L, 2005A&A...434..385G} have found
intermediate targets filling the gap, but even these samples may be
biased by the shallowness of the respective surveys and do not give
the full picture of the BL Lac population. Nowadays, it is clear that
the distribution of synchrotron peak frequencies is not bimodal 
\citep[e.g., ][]{2010ApJ...716...30A}, also resulting in
continuous coverage in the $\alpha_{ox}$ - $\alpha_{ro}$ plane.  BL
Lacs are commonly dubbed as low-frequency peaked (LBL),
intermediate-frequency peaked (IBL), and high-frequency peaked (HBL) BL
Lacs depending on the frequency of the synchrotron peak
$\nu_{peak}$. Given the continuous distribution of $\nu_{peak}$, the
limits for different classes are somewhat arbitrary. In this paper we
use the common convention to classify a BL Lac as an LBL if
$\log{\nu_{peak}} < 14$, IBL if $14 \leq
\log{\nu_{peak}} < 15$, and HBL if $\log{\nu_{peak}} \geq 15$.

With the advent of large optical surveys (such as SDSS), it became
possible to obtain samples that potentially populate the IBL
region and may be more representative of the BL Lac class as a whole. 
\cite{2005AJ....129.2542C} (C05 hereafter) extracted a sample
of 240 probable BL Lac candidates from the SDSS survey
DR2 \citep{2000AJ....120.1579Y} in order to find an IBL sample not
biased by X-ray or radio properties (see note 1 in
\cite{2011A&A...529A.162H}, hereafter Paper I). A more recent
selection, albeit with a slightly different approach, was presented by
\cite{2010AJ....139..390P}. They recovered the majority of candidates
found by C05 and enlarged the probable candidate list to over 700
objects using SDSS DR7. Despite the careful selection process,
possible confusion by stellar (e.g., DC white dwarfs, see
\citealt{1978ARA&A..16..487A}) or extragalactic (e.g., weak-lined
quasars, see \citealt{1999ApJ...526L..57F}) sources may be present and
cannot be ruled out without further observations.

In Paper I we presented the polarization properties of 182/204
BL Lac candidates from C05. We found 124 out of 182 targets (68\%) to be 
polarized and 95 of the polarized targets (77\%) to be highly polarized 
($>$ 4\%). This indicates that the C05 sample of probable BL Lac objects
indeed contains a large number of bona fide BL Lacs. 

With the present paper we enlarge our study of the properties of this sample.
Using our polarimetric data in combination with the SDSS measurements, we 
look for optical variability and study the host galaxy properties of our 
182 objects. In addition, using the data available in the literature,
we constructed broad-band spectral energy distributions (SEDs). 
They are used to fit simple synchrotron models 
to them in order to derive peak frequencies and 
to determine their LBL/IBL/HBL nature. Finally, new optical spectra for 
27/182 objects are presented. In section 2 we briefly summarize the 
resources used for our data extraction followed by a description of the
analysis in section 3. The results are discussed in section 4 and 
summarize in section 5. For a detailed discussion of the global properties 
of our sample, a comparison to other samples and a potential revision 
of the defining criterion of a BL Lac, we refer to Paper III (Nilsson 
et al., in prep). 

Throughout this paper we use standard cosmology 
($H_0 = 70$ km s$^{-1}$
Mpc$^{-1}$, $\Omega_{M}$ = 0.3, and $\Omega_{\Lambda}$ = 0.7).
When discussing the results of a K-S test, we denote two distributions
that are significantly different if the null hypothesis that they are drawn from
the same parent population can be rejected with p $<$ 1\%.

\section{Data acquisition and reduction}

\subsection{Variability and host galaxies}

The data that we use for variability and host galaxy analysis 
were presented in Paper I, where a detailed log of the observations 
can be found.
Here we briefly repeat the main characteristics of the
observations.

Alltogether, 123 targets were observed at the ESO New Technology
Telescope (NTT) on La Silla, Chile during Oct. 2-6, 2008 and Mar. 28 -
Apr. 1, 2009. The observations were made with the EFOSC2 instrument
through a Gunn-r filter (\#786).  We used a 2k Loral chip with a gain of
0.91 e$^-/$ADU, readout noise of 7.8 e$^-$ and pixel scale of
0\farcs24/pixel in 2x2 binning mode.  The total field of view was
4'$\times$4'.  The observations were made in polarimetric mode; i.e., a
Wollaston prism and a half-wave plate were inserted into the beam,
resulting in two images of each target on the CCD, separated by 10
\arcsec. One polarization observation consisted of four exposures of
10-1000s each at different polarization angles (0, 22.5, 45, and 67.5 
degrees) of the half-wave plate, resulting in eight images per target 
per polarization observation. In most cases a single sequence was
obtained, but two to three sequences were made for fainter targets. 
Seeing was generally good (0\farcs6-1\farcs2) during the NTT
runs and the weather was photometric, except for the last half night 
of NTT run in March, when thin clouds increasingly covered the sky
toward the morning.

Another set of 47 targets was observed in service mode at the Calar
Alto (CA) 2.2 m telescope using the CAFOS instrument on Feb. 18-24,
2009.  The observations were made through a Gunn-r filter using the
central 1000$\times$1000 pixels of the Site-CCD with a gain of 2.3
e$^-$/ADU, readout noise of 5.1 e$^-$, and pixel scale of
0\farcs51/pixel, resulting in a field of view of 7'$\times$7'. The
observations were made using a polarimetric setup similar to the NTT,
except that the separation of the images was 19\arcsec. Exposure times
varied from 30 to 1000 s per half-wave plate position.  Seeing was
better than 1\farcs5 throughout the run and weather was photometric.

Finally, 25 objects were observed using the ALFOSC instrument at the
Nordic Optical Telescope (NOT). Observations were made through the
SDSS-r' (\#84) filter using the central 1500$\times$650 pixels of an
E2V-CCD with a gain of 0.736 e$^-$/ADU, readout noise of 5.3 e$^-$, and
pixel scale of 0\farcs19/pixel, giving a field of view of
4\farcm7$\times$2'. A similar polarimetric mode was used here also,
except that a calcite plate was inserted into the beam to provide a beam
separation of 15\arcsec. Exposure times varied from 150 to 1000 s per
half-wave plate position. Seeing varied between 0\farcs7-1\farcs5, and
the weather was mostly photometric throughout the run, except for some
low-altitude cirrus on the last of the three nights.

All in all, we have 195 observations of 182 targets, i.e. 13 targets
were observed twice. The images were reduced by first subtracting the
bias frame and then were divided the frames by a flat-field, which was
obtained either from the twilight sky (NTT and NOT) of from an evenly
illuminated screen inside the dome (CA). Dark current was negligible
in all cases.

\subsection{Optical photometry (SDSS)}

For estimating the zero points of the various science fields, the photometry 
performed by the Sloan Digital Sky Survey (SDSS) Data Release 2 
\citep{2004AJ....128..502A} was used.
The SDSS is the deepest and most complete optical survey to use a dedicated 
2.5m telescope at the Apache Point observatory. A detailed description of the 
instrument can be found in \citet{1998AJ....116.3040G}. 

\subsection{Optical spectra}

Using the analysis of Paper I, candidates with unknown redshift
and high polarization ($>$4\%) were targeted for spectroscopy. In total 
27 of 87 targets fulfilling this requirement were observed.
The observations were performed at the Calar Alto 2.2 m telescope
during eight nights between March 17 and April 27, 2011. We used the CAFOS
instrument in the long-slit mode with the G-200 grism (Nr.9) and a
slit with of 2-3\arcsec (10\arcsec for standards).  With the chosen
grism, the spectral range was from 440 to 850 nm with a resolution of
$\lambda/\Delta\lambda \approx 400$ at 500 nm. Every night at least one
spectrophotometric 
standard star (Hiltner102, HZ44, BD+33$\_$2642) was observed, and a continuum 
lamp flat field and an arc exposure was made after every observation. 
In order to reach a high S/N ratio and to be able to correct for
cosmic ray hits, three 2400 s exposures were taken for each object.

The spectra were reduced by subtracting the average bias images taken
at the beginning and at the end of each night. Then every image was
divided by its adjacent bias-subtracted flat using standard
IRAF\footnote{Image Reduction and Analysis Facility
  (www.iraf.net) is distributed by the National Optical
  Astronomy Observatories, which are operated by the Association of
  Universities for Research in Astronomy, Inc., under cooperative
  agreement with the National Science Foundation.} tasks.  The
deflection of the spectrum perpendicular to the dispersion axis
(centroid shift) was fitted by a third-order polynomial. After 
background determination and subtraction the spectra were extracted
using an appropriate aperture. The spectra were then wavelength-calibrated 
using the arc-lamp-spectra and the night-sky and finally
flux-calibrated using the standard star observed the same night.

\subsection{Data for SED fits}

To construct broad-band SEDs we used archival data from 
FIRST (Faint Images of the Radio Sky at 
Twenty-Centimeters, \citealp{1995ApJ...450..559B}), 
the NVSS (NRAO VLA Sky Survey, \citealp{1998AJ....115.1693C}),
WISE (Wide-Field Infrared Survey Explorer, \citealp{2010AJ....140.1868W}), 
UKIDSS (the United Kingdom
Infrared Deep Sky Survey DR9, \citealp{2007MNRAS.379.1599L}), 
the SDSS (Sloan Digital Sky Survey DR5, 
\citealp{2007ApJS..172..634A}), 
GALEX (Galaxy Evolution Explorer, \citealp{2005ApJ...619L...1M}), 
and the RASS (ROSAT All Sky Survey, \citealp{2000yCat.9029....0V}).
Some more data points were derived using the NED 
(NASA extragalactic database - http://ned.ipac.caltech.edu/).
Broad-band SEDs could be retrieved for 104 out of our 182 objects.

\section{Analysis}

\subsection{Flux calibration}

We used the data set from Paper I for the test of variability and for
the host galaxy study. Both studies require accurate photometric
calibration of the frames, which we describe in this section.
For the flux calibration we used stars on the CCD frames with SDSS
r-band modelMag magnitudes and
$g - r$ colors available from the SDSS DR2. Each observing run, NTT March, NTT 
October, NOT, CA was calibrated as a separate block with the exception 
of NTT March, where
the last night of this observing run was treated separately due to
increasing cloud coverage toward the end of the night. The rest of
the data were obtained in photometric conditions.

\begin{figure}
\centering
\includegraphics[width=8.5cm]{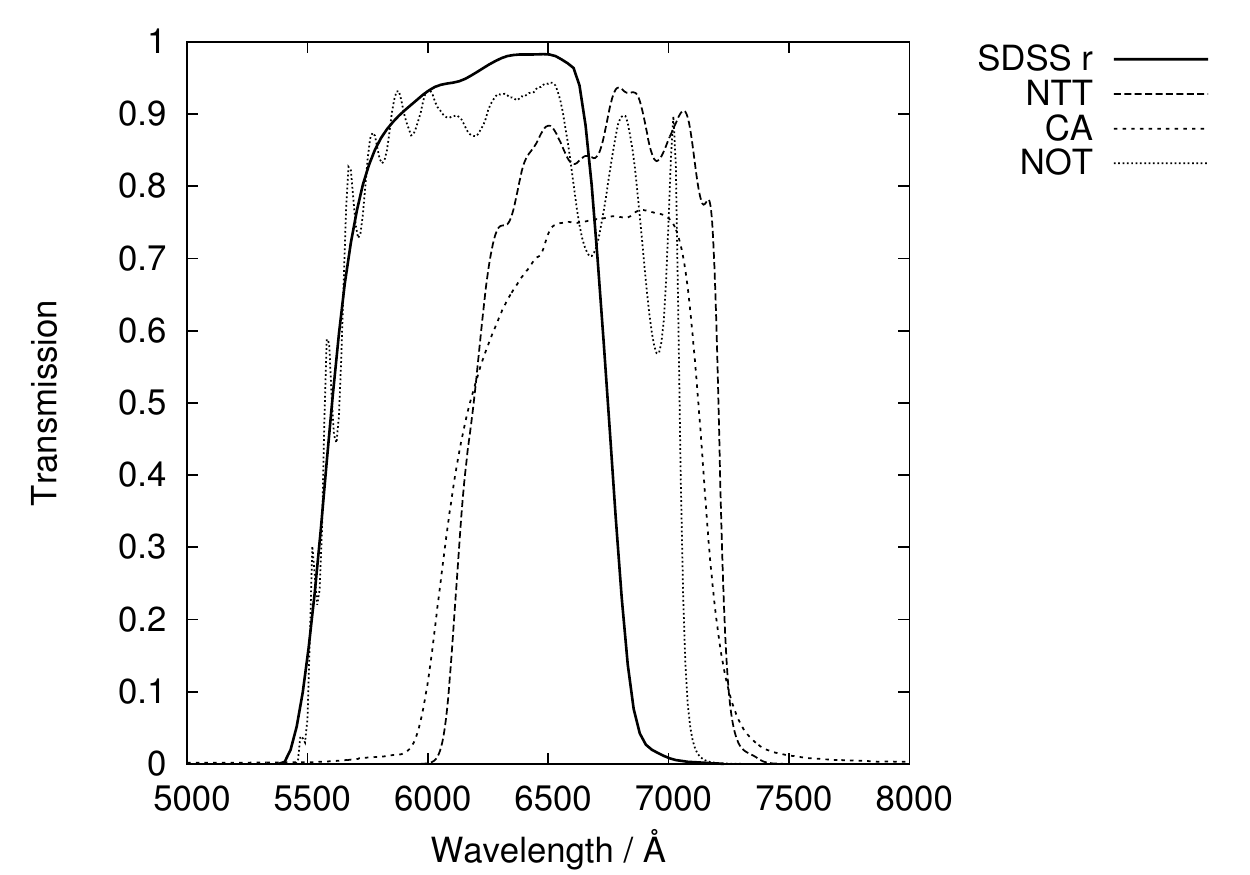}
\caption{\label{filtertrans} 
Transmission curves of the filters used for our observations together with the
SDSS r-band filter.
}
\end{figure}

Since our filter band-passes differ from the SDSS r-band
(Fig. \ref{filtertrans}), a color dependence in the calibration is
expected. Furthermore, the spectra of the BL Lac nuclei are dominated
by a power-law continuum, which differs significantly from the spectra
of the stars used for calibration, whereas the host galaxies have a
SED closer to the stars. Based on these considerations we used two
different equations for the calibration, one for the stars and host
galaxies and another for the BL Lac nuclei. The general form of this
equation can be written as
\begin{equation}
\label{starcolordep}
r = \overline{K_0} - 2.5 \log N + C
\end{equation}
where $r$ is the SDSS r-band magnitude, $\overline{K_0}$ the
magnitude zero point of the run, $N$ the measured counts/s from the
target (ADU/s), and $C$ a color-correction term. For the
stellar/host galaxy correction we used a linear color correction
$C=s\cdot(g-r)$ with $s$ the slope. 
The color correction 
for the active nucleus that is color-independent constant which 
will be discussed extensively in the Section
\ref{VariabilityAna}. The linear equation 
was fit to each run separately to obtain $\overline{K_0}$ and $s$. The
count rate $N$ for each star was determined by performing aperture
photometry to all eight images in the polarization sequence, thus
eliminating the modulation by the polarization optics.  We did not
correct for atmospheric absorption since the average absorption is
included in $\overline{K_0}$, and the RMS scatter of the zero points
$K_0$ of individual images is much higher than the extinction by the
atmosphere; i.e., no dependence of the zero point as a function of
airmass was found.

\begin{figure}
\begin{center}
\includegraphics[width=8.5cm]{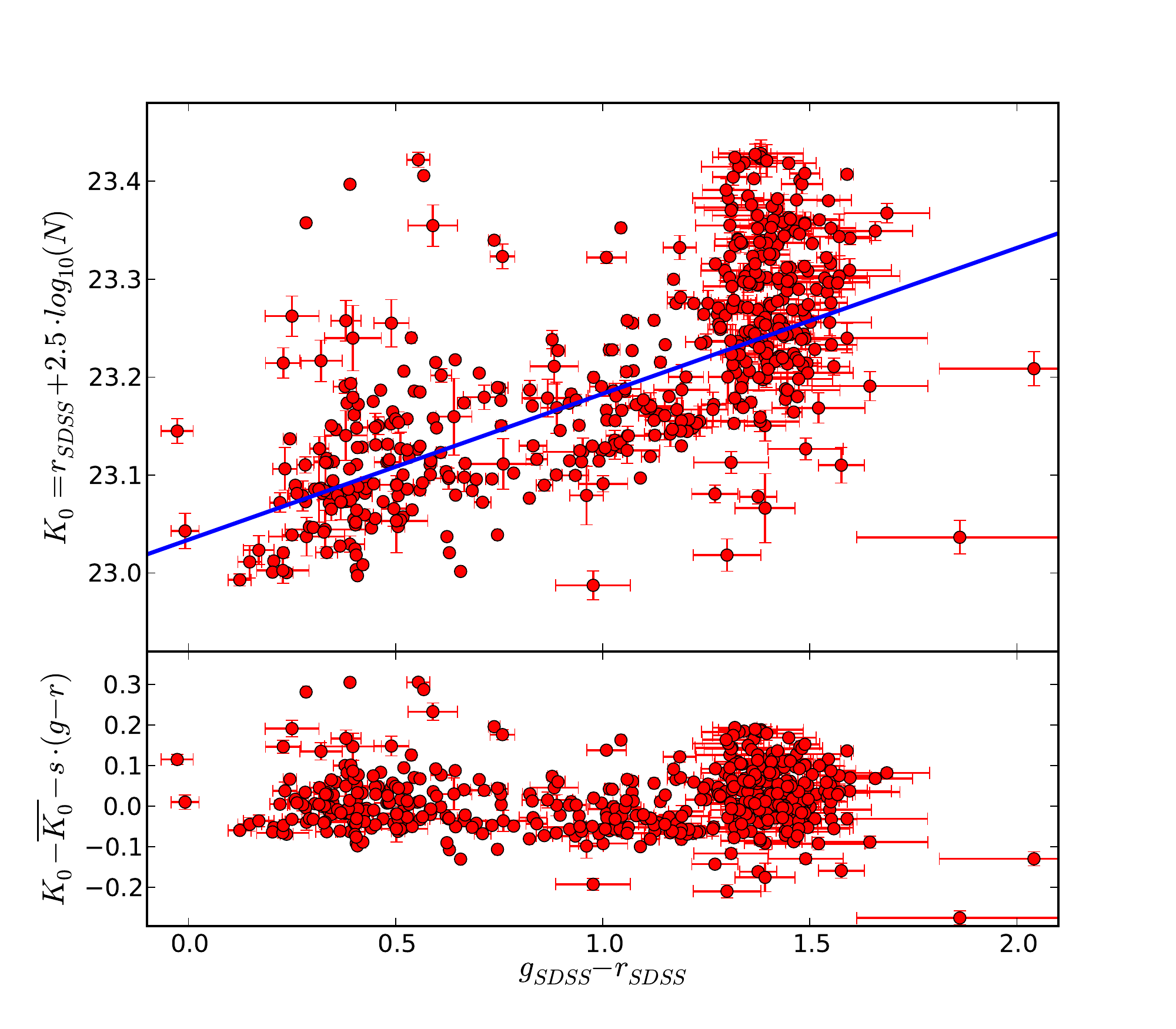}
\end{center}
\caption{\label{color} \emph{Upper panel}: The linear dependence of the
   ZP as function of the SDSS color (for 412 stars) for Calar Alto data. 
   The vertical branch at $g - r > 1.3$ is discussed in the text. The 
   blue line is the fitted behavior found by minimizing $\chi^2$. 
  \emph{Lower panel}: The color
  dependence after subtraction of a linear dependence.}
\end{figure}

Figure \ref{color} shows an example of the color dependence in the data.
There is an approximately linear dependence of $K_0=r+ 2.5 \log N$
on the color of the star. Two things are apparent in this plot: some
stars reside significantly below or above the main body of the data, and
there is a conspicuous clustering of points in the upper righthand
corner of the plot. The former property is probably due to measurement
errors and variability of the stars. The second property, present in
all but the NOT data, resides approximately at $g - r > 1.3$, which
corresponds to stars with temperatures of less than $\sim 3500$K,
indicating that the branch is mainly populated by M dwarfs and red
giants. While M dwarfs show a more or less linear color dependence in
the optical \citep{1995ApJ...445..433A}, red giants show broad
absorption bands in the SDSS r-band \citep{1980ApJS...42..501J},
leading to an increase of $r$ (decrease of $g - r$). Since the depth
of the absorption band increases with decreasing temperature
(increasing $g - r$), the color $g - r$ only weakly depends on the
temperature, so that colder giants (down to $\sim 2000$K) reside in the same
color range. This explains the vertical structure of the color plot in an 
excellent way. 
The reason for not seeing the 
vertical branch in the NOT data set is obviously
that nearly identical filters were used, and furthermore the small
number statistics of calibration objects, due to the small FOV,
prohibits the detection of such a possible branch (called red branch 
hereafter).

To exclude extreme outliers and to get rid of highly
variable stars, we subsequently performed a Kappa-Sigma clipping of the
data. We first fitted the color dependence with a linear dependence
and subtracted the fit from the data. Then we computed the standard
deviation of the residuals and clipped away the most extreme outsiders
(threshold $\sim 3\sigma$). This clipping was iterated twice to obtain
the final fit. The fit was then subtracted from the individual $K_0$'s
and a histogram of the residuals created. To this histogram, a Gaussian
distribution is fitted by minimizing $\chi^2$. The center of the
distribution is then the $\overline{K_0}$ and the Gaussian width of 
the distribution $\sigma_{K_0}$ its error.

At this point the red branch introduces another problem. The
clustering of points at $g - r > 1.3$ may bias the fitted $s$ value
upward, resulting in a double-peaked or skewed Gaussian distribution
of the residuals. Thus another possibility for fitting the color dependence
is to minimize the width of the resulting histogram by iteratively
fitting different lines to the color dependence and fitting a Gaussian
to the resulting residual distribution. Fortunately only for the 
NTT(Oct) run does the branch have a significant effect on the resulting
distribution.  For these data the difference in the slopes
$s_{\sigma}$, obtained by minimizing the width of the Gaussian and
$s_{\chi^2}$, obtained by minimizing the $\chi^2$ in the ZP-color
plot, is treated as an additional error (see below for details).  The
fitted color terms $s$ can be found in Table \ref{tab:hepp}.

\subsection{Variability \label{VariabilityAna}}

After performing the calibration we measured the target brightnesses by
aperture photometry in a similar way to the calibration stars. The
light from disturbing adjacent objects was subtracted if there was any
leakage into the aperture. Since the total flux of our targets is a
sum of two components, the AGN nucleus and the host galaxy, with the
former definitely exhibiting a non-stellar spectrum, the color term
$C=s\cdot(g-r)$ in Eq. \ref{starcolordep} derived from stars is valid 
only for host-galaxy-dominated targets. A different $C$ should in principle 
be used for power-law dominated targets.

\begin{table}[]
\caption{Computed parameters for testing variability.}
\centering{}
\begin{tabular}{cccccc}
\hline
\hline
\multicolumn{1}{c}{Parameter} & \multicolumn{2}{c}{NTT(Mar)} & \multicolumn{1}{c}{NTT} & \multicolumn{1}{c}{CA} & \multicolumn{1}{c}{NOT} \tabularnewline
 & N 1-3 & N 4 & (Oct) & & \tabularnewline
\hline
\# calibrators & 344 & 220 & 306 & 412 & 40 \tabularnewline 
s & 0.14 & 0.14 &0.14 
 & 0.15 & -0.003\tabularnewline
$C_{AGN}$ & 0.10 & 0.10 &0.10 & 0.09 & 0.02 \tabularnewline
$\sigma_{K_0}$ & 0.023 & 0.053 & 0.027 & 0.062 & 0.038 \tabularnewline
$\sigma_{c}$ & 0.019 & 0.014 & 0.038 & 0.034 & 0.001 \tabularnewline
$\langle VL \rangle$ & 0.101 & 0.182 & 0.208 & 0.275 & 0.152 \tabularnewline
\hline
\end{tabular}
\label{tab:hepp}
\end{table}

To estimate the color correction for power-law-dominated targets,
denoted $C_{AGN}$ here, we created a power-law spectrum with spectral
slope of $\alpha_{\nu}=1.16$ $(F_{\nu} \propto \nu^{-\alpha})$, which 
is typical
of optically selected BL Lac candidates \citep{2010AJ....139..390P},
and used synthetic photometry with the SDSS r-band filter bandpass
curve and the NTT, CA, and NOT filter band passes to derive
$C_{AGN}$. The derived color-correction values $C_{AGN}$ can be seen
in Table \ref{tab:hepp}.  From this table we see that the color
correction for power-law dominated targets is not very different from
stellar targets. A power-law index of $\alpha_{\nu} = 1.16$ corresponds
to g - r = 0.36, for which the stellar color correction is $\sim 0.05$
mag in the case of the NTT and CA data. 

Since the total light from our targets is a superposition of two
components, the power-law nucleus and the host galaxy with a
stellar-type spectrum, the SDSS r-band magnitude can be calculated from
our data only if the power-law slope and the host galaxy fraction are
known precisely. As described in the next section, we were able to
resolve the host galaxy in only about one third of the targets, so even
though the host galaxy fraction is known for a significant part of our
sample, it is uncertain for the major part.  Additionally, the
power-law index of the optical nucleus is uncertain for most of the
targets. Because of these uncertainties and in order to treat the whole
sample homogeneously, we treated the entire sample statistically using
an average $\alpha_{nu}$ and nucleus/host galaxy ratio. This obviously
introduces errors to the derived r-band magnitudes, and we
propagated this error into the final magnitude errors as described
below.

Based on the discussion above, the SDSS r-band magnitudes were
computed as follows. We first computed the average of the two 
extreme color corrections (pure AGN and pure host galaxy), 
averaged over each run:
\begin {equation}
\langle C \rangle = 
\frac{1}{N}
\sum\limits_{i=1}^{N}~\left(\frac{s\cdot(g-r)+C_{AGN}}{2}\right)
=
\frac{1}{N}
\sum\limits_{i=1}^{N}~C_i\ ,
\end{equation}
where $N$ is the
number of targets in the run.  The $\langle C \rangle$ gives the
typical correction between the two SEDs (stellar and power law)
through the different filters used. This value is then added to
the ``raw'' magnitude (Eq. \ref{starcolordep} without 
color correction), and the RMS
scatter of this quantity, $\sigma_C$ is treated as the error of our
color correction.  This RMS scatter depends on the sample and even on color 
in the case of NTT(Oct), for the reasons explained below.

An object is called variable if the difference between our magnitude
and the SDSS magnitude, hereafter called {\em variability amplitude}, is
greater than the variation limit $VL$, which is computed by
\begin{equation}
VL = 3 \sqrt{\sigma_{K_0}^2+\sigma_{C}^2+\sigma_{SDSS}^2+\sigma_{phot}^2}\ ,
\end{equation}
where $\sigma_{K_0}$ and $\sigma_{C}$ have been discussed above and
$\sigma_{SDSS}$ and $\sigma_{phot}$ are the photometry errors of the SDSS
and our data, respectively. The errors of the SDSS and our photometry 
are generally small compared to $\sigma_{K_0}$ and $\sigma_{C}$.

Slight adjustments to this scheme were made owing to complications in the
data. To compensate for the clouds in the second half of the fourth
night of the NTT March run, the $K_0$ of each image was derived, and a
correction due to clouds was computed using the average $K_0$ of the
first half of the night. Even though this was done for each image
individually, the overall distribution of $K_0$ of the fourth night was
broadened significantly so that we decided to evaluate the last night
separately.

For the NTT October data, the red
giant branch had such a strong influence on the resulting standard
deviation of the Gaussian distribution that the $\chi^2$ fit could not be
applied to fit the color dependence. The difference in the color term
$s$ between the $\chi^2$ ($s_{\chi^2}$) and the "best Gauss" fit
($s_{\sigma}$) was added as an additional (color-dependent) error to
the variation limit:
\begin{equation}
\label{vlequation}
VL^2_{NTT(Oct)}=VL^2+\left[\left(s_{\chi^2}-s_{\sigma}\right) \left( g-r \right)_{S}\right]^2\,\,.
\end{equation}
This additional error affected the variation statement of only two
objects.  The CA data also showed a rather extreme red branch,
but owing the broadening of the distribution to the
smaller primary mirror, the effect was negligible so that $s_{\chi^2}
\approx s_{\sigma}$. Because of the small mirror and the use of the Gunn r
filter, the CA data resulted in the worst sensitivity for testing
variability.

The NOT data set was affected by a very small FOV so that the number
of calibration objects was very low. Of the 59 chosen objects, 19 were
cut away by the two clippings, resulting in a Gaussian plot with
fairly low number statistics, so instead of employing a Gaussian fit
to the data, we used the standard deviation of the ZP distribution to
estimate $\sigma_{K_0}$.  The NOT data benefit from the available filter,
which is nearly identical with the SDSS r-band, causing the color
correction for calibration stars and BL Lac candidates to be very
small. This shows the importance of the choice of the filter since the
only parameters affecting the precision of the photometry, in addition
to photometric errors and clouds, is the collecting area of the
telescope.  We therefore would expect a much higher precision at NTT,
which was frustrated by the color correction.

For four objects no statement of variability could be made
as those showed such extreme spectra that our color
correction cannot yield reliable results. For instance, in
SDSS J004054.65-091526.8, the Ly$\alpha$ edge lies exactly between the
Gunn r and the SDSS-r filter. These objects were excluded from the
variability analysis.

\subsection{Host galaxies \label{HostGalaxyAna}}

The host galaxy analysis was performed with the model fitting software
we have used extensively during our previous studies of BL Lac host
galaxies
\citep[e.g.][]{1999A&A...341..683H,2003A&A...400...95N,2007A&A...475..199N}.
A more detailed description of the fitting procedure can be found in
\cite{1999PASP..111.1223N}, here we briefly describe the main features of
the program and its application to present data.

\begin{figure}
\begin{center}
\includegraphics[width=8.5cm]{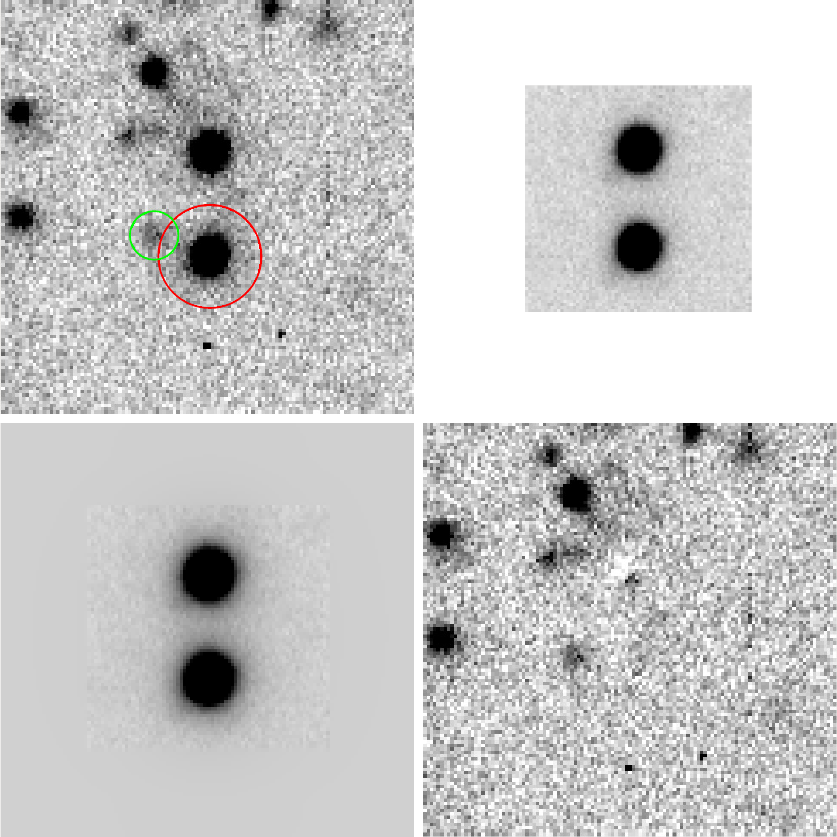}
\end{center}
\caption[]{\label{esimkuva} {\em Upper left panel}: Gunn r-band image of
  SDSS J215051.73+111916.6 obtained at the NTT. Note the double image of the
  target created by the polarization optics. Field size is 
29$\times$29\arcsec, north is up and east to the left. The red circle marks
  the area included in the fit and the green circle an area masked out
  due to an overlapping target. {\em Upper right panel}: PSF image
  extracted from a field star. {\em Lower left panel}: Best fit model
(core + host galaxy) of the target. {\em Lower right panel}: Residuals
after subtracting the model.}
\end{figure}

Polarimetric imaging differs from conventional imaging in a few
important aspects. Firstly, each target produces two images on the CCD
corresponding to the two orthogonally polarized beams (see
Fig. \ref{esimkuva}). In our case the beams were separated vertically
by 10-19\arcsec on the CCD depending on the instrument. Secondly, the
intensity of polarized targets is modulated by the position angle of
the half-wave plate, and therefore some BL Lacs exhibit varying peak
intensity in both beams over the four image polarimetric
cycles. Furthermore, since the two beams go through different optical
paths, the PSF shapes of the two images are different from each other.
The first property means that great care must be exercised to identify
and mask out overlapping images from nearby targets. Furthermore, if
the diameter of the target is larger than the beam separation, the two
images start to overlap, which must be taken into account
in the analysis.  The second property can be circumvented by summing
the four images of the polarimetric cycle, which effectively removes
the modulation. The third property does not cause problems if an
empirical PSF (i.e., a PSF derived from field stars) is used.

The model fitted to the observed images consists of two components,
the unresolved core, parameterized by position $x_c$,$y_c$ and
magnitude $m_c$, and the host galaxy parameterized by position
$x_g$,$y_g$, magnitude $m_g$, effective (half-light) radius $r_e$,
ellipticity $\epsilon_g$, and position angle of the host galaxy
$\theta_g$. The ellipticity and position angle were free parameters
only for well-resolved targets and were fixed at 0.0 for the rest.
To test for host galaxy type, we fitted two different host galaxy
models: a bulge model represented by deVaucouleurs profile with
$\beta = 0.25$ and a disk model with $\beta = 1.0$, where $\beta$ is
the profile slope in
\begin{equation}
I(r) = I(r_e)\ {\rm dex} \left\{
-b_{\beta}
\left[
\left(
\frac{r}{r_e}
\right)^{\beta}
- 1
\right]
\right\}\ ,
\end{equation}
where $b_{\beta}$ is a $\beta$-dependent constant so that $r_e$ always
encircles half the host galaxy light. The model fit was made using
an iterative Levenberg-Marquardt loop, which finds the set of parameters
minimizing the chi squared between the data and the model.

The model was convolved with the PSF, which was obtained from a
suitably bright field star. We first tried the fit using both images
simultaneously; i.e., the PSF consisted of a double image of a field
star, and both target images were used for the fit. This has the
advantage that even cases where the two images partly overlap can be
fit accurately since the overlap is included in the model.  However,
our simulations (see below) showed that the results are sometimes very
noisy in this case since the distance between the two images was not
constant over the field of view, leaving strong residuals after the
fit. The other disadvantage was that a PSF that consists of both
images of a star includes lots of pure sky, especially at the CA and
NOT where image separation was large, resulting in noisier PSF
scaling and consequently noisier results.  We thus decided to only use
one of the images for fitting, the one with fewer overlapping
targets and/or rounder PSFs.  This effectively means sacrificing
half of the signal for better fitting results and simplified error
analysis.

The fitting procedure thus progressed as follows. First all 4 to 12
images in the polarimetric sequence were summed, and of the two target
images on the CCD, the one better suited to fitting was selected.  All
overlapping targets were masked out and the background was subtracted
by measuring 4 to 8 sky regions around the target. Next, a suitable
PSF star was selected and extracted. For the NTT images we extracted a
``double'' PSF, but used only one of the PSF images for fitting. This
enabled us to model any leakage from one component of the double image
to the other.  For the CA and NOT images, only the half of the PSF
corresponding to the selected target image was extracted. In the few
cases where the two images overlapped, we took great care to mask the
regions affected by the overlap. In most cases, however, the two
images were clearly separated, and as stated above, for the 123 NTT
images the overlap was included in the model, so image overlap had no
major effect on our results.  Next we fitted the image with a model
consisting of only the core component; i.e., the fit had three free
parameters, $x_c$, $y_c$, and $m_c$. If the residuals showed any hint
of a host galaxy, we continued by fitting the $\beta = 0.25$ and
$\beta = 1.0$ models to the observed image.  During these fits the
position of the core and host galaxy were held constant at the values
obtained from the pure core fit and $m_c$, $m_g$, and $r_e$ (and
$\epsilon_g$ and $\theta_g$ for the largest targets) were allowed to
change freely.

Calibration of the data was made using Eq. \ref{starcolordep} and $C =
s\cdot(g-r)$ for host galaxies and the $C_{AGN}$ values in Tab.
\ref{tab:hepp} for the AGN. The redshift-dependent $g - r$ color of
the host galaxies was taken from \cite{1995PASP..107..945F} using the
curve for elliptical galaxies. For the galaxies with no z, we used z =
0.5.

In addition to the model fits, we performed Monte Carlo simulations to
determine the errors of fitted parameters, to decide if a host galaxy
was detected and to determine if one of the host galaxy models
(bulge or disk) is clearly preferred. We created 100 simulated images
of each target corresponding to the best-fit parameters and including
properly scaled photon and readout noise, sky determination error, and
PSF variability and performed the fits on the simulated images in
exactly the same way as for the real images. The PSF variability was
introduced by producing two slightly different PSFs for each
simulation.  The first PSF was used to convolve the simulated model,
and the second PSF was used in the fit as the PSF model. Both PSFs
were represented by a Moffat profile, but they differed with respect
to their ellipticity and position angle by an amount that
quantitatively reproduced the peak-to-peak residuals seen in the
data. All fits on simulated images were made keeping $\beta$ constant
at 0.25 or 1.0, depending on which model was preferred by the actual
fit on the data. However, the $\beta$ value for the simulated host
galaxy was drawn from a Gaussian distribution with average 0.25 or 1.0
and 10\% standard deviation to simulate the natural variability of
galaxy profiles.

After the simulations we computed the standard deviations $\sigma$ of
the fitted parameters. To consider the host galaxy detected, we
required that $\sigma_{m_g}$ is $< 0.3$. Furthermore, we considered that the
host galaxy type was determined if the chi squared value of one model
(e.g. bulge) was significantly better than the other (disk),
significantly meaning that $|\chi^2_{bulge} - \chi^2_{disk}| <
3\ \sigma_{\chi^2}$, where $\sigma_{\chi^2}$ is the standard deviation
of the chi squared in the simulations.

\subsection{Optical spectra}

After inspecting all spectra for cosmic ray hits and correcting by
interpolation, the S/N ratio of the individual spectra was
determined (cf. Table \ref{specsum}). This was found to be  50\% higher 
on average than in the SDSS spectra, as expected from comparable
mirror sizes and an exposure time at least a factor of two higher. 
All possible absorption and emission features were tested individually 
for their
reliability. Any feature was deemed reliable if it 
was present in at least two thirds of the individual spectra
10$\sigma$ above the background, or in all three spectra 5$\sigma$
above the background. The confirmed spectral features were then
compared to the most prominent features typically seen in BL Lacs
surrounded by an elliptical galaxy in order to determine a 
redshift.

\subsection{Broad-band SEDs}

One powerful tool for separating BL Lac objects, say 
from thermal sources, is the inspection of their broad-band 
SED. 
In AGN, the distribution is a superposition 
of thermal emission 
from the accretion disk (power-law with exponential drop-off),
a thermal component from the dusty torus, and synchrotron emission 
from a jet, as well as host galaxy emission, while for thermal sources 
(e.g., stars) the spectrum is dominated by 
blackbody emission in the NIR-UV range alone.
BL Lacs should be entirely dominated by synchrotron 
emission at low frequencies and synchrotron-self-Compton processes
at higher frequencies. Since the flux spans a range of five orders of 
magnitude, with variability across all bands, the errors 
of the fluxes, as well as non-simultaneity were
not taken into account for the SED fits. Once the SEDs are 
fitted and a peak frequency is obtained, their rest-frame 
frequencies are derived using the spectroscopic redshift 
(including uncertain ones) given by SDSS.
The surveys, along with the bands and their central wavelengths, 
are listed in Table \ref{tab:dbQueries}. 
\begin{table}[htb!]
\caption{\label{tab:dbQueries}Surveys used for extracting of the SEDs.}
\centering
\begin{tabular}{c|c|c}
Survey & Band & $\mathrm{log}_{10}\left(\nu_{\mathrm{cen}}\right)$ \tabularnewline
\hline
\hline
ROSAT & 1.2 - 2.0 keV & 17.512\tabularnewline
\hline
\multirow{2}{*}{GALEX} & NUV & 15.115\tabularnewline
 & FUV & 15.287\tabularnewline
\hline
 & u & 14.928\tabularnewline
 & g & 14.800\tabularnewline
SDSS & r & 14.683\tabularnewline
 & i & 14.595\tabularnewline
 & z & 14.521\tabularnewline
\hline
\multirow{4}{*}{UKIDSS} & y & 14.468\tabularnewline
 & J & 14.380\tabularnewline
 & H & 14.265\tabularnewline
 & K & 14.135\tabularnewline
\hline
 & W1 & 13.946\tabularnewline
WISE & W2 & 13.814\tabularnewline
 & W3 & 13.398\tabularnewline
\hline
FIRST/NVSS & 1.42GHz & 9.152\tabularnewline
\end{tabular}
\end{table}

The magnitudes are converted to spectral fluxes (Jy) 
using the standard zero points. For FIRST, NVSS, and GALEX, they are  
already tabulated in Jy, while for WISE, UKIDSS, and SDSS, we used the conversion given 
in \citet{2010AJ....140.1868W}, \citet{2006MNRAS.367..454H}, and
\citet{1996AJ....111.1748F}, respectively.
The ROSAT fluxes are given in $erg/cm^2/s$ and are converted to Jy by
$$F_{\nu}[Jy]=F[erg/cm^2/s]\cdot 10^{23}/\nu_{\mathrm{cen}}[\mathrm{Hz}]\, .$$
The spectral fluxes are then multiplied with the frequency to obtain 
flux-energy densities.

To derive peak frequencies we did not fit a full
synchrotron model starting from  
a given electron distribution. Instead, we followed the approach by 
\citet{2008A&A...488..867N} and applied a second-order
polynomial to the data in log-log space. This avoids over-fitting 
of the mostly poorly populated SEDs by tuning too many free parameters.

In addition, we did not include the ROSAT data in our fits since we
are only interested in the peak frequency of the synchrotron emission .
Moreover, flux densities at keV energies are available for only a few objects 
and are completely absent at higher frequencies. On the other hand, 
the X-ray data at least allow the reliability of a fit to be judged.

Because we are fitting global SEDs, we restrict ourselves further 
to objects where measurements in at least 12 bands are available.
Since our measurements are heavily skewed toward IR-optical frequencies,
we separate our fits into three categories. Fits to objects with fewer 
than two radio data points are flagged  ``Uncertain''. 
If the number of radio data points exceeds two, but the total number of 
SED points is less than 14, the objects are flagged ``OK'', and
the rest of the objects are flagged as ``Good''. 

The SED fits for each object are displayed 
in the Appendix (\ref{SEDfits}). The fits give a good indication of the peak 
frequency for most of the objects, even though the fits are not satisfactory 
for some of the SEDs, for the reasons mentioned above.
For three targets, we derived unreasonably high 
peak frequencies ($\mathrm{log10}\left(\nu_{peak}\right) > 30$). 
They were not taken into account any further but are 
shown for completeness. The reliability of the SED fits for targets
with only one or no radio data point is somewhat questionable
since the low-frequency part of the SED is strongly 
underpopulated here. Also targets where the host galaxy contribution 
outweighs the flux originating in the nucleus (i.e., core fraction $<$ 0.5) 
should be considered as rather uncertain.

\section{Results}

\subsection{Variability}

\begin{figure}
\begin{center}
\includegraphics[width=8.5cm]{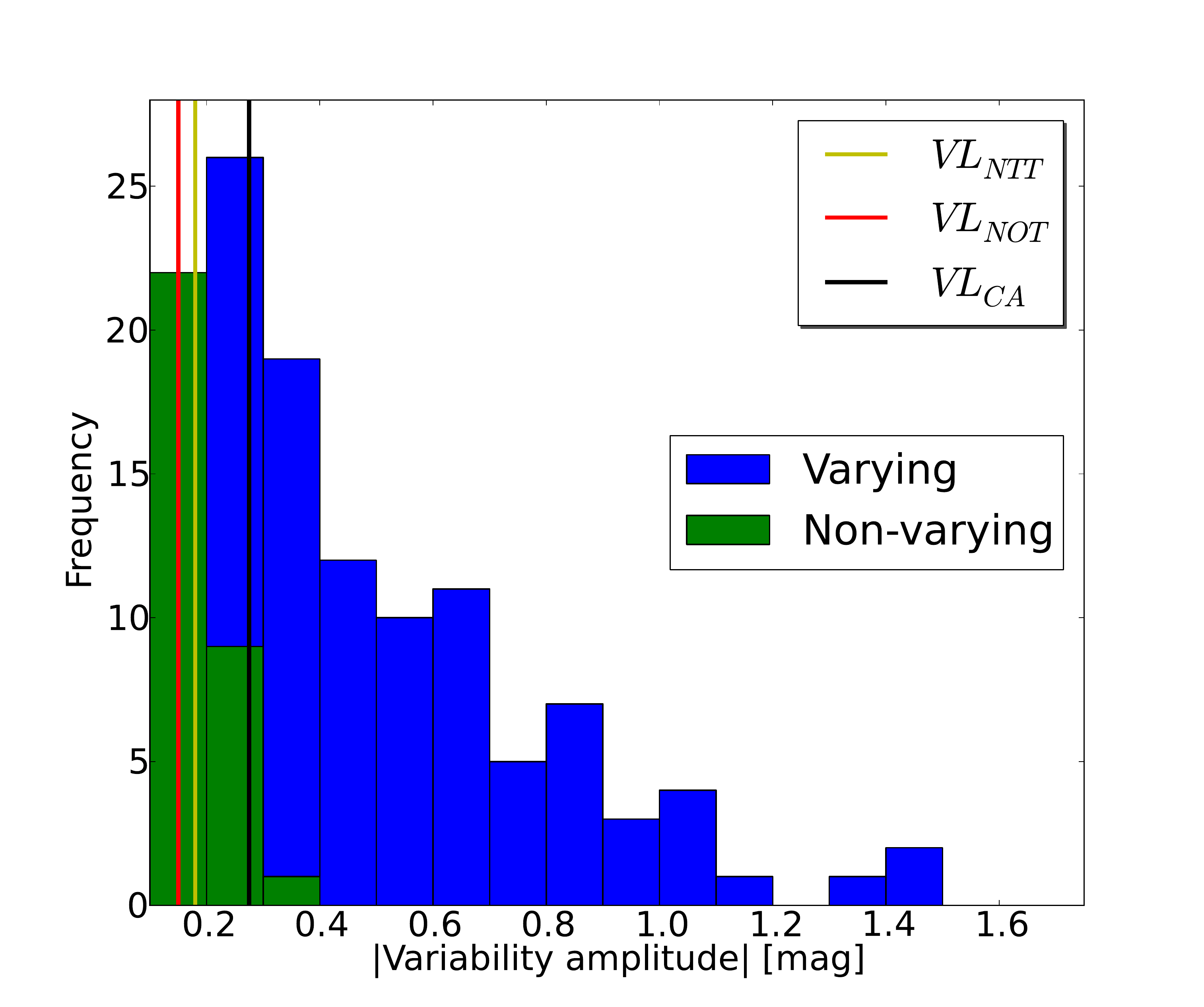}
\end{center}
\caption{\label{HistoPlot} Histogram of variability amplitudes.}
\end{figure}

After evaluating all the objects described above, 107 of the 182
targets (59\%) showed variations according to our definition. Of
the 13 objects measured at two different epochs, two showed variations
at one epoch but no variability in the other with respect to the 
SDSS photometry. This is likely to be
partly due to the lower sensitivity of the CA data and partly due to our
inability to detect variability using very few data points, so both
objects were marked as variable. The distribution of variability
amplitudes in Fig. \ref{HistoPlot} shows that extreme variations up to
2 mag occur, but they are very rare, and most objects vary within the
range of 0.2-0.4 mag.

We checked to what extent the above result depends on observational errors
by creating 1000 mock samples of 178 targets (182 minus the 4 targets
with complex spectra). Simulated magnitudes for both the SDSS and our
photometry were drawn from a Gaussian distribution with a mean
corresponding to the observed value and $\sigma = \sigma_{SDSS}$ for
the simulated SDSS magnitudes and $\sigma = \sqrt{\sigma_{K_0}^2 +
  \sigma_{C}^2 + \sigma_{phot}^2}$ for simulated photometry of this
paper. We also included the additional error due to the red branch 
in the simulation and that 13 targets were observed twice by
us. We then classified the targets into variable and non-variable using
the same criteria as for the real data and computed $N_{var}$, the
number of variable targets. The resulting distribution of $N_{var}$ is
close to Gaussian with the center at $N_{var}$ = 108.8 and $\sigma =
3.4$.  The 3$\sigma$ confidence interval of $N_{var}$ is thus
(99,119), meaning that up to 8 targets marked as variable in our
sample could be non variable in reality. On the other hand, we may have
failed to detect variability in up to 12 targets. These numbers do not
take into account the very sparse sampling. Having more sampling
points would increase $N_{var}$, so our fraction of variable targets
should be considered a lower limit.

\begin{figure}
\begin{center}
\includegraphics[width=8.5cm]{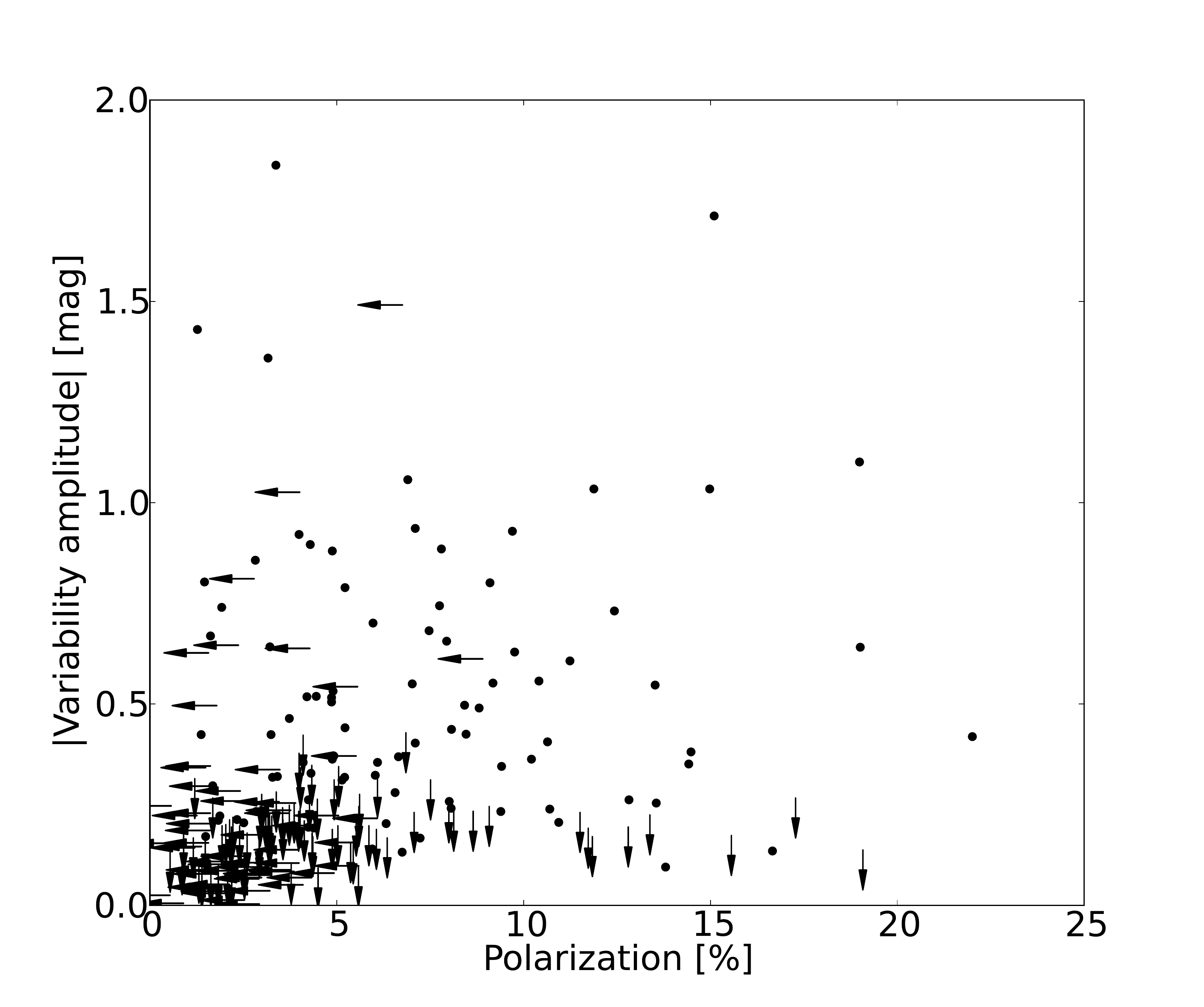}
\end{center}
\caption{\label{PolPlot}
Polarization versus variability amplitude.}
\end{figure}

The correlation between polarization discussed in Paper I 
and variability amplitude is
shown in Fig. \ref{PolPlot}. Out of the 107 variable objects, 83
($78\%$) are polarized as well. The polarization fraction of
non-variable objects is significantly lower ($<55\%$), but neither a
K-S test nor a linear regression fitted to the
variability-polarization plane with subsequent Kappa-Sigma-clipping
yielded any correlation between polarization and variability.  Since we 
are comparing an absolute (one-epoch) to a relative (two-epoch) 
measurement, we do not expect to see any simple correlation.

\begin{figure}
\begin{center}
\includegraphics[width=8.5cm]{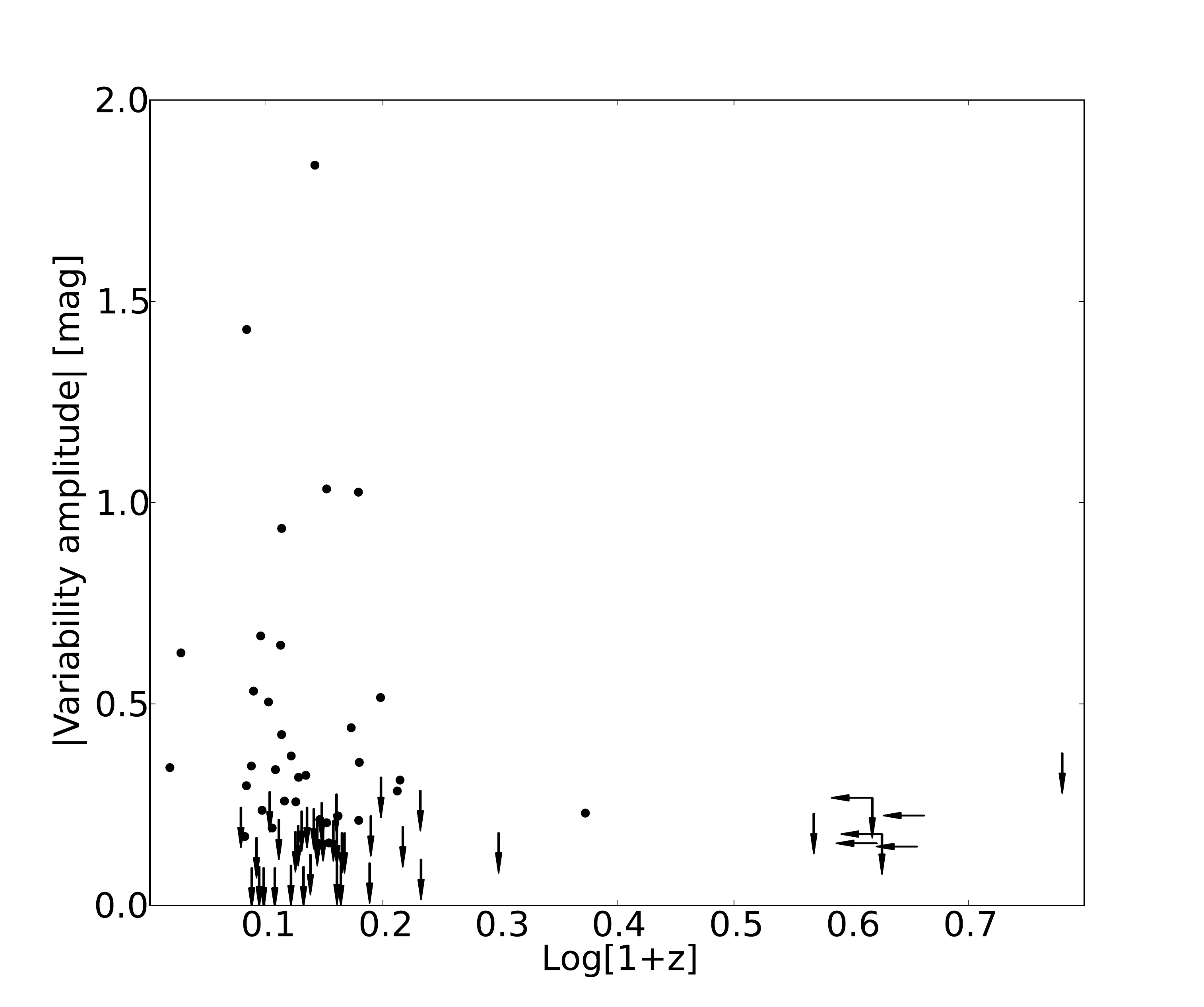}
\end{center}
\caption{\label{RedPlot}
Correlation between redshift and variability amplitude. Uncertain 
redshifts as denoted from the SDSS pipeline flags are marked with a 
horizontal arrow.}
\end{figure}

Out of the 107 varying objects, only 37 ($35\%$) have a secure redshift
while further 31 have lower limits and/or uncertain
redshifts. Correspondingly, out of 72 non-varying objects 38 have
secure redshifts and additional 12 lower limits/uncertain
redshifts. This highlights the importance of high S/N
spectroscopy. The dependence of variation on redshift (only including
certain ones) is shown in Fig.  \ref{RedPlot}. One still has to take
into account that for low-redshift targets, the variability is
underestimated because host galaxy light might have a significant 
impact on the photometry.  

Figure \ref{RedPlot} shows two interesting features. First of all, 
low-redshift BL Lac candidates show a wide range of variability amplitudes
up to 2 mag, while all objects at z $>$ 0.6 show variability amplitudes
of 0.3mag at most or did not show variability at all. Second,
the redshift distribution is far from being continuous. The majority of 
objects are at redshifts $\leq$ 0.8, only two are between z = 0.8
and 2.7, while seven of them are at z = 2.7...5.0. 
There could be two reasons (or a mixture of both) for this behavior. 
The high-redshift targets belong to a different class of objects; e.g., 
weak-lined QSOs (e.g., \citealp{2009ApJ...696..580S}), 
where one a priori would not expect large variability 
amplitudes. Alternatively, the redshifts for some of these objects are 
not correct. In fact, the redshift of three objects at z $>$ 2.7 
is flagged with a small delta-$\chi^2$ and two others are marked as 
uncertain since they exhibit negative emission features in SDSS DR10.
It is also worth noting that in the Roma-BZ catalog 
\citealp{2009A&A...495..691M} the number of known BL Lac objects with 
proper redshifts exceeding 1 is about a dozen 
with none of them exceeding a redshift of two. Whether these ``high-redshift'' 
targets belong to a different class of objects or whether their redshifts
are not correct will be adressed in Paper III.

Only 6 of the 107 ($\sim 5.6\%$) varying sources have neither a radio
nor X-ray counterpart. Three out of those six have a reliable redshift so
that only three objects are only variable and do not have any other of the
properties discussed in this paper. That only 38 of the varying
objects have a radio and X-ray counterpart reflects the shallowness of 
the X-ray data used.


\subsection{Host galaxies}

We were able to resolve the host galaxy in 66 targets; i.e., in 36\% of
the sample. The results for resolved targets are summarized in Table
\ref{hostit}. Column 1 gives the target name, Col. 2 the telescope
used, Col. 3 the redshift, listed only when the redshift determination
is secure, Col. 4 the SDSS r'-band magnitude of the core, Col. 5 the
SDSS r'-band magnitude of the host galaxy, Col. 6 the effective radius in
arcsec, Col. 7 the ellipticity of the host galaxy, Col. 8 the position
angle of the host galaxy, Col. 9 host galaxy type (U = undefined, B = bulge),
Col. 10 the fraction of the core flux of the total flux, Col. 11
SDSS r'-band absolute magnitude, and Col. 12 the effective radius in kpc.
In Appendix \ref{SEDfits}, we show the surface brightness profiles of the 
resolved host galaxies, as well as the fits to them. In addition, one example 
of an unresolved and marginally resolved host galaxy based on the observations 
on each of the three telescopes used is displayed as well.

\begin{table*}
\caption{\label{hostit}Host galaxy results for the bulge model.}
\centering
\begin{tabular}{lllccccccccc}
SDSS & Tel. & z & $m_c$ & $m_g$ & $r_e$ & $\epsilon_g$ & 
$\theta_g$ & Host & Core & $M_{r'}$ & $r_e$\\
& & & (mag) & (mag) & (arcsec) & & (deg) & type & frac & (mag) & (kpc)\\
\hline
000121.47-001140.3 & NTT & 0.4620 & 19.88 $\pm$ 0.15 & 19.78 $\pm$ 0.10 &  2.1 $\pm$ 0.6 & 0.39 & -40  &   U & 0.60 $\pm$ 0.07 & -22.92& 12.2\\
002200.95+000658.0 & NTT & 0.3057 &          -       & 18.53 $\pm$ 0.07 &  1.5 $\pm$ 0.2 & 0.10 &  61  &   B & (0.16 $\pm$ 0.07) & -22.74&  6.7\\
002839.77+003542.2 & NTT &   -    & 19.76 $\pm$ 0.16 & 20.42 $\pm$ 0.26 &  2.4 $\pm$ 3.9 & 0.00 &   0  &   U & 0.85 $\pm$ 0.10 &      -&    -\\
010326.01+152624.8 & NTT & 0.2461 & 19.59 $\pm$ 0.22 & 17.04 $\pm$ 0.08 &  4.5 $\pm$ 0.5 & 0.30 & -19  &   B & 0.25 $\pm$ 0.04 & -23.81& 17.3\\
011012.66-004746.9 & NTT & 0.5477 & 20.16 $\pm$ 0.11 & 19.23 $\pm$ 0.18 &  8.2 $\pm$ 3.3 & 0.37 & -45  &   U & 0.69 $\pm$ 0.05 & -24.15& 52.4\\
012155.87-102037.2 & NTT & 0.4695 &          -       & 19.26 $\pm$ 0.09 &  0.7 $\pm$ 0.1 & 0.00 &   0  &   U & (0.20 $\pm$ 0.08) & -23.52&  4.0\\
012750.83-001346.6 & NTT & 0.4376 & 20.36 $\pm$ 0.30 & 19.37 $\pm$ 0.10 &  1.6 $\pm$ 0.3 & 0.00 &   0  &   U & 0.39 $\pm$ 0.09 & -23.13&  9.1\\
020106.18+003400.2 & NTT & 0.2985 & 19.27 $\pm$ 0.36 & 18.39 $\pm$ 0.09 &  2.0 $\pm$ 0.4 & 0.00 &   0  &   U & 0.49 $\pm$ 0.09 & -22.81&  8.8\\
023813.68-092431.4 & NTT & 0.4188 &          -       & 19.27 $\pm$ 0.06 &  1.2 $\pm$ 0.1 & 0.18 & -22  &   B & (0.07 $\pm$ 0.06) & -23.05&  6.5\\
024302.93+004627.3 & NTT & 0.4089 & 20.55 $\pm$ 0.39 & 19.35 $\pm$ 0.09 &  1.2 $\pm$ 0.3 & 0.38 & -32  &   U & 0.33 $\pm$ 0.09 & -22.89&  6.3\\
024752.13+004106.3 & NTT & 0.3929 & 20.59 $\pm$ 0.23 & 19.72 $\pm$ 0.08 &  1.6 $\pm$ 0.3 & 0.47 &   7  &   B & 0.44 $\pm$ 0.07 & -22.41&  8.7\\
025612.47-001057.8 & NTT & 0.6302 & 20.53 $\pm$ 0.16 & 19.98 $\pm$ 0.10 &  3.1 $\pm$ 0.9 & 0.25 & -33  &   U & 0.62 $\pm$ 0.07 & -24.11& 21.5\\
030433.96-005404.7 & NTT & 0.5112 & 18.69 $\pm$ 0.25 & 19.43 $\pm$ 0.23 &  1.8 $\pm$ 1.9 & 0.00 &   0  &   U & 0.80 $\pm$ 0.14 & -23.80& 11.0\\
032356.64-010829.6 & NTT & 0.3923 & 20.67 $\pm$ 0.17 & 19.65 $\pm$ 0.09 &  2.7 $\pm$ 0.6 & 0.38 &  76  &   U & 0.50 $\pm$ 0.07 & -22.62& 14.3\\
083918.75+361856.1 & NOT & 0.3343 & 20.14 $\pm$ 0.26 & 19.15 $\pm$ 0.11 &  1.5 $\pm$ 0.5 & 0.00 &   0  &   U & 0.42 $\pm$ 0.09 & -22.44&  7.1\\
084225.52+025252.7 & NTT & 0.4251 & 19.94 $\pm$ 0.11 & 19.04 $\pm$ 0.07 &  2.4 $\pm$ 0.4 & 0.13 &  59  &   B & 0.45 $\pm$ 0.04 & -23.33& 13.1\\
085638.50+014000.7 & NTT & 0.4479 & 19.83 $\pm$ 0.13 & 20.07 $\pm$ 0.11 &  1.2 $\pm$ 0.5 & 0.00 &   0  &   U & 0.66 $\pm$ 0.07 & -22.53&  6.6\\
085749.80+013530.3 & NTT & 0.2812 & 18.79 $\pm$ 0.14 & 17.61 $\pm$ 0.07 &  1.8 $\pm$ 0.2 & 0.32 &  77  &   B & 0.39 $\pm$ 0.05 & -23.46&  7.5\\
094432.33+573536.2 &  CA &   -    & 19.69 $\pm$ 0.17 & 20.40 $\pm$ 0.26 &  1.5 $\pm$ 3.3 & 0.00 &   0  &   U & 0.73 $\pm$ 0.10 &      -&    -\\
094542.24+575747.7 & NOT & 0.2289 & 16.92 $\pm$ 0.21 & 16.89 $\pm$ 0.14 &  4.3 $\pm$ 1.9 & 0.00 &   0  &   U & 0.72 $\pm$ 0.11 & -23.51& 15.7\\
094620.21+010452.1 & NTT & 0.5775 & 19.81 $\pm$ 0.24 & 20.35 $\pm$ 0.22 &  1.6 $\pm$ 2.6 & 0.00 &   0  &   U & 0.74 $\pm$ 0.13 & -23.60& 10.7\\
095127.82+010210.2 & NTT &   -    & 19.97 $\pm$ 0.36 & 19.86 $\pm$ 0.17 &  0.9 $\pm$ 0.3 & 0.31 & -19  &   U & 0.55 $\pm$ 0.12 &      -&    -\\
102013.78+625010.1 &  CA & 0.2495 & 19.28 $\pm$ 0.19 & 17.67 $\pm$ 0.09 &  3.0 $\pm$ 0.4 & 0.00 &   0  &   B & 0.35 $\pm$ 0.05 & -22.97& 11.6\\
102523.04+040229.0 & NTT & 0.2078 & 19.29 $\pm$ 0.11 & 17.87 $\pm$ 0.09 &  3.0 $\pm$ 0.5 & 0.12 & -30  &   B & 0.44 $\pm$ 0.04 & -22.33& 10.4\\
103220.29+030949.2 & NTT & 0.3233 & 18.86 $\pm$ 0.16 & 19.89 $\pm$ 0.22 &  1.2 $\pm$ 1.3 & 0.00 &   0  &   U & 0.81 $\pm$ 0.09 & -21.58&  5.8\\
103940.70+053609.3 & NTT & 0.5103 & 20.31 $\pm$ 0.25 & 20.02 $\pm$ 0.10 &  0.9 $\pm$ 0.5 & 0.00 &   0  &   U & 0.52 $\pm$ 0.10 & -23.04&  5.6\\
105151.84+010310.7 & NTT & 0.2654 & 18.51 $\pm$ 0.11 & 19.40 $\pm$ 0.13 &  2.1 $\pm$ 1.0 & 0.28 &  67  &   U & 0.84 $\pm$ 0.07 & -21.52&  8.4\\
105606.62+025213.5 & NTT & 0.2360 & 19.67 $\pm$ 0.21 & 18.11 $\pm$ 0.08 &  1.8 $\pm$ 0.4 & 0.10 &  73  &   B & 0.36 $\pm$ 0.06 & -22.46&  6.7\\
105752.79-005908.3 & NTT &   -    & 20.11 $\pm$ 0.24 & 20.87 $\pm$ 0.28 &  0.8 $\pm$ 1.8 & 0.00 &   0  &   U & 0.72 $\pm$ 0.13 &      -&    -\\
110356.15+002236.4 & NTT & 0.2747 & 19.47 $\pm$ 0.12 & 17.87 $\pm$ 0.08 &  3.4 $\pm$ 0.4 & 0.22 &  84  &   B & 0.40 $\pm$ 0.03 & -23.12& 14.2\\
110704.78+501037.9 &  CA & 0.7061 & 20.27 $\pm$ 0.26 & 20.12 $\pm$ 0.22 &  0.7 $\pm$ 0.3 & 0.00 &   0  &   U & 0.48 $\pm$ 0.11 & -24.38&  4.9\\
111717.55+000633.6 & NTT & 0.4511 & 19.19 $\pm$ 0.14 & 19.56 $\pm$ 0.12 &  1.0 $\pm$ 0.3 & 0.22 &  36  &   U & 0.67 $\pm$ 0.08 & -23.07&  5.8\\
115404.54-001009.9 & NTT & 0.2535 & 18.82 $\pm$ 0.15 & 18.31 $\pm$ 0.09 &  1.6 $\pm$ 0.4 & 0.00 &   0  &   U & 0.55 $\pm$ 0.07 & -22.39&  6.3\\
120303.50+603119.1 &  CA & 0.0653 & 16.50 $\pm$ 0.10 & 14.82 $\pm$ 0.09 &  7.4 $\pm$ 0.8 & 0.02 &  77  &   B & 0.46 $\pm$ 0.03 & -22.56&  9.3\\
121758.72-002946.2 & NTT & 0.4188 & 19.39 $\pm$ 0.15 & 19.32 $\pm$ 0.22 &  1.6 $\pm$ 1.8 & 0.00 &   0  &   U & 0.59 $\pm$ 0.08 & -22.99&  9.0\\
121944.98+044622.4 & NTT & 0.4891 & 18.54 $\pm$ 0.12 & 19.83 $\pm$ 0.22 &  1.0 $\pm$ 1.2 & 0.00 &   0  &   U & 0.84 $\pm$ 0.08 & -23.05&  6.2\\
122300.31+515313.9 &  CA & 0.3650 & 19.76 $\pm$ 0.12 & 19.46 $\pm$ 0.10 &  2.0 $\pm$ 0.5 & 0.00 &   0  &   U & 0.57 $\pm$ 0.05 & -22.37& 10.2\\
122809.13-022136.1 & NTT & 0.3227 & 20.73 $\pm$ 0.29 & 19.36 $\pm$ 0.07 &  0.8 $\pm$ 0.1 & 0.18 & -58  &   B & 0.27 $\pm$ 0.06 & -22.09&  3.6\\
124225.39+642919.1 &  CA & 0.0424 &          -       & 15.51 $\pm$ 0.08 &  4.7 $\pm$ 0.4 & 0.24 &  66  &   B & (0.05 $\pm$ 0.05) & -20.91&  3.9\\
124425.30+044459.7 & NTT & 0.3999 &          -       & 19.46 $\pm$ 0.07 &  1.1 $\pm$ 0.1 & 0.40 & -34  &   U & (0.16 $\pm$ 0.07) & -22.69&  5.9\\
124834.30+512807.8 &  CA & 0.3508 & 18.19 $\pm$ 0.20 & 18.86 $\pm$ 0.26 &  1.0 $\pm$ 0.6 & 0.00 &   0  &   U & 0.71 $\pm$ 0.11 & -22.81&  4.7\\
125820.79+612045.6 &  CA & 0.2235 &          -       & 18.09 $\pm$ 0.07 &  0.6 $\pm$ 0.1 & 0.00 &   0  &   B & (0.00 $\pm$ 0.00) & -22.27&  2.1\\
131330.15+020105.9 & NTT & 0.3558 & 18.85 $\pm$ 0.07 & 18.50 $\pm$ 0.20 &  5.3 $\pm$ 2.9 & 0.21 &  69  &   U & 0.73 $\pm$ 0.04 & -23.25& 26.2\\
132301.01+043951.4 & NTT & 0.2244 & 18.41 $\pm$ 0.11 & 17.65 $\pm$ 0.09 &  3.0 $\pm$ 0.8 & 0.12 &   9  &   U & 0.58 $\pm$ 0.05 & -22.75& 10.8\\
132759.76+645811.3 &  CA & 0.4468 &          -       & 18.71 $\pm$ 0.08 &  2.1 $\pm$ 0.3 & 0.23 & -69  &   B & (0.04 $\pm$ 0.06) & -23.84& 12.2\\
133105.71-002221.2 & NTT & 0.2426 & 20.80 $\pm$ 0.33 & 18.37 $\pm$ 0.05 &  1.1 $\pm$ 0.1 & 0.11 & -54  &   B & 0.14 $\pm$ 0.04 & -22.23&  4.1\\
134037.59-014847.6 & NTT & 0.5130 & 20.11 $\pm$ 0.16 & 20.25 $\pm$ 0.09 &  1.5 $\pm$ 0.8 & 0.00 &   0  &   U & 0.69 $\pm$ 0.09 & -22.87&  9.2\\
141003.92+051557.7 & NTT & 0.5440 & 20.25 $\pm$ 0.09 & 19.69 $\pm$ 0.08 &  2.1 $\pm$ 0.4 & 0.24 & -24  &   B & 0.55 $\pm$ 0.04 & -23.65& 13.6\\
141030.84+610012.8 &  CA & 0.3833 & 20.08 $\pm$ 0.20 & 19.18 $\pm$ 0.09 &  1.3 $\pm$ 0.2 & 0.00 &   0  &   U & 0.36 $\pm$ 0.06 & -22.80&  6.7\\
145111.69+580003.0 &  CA & 0.4053 & 20.13 $\pm$ 0.31 & 18.84 $\pm$ 0.10 &  2.4 $\pm$ 0.5 & 0.34 &  21  &   U & 0.33 $\pm$ 0.08 & -23.32& 13.2\\
150006.49+012956.0 & NTT & 0.7083 & 20.18 $\pm$ 0.19 & 21.38 $\pm$ 0.27 &  1.0 $\pm$ 2.5 & 0.00 &   0  &   U & 0.83 $\pm$ 0.11 & -23.21&  7.1\\
161541.22+471111.8 &  CA & 0.1986 & 18.09 $\pm$ 0.13 & 17.32 $\pm$ 0.09 &  2.8 $\pm$ 0.4 & 0.00 &   0  &   U & 0.53 $\pm$ 0.05 & -22.70&  9.0\\
162115.21-003140.4 & NTT &   -    & 19.73 $\pm$ 0.10 & 20.34 $\pm$ 0.13 &  0.6 $\pm$ 0.1 & 0.00 &   0  &   U & 0.68 $\pm$ 0.06 &      -&    -\\
165109.18+421253.5 &  CA & 0.2686 &          -       & 18.85 $\pm$ 0.12 &  1.0 $\pm$ 0.2 & 0.00 &   0  &   U & (0.15 $\pm$ 0.11) & -22.04&  4.1\\
165808.33+615001.9 & NOT & 0.3742 & 18.64 $\pm$ 0.23 & 18.42 $\pm$ 0.16 &  2.0 $\pm$ 1.0 & 0.00 &   0  &   U & 0.64 $\pm$ 0.10 & -23.56& 10.3\\
205523.36-050619.3 & NTT & 0.3426 & 19.65 $\pm$ 0.16 & 18.52 $\pm$ 0.08 &  2.2 $\pm$ 0.4 & 0.22 & -24  &   U & 0.38 $\pm$ 0.05 & -23.25& 10.5\\
205938.57-003756.0 & NTT & 0.3354 & 20.17 $\pm$ 0.22 & 18.72 $\pm$ 0.08 &  2.2 $\pm$ 0.4 & 0.21 &  31  &   U & 0.37 $\pm$ 0.06 & -22.97& 10.5\\
211611.89-062830.4 & NTT & 0.2916 & 19.51 $\pm$ 0.44 & 18.64 $\pm$ 0.13 &  0.8 $\pm$ 0.2 & 0.00 &   0  &   U & 0.37 $\pm$ 0.11 & -22.78&  3.7\\
213950.32+104749.6 & NTT & 0.2960 & 21.55 $\pm$ 0.49 & 17.68 $\pm$ 0.09 &  8.4 $\pm$ 1.2 & 0.28 &  89  &   B & 0.11 $\pm$ 0.04 & -23.59& 37.2\\
215051.73+111916.5 & NTT &   -    & 19.47 $\pm$ 0.19 & 20.04 $\pm$ 0.18 &  1.2 $\pm$ 0.8 & 0.00 &   0  &   U & 0.70 $\pm$ 0.11 &      -&    -\\
215305.36-004230.7 & NTT & 0.3416 & 18.98 $\pm$ 0.34 & 18.85 $\pm$ 0.24 &  0.5 $\pm$ 0.3 & 0.00 &   0  &   U & 0.50 $\pm$ 0.14 & -23.02&  2.3\\
221108.34-000302.5 & NTT & 0.3619 & 18.97 $\pm$ 0.29 & 19.12 $\pm$ 0.18 &  0.6 $\pm$ 0.2 & 0.33 &  57  &   U & 0.58 $\pm$ 0.11 & -22.78&  3.3\\
221109.88-002327.5 & NTT & 0.4476 & 20.17 $\pm$ 0.23 & 19.08 $\pm$ 0.07 &  2.0 $\pm$ 0.4 & 0.20 &  54  &   U & 0.43 $\pm$ 0.07 & -23.64& 11.6\\
221456.37+002000.1 & NTT &   -    & 19.69 $\pm$ 0.16 & 20.44 $\pm$ 0.21 &  1.7 $\pm$ 1.7 & 0.00 &   0  &   U & 0.77 $\pm$ 0.10 &      -&    -\\
224819.44-003641.6 & NTT & 0.2123 & 19.87 $\pm$ 0.24 & 17.15 $\pm$ 0.07 &  4.3 $\pm$ 0.6 & 0.11 &  37  &   B & 0.19 $\pm$ 0.04 & -23.28& 14.8\\
235604.03-002353.8 & NTT & 0.2830 & 19.72 $\pm$ 0.23 & 19.05 $\pm$ 0.09 &  1.2 $\pm$ 0.3 & 0.22 &  50  &   U & 0.46 $\pm$ 0.08 & -22.01&  5.2\\

\hline
\end{tabular}
\end{table*}

The absolute magnitudes $M_{r'}$ in Table \ref{hostit} were computed from
\begin{equation}
M_{r'} = m_g - DM - K_{r'} - A_{r'} + E(z)\ ,
\end{equation}
where $DM$ is the distance modulus, $K_{r'}$ is the K correction
\citep{1995PASP..107..945F}, $A_{r'}$ is the galactic extinction
\citep{1998ApJ...500..525S}, and $E(z) = 0.93 \cdot z$ is the evolution
correction. The latter was computed using the PEGASE code
\citep{1997A&A...326..950F} by assuming initial ISM metallicity $Z_0 =
0.004$, single starburst 11 Gyr ago (z = 2.6), and passive evolution
thereafter.

None of the targets were unambiguously associated with a disk type
galaxy. In seven cases out of 66 a disk host galaxy formally gave a better
fit, but the simulations showed that in none of these cases the host
galaxy type was secure. In all 19 cases where our simulations showed
the host galaxy type to be well determined the bulge model was
preferred. This result is in line with previous BL Lac host galaxy
imaging surveys
\citep[e.g.,][]{2000ApJ...532..816U,2003A&A...400...95N}, which show
that BL Lacs are almost exclusively found in ellipticals.  Given this
result we use the bulge fit results for all targets from this point
on.

As Table \ref{hostit} indicates, in eight cases we were not able to detect
an optical core. In seven cases out of these eight the fit returned a value
for the core magnitude $m_c$, but subsequent error simulations
indicated that the error of $m_c$ was $>$ 0.5 mag. We mark these cores as 
undetected, although it is possible that weak cores below our
detection limit are present in these targets or the objects were in a 
very low state. Column 10 on Table
\ref{hostit} gives the core fraction; i.e., the ratio between the core
flux and total flux within the aperture used for polarimetry in Paper
I. This was measured directly from the two-dimensional model images of
the core and the host galaxy using aperture photometry. We list in
parenthesis the core fraction also for the eight targets with formally
undetected cores, computed from the core magnitude returned by the
fit. The formal core fractions of the non-detected cores are very low,
0.04-0.20, as expected. The only case in which the fit converged
toward $m_c \rightarrow \infty$, i.e. core fraction of 0.0, is
SDSS125820.79+612045.6, which is the only target where we have no
evidence of an optical core. 

{\bf SDSS012155.87-102037.2}: A disk fit formally gives a better fit,
but both bulge and disk fits leave strong residuals, which are unlikely
to be due to PSF verifiability and give an impression of a tight
gravitational lens system or a merger of several galaxies. A weak
core is indicated by the fits, but not significantly detected.

{\bf SDSS094432.33+573536.2}: There are two compact objects within 
3.0 arcsec from the core.

{\bf SDSS100050.22+574609.1}: This is one of the targets with only
marginal detection of the host galaxy. The nucleus is surrounded
by a very elongated feature giving an impression of a nearly edge-on
disk galaxy.

\subsection{Optical spectra}

All the obtained and reduced spectra can be found in the Appendix 
\ref{spectralPlots}.
In spite of our high S/N spectra for only 
one out of our 27 objects a unique redshift could be assigned.
SDSS J105829.62+013358.8 shows a very broad
emission feature at 5289\AA, which was identified with
MgII yielding a redshift of z=0.89$\pm$0.02.
In the SDSS-pipeline (DR8 and earlier), the broad feature was identified 
as CIII (z=1.78$\pm$0.003), but in the following data releases, the 
feature was also attributed to MgII emission
at z = 0.8933$\pm$0.0004. Out of the remaining 26 objects,
17 have at least one absorption/emission line in their spectra, 
while the spectra of the remaining 9 objects appear featureless.
No unique redshift could be assigned for the 17 objects with a 
single emission/absorption line,
.

Table \ref{specsum} with our analysis of the 
spectroscopic data  as well as the median-combined spectra for all 
objects, can be found the Appendix. In Table \ref{specsum}
the target name, the r-band 
magnitude from SDSS, the S/N per resolution element at 6500\AA~
in the spectra and the central wavelength for the emission/absorption 
lines detected is given . 
We also indicate their equivalent widths. 

\subsection{Broadband SEDs}
\begin{figure}[htb!]
\includegraphics[width=1\textwidth]{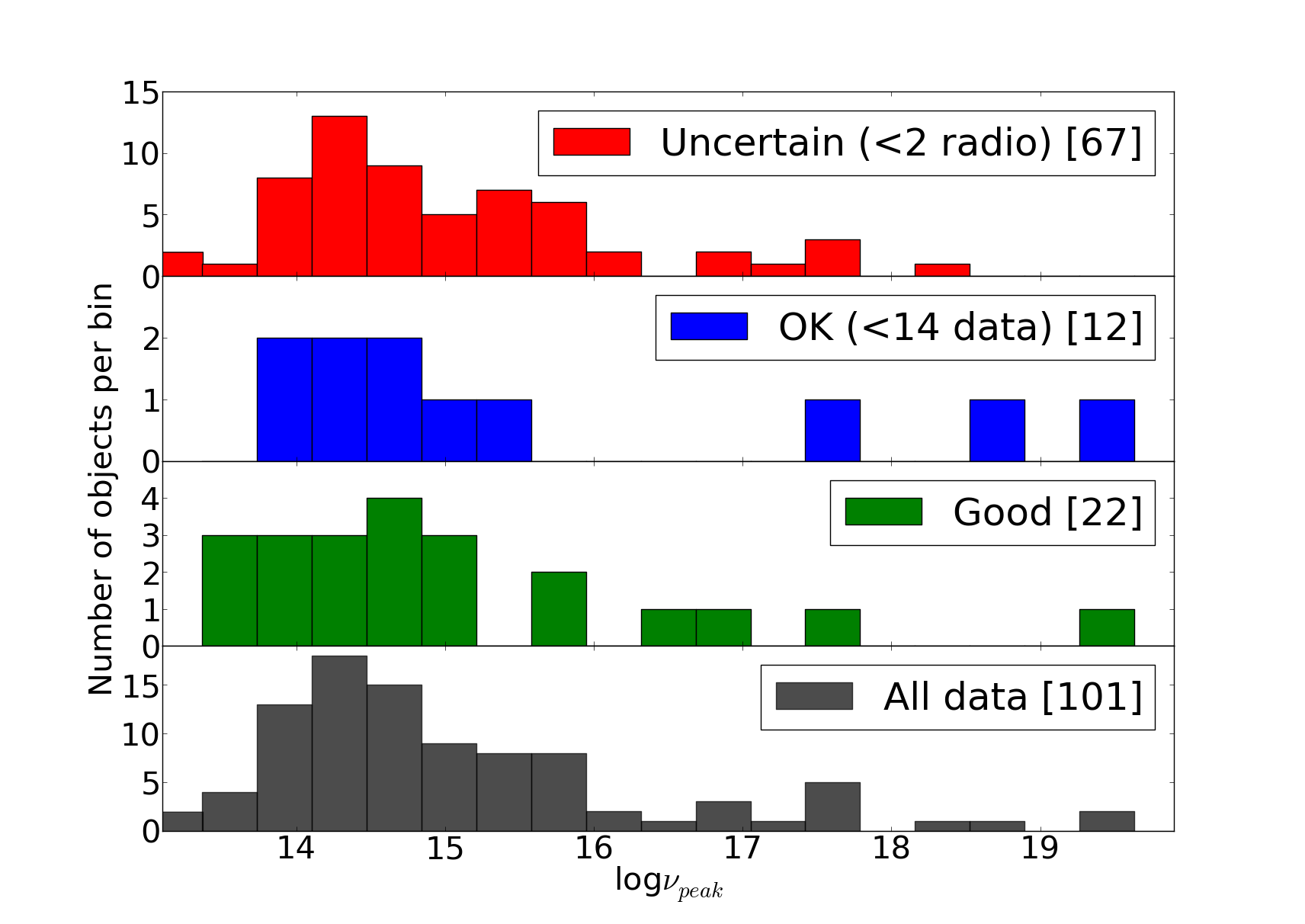}
\caption{Distribution of frequency peaks for the different quality classes. 
For better clarity the 6 (5 uncertain, 1 OK) objects above 
$\mathrm{log10}\left(\nu_{\mathrm{peak}}\right) > 20$ are not shown.
\label{peakHist}}
\end{figure}

The main scope of polynomial fits to the SEDs was to derive the 
synchrotron peak frequencies for the objects in our sample. 
As shown in  Fig. \ref{peakHist},
 most of the objects have peak frequencies between  
$13.5 \leq \mathrm{log}_{10}\left(\nu_{\mathrm{peak}}\right) \leq 16$ with a 
faint tail towards higher frequencies. This is a similar range to the one 
found by \citet{2008A&A...488..867N}, although the peak frequencies they derived 
are shifted toward lower frequencies. This is due to their sample 
selection of radio-bright blazars. The distribution is not homogeneous
but rather peaked at $\mathrm{log}_{10} \sim 14.5$. Thus the sample 
seems to contain a substantial fraction of IBL.

The derived synchrotron peak frequencies are affected by
variability, temporal evolution of the peak frequency, shallowness of 
the data, and a significant host galaxy component in some cases.
We made Monte Carlo simulations
where the photometric data points were varied within their errors,
which showed that photometric errors alone can cause
shifts in the peak frequency up to 0.2 in the logarithmic scale.
Variability and peak shifts are difficult to account, for and we also did 
not subtract the host galaxy light since we expect that the influence to the 
general distribution is very low, although in some individual cases
there might be larger shifts because of the host galaxy light.

For three of our objects, the SED fits yielded extremely high peak
frequencies. The SED of SDSS J094432.34+573536.15 shows almost a
linear slope while SDSS J104523.87+015722.09 seems to be dominated
by black body radiation. The latter source was identified by
\citet{2004ApJ...607..426K} as a DC white dwarf but not confirmed by
\citet{2006ApJS..167...40E}. This source is apparently a
radio source \citep{2005AJ....129.2542C}. FIRST lists a 2.7mJy
radio source about 1.3'' south of the SDSS position,
where a faint optical counterpart is present on our NTT image.
Both objects may have entered the 3\arcsec SDSS fiber,
and the resulting spectrum be a superposition from both components.
SDSS J140450.91+040202.16 would finally yield a synchrotron peak between
$15 \leq \mathrm{log10}\left(\nu_{\mathrm{peak}}\right) \leq 16$ when
X-ray data would have been taken into account.

In Table \ref{summaryAllProps} in the Appendix, we summarize the
global properties of our targets derived in Paper I and in this work. 
This includes the redshifts from SDSS, the spectral indices 
$\alpha_{ox}$ and $\alpha_{ro}$, the polarization properties, the
variability limit and variabiliy amplitudes, the core fraction 
derived from the host galaxy and the peak-frequencies from the 
SED fits.

\section{Summary}

We presented a detailed analysis of the properties
of 182 probable BL Lac candidates from the SDSS extracted by 
\cite{2005AJ....129.2542C}. Particular emphasis was given to their variability
characteristics and their broad-band radio-UV SEDs. We also examined
our data for the presence of a host galaxy of the targets.
In addition, we presented new optical spectra of 27 targets with
improved S/N with respect to the SDSS spectra. 
Our main results can be summarized as follows:
\begin{itemize}
\item About 60\% (107/182) of the objects show variability
  on long timescales between SDSS DR2 and our observations in 2008-09.
  The 3$\sigma$ confidence interval for the number of variable
  targets, when taking only observational errors into account, is (99~119).
\item Using two-dimensional model fits, we were able to resolve the
  host galaxy in 66 targets, 7 of them without a secure SDSS redshift.
  None of the host galaxies is unambiguously associated with a disk
  galaxy. In the 19 cases where the host galaxy classification 
  is unique, a deVaucouleurs model is preferred. The luminosity
  distribution is consistent with earlier results of BL Lac host
  galaxies if a bias in the sample is taken into account.
\item We analyzed 104 broad-band radio-UV SEDs and determined 
  the synchrotron peak frequency. The objects have peak frequencies between  
$13.5 \leq \mathrm{log}_{10}\left(\nu_{\mathrm{peak}}\right) \leq 16$ with a 
faint tail toward higher frequencies. The distribution is not homogeneous
but instead peaked at $\mathrm{log}_{10} \sim 14.5$. Thus the sample 
seems to contain a substantial fraction of IBL.
\item Our new optical spectra did not reveal any new redshift for any of 
our objects. For SDSS J105829.62+013358.8, 
we could confirm the SDSS redshift of z = 0.89$\pm$0.02.
\item There is potentially a population of high-redshift 
BL Lacs as indicated in Fig. \ref{RedPlot} with low variability 
amplitudes. This could alternatively be a different class of objects 
(high-redshift weak-lined QSOs) or an artifact due to wrong redshift 
assignments.
   
\end{itemize}


Overall, our results, including the analysis of the polarimetric properties
imply  that the \cite{2005AJ....129.2542C} 
sample, is only marginally contaminated by stellar sources and is likely 
to contain a high fraction of bona fide BL Lacs. It potentially contains 
a high fraction of IBL. The detailed discussion of these properties,
a comparison to BL Lac samples determined by other selection criteria
and a potential revision of the defining criteria of a BL Lac will
be presented in a forthcoming paper (Nilsson et al., in prep).


\begin{acknowledgements}
JH acknowledges support by the Deutsche Forschungsgemeinschaft (DFG)
through grant HE 2712/4-1.  Part of this work was supported by the
COST Action MP1104 "Polarization as a tool to study the Solar System
and beyond".

The data presented here were obtained with ALFOSC, which is provided
by the Instituto de Astrofisica de Andalucia (IAA) under a joint
agreement with the University of Copenhagen and NOTSA. This research 
made use of NASA's Astrophysics Data
System Bibliographic Services. This research made use of the
NASA/IPAC Extragalactic Database (NED), which is operated by the Jet
Propulsion Laboratory, California Institute of Technology, under
contract with the National Aeronautics and Space Administration.
\end{acknowledgements}
\bibliographystyle{aa}
\bibliography{paper.bib}
\clearpage
\onecolumn
\begin{appendix}
\section{Appendix}
\begin{table*}[h!]
\centering{}
\caption{\label{specsum}Results of the analysis of the optical spectra.
Negative EW indicate absorption lines.}
\begin{tabular}{lllcc}
\hline
\hline
Nr	& Object name	        & $r_{SDSS}$ &		S/N per RE		& Lines \\
	& [SDSS]    	        & [mag]     &	at $\lambda$ 6500$\mathring{A}$	& CWL / EW [$\mathring{A}$]\\
\hline
I	& J003514.72+151504.1	& 16.59	    &	174 &   5721 / -0.42 \\
II	& J011452.77+132537.5	& 17.03	    &	186 &	7545 / 0.34  \\
III	& J022048.46$-$084250.4	& 18.27	    &	62  &	-	    \\
IV	& J032343.62$-$011146.1	& 16.81     &	174 &	7188 / -3.02, 8168 / -2.44 \\
V	& J074054.60+322601.0	& 18.67     &	62  &	7536 / 4.10 \\
VI	& J085920.56+004712.1	& 18.76     &	37  &	5049 / -0.33 \\
VII	& J091848.57+021321.8	& 18.55     &	37  &	-	\\
VIII	& J100612.23+644011.6	& 18.87     &	62  &   7127 / 3.52 \\
IX	& J101858.55+591127.8	& 17.75     &	87  &	-	\\
X	& J105829.62+013358.8	& 17.86     &	112 &	5289 / 9.27 , 8175 / -1.08 (z=0.89$\pm$0.02) \\
XI	& J110735.92+022224.5	& 18.59     &	37  &	-	\\
XII	& J113245.61+003427.7	& 17.44     &	99  &	-	\\
XIII	& J113523.70+660941.0	& 18.94     &	37  &	8082 / -3.52, 8146 / -3.33 \\
XIV	& J114312.11+612210.8	& 17.93     &	99  &	5001 / 0.42	\\
XV	& J114926.13+624332.5	& 18.93	    &	25  &	8288 / 5.24	\\
XVI	& J121300.80+512935.6	& 18.45     &	62  &	-	\\
XVII	& J121500.80+500215.6	& 17.41     &	74  &	-	\\
XVIII	& J121834.93$-$011954.3	& 17.55     &	136 &   6970 / 2.8 , 7196 / -0.53	\\
XIX	& J123341.33$-$014423.7	& 18.31     &	50  &	5995 / 1.2	\\
XX	& J131106.48+003510.0	& 17.86     &	99  &	5419 / 0.40	\\
XXI	& J135738.70+012813.6	& 17.82     &	74  &	-	\\
XXII	& J140450.91+040202.2	& 16.31     &	186 &	7188 / -1.89, 8160 / -1.98	\\
XXIII	& J141004.65+020306.9	& 18.15     &	87  &	8384 / 3.40	\\
XXIV	& J141826.33$-$023334.1	& 16.64     &	174 &	7193 / -2.88	\\
XXV	& J141927.50+044513.8 	& 18.18     &	50  &	6336 / -6.13	\\
XXVI	& J143657.71+563924.8	& 18.42     &	50  &	7518 / 3.6	\\
XXVII	& J170124.64+395437.1 	& 16.88     &	223 &	-	\\
\hline
\end{tabular}
\end{table*}

\onltab{2}{
\thispagestyle{empty}
\begin{figure*}[htb]
\caption{Flux-calibrated spectra. 
The green lines are the binned (factor of 2) spectra with the 
red lines representing the typical error (obtained by IRAF - routines) of 
the respective data point.
Features stemming from an imperfect sky-subtraction 
are marked in light gray, significant (5$\sigma$ in all 3 spectra or 10$\sigma$ in 2/3 spectra) 
absorption/emission features originating in the object are marked 
light magenta/blue.\label{spectralPlots}}
\centering
\includegraphics[width=\textwidth]{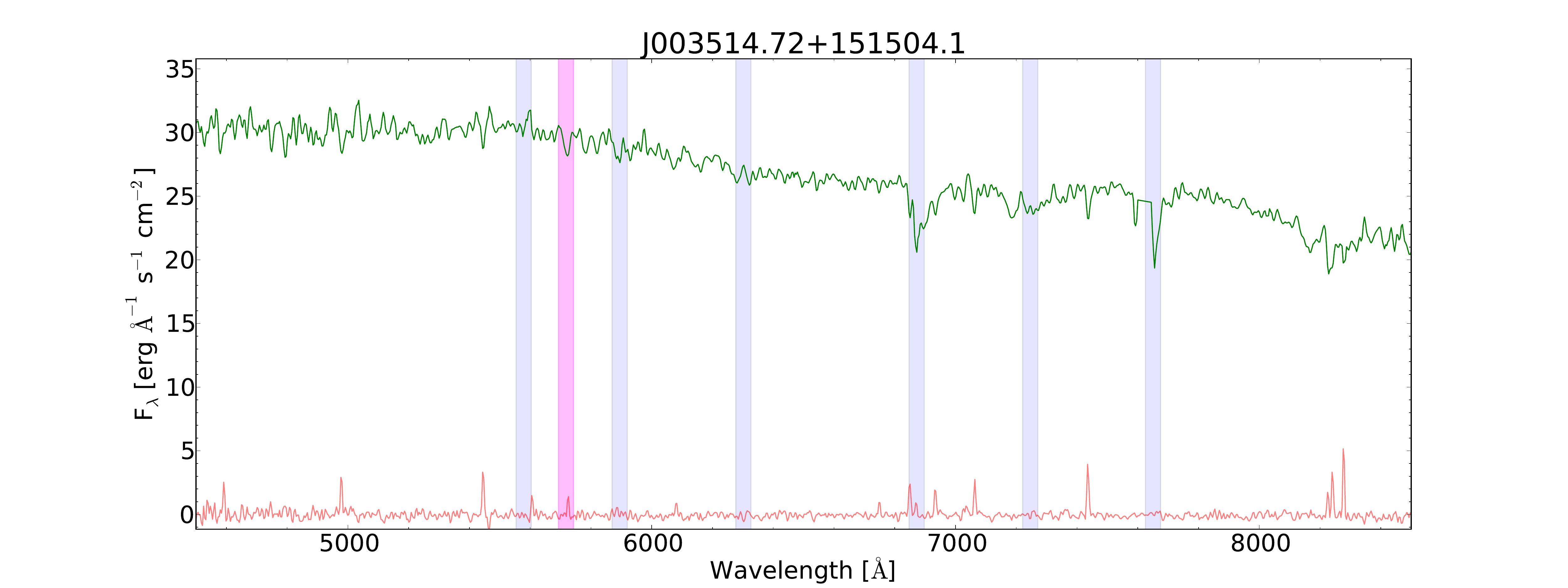}
\includegraphics[width=\textwidth]{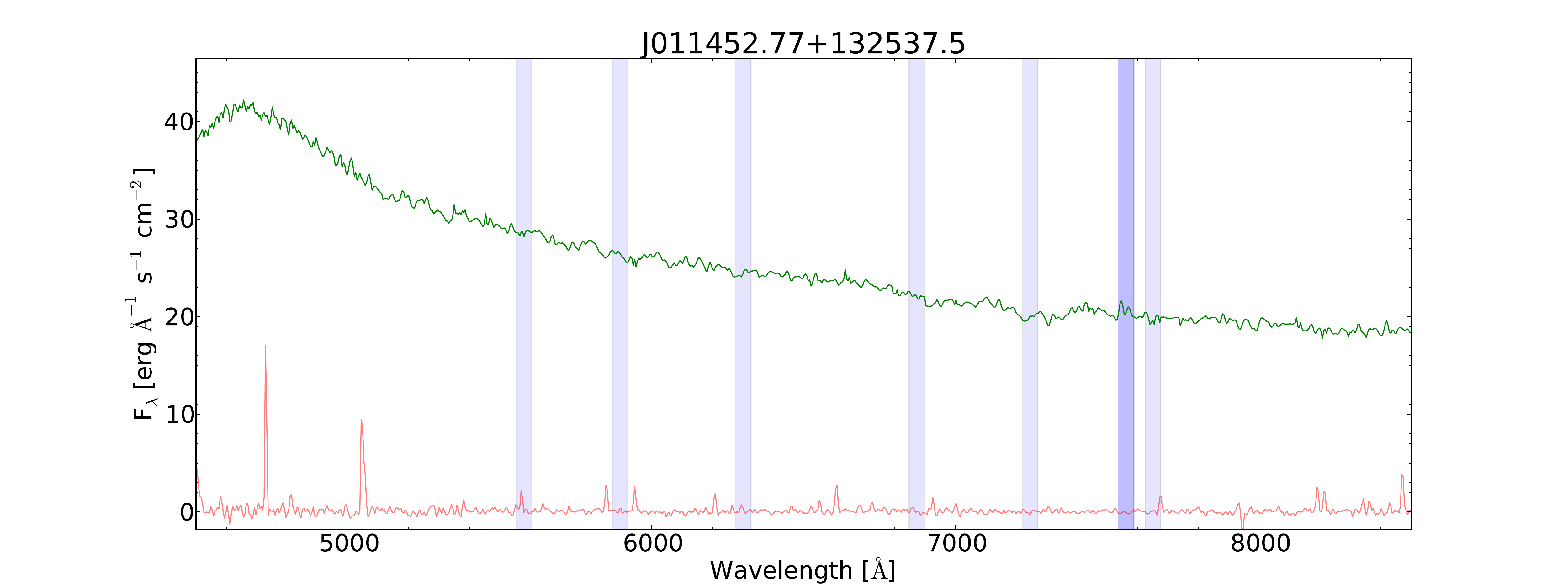}
\end{figure*}
\begin{figure*}[htb]
\centering
\includegraphics[width=\textwidth]{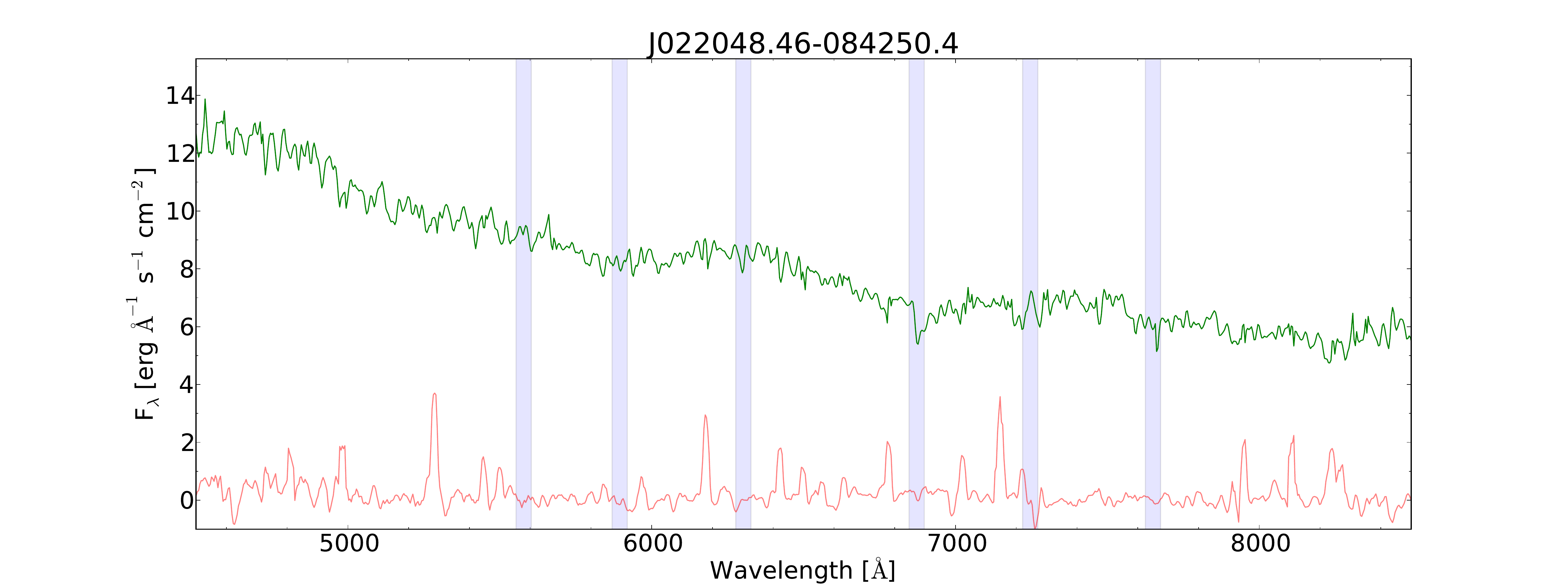}
\includegraphics[width=\textwidth]{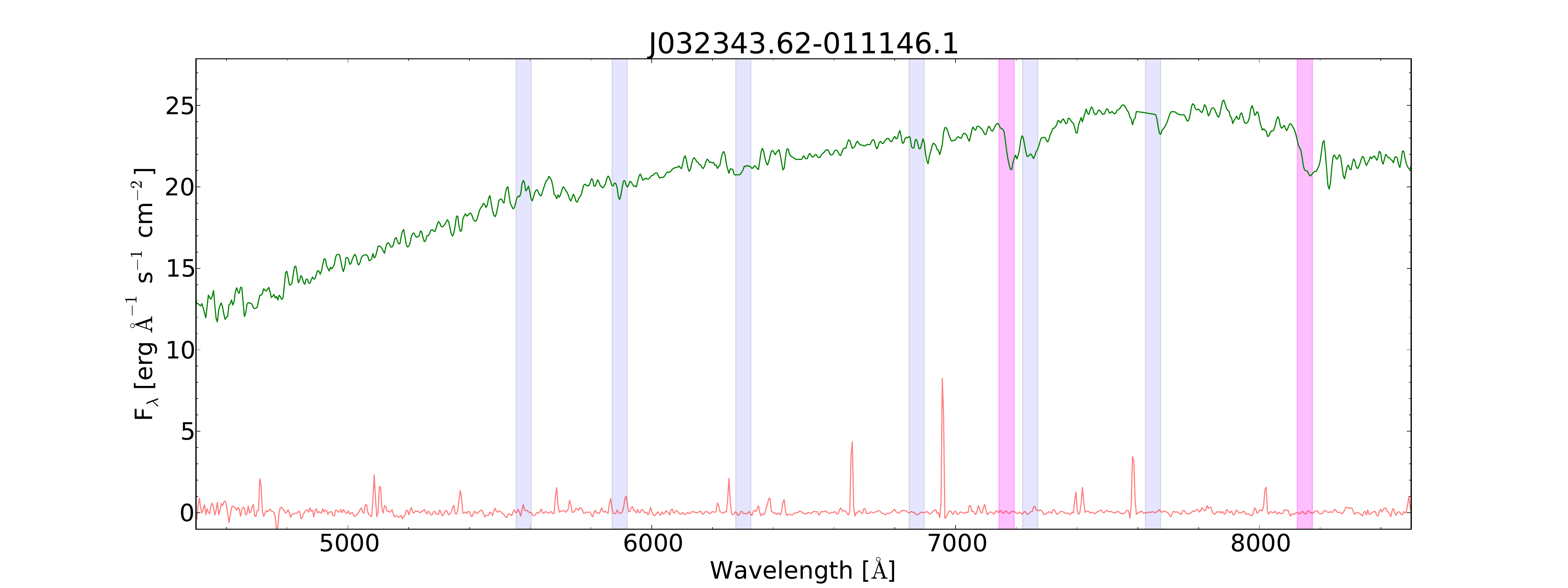}
\includegraphics[width=\textwidth]{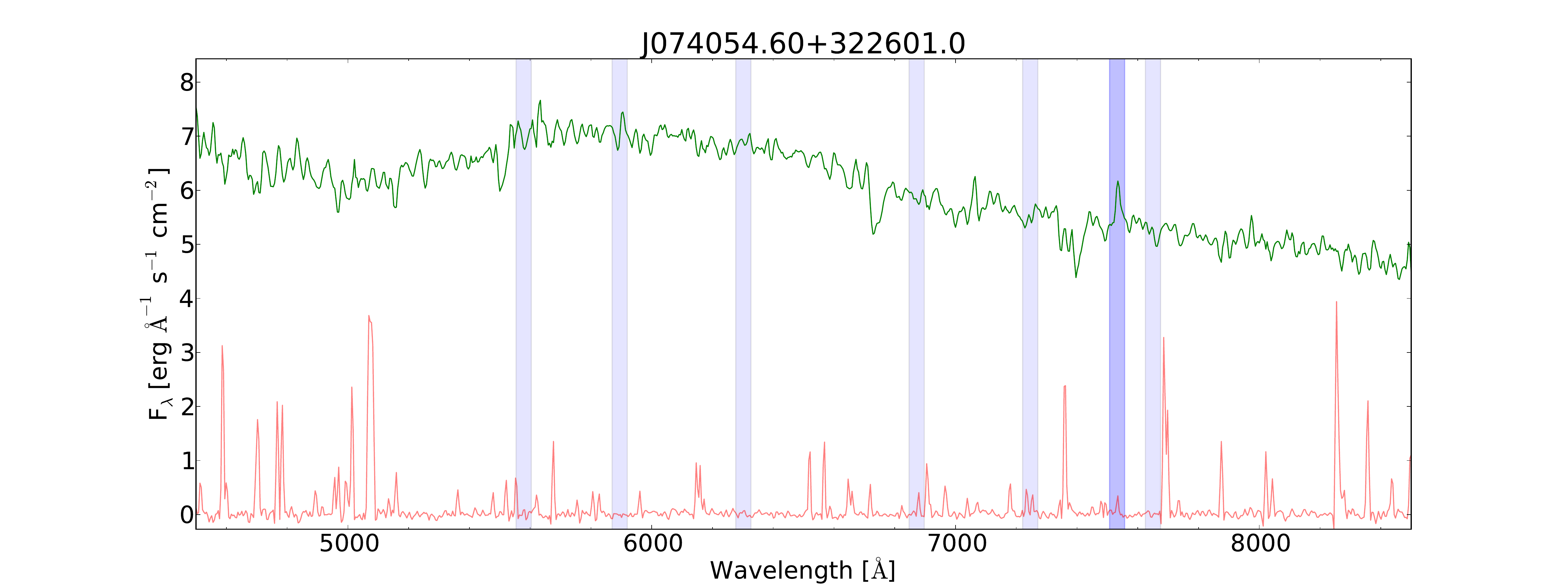}
\end{figure*}
\begin{figure*}[htb]
\centering
\includegraphics[width=\textwidth]{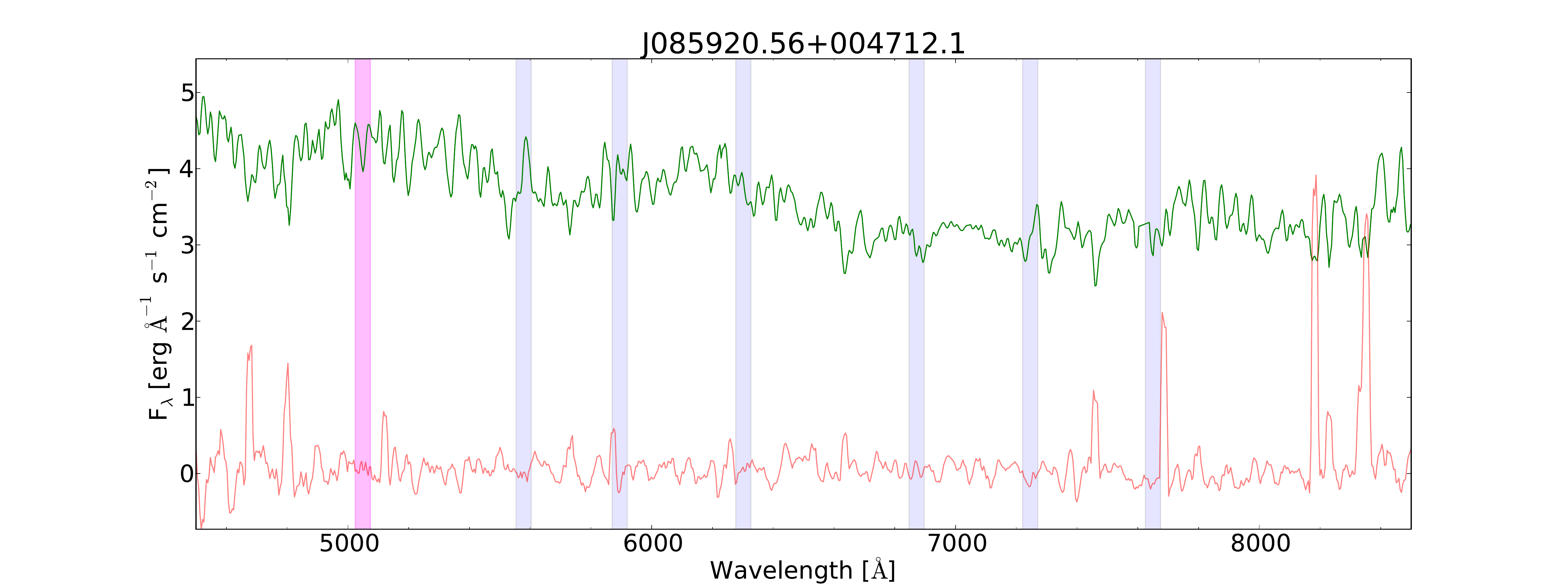}
\includegraphics[width=\textwidth]{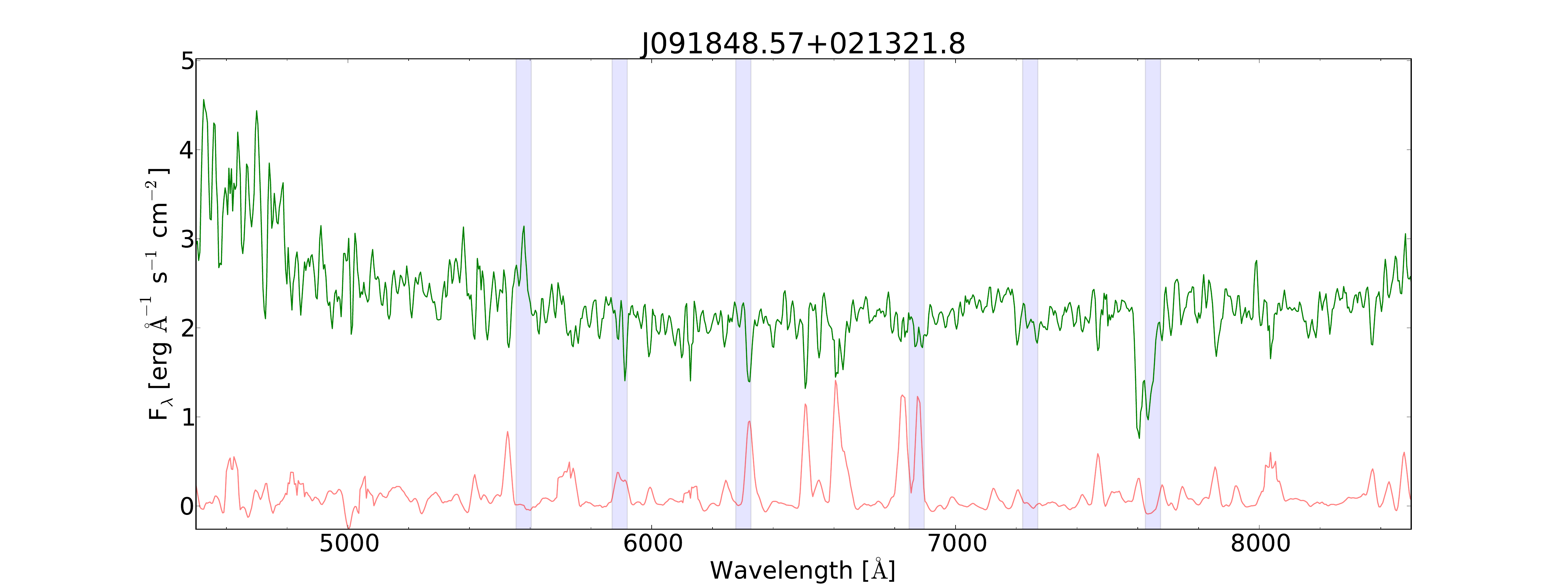}
\includegraphics[width=\textwidth]{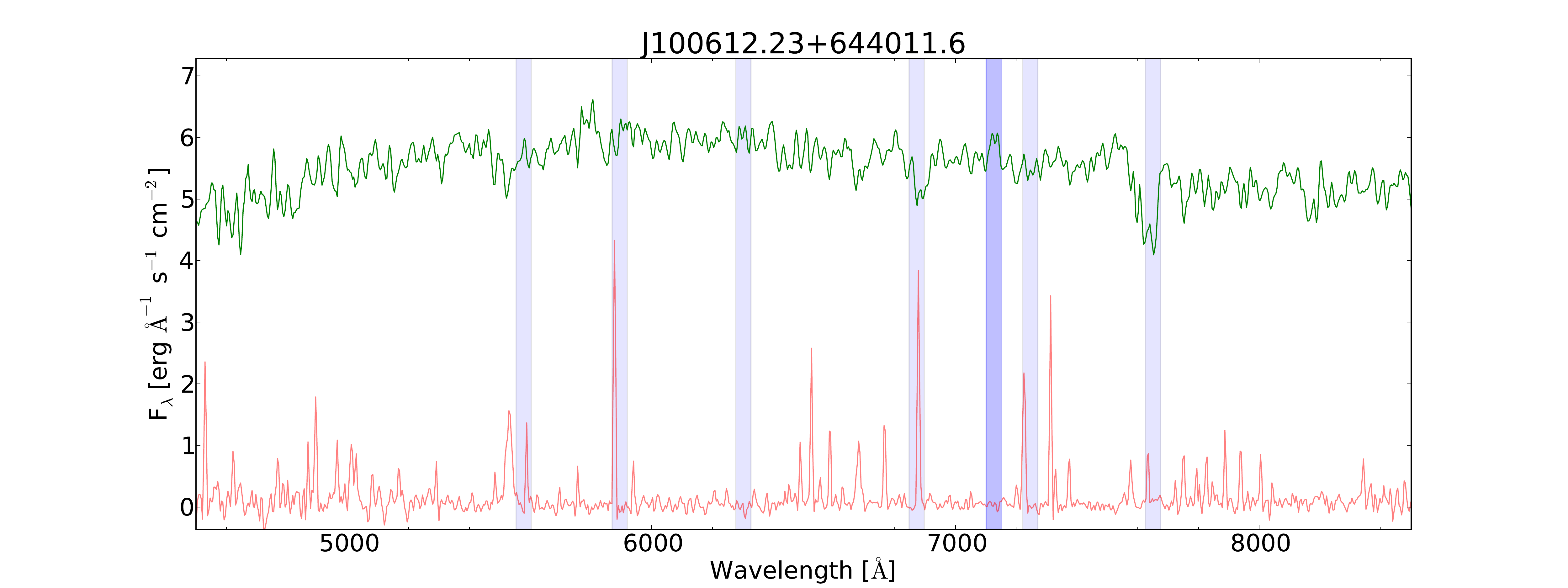}
\end{figure*}
\begin{figure*}[htb]
\centering
\includegraphics[width=\textwidth]{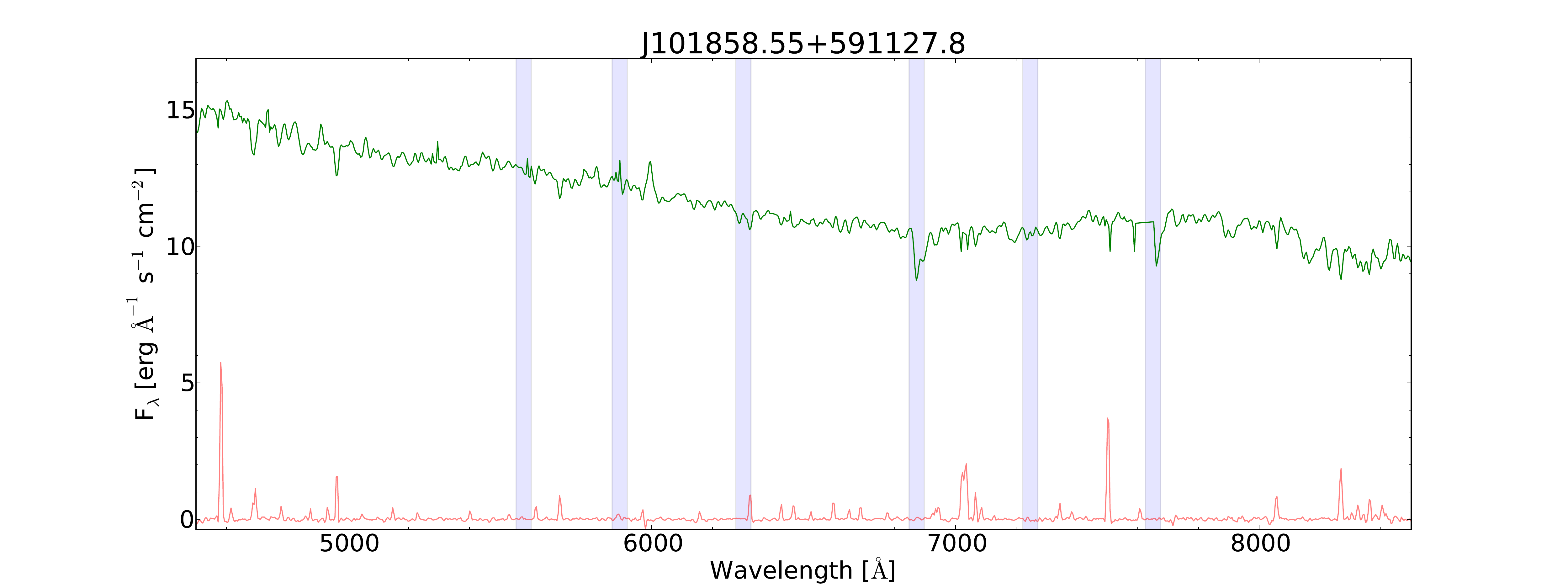}
\includegraphics[width=\textwidth]{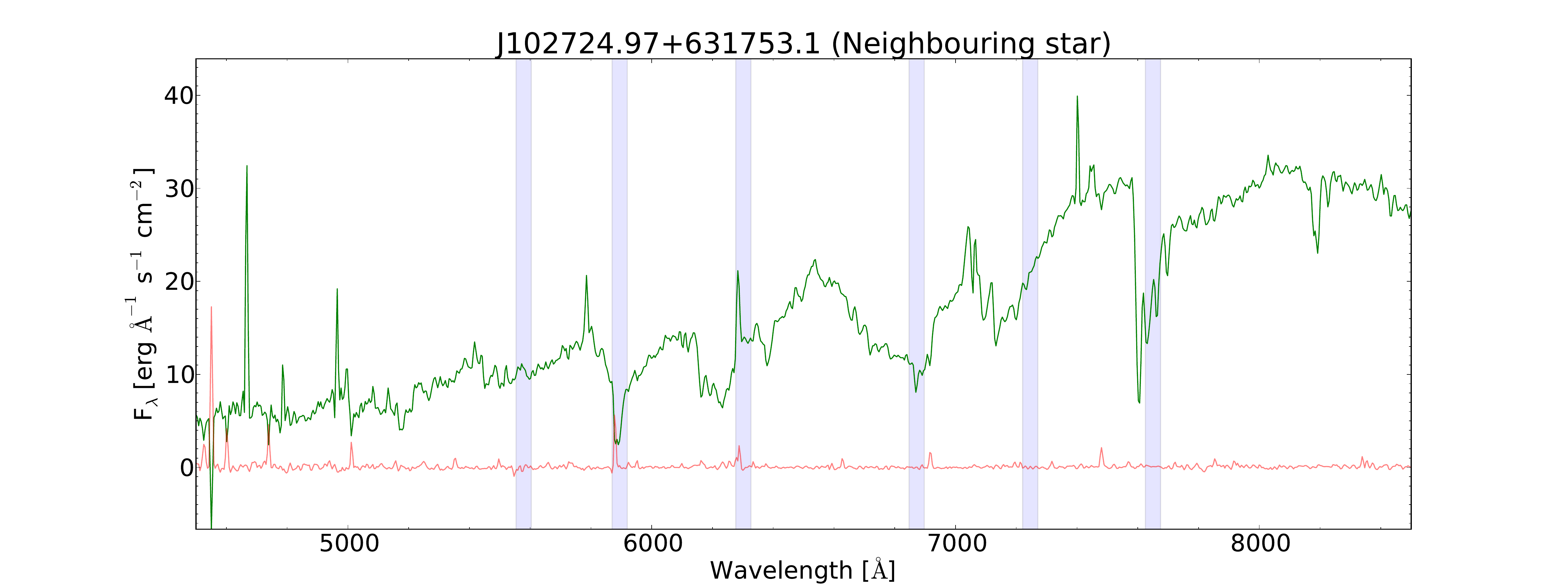}
\includegraphics[width=\textwidth]{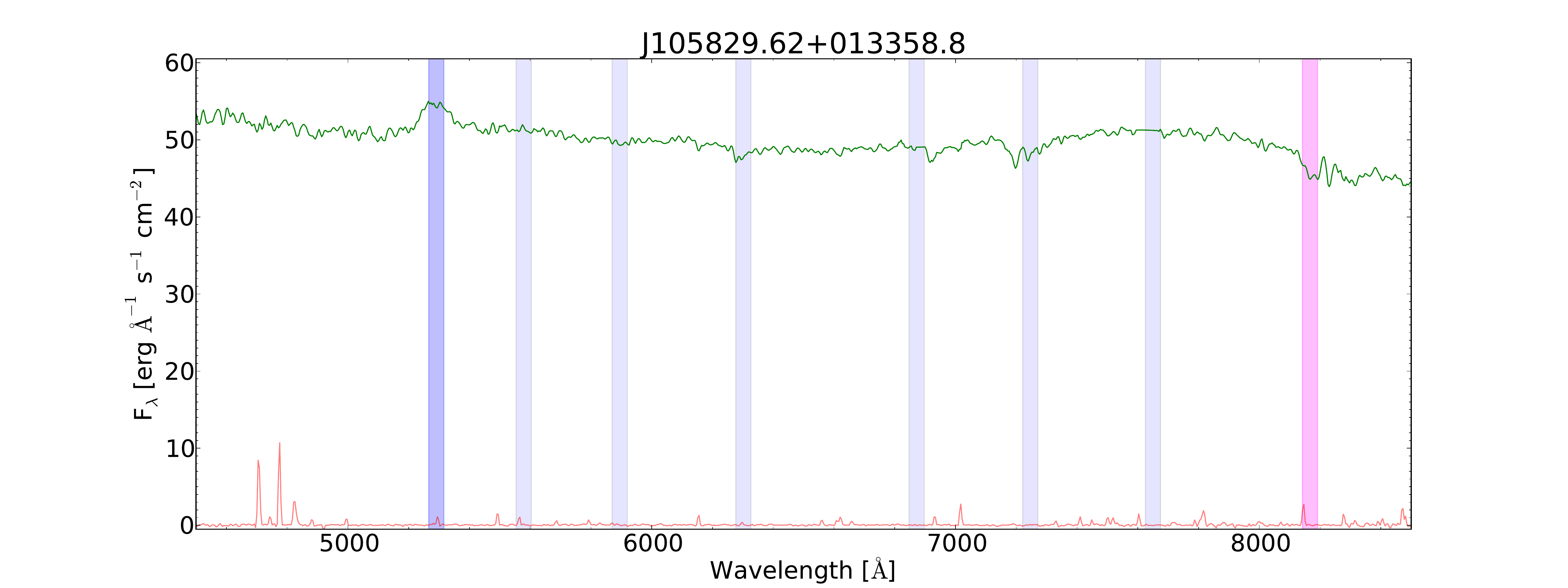}
\end{figure*}
\begin{figure*}[htb]
\centering
\includegraphics[width=\textwidth]{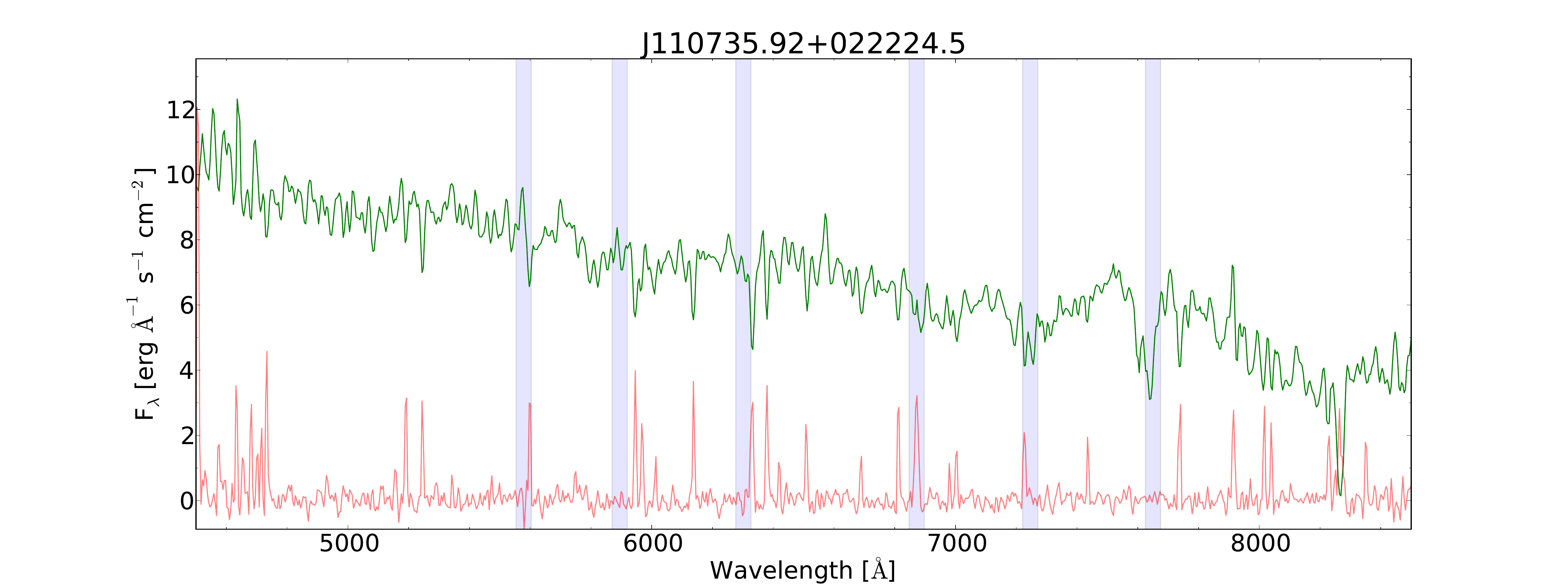}
\includegraphics[width=\textwidth]{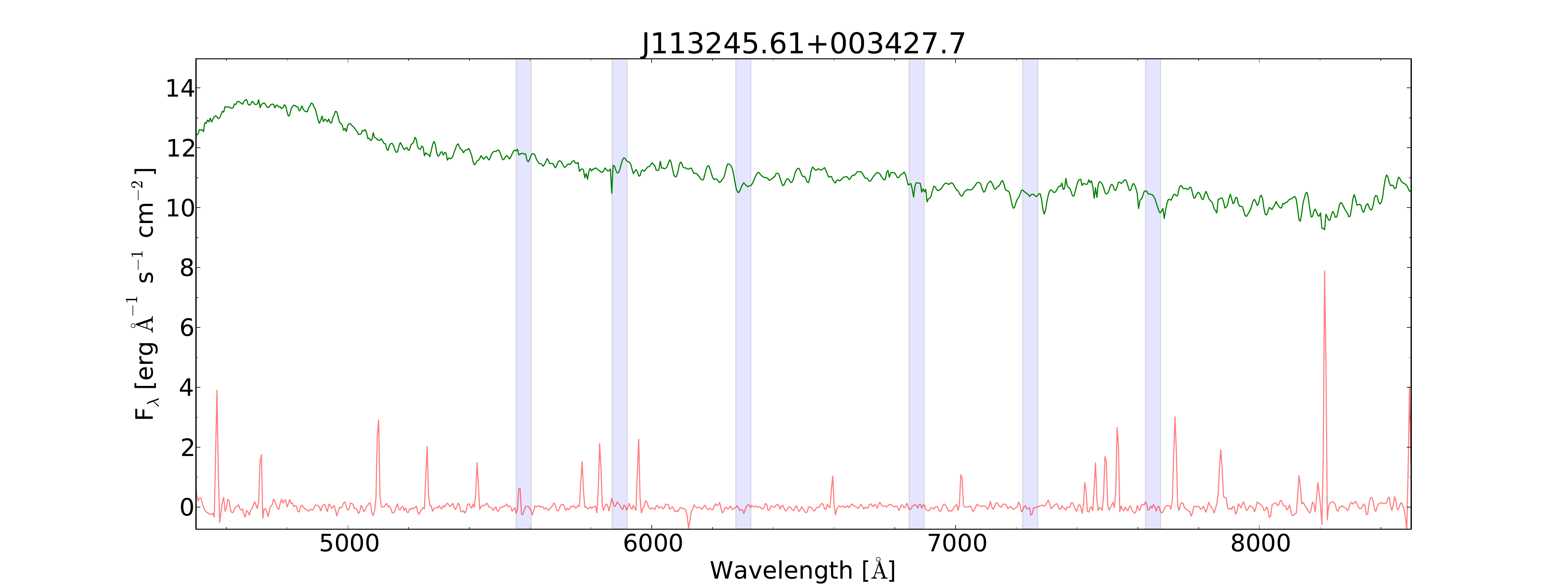}
\includegraphics[width=\textwidth]{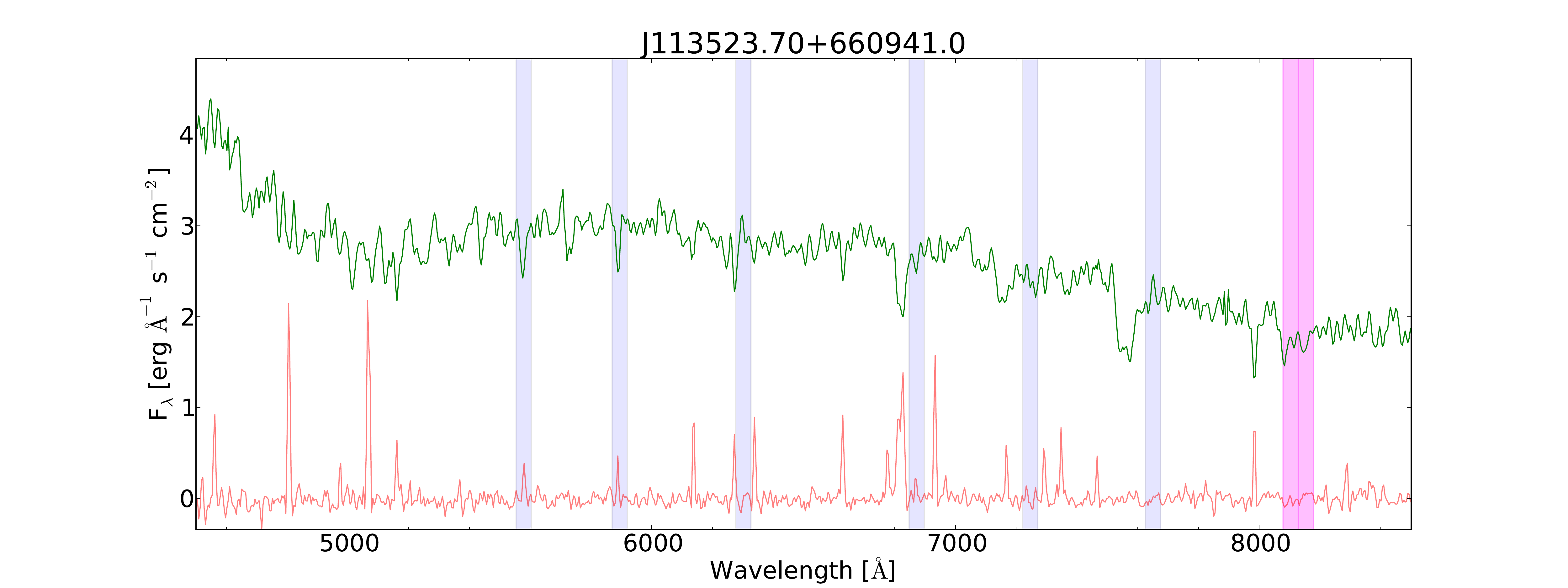}
\end{figure*}
\begin{figure*}[htb]
\centering
\includegraphics[width=\textwidth]{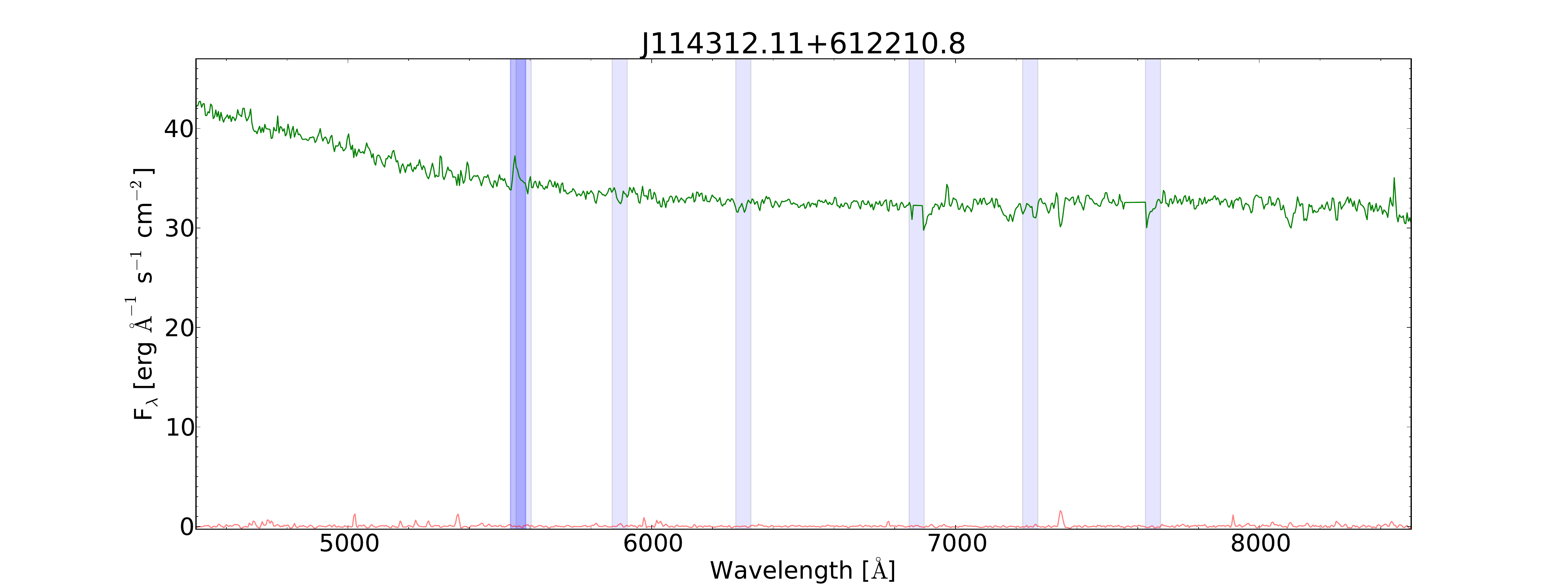}
\includegraphics[width=\textwidth]{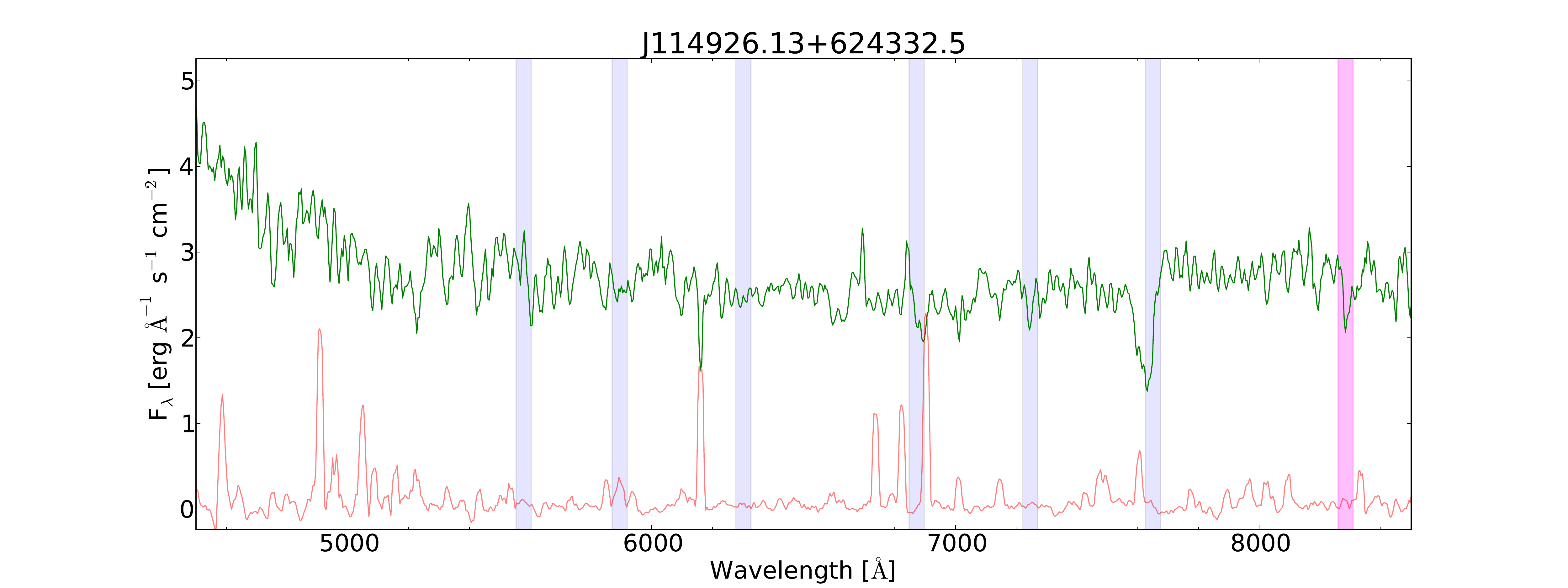}
\includegraphics[width=\textwidth]{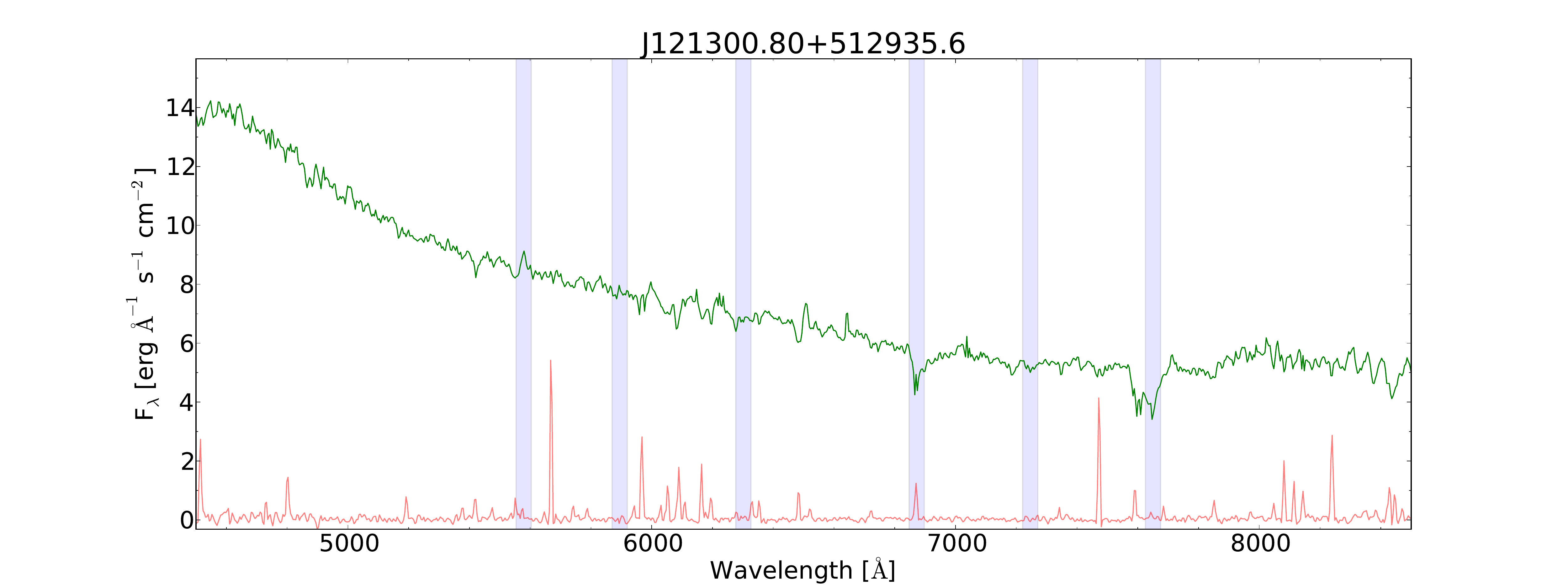}
\end{figure*}
\begin{figure*}[htb]
\centering
\includegraphics[width=\textwidth]{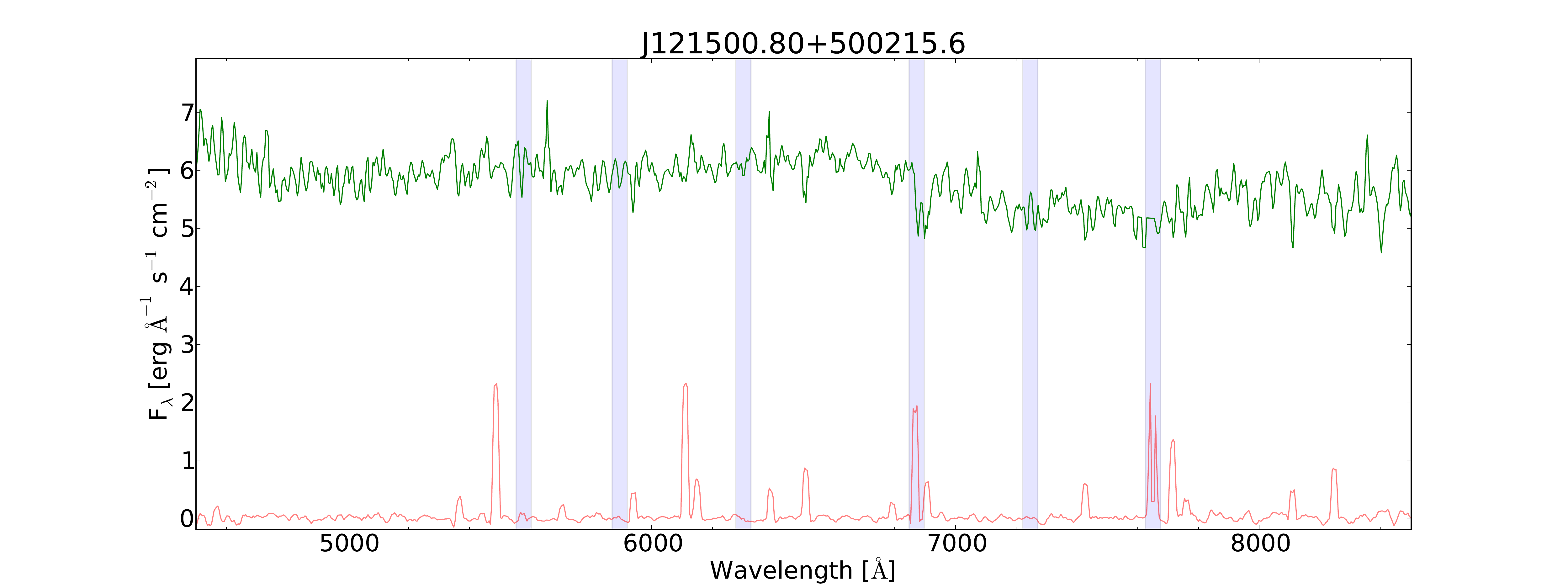}
\includegraphics[width=\textwidth]{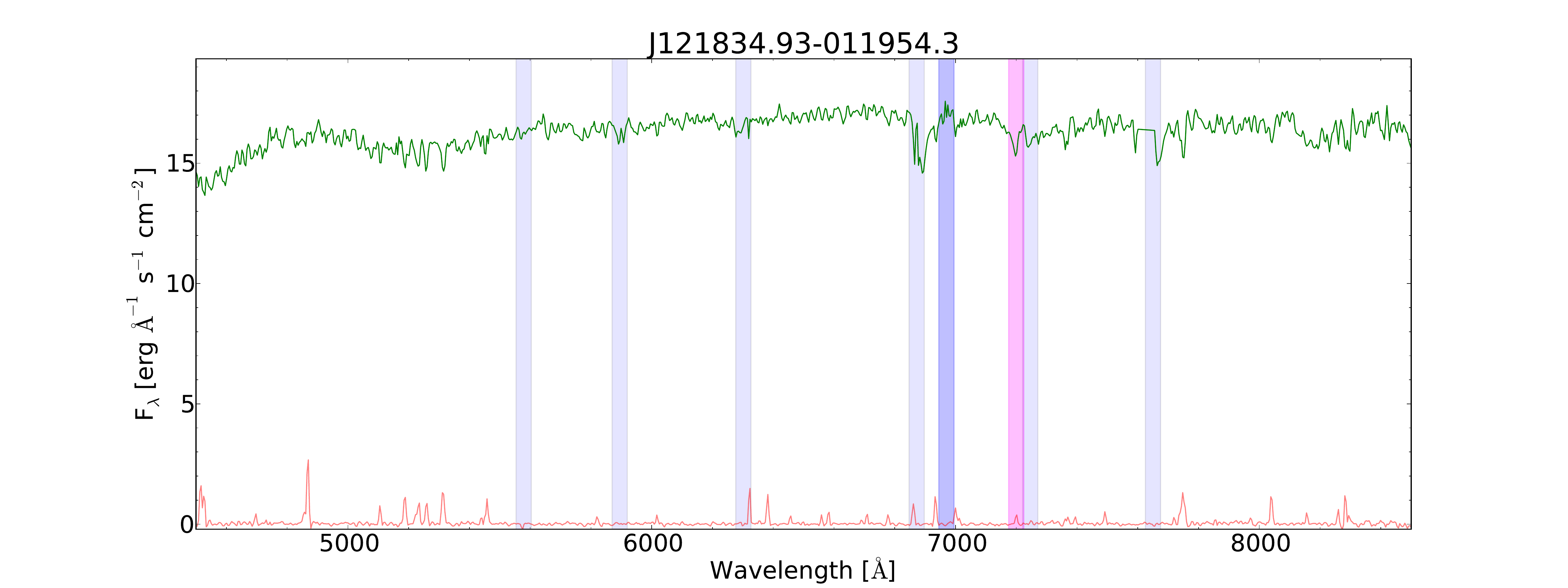}
\includegraphics[width=\textwidth]{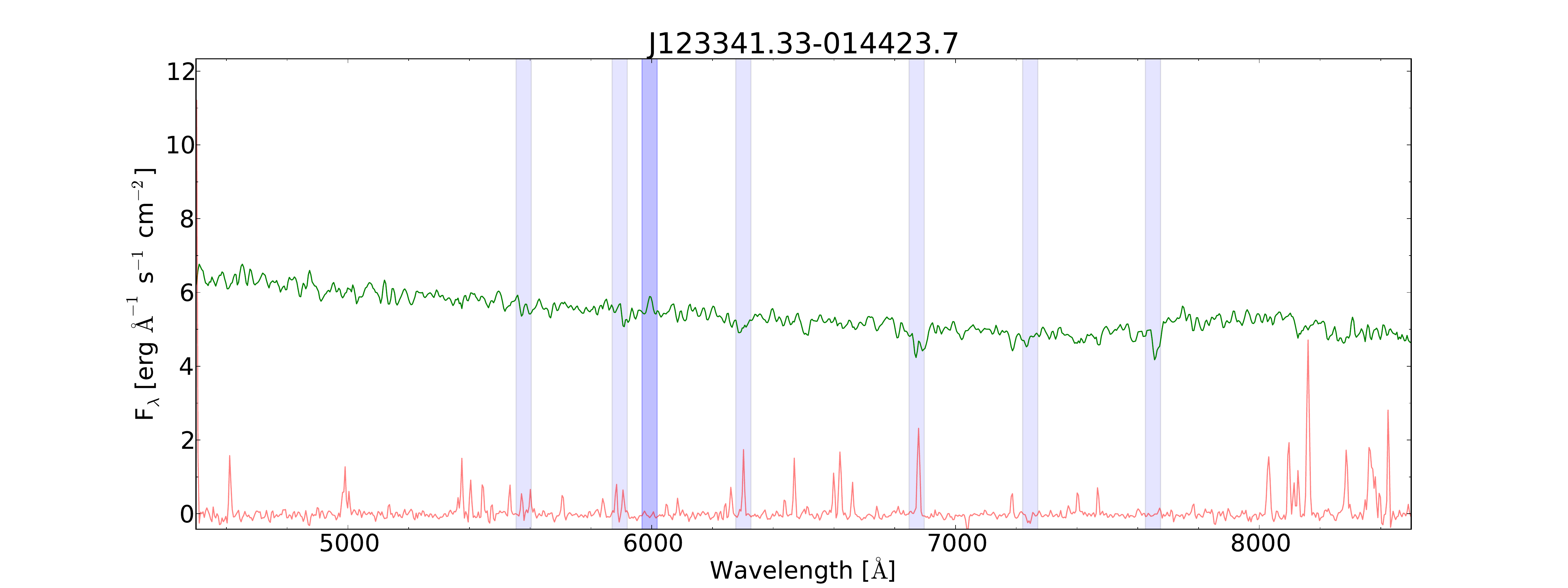}
\end{figure*}
\begin{figure*}[htb]
\centering
\includegraphics[width=\textwidth]{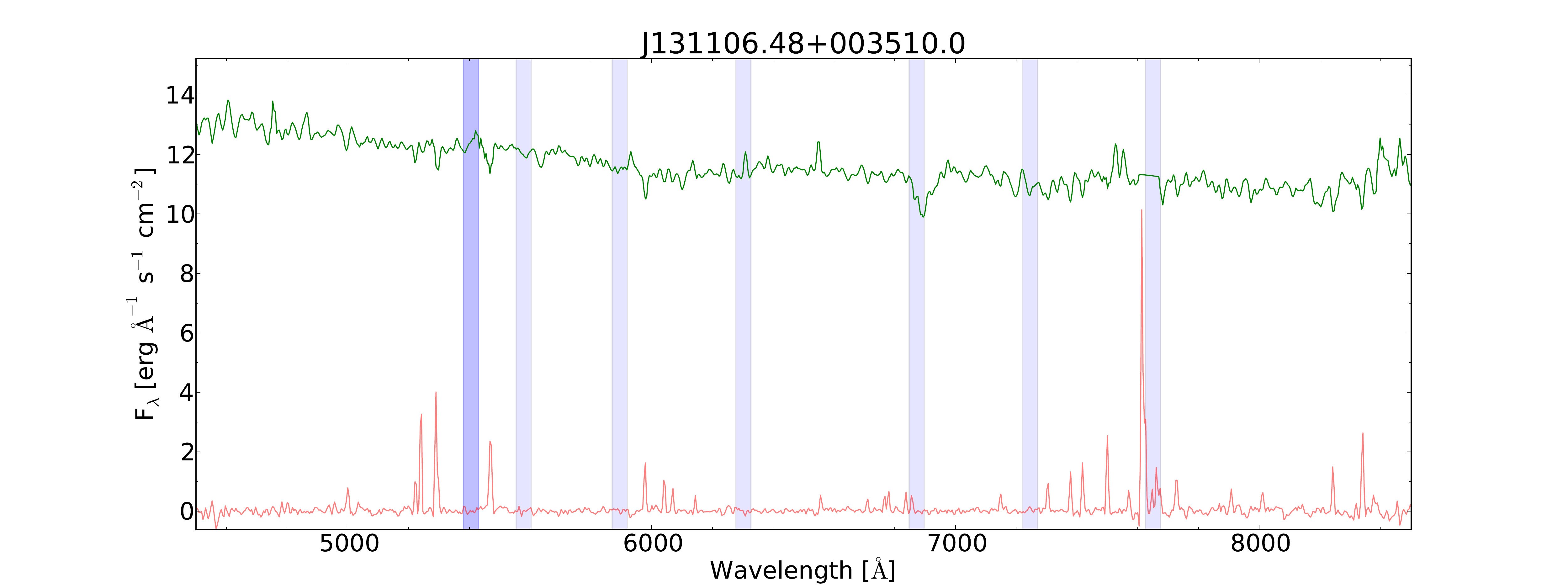}
\includegraphics[width=\textwidth]{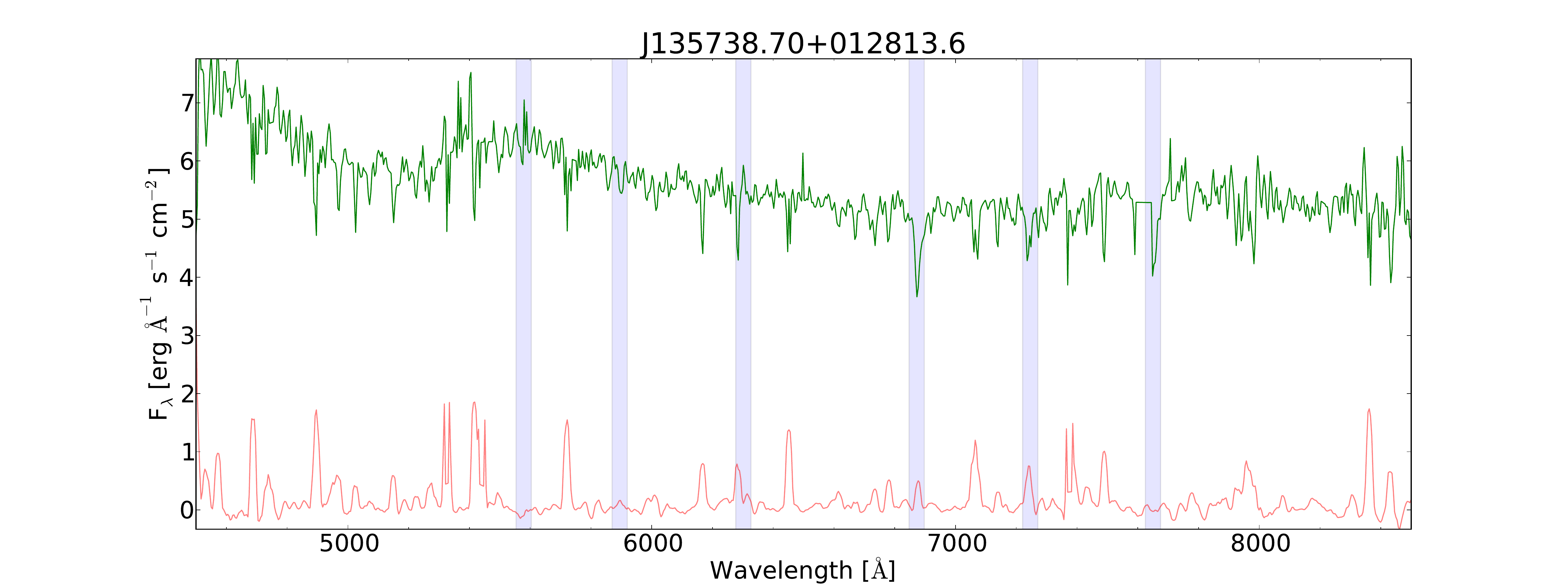}
\includegraphics[width=\textwidth]{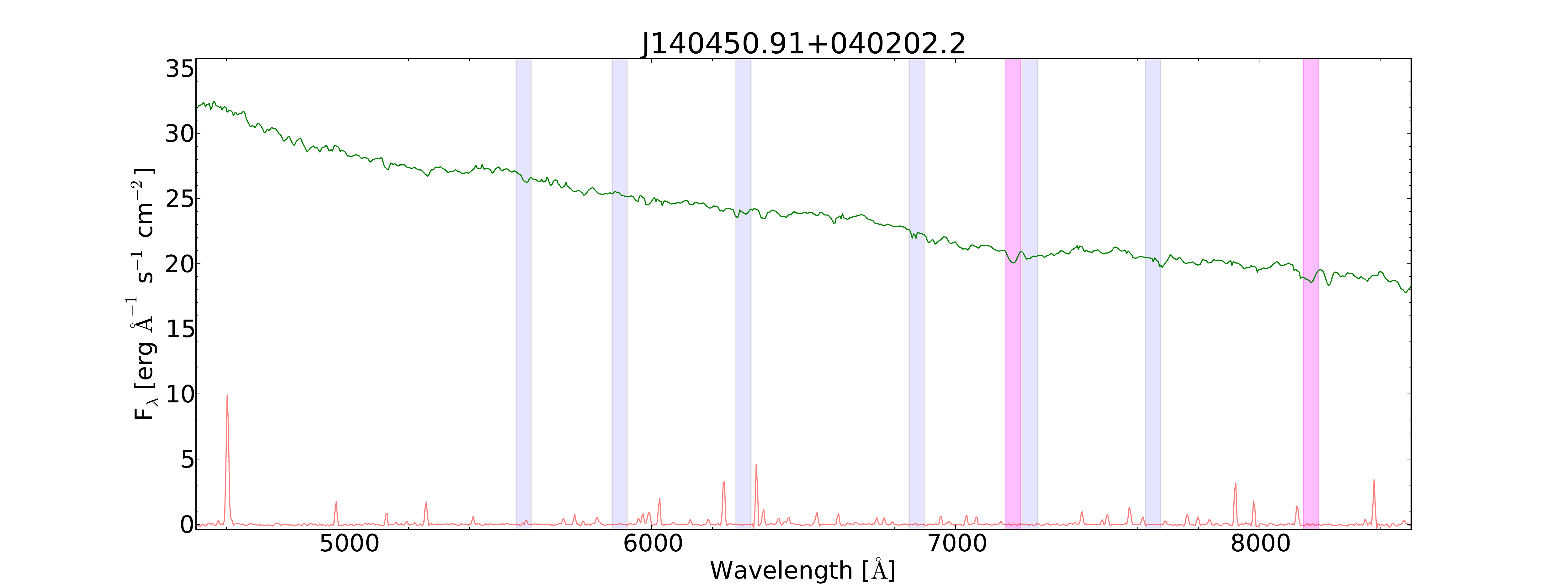}
\end{figure*}
\begin{figure*}[htb]
\centering
\includegraphics[width=\textwidth]{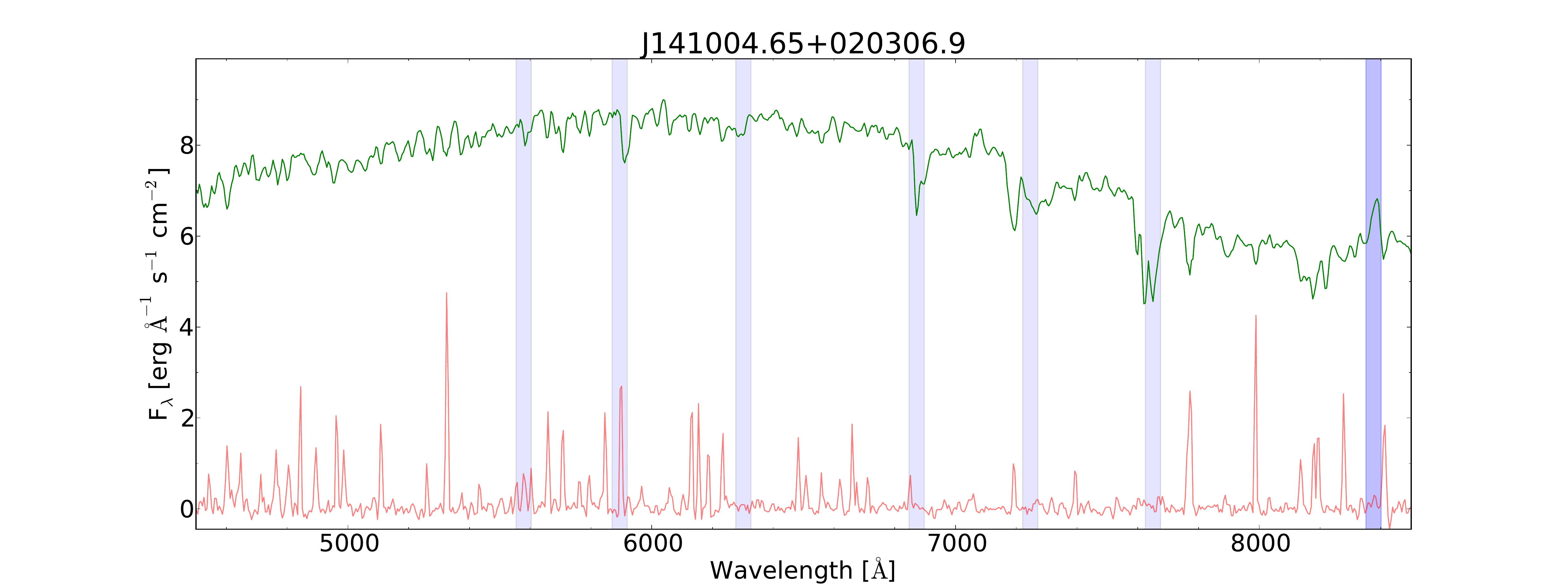}
\includegraphics[width=\textwidth]{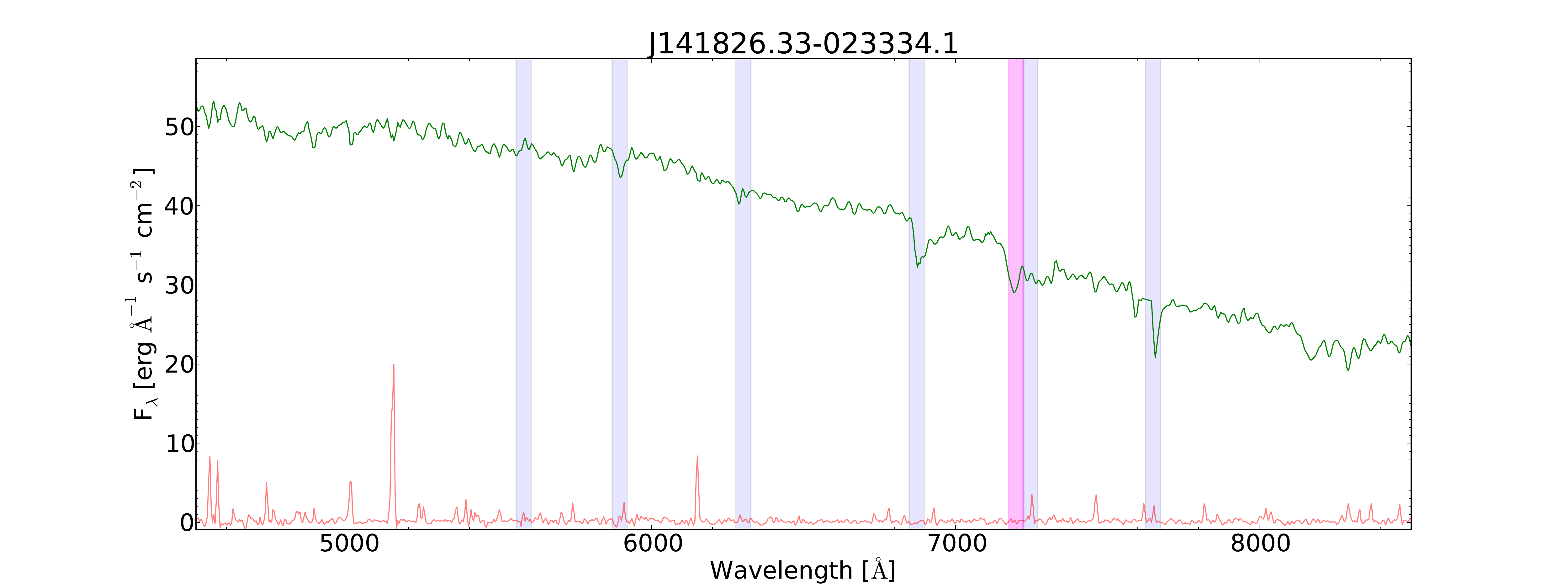}
\includegraphics[width=\textwidth]{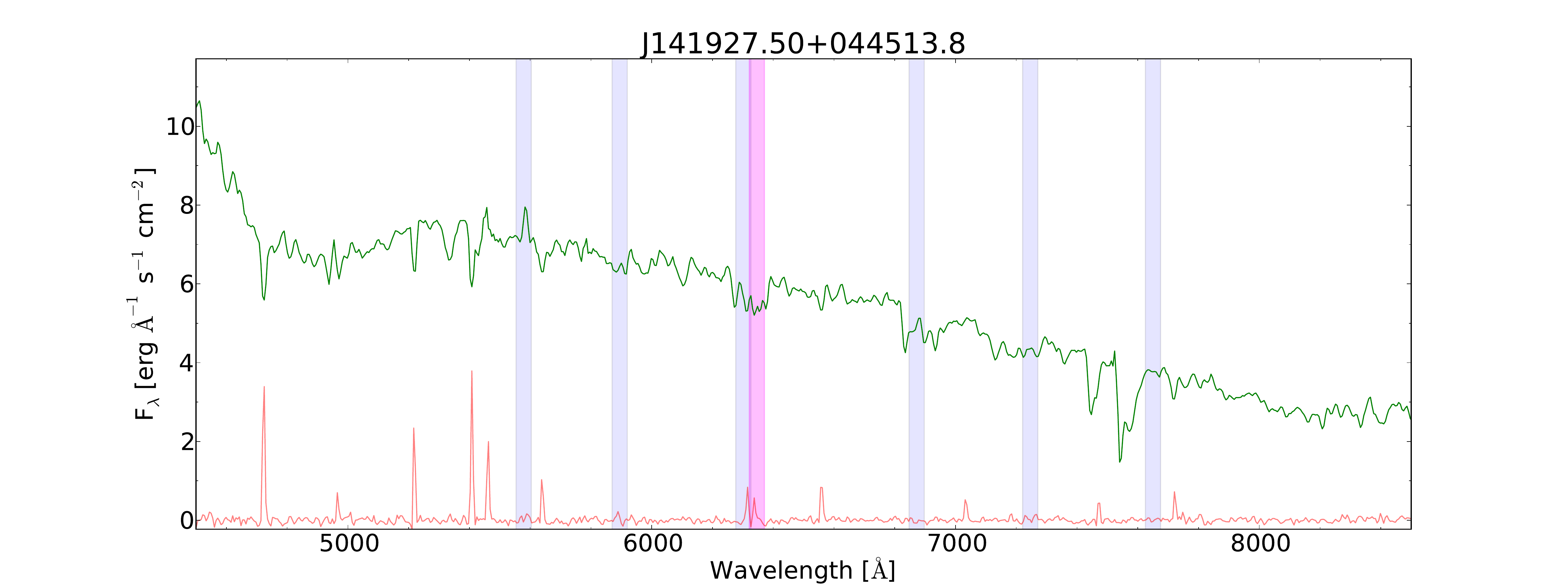}
\end{figure*}
\begin{figure*}[htb]
\centering
\includegraphics[width=\textwidth]{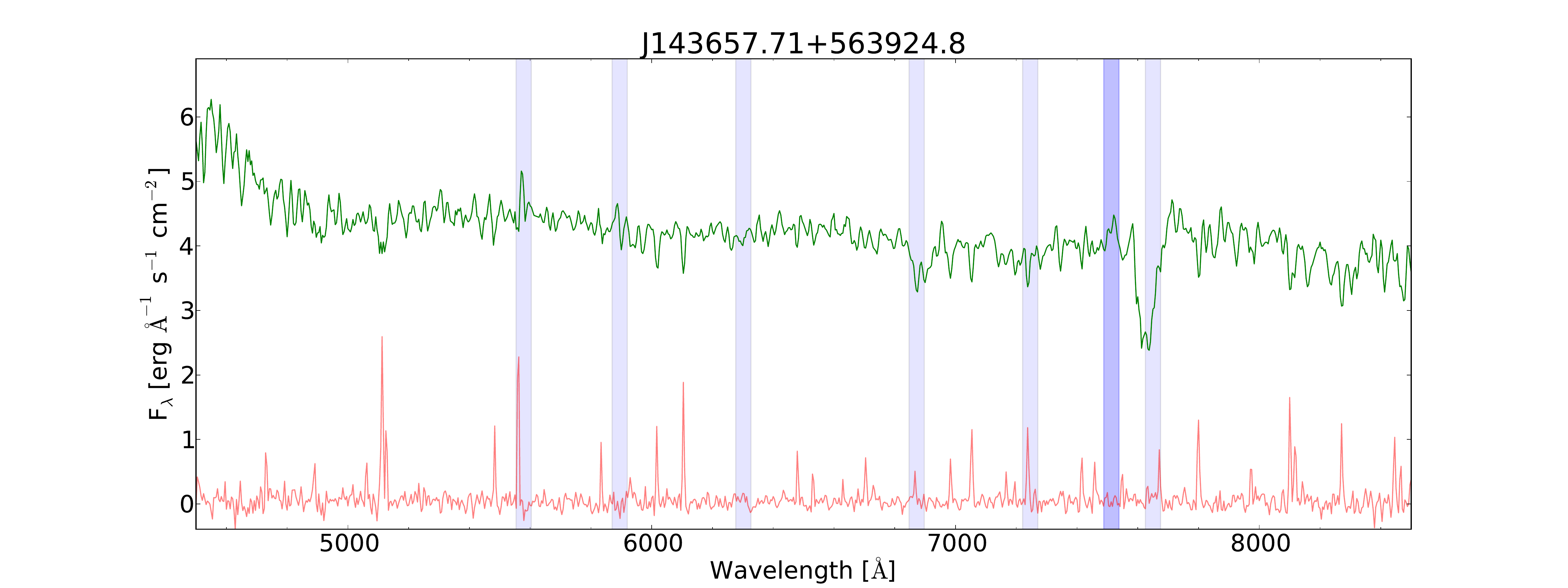}
\includegraphics[width=\textwidth]{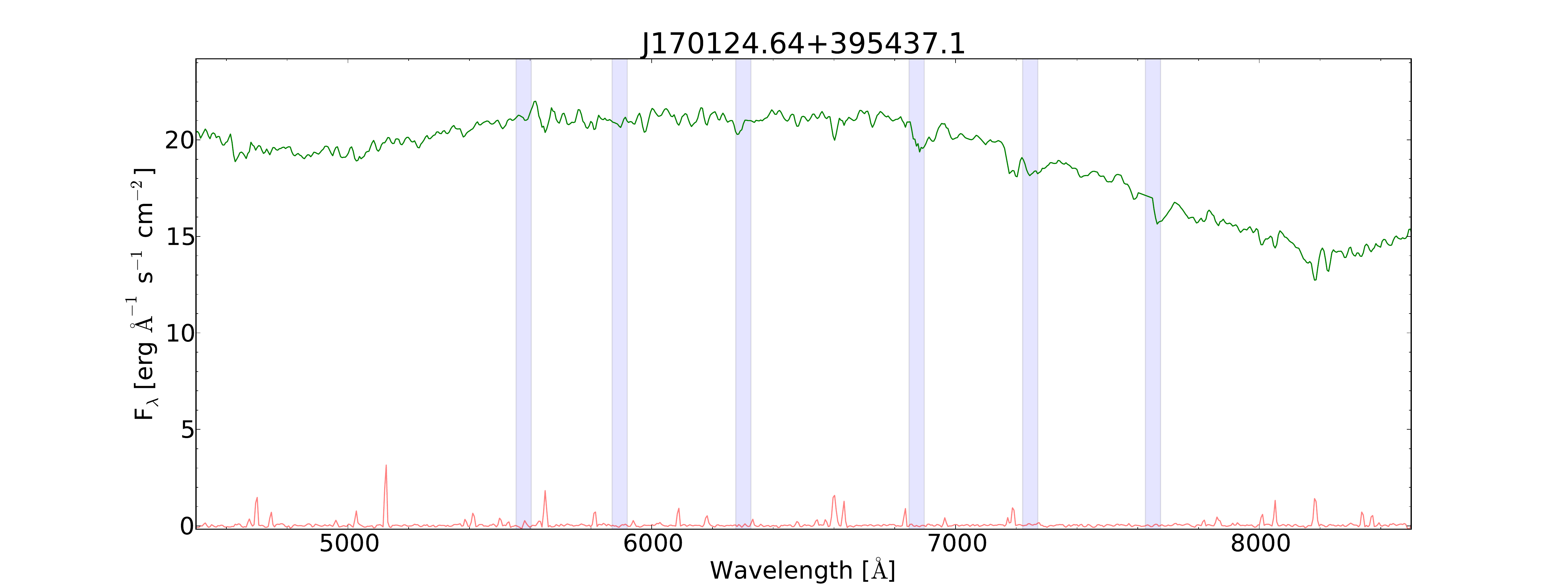}
\end{figure*}
}

\newpage
\begin{figure*}[htb!]
\caption{SED fits, and peak frequencies 
of all objects with more than 12 data points in the SED. 
All objects with fewer than 2 radio measurements (uncertain) are labeled 
with a ``[?]'' in the figure caption. 
The three objects where the fit did not yield any reasonable results 
are emphasized with a ``*'' and objects with a 
significant influence of the host galaxy (core fraction $<$ 0.5) on the 
SED are highlighted with a ``$\dagger$'' in the title.
\label{SEDfits}}
\includegraphics[width=0.3\textwidth]{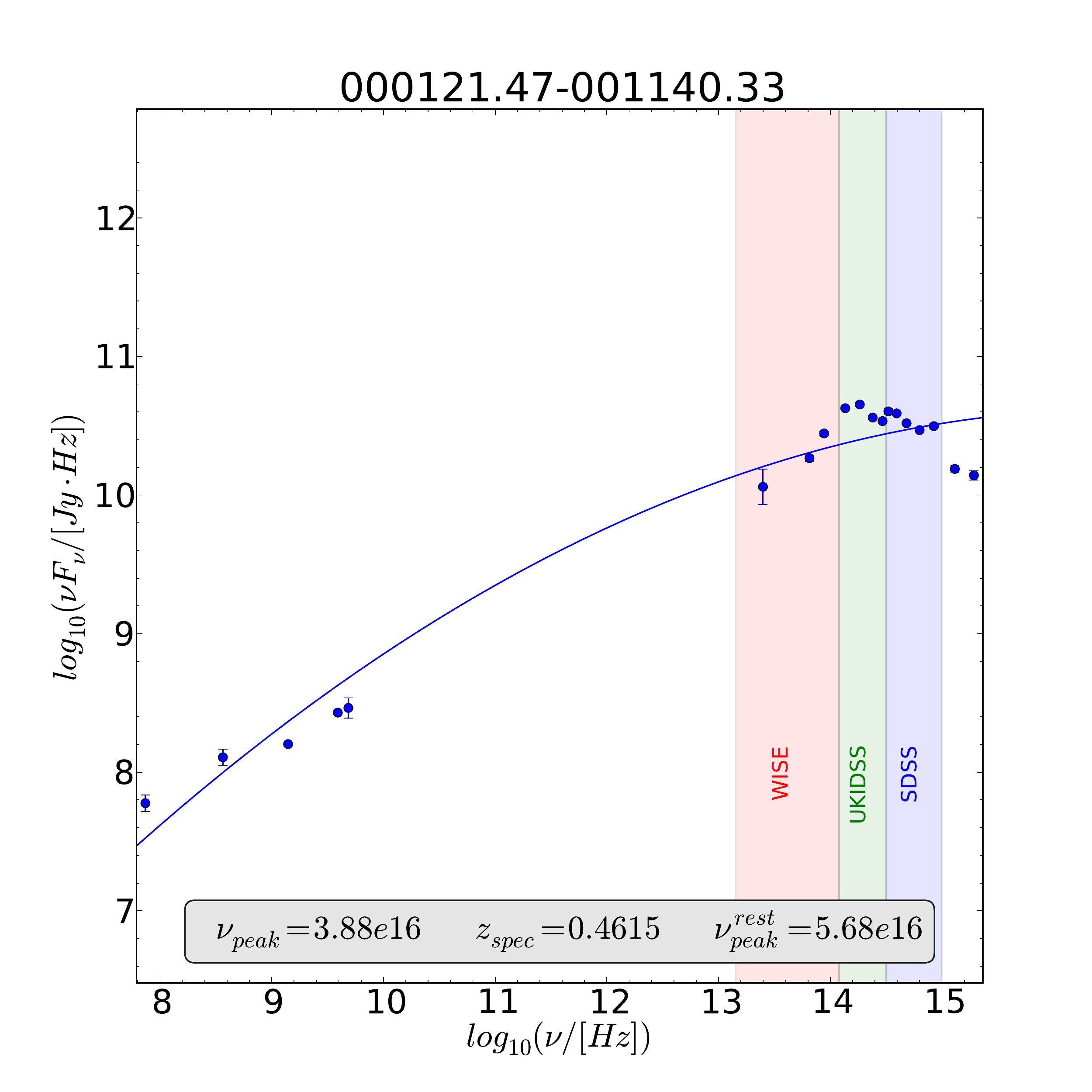}
\includegraphics[width=0.3\textwidth]{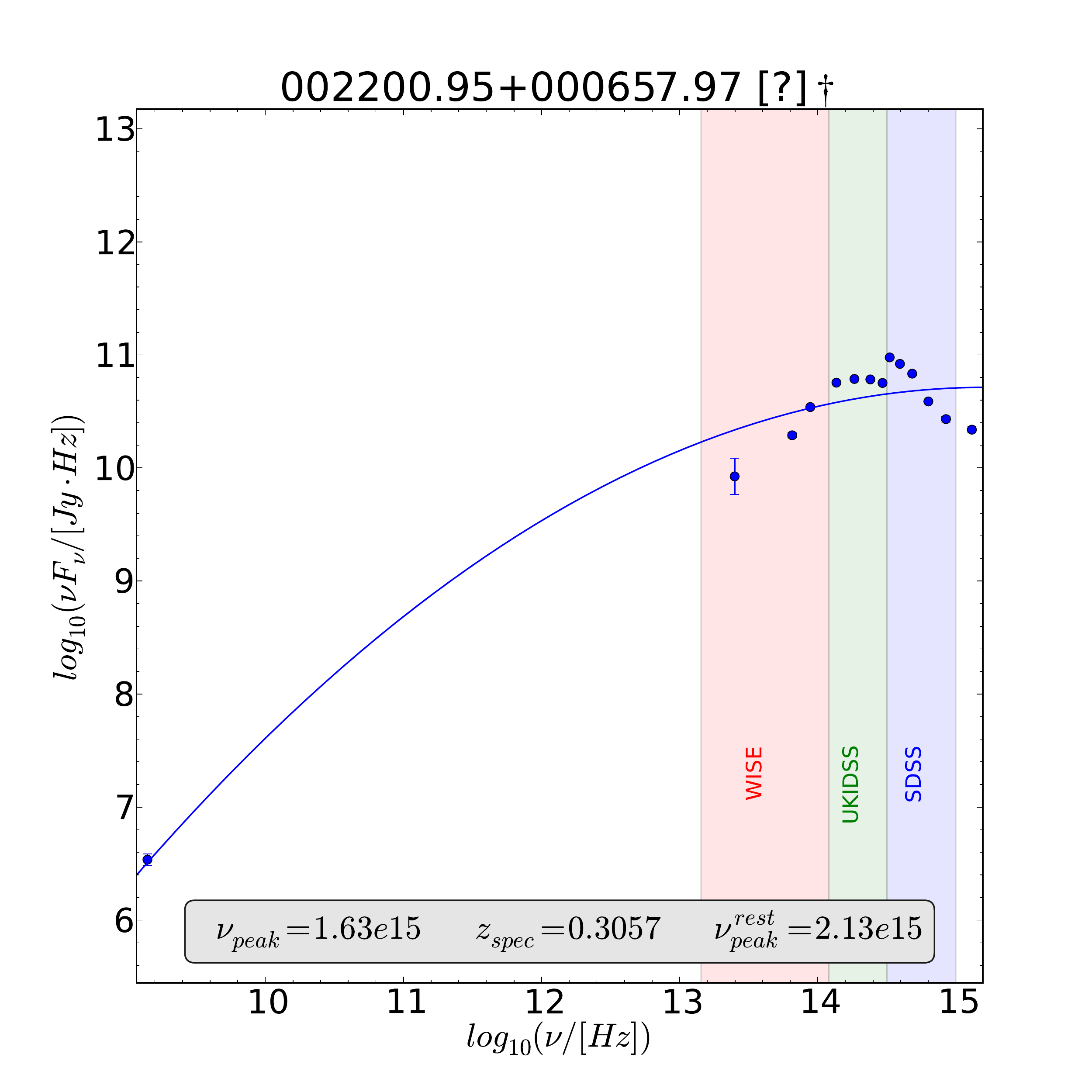}
\includegraphics[width=0.3\textwidth]{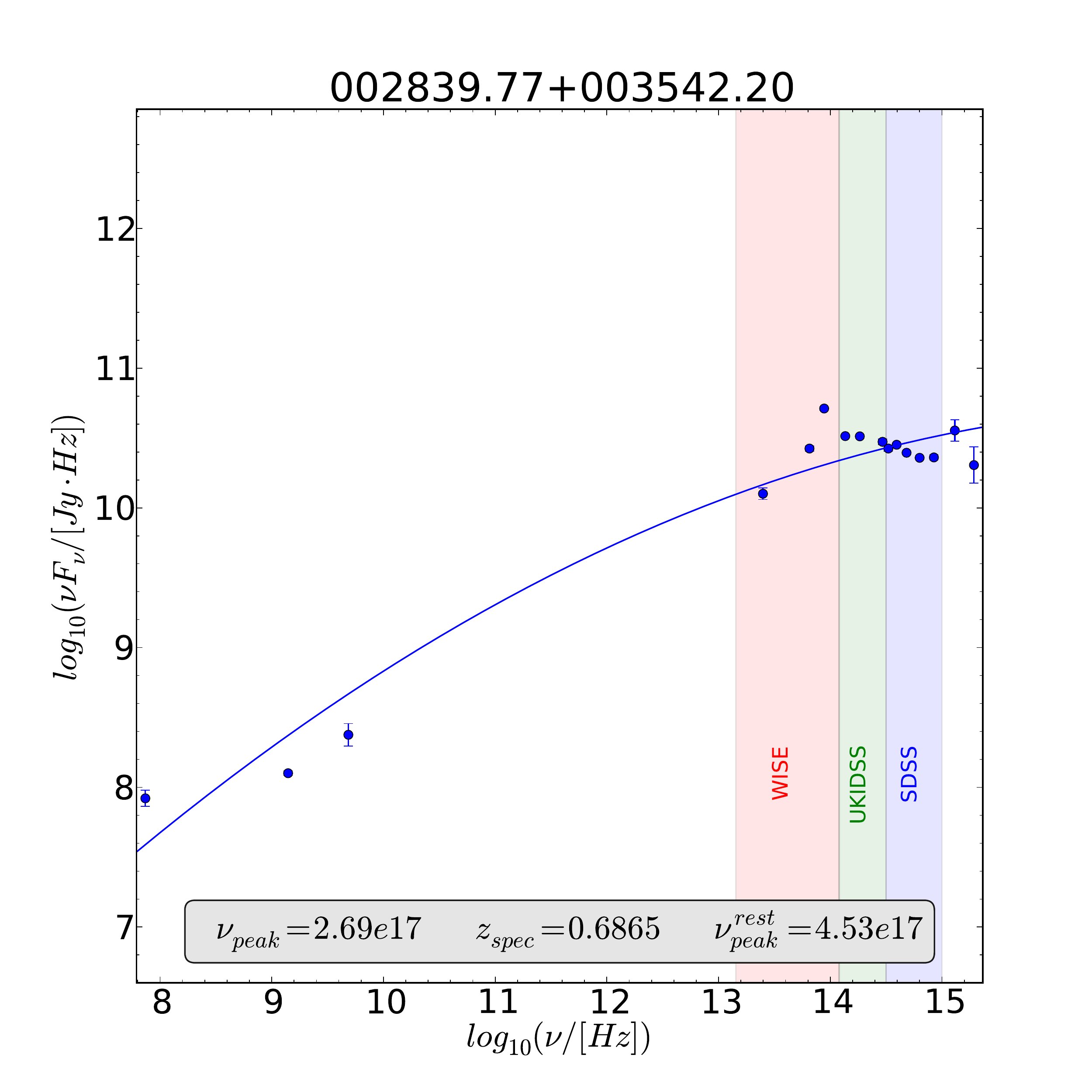}\\

\includegraphics[width=0.3\textwidth]{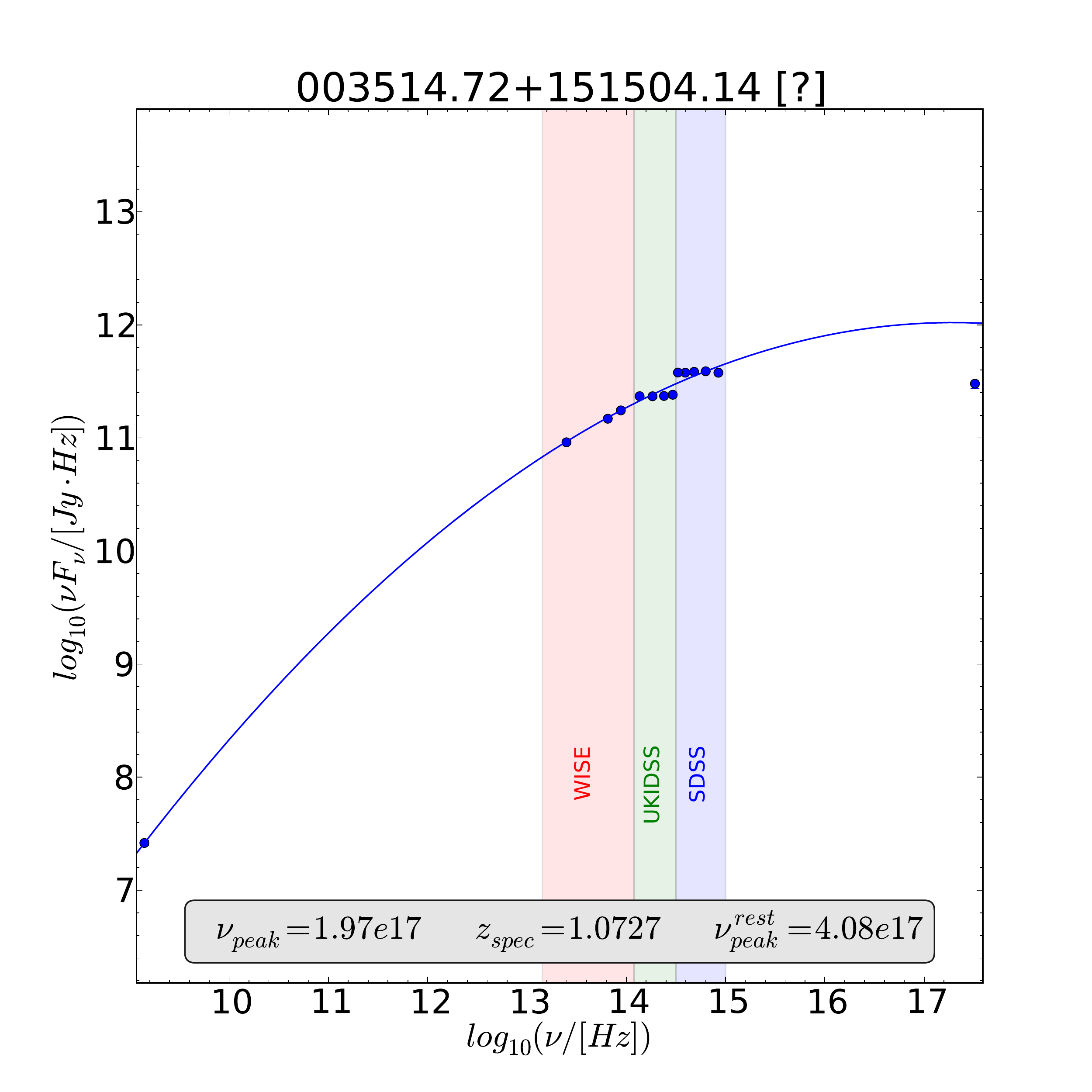}
\includegraphics[width=0.3\textwidth]{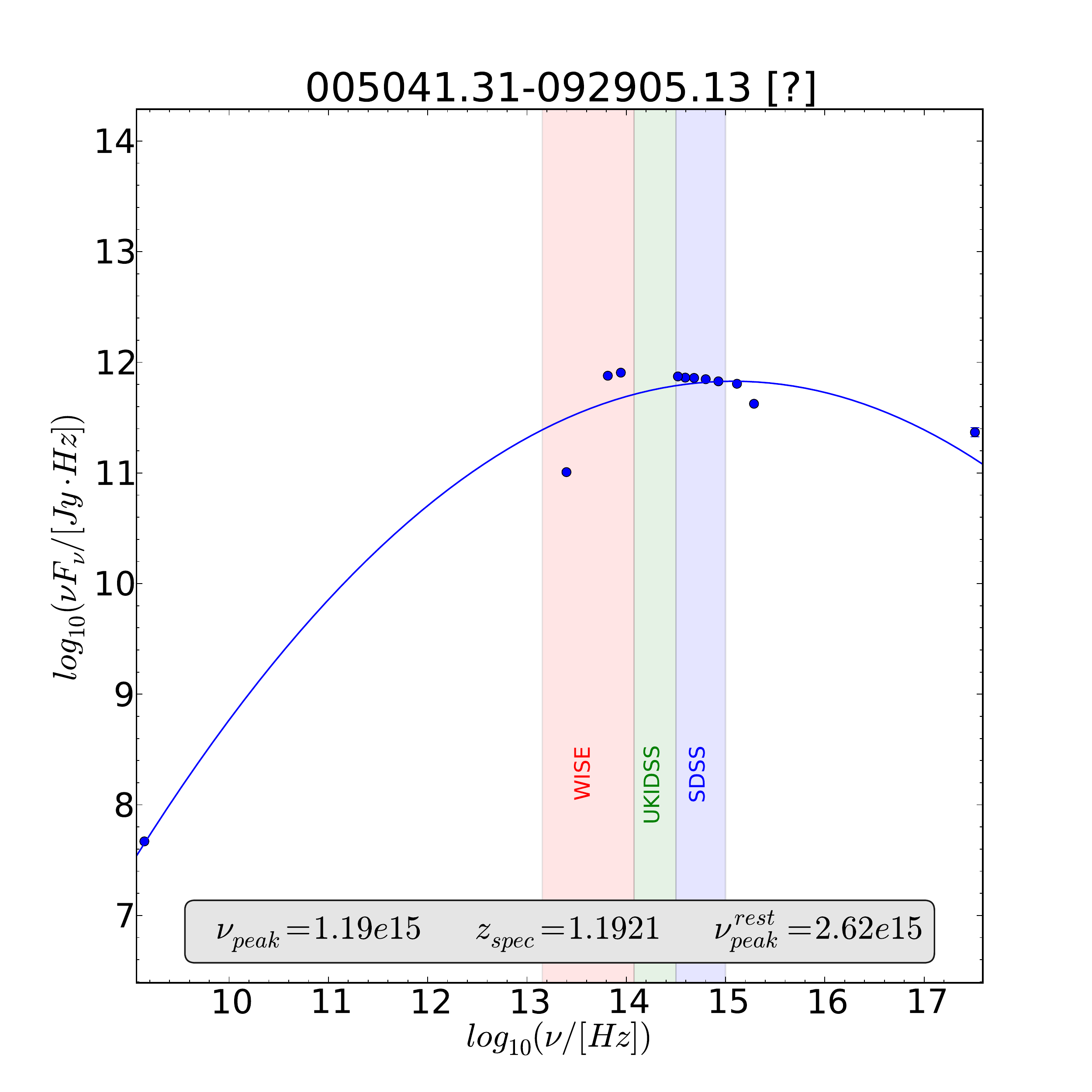}
\includegraphics[width=0.3\textwidth]{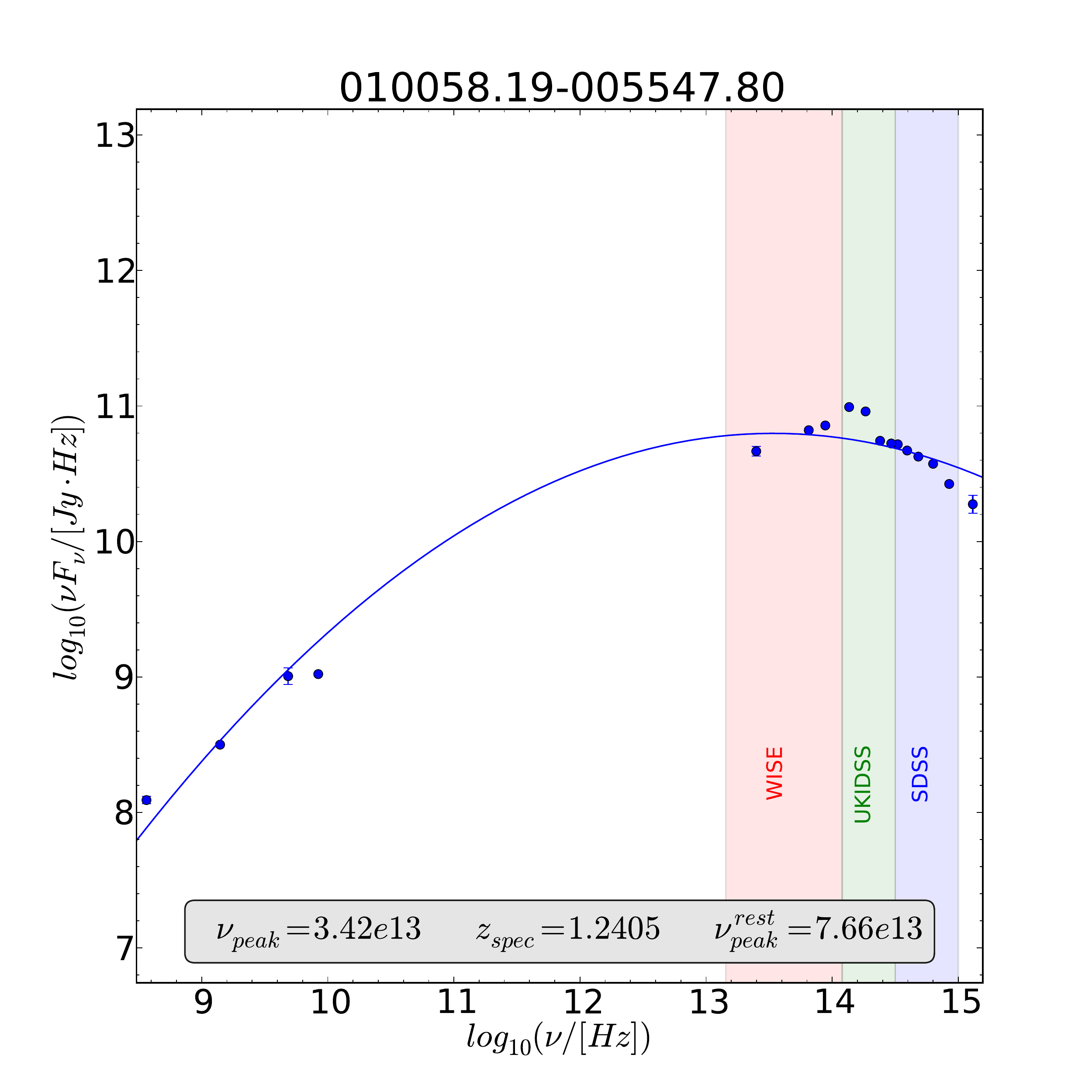}\\

\includegraphics[width=0.3\textwidth]{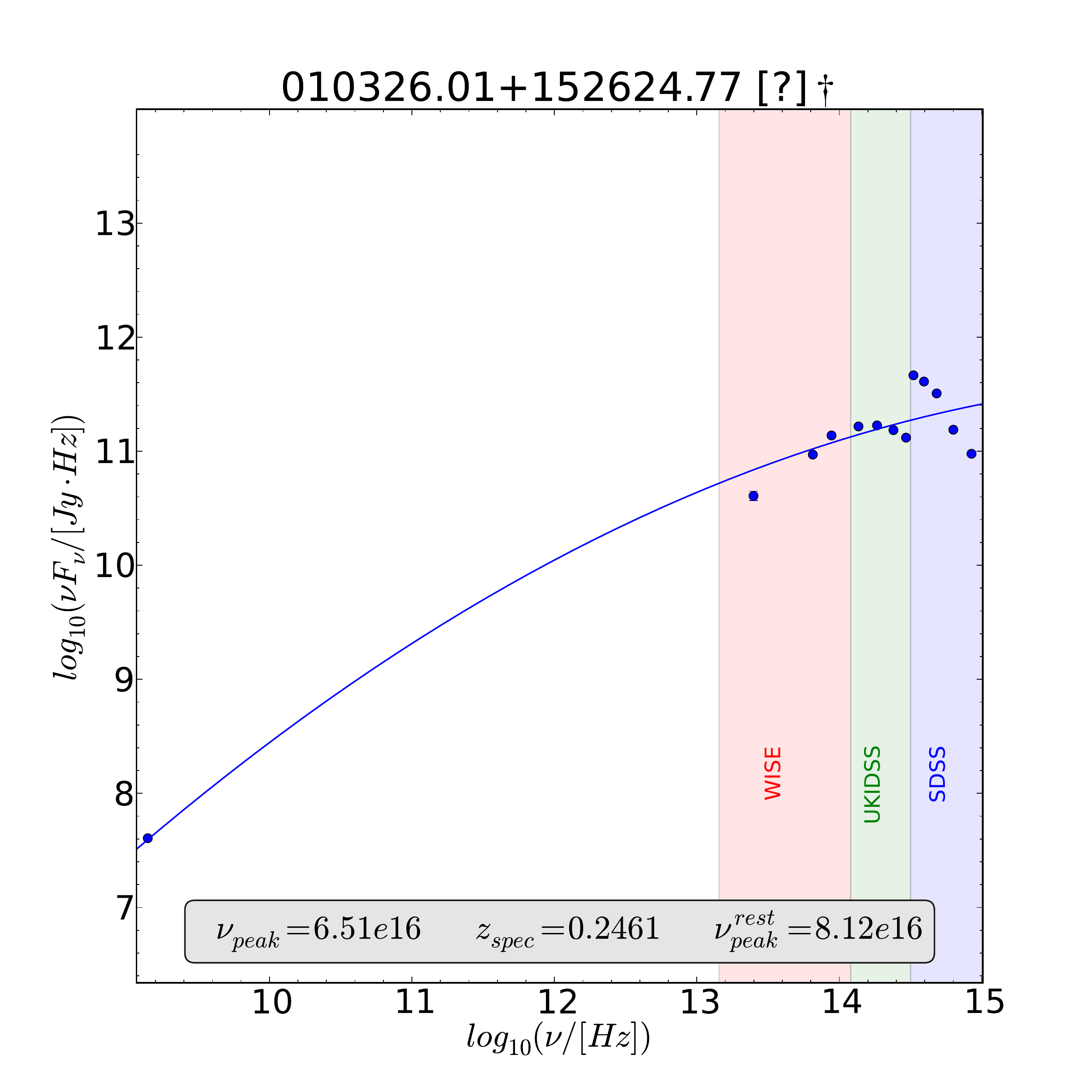}
\includegraphics[width=0.3\textwidth]{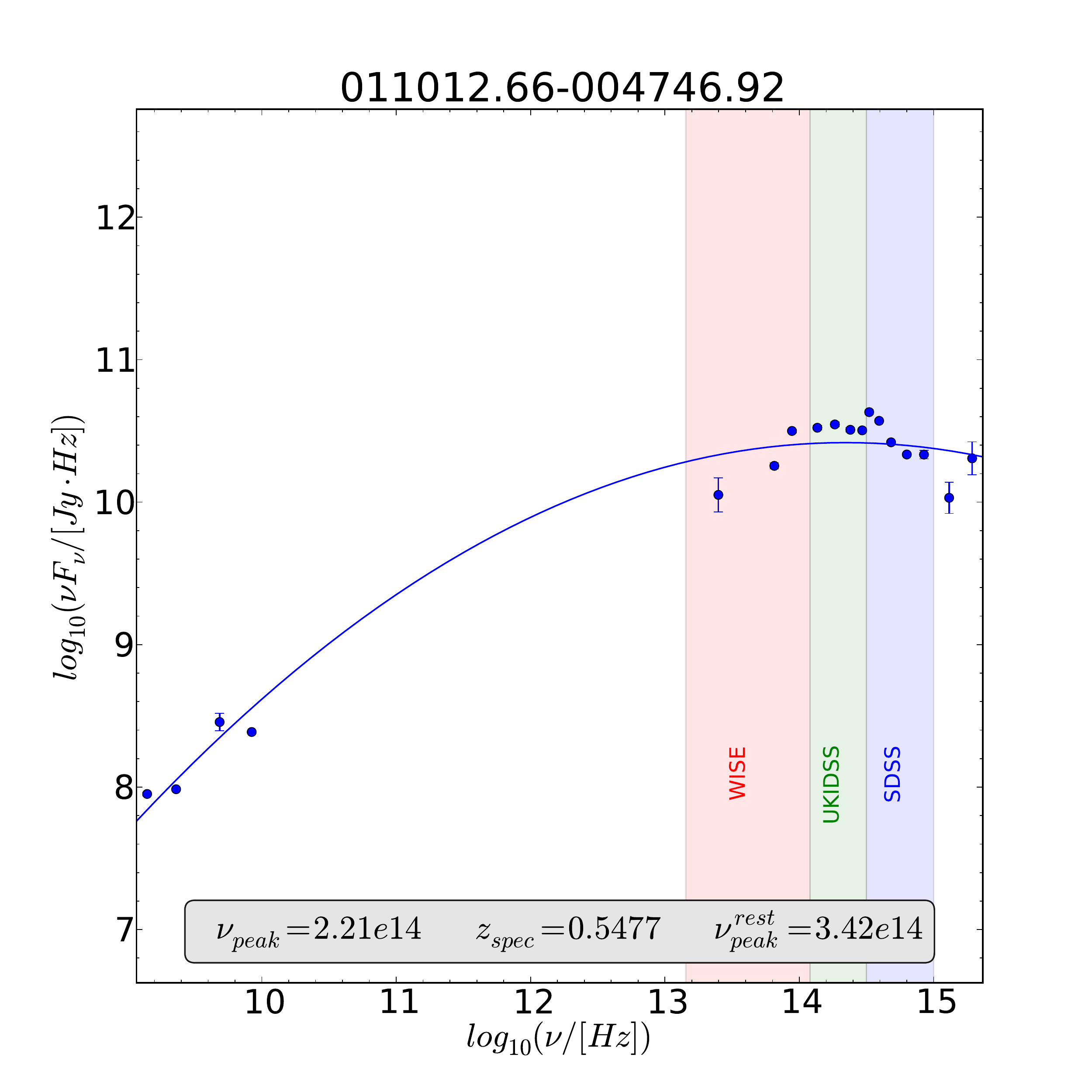}
\includegraphics[width=0.3\textwidth]{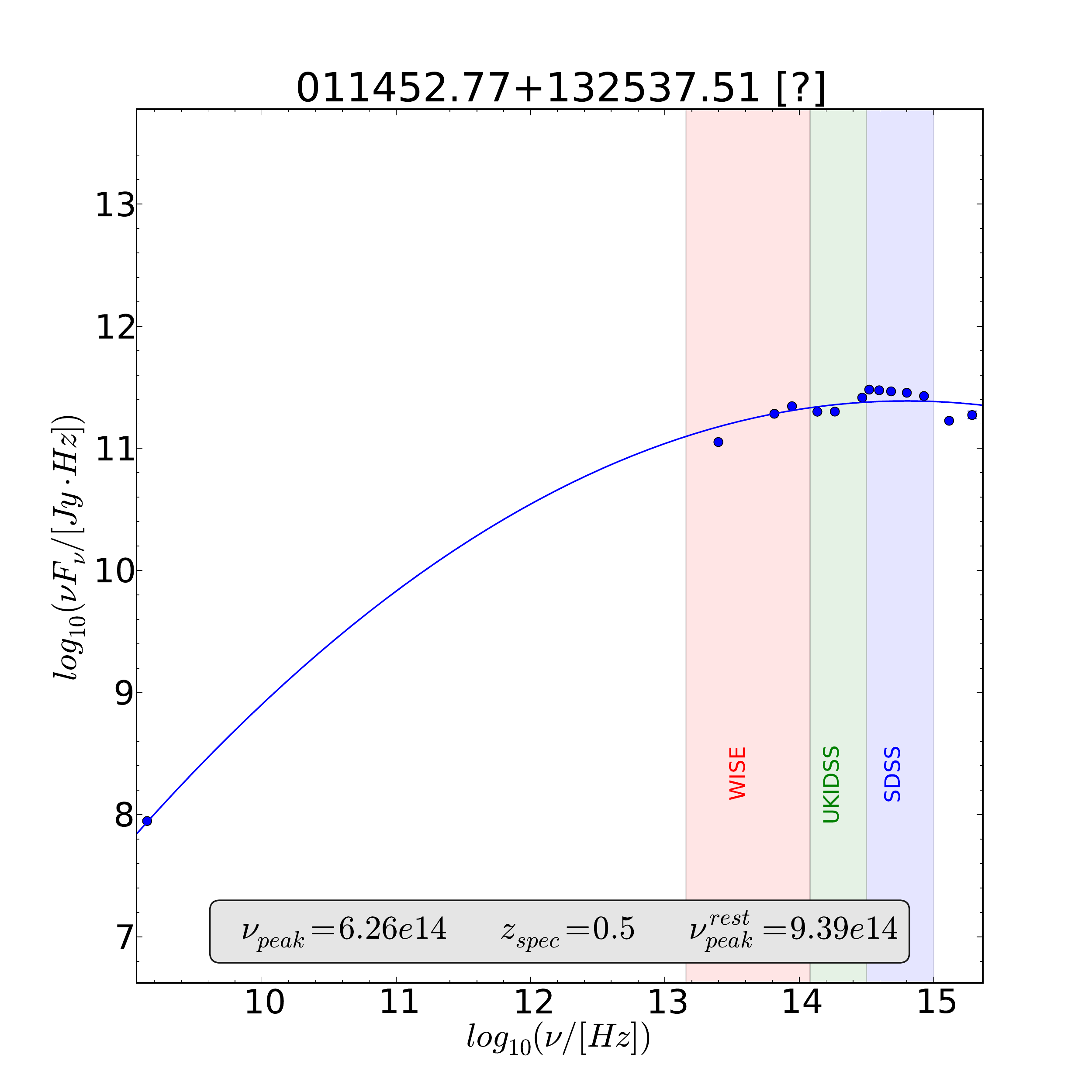}\\

\includegraphics[width=0.3\textwidth]{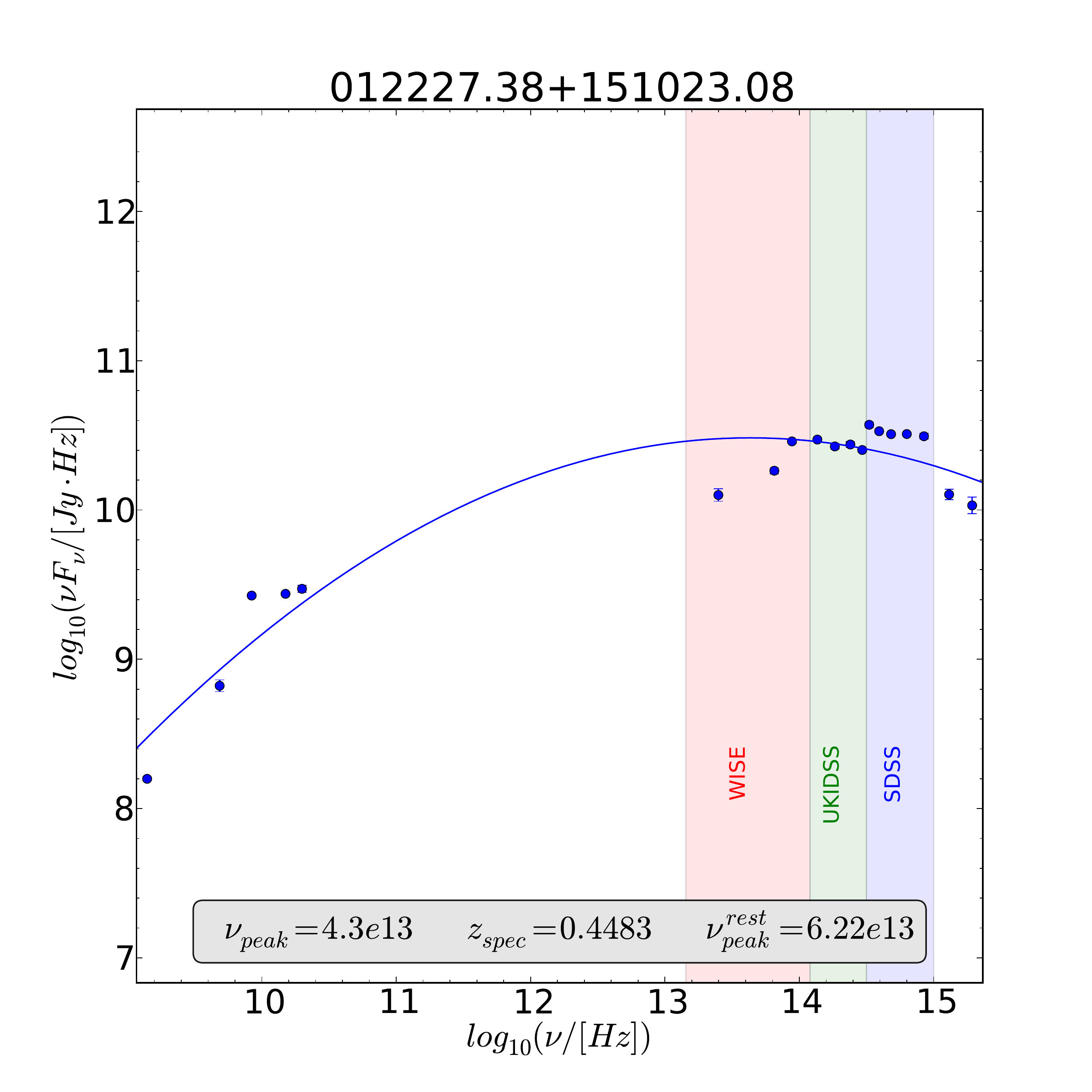}
\includegraphics[width=0.3\textwidth]{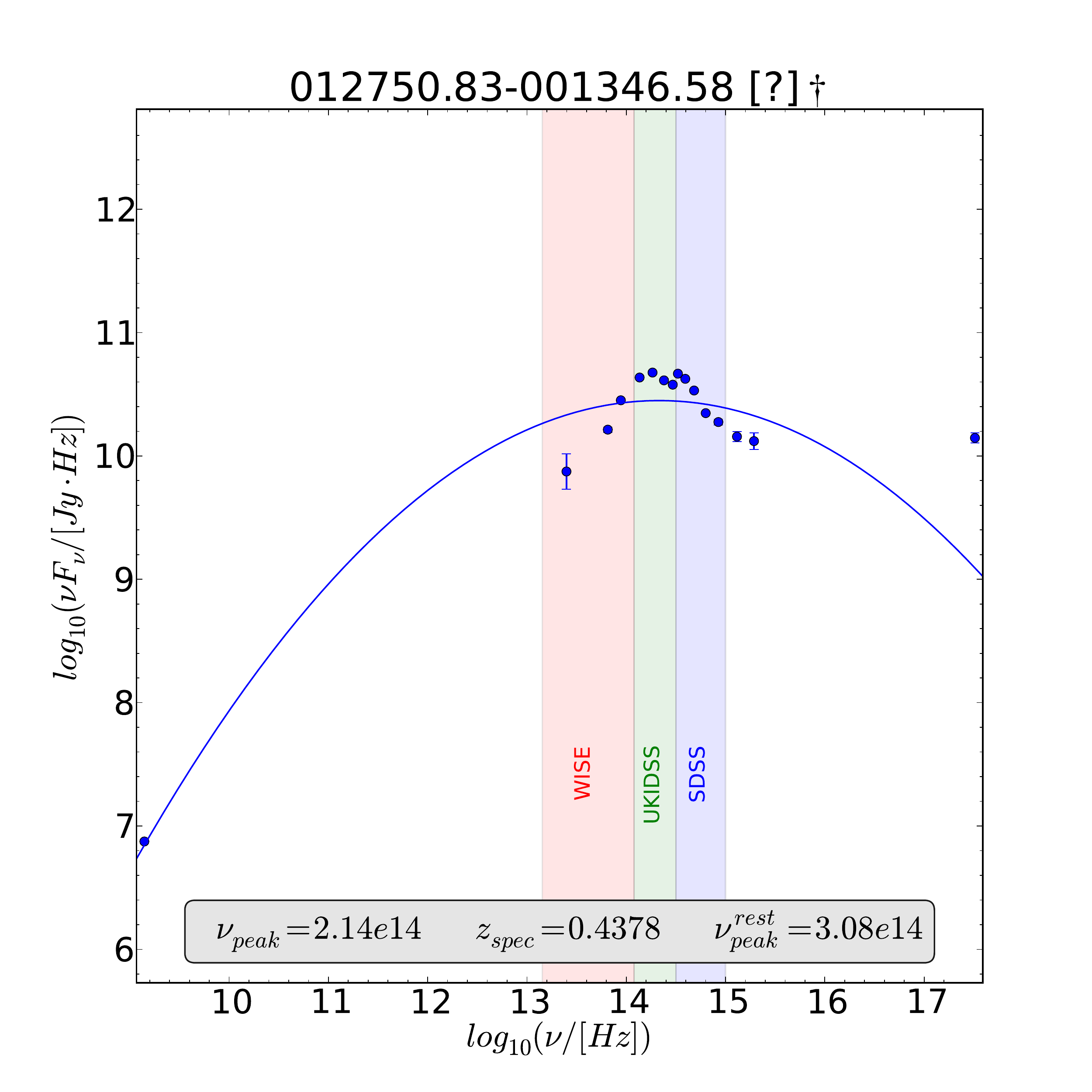}
\includegraphics[width=0.3\textwidth]{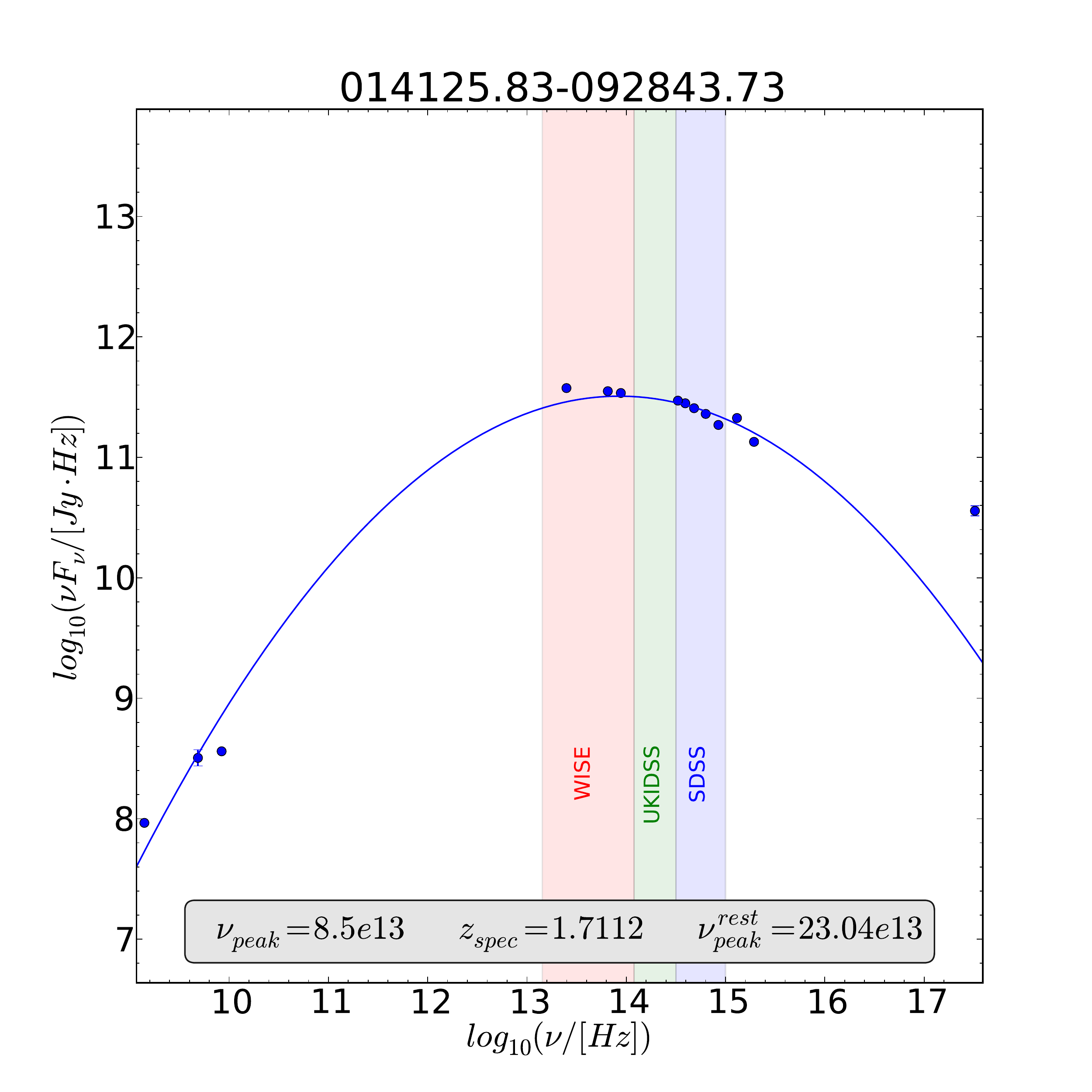}\\

\end{figure*}
\setcounter{figure}{0}
\begin{figure*}[htb!]
\caption{--Continued.}

\includegraphics[width=0.3\textwidth]{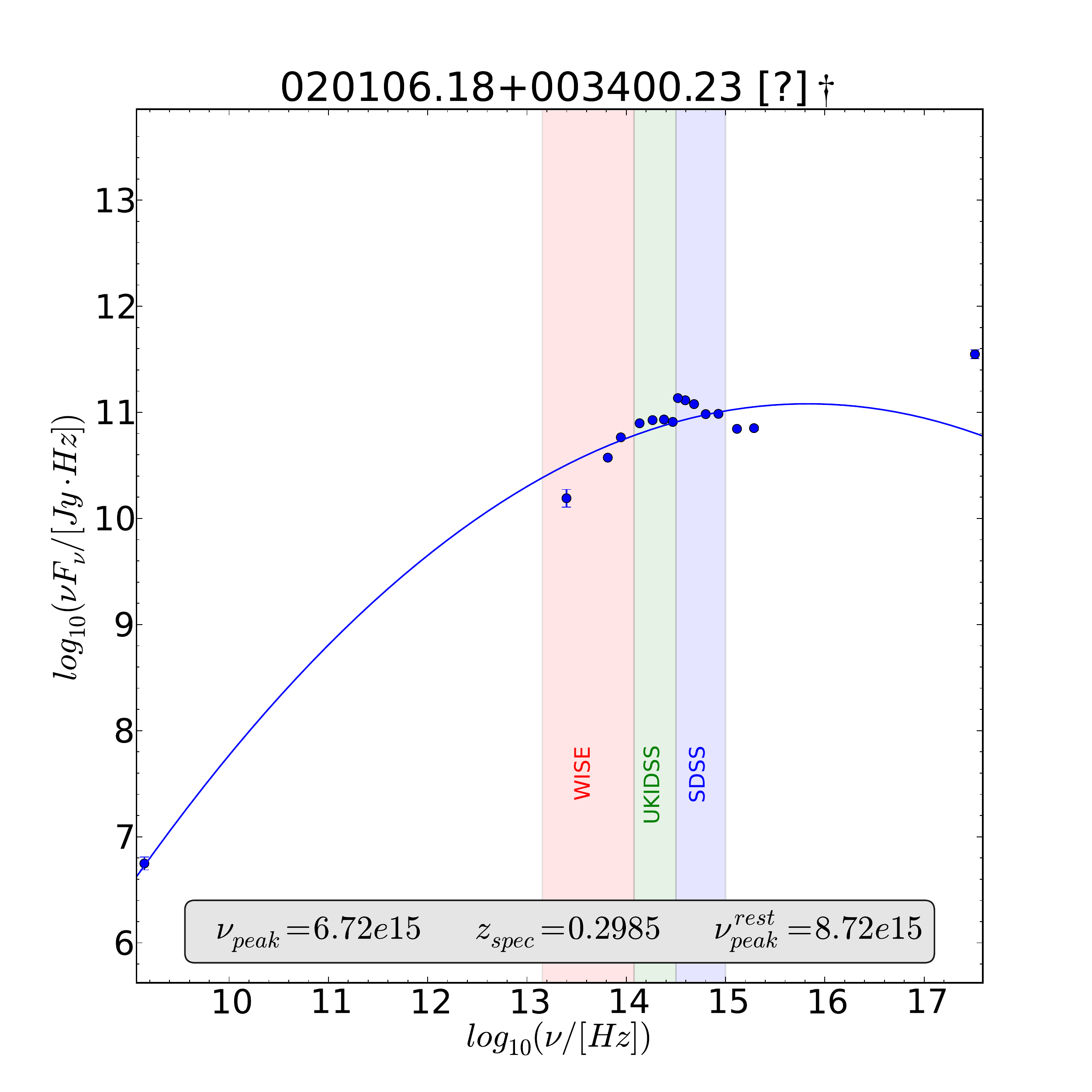}
\includegraphics[width=0.3\textwidth]{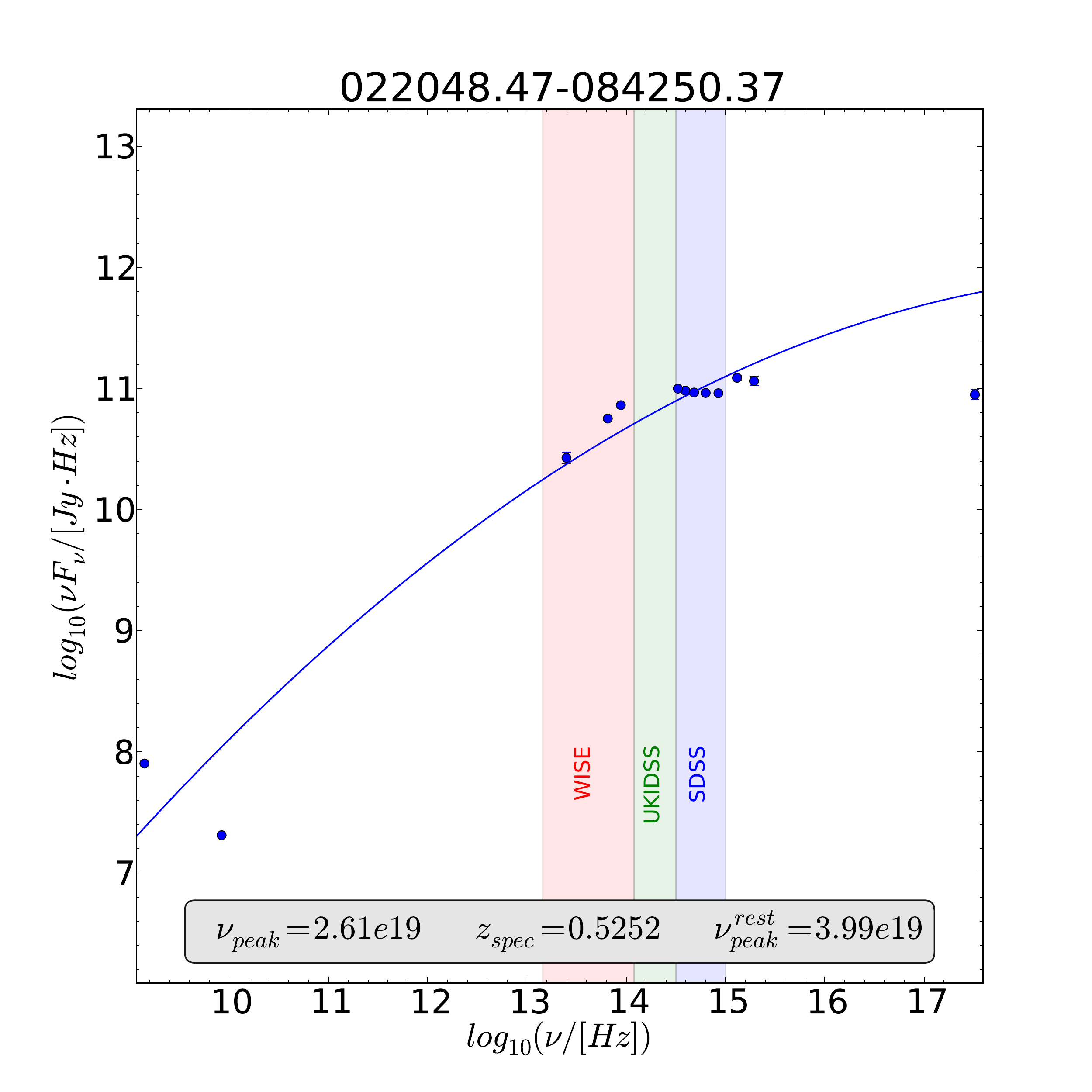}
\includegraphics[width=0.3\textwidth]{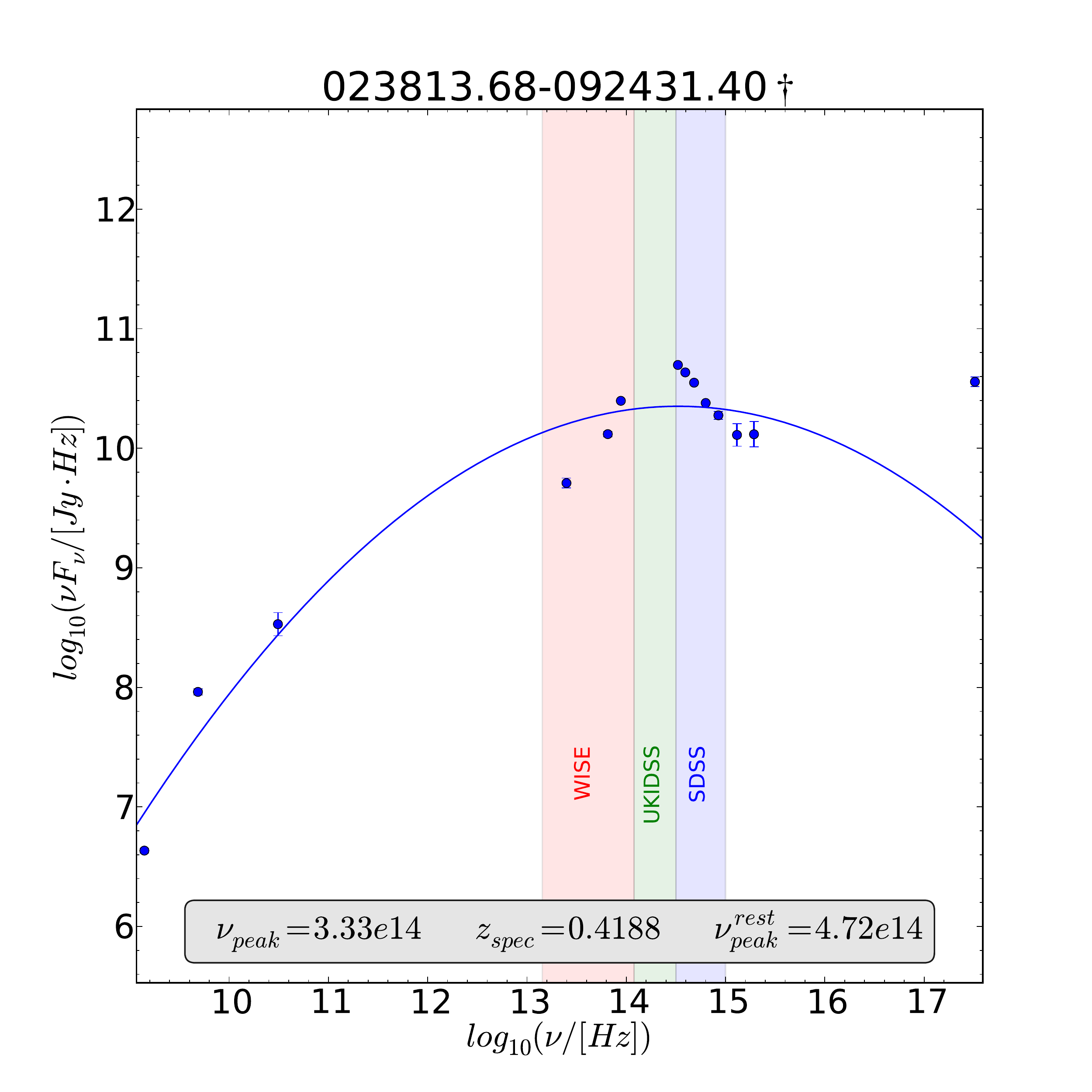}\\

\includegraphics[width=0.3\textwidth]{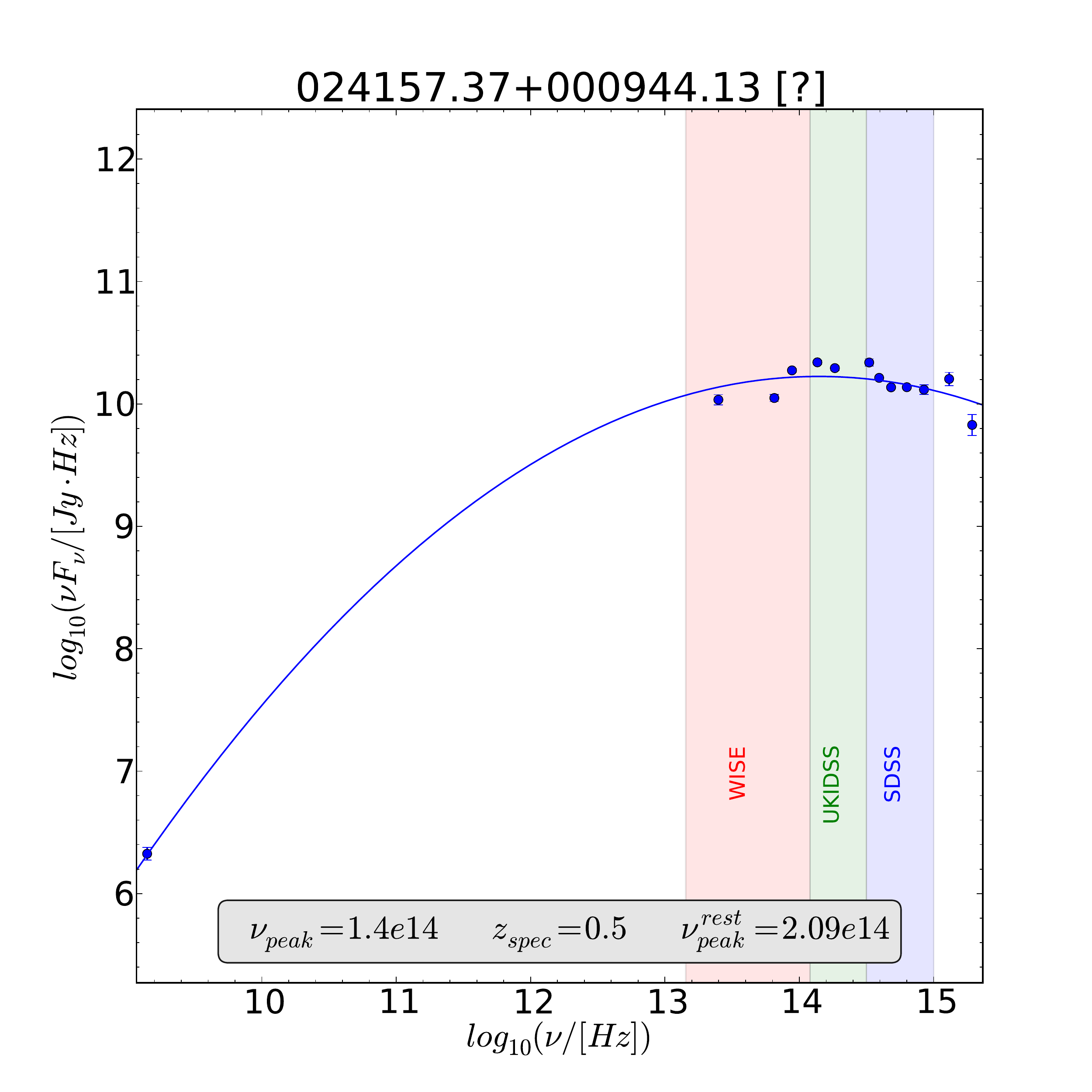}
\includegraphics[width=0.3\textwidth]{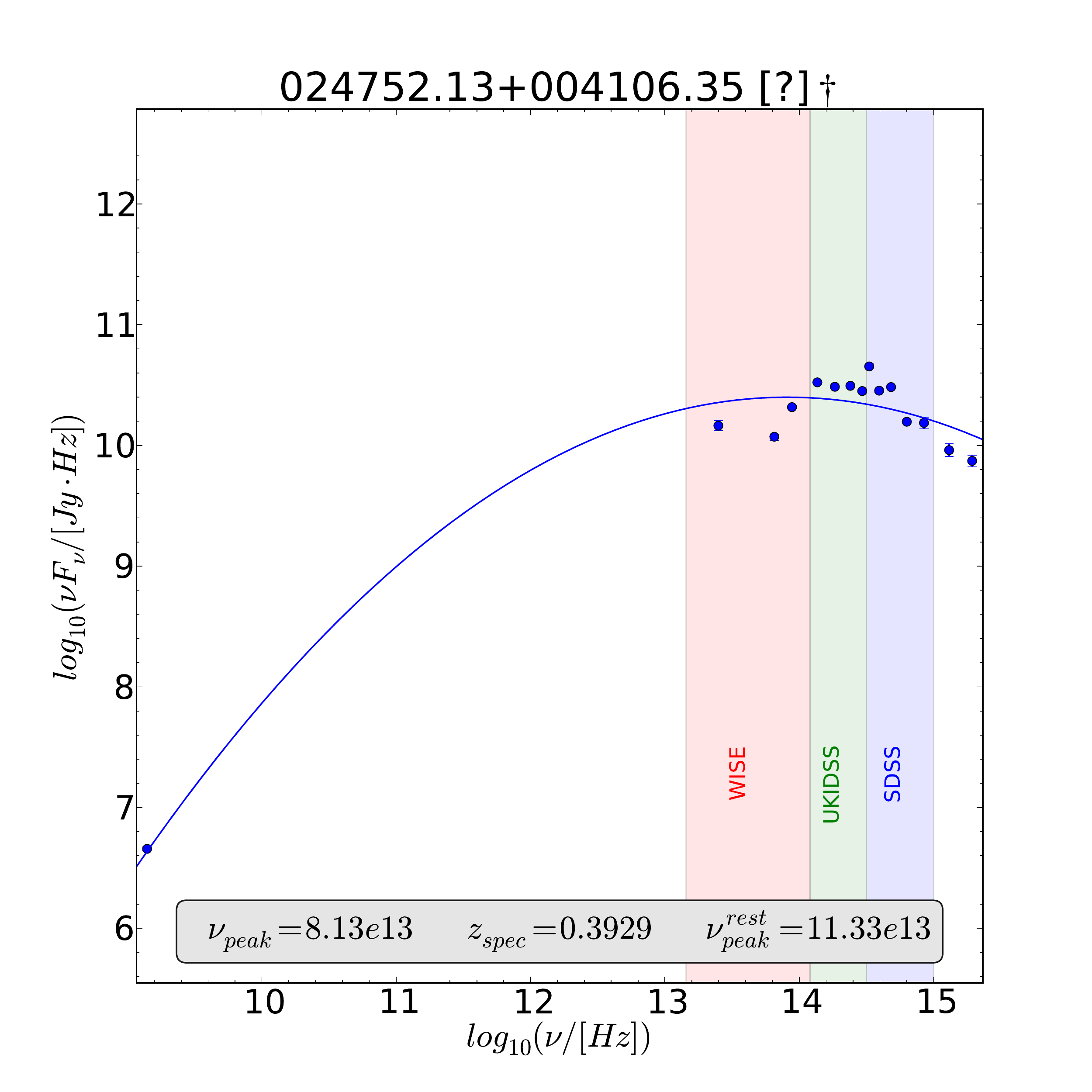}
\includegraphics[width=0.3\textwidth]{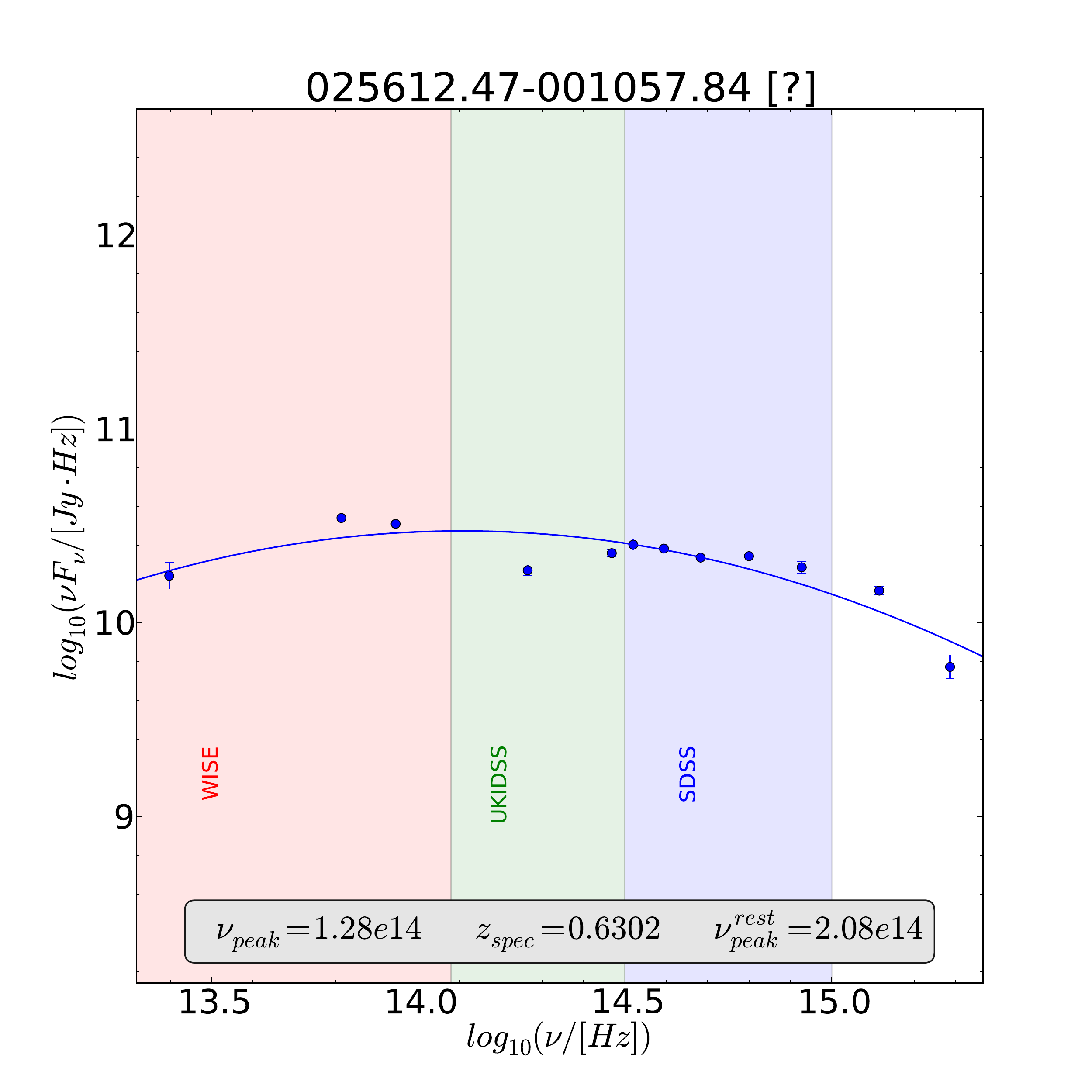}\\

\includegraphics[width=0.3\textwidth]{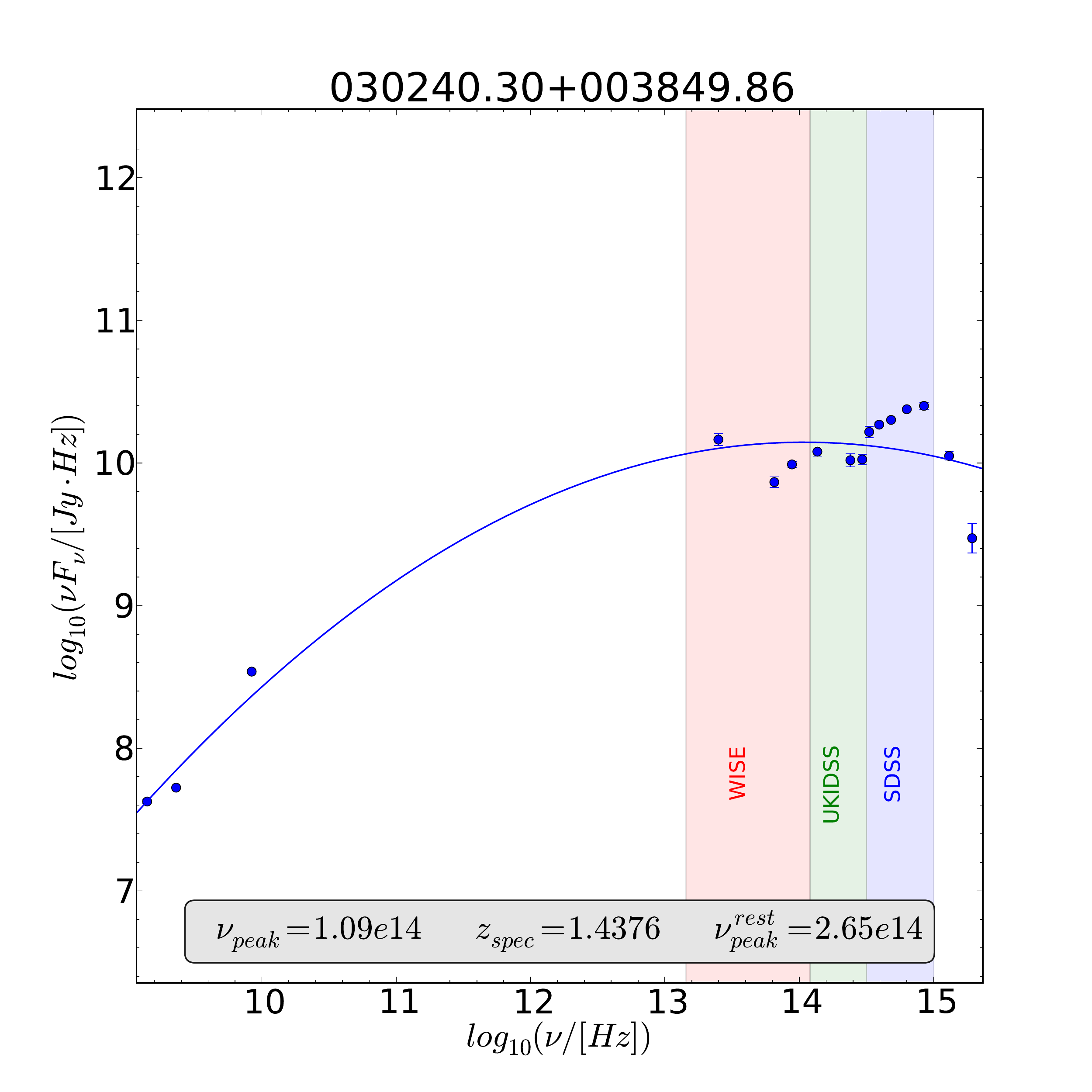}
\includegraphics[width=0.3\textwidth]{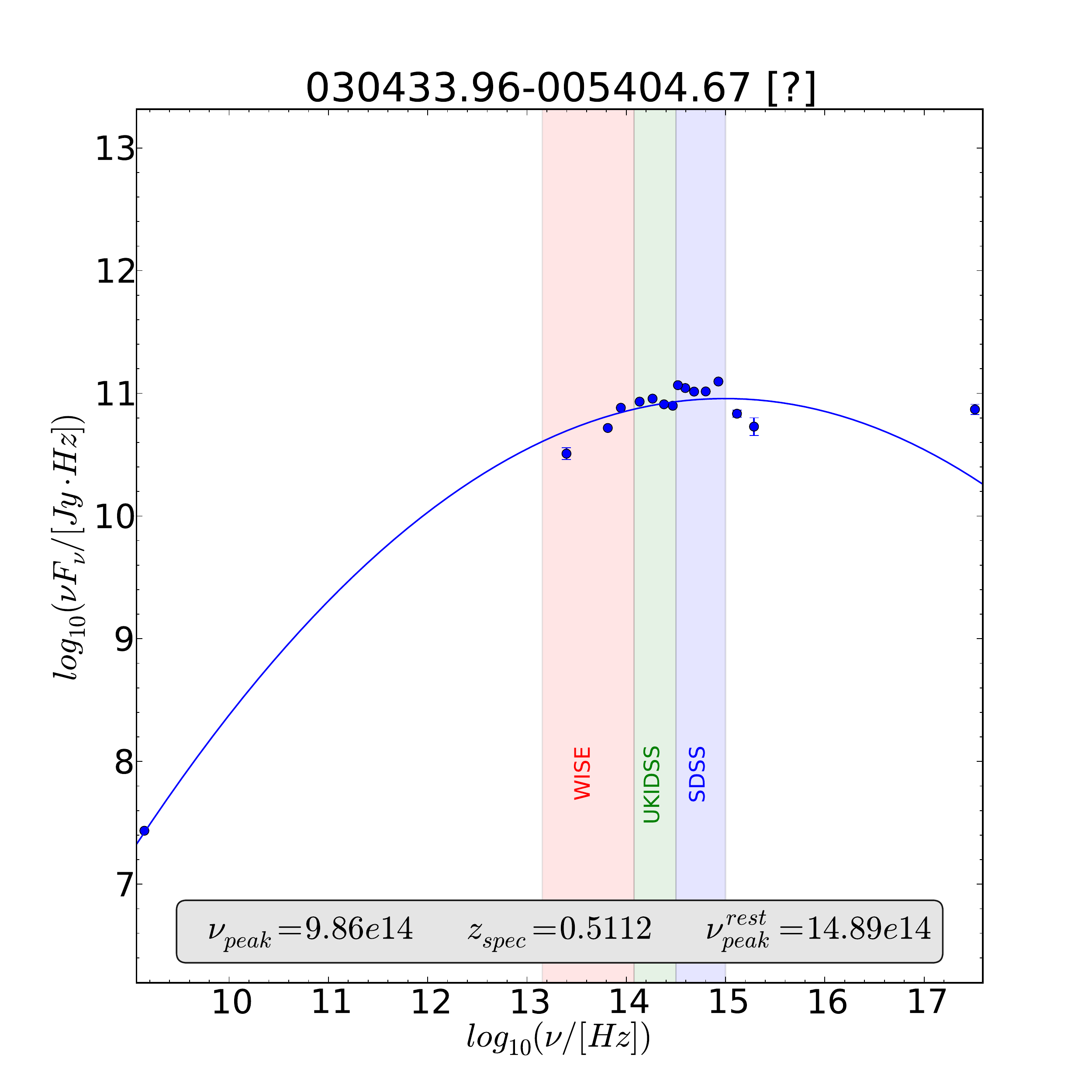}
\includegraphics[width=0.3\textwidth]{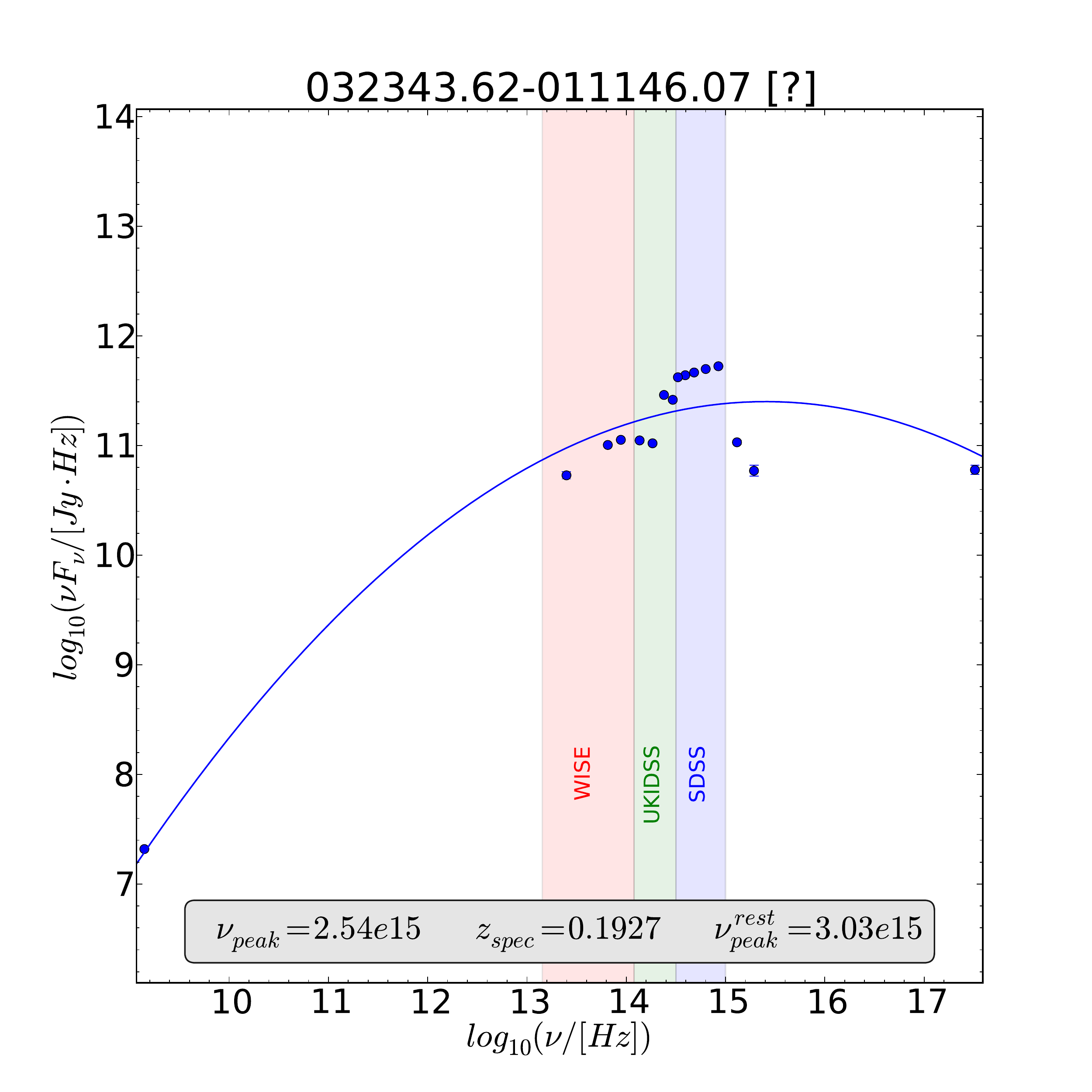}\\

\includegraphics[width=0.3\textwidth]{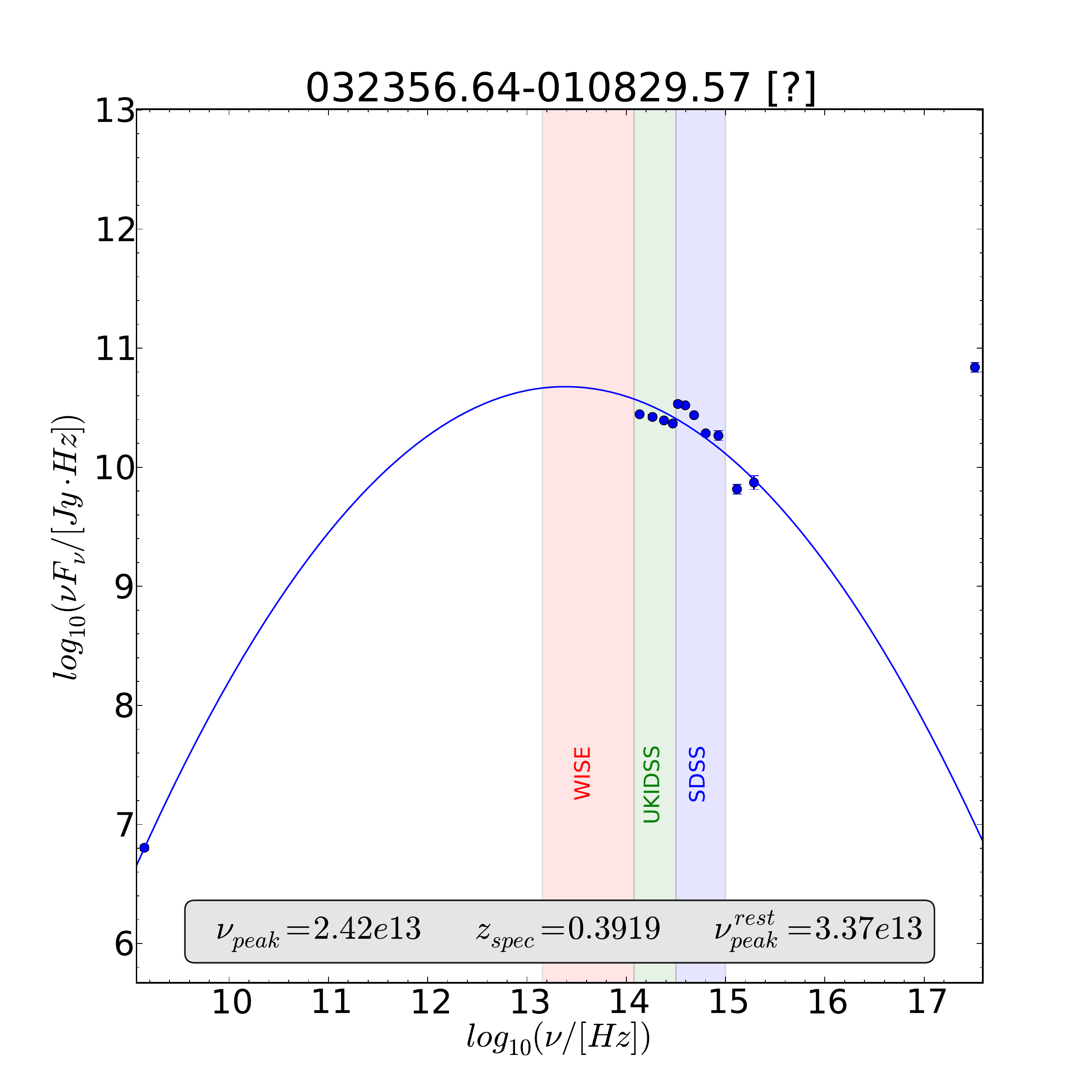}
\includegraphics[width=0.3\textwidth]{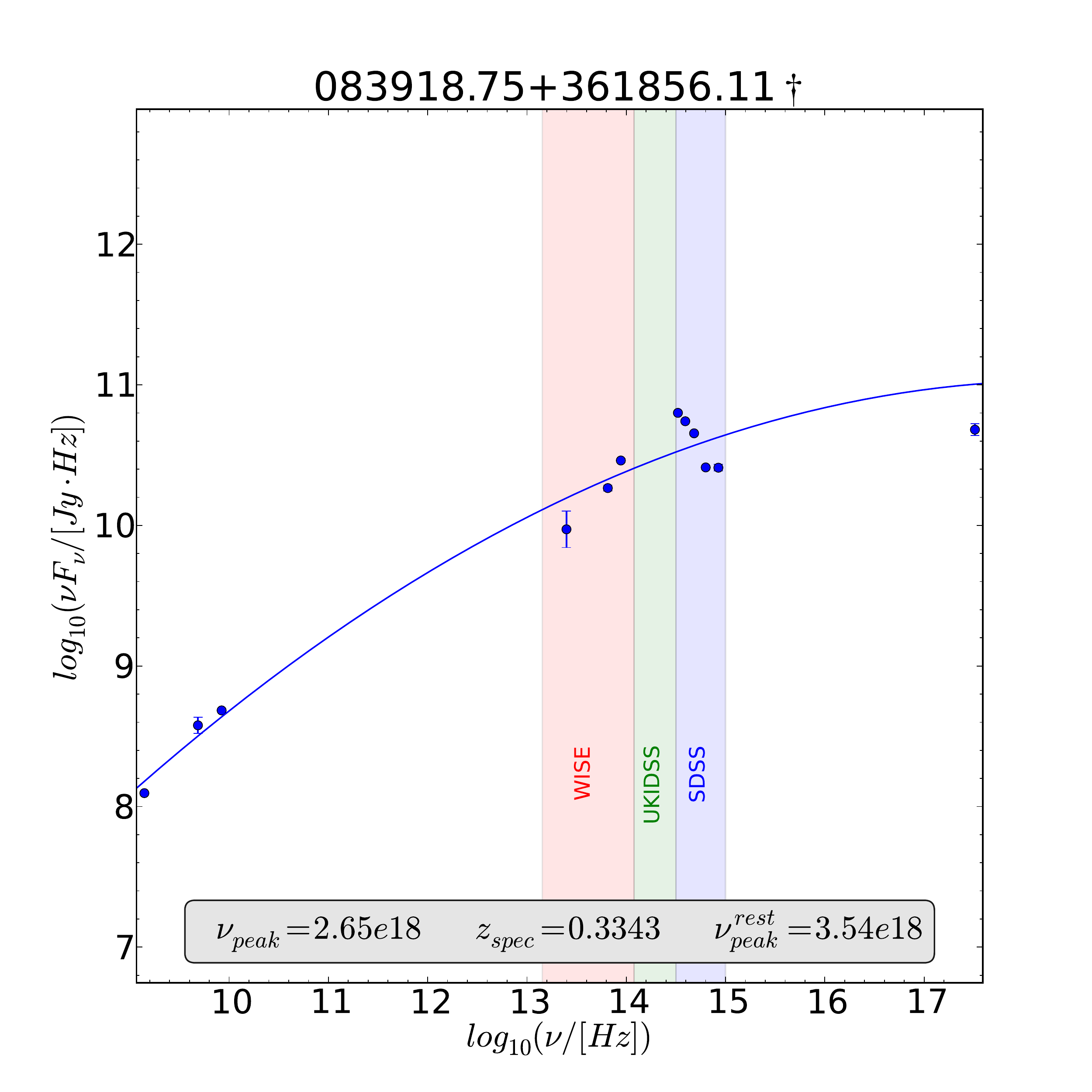}
\includegraphics[width=0.3\textwidth]{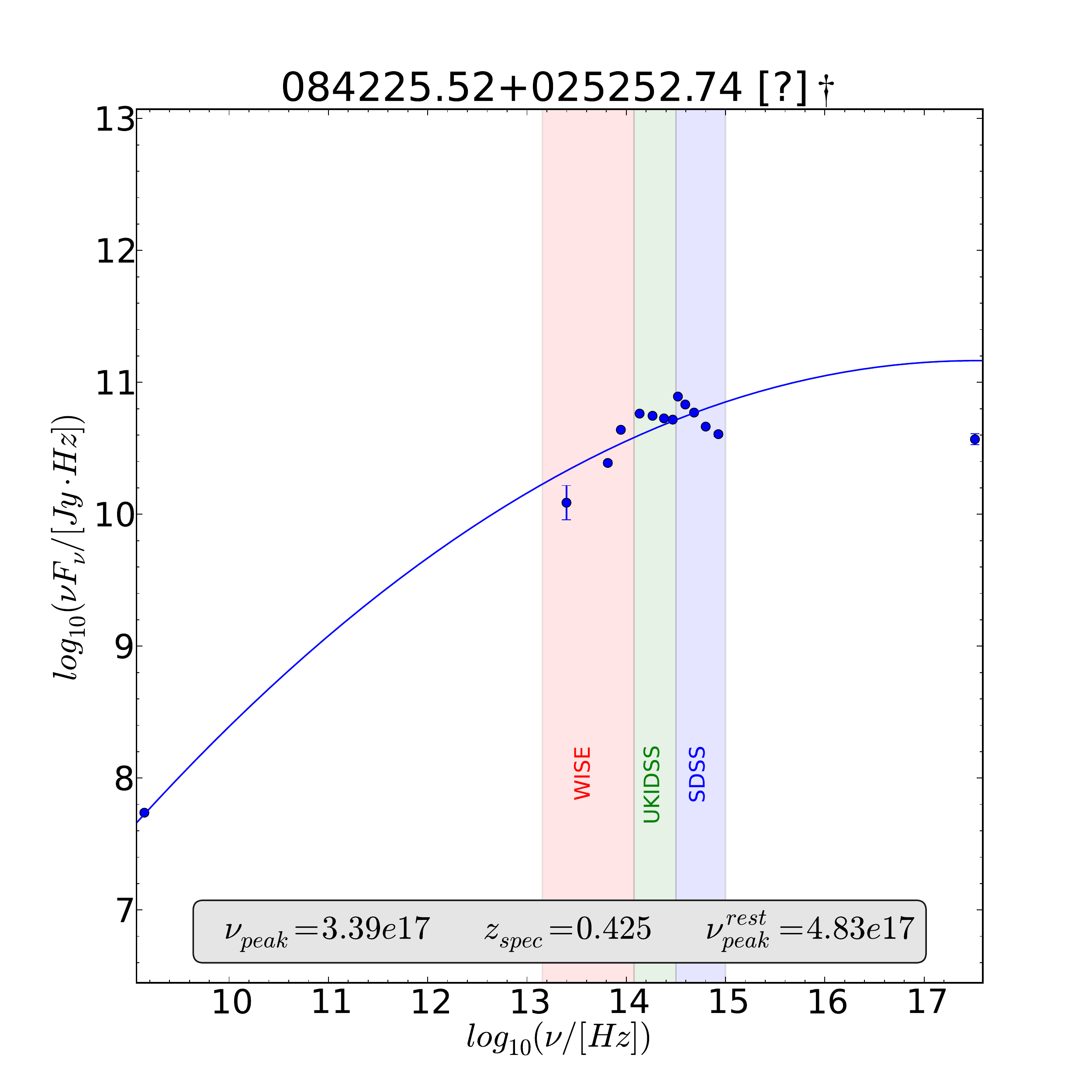}\\

\end{figure*}
\setcounter{figure}{0}
\begin{figure*}[htb!]
\caption{--Continued.}

\includegraphics[width=0.3\textwidth]{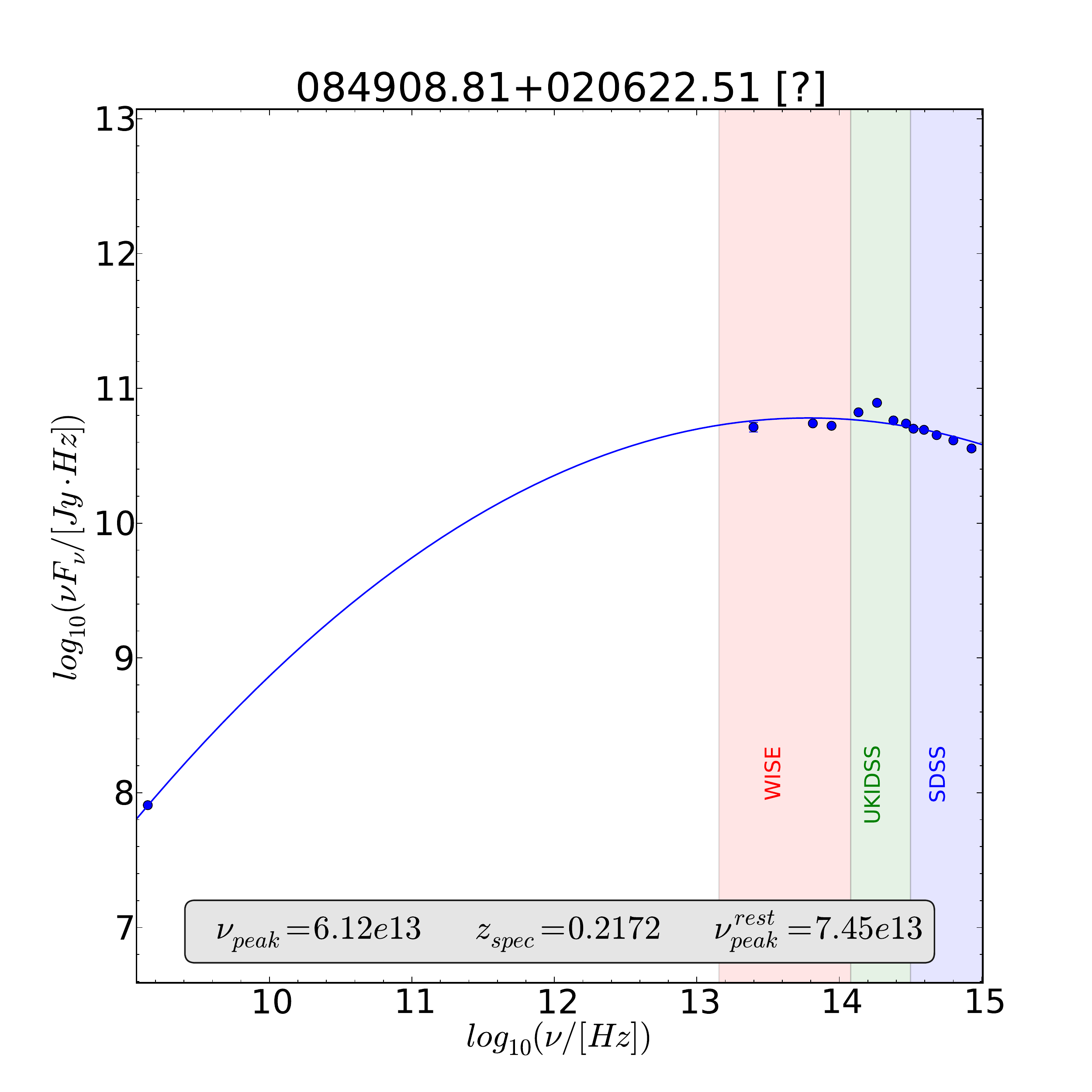}
\includegraphics[width=0.3\textwidth]{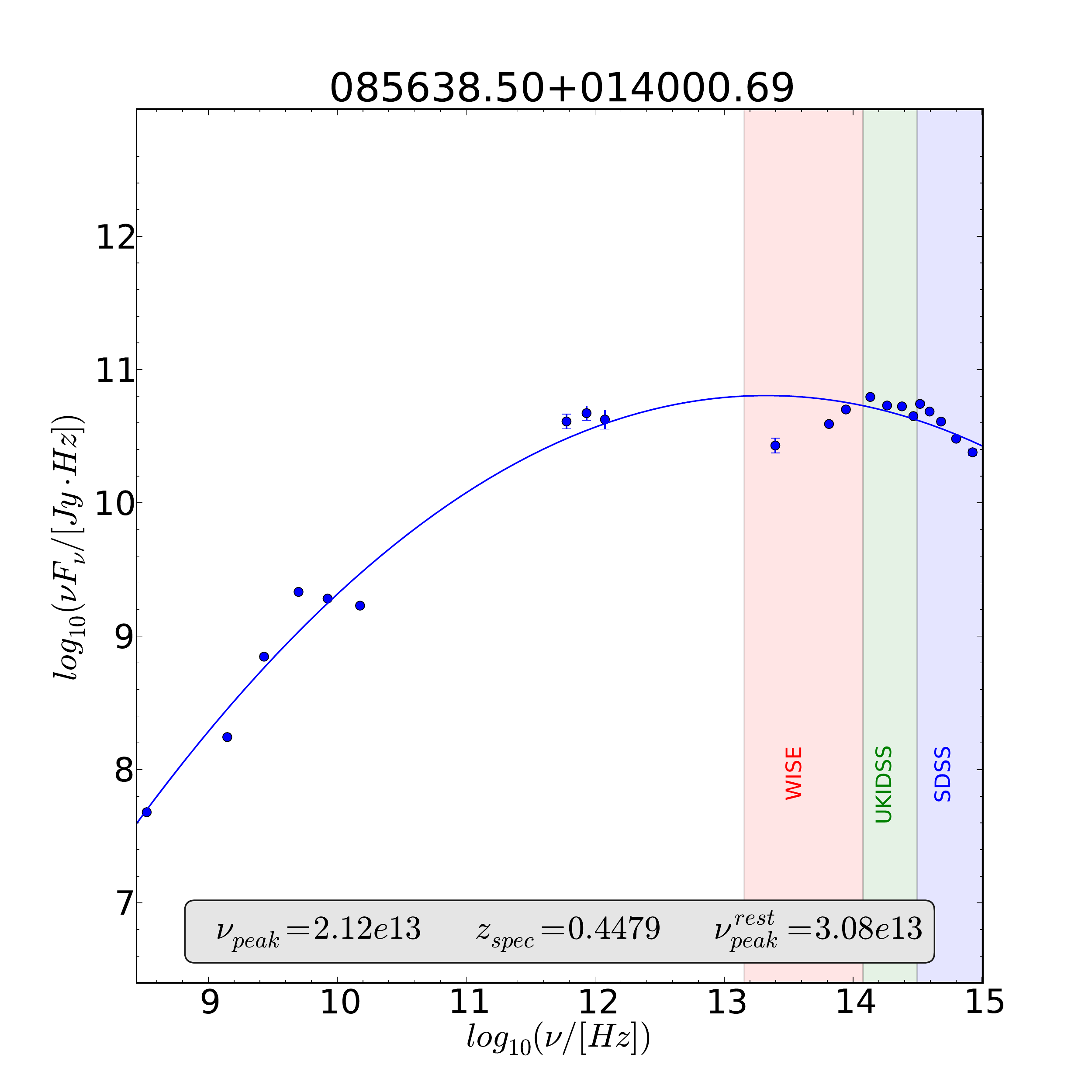}
\includegraphics[width=0.3\textwidth]{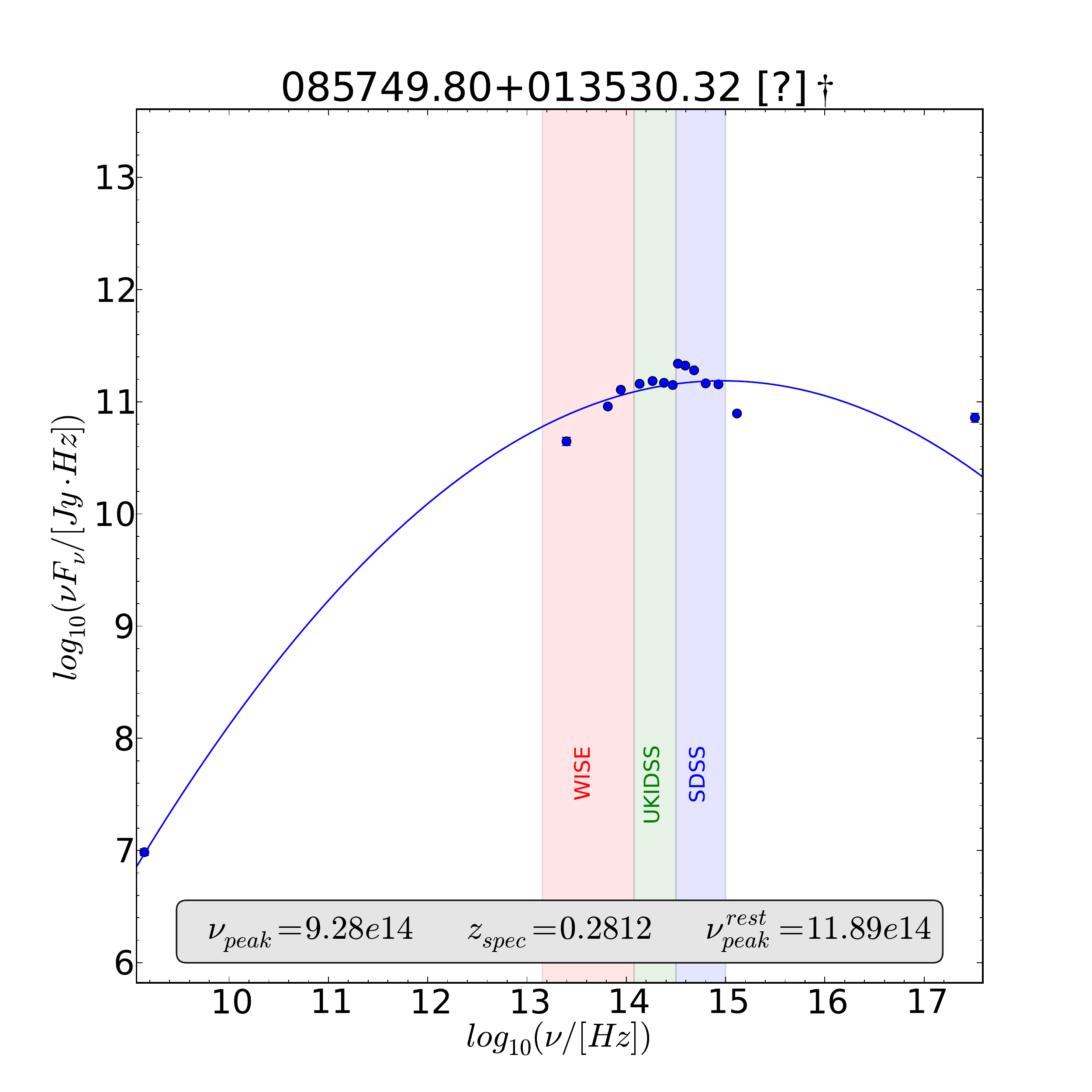}\\

\includegraphics[width=0.3\textwidth]{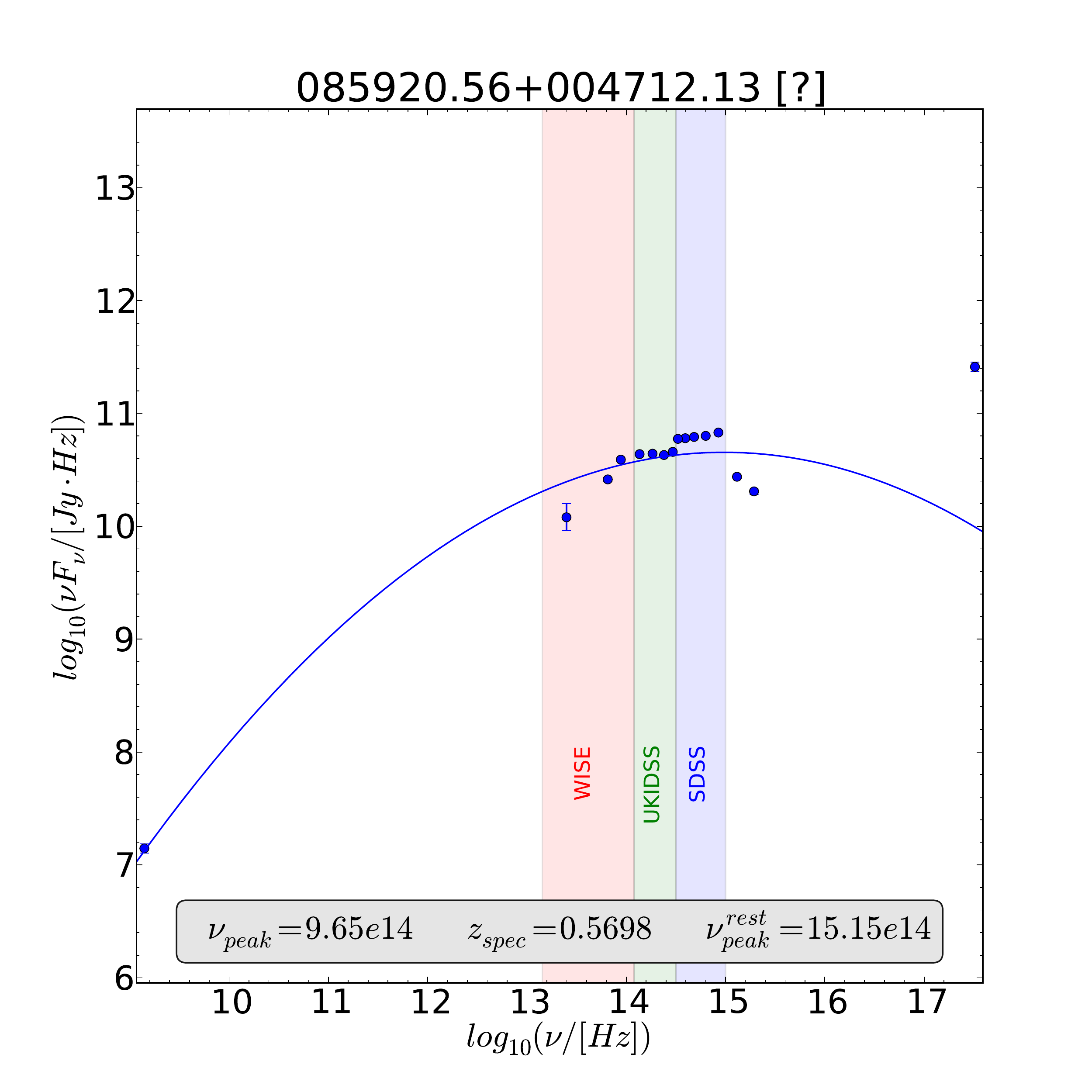}
\includegraphics[width=0.3\textwidth]{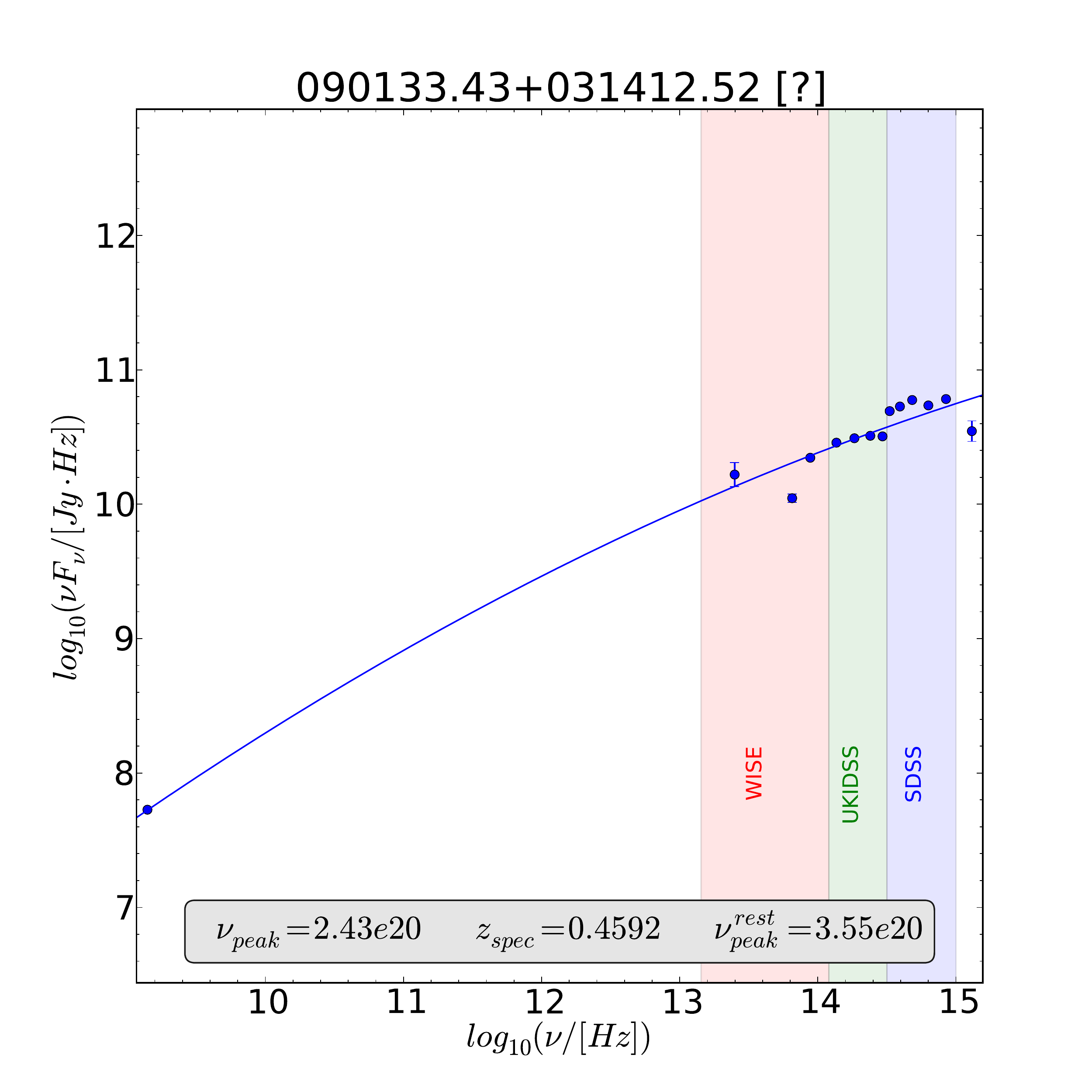}
\includegraphics[width=0.3\textwidth]{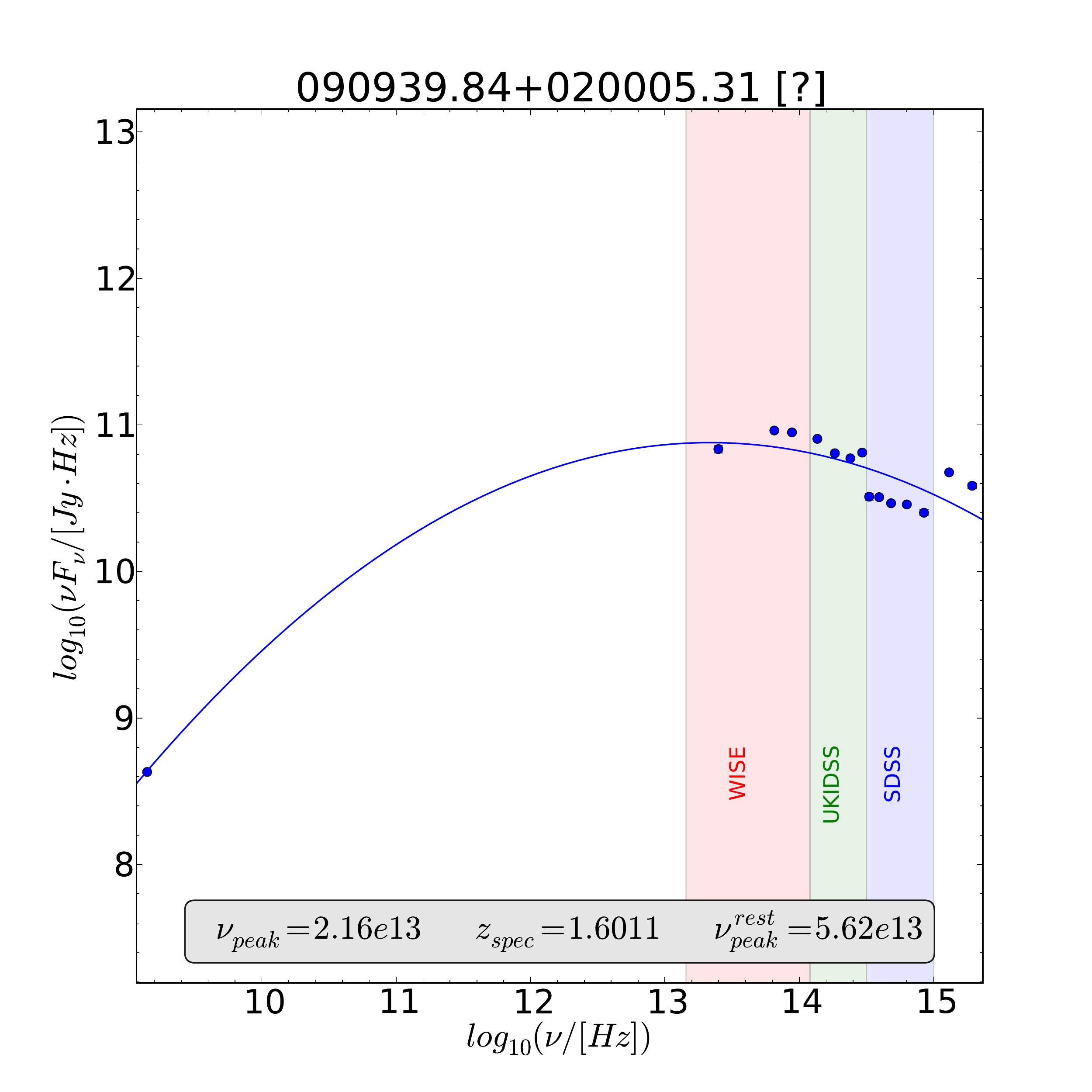}\\

\includegraphics[width=0.3\textwidth]{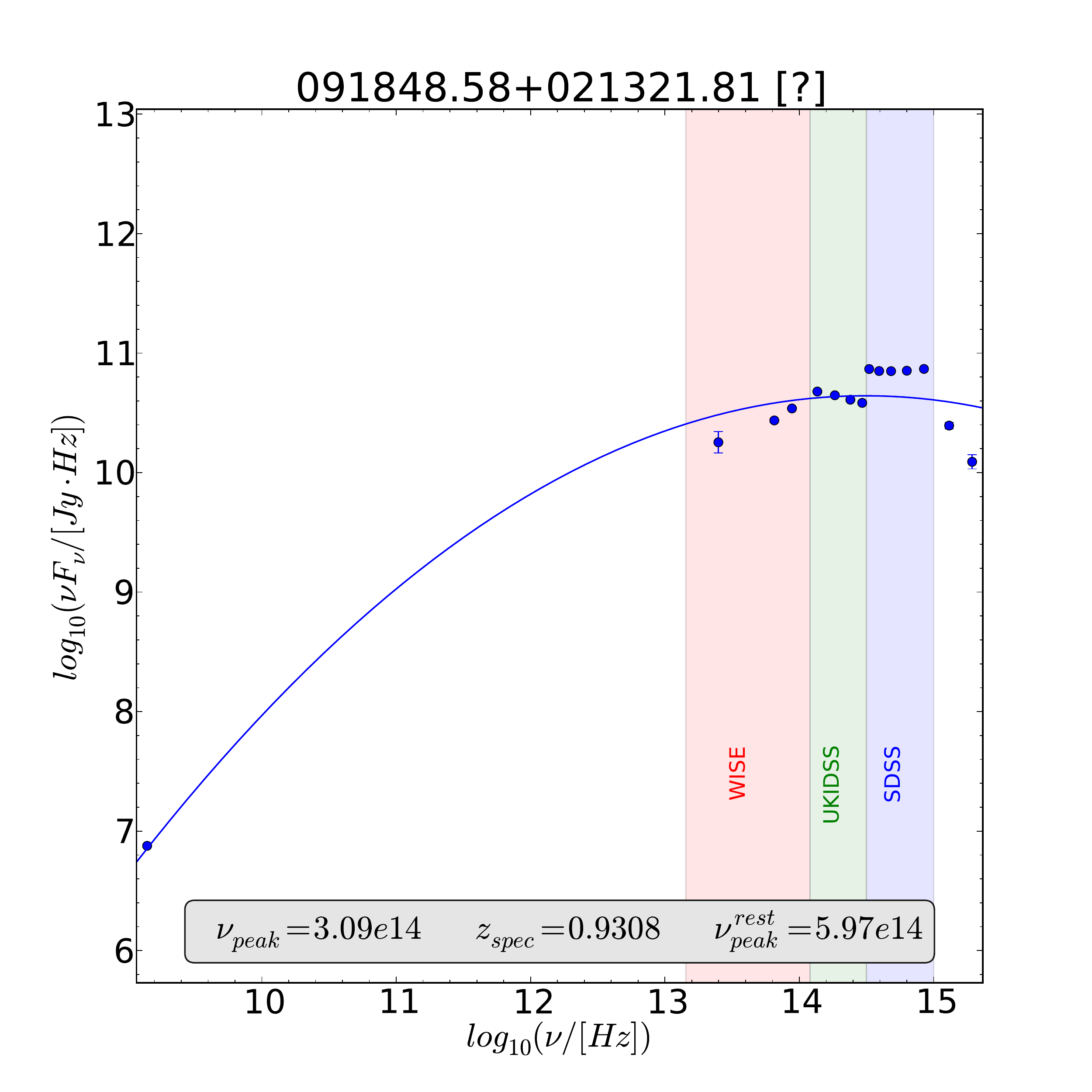}
\includegraphics[width=0.3\textwidth]{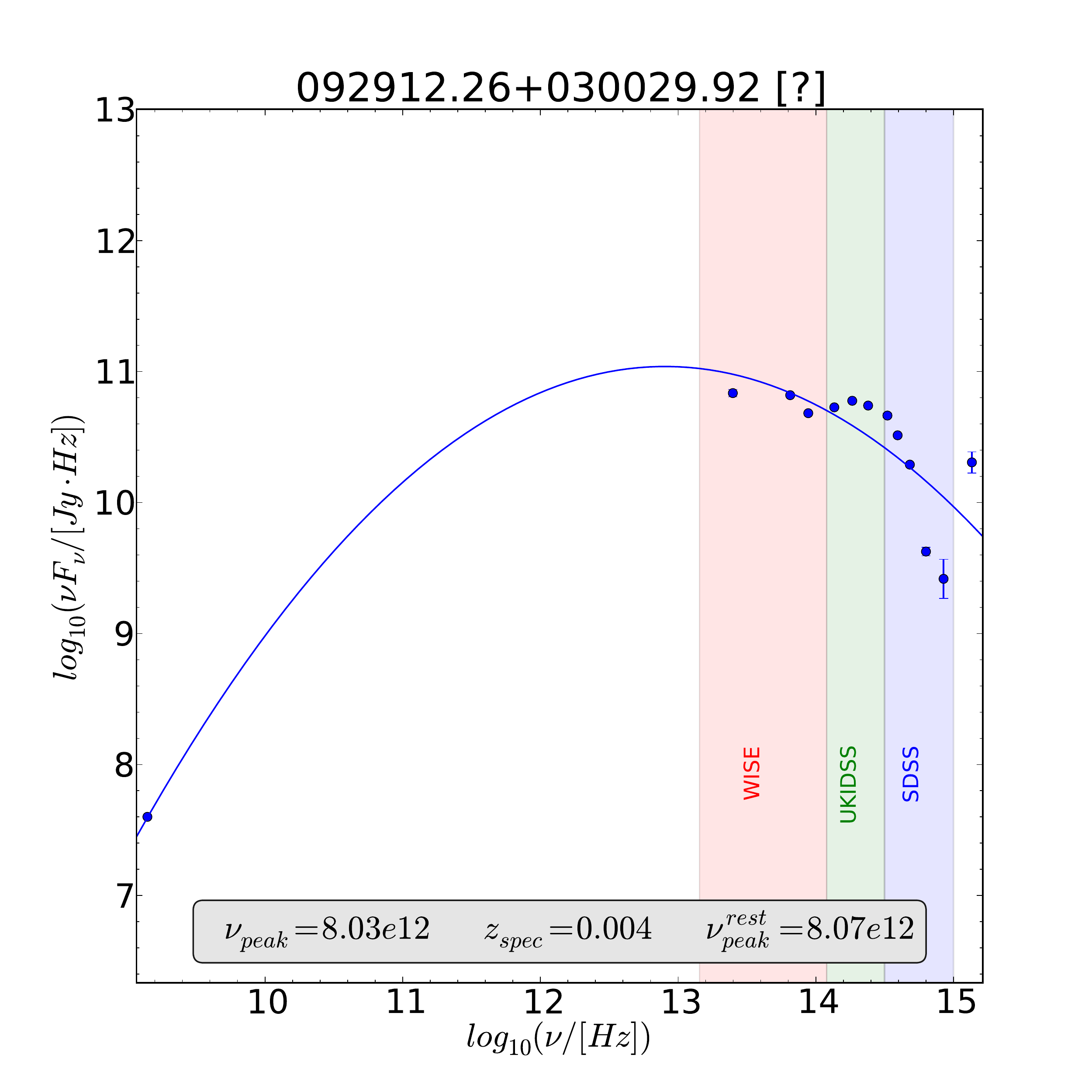}
\includegraphics[width=0.3\textwidth]{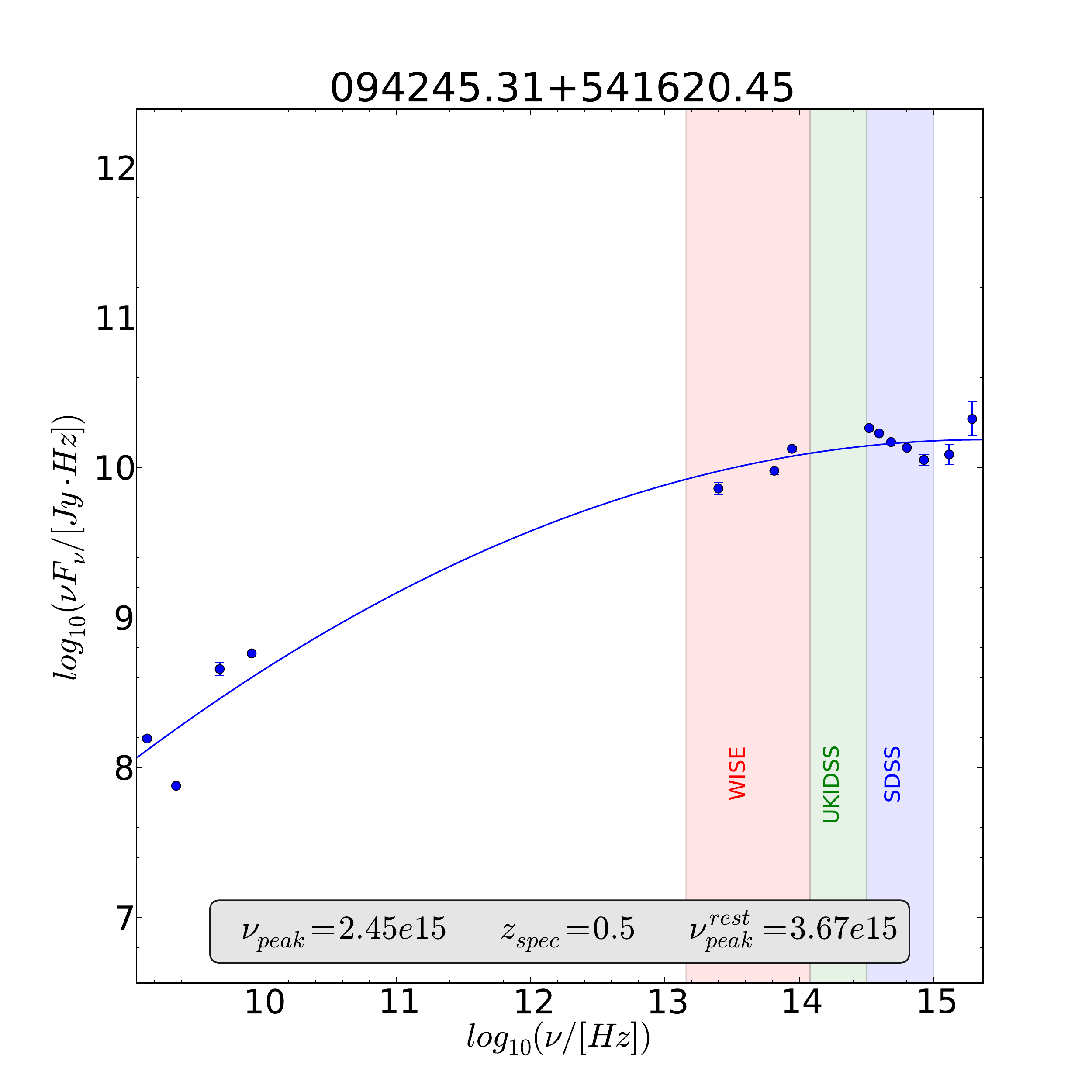}\\

\includegraphics[width=0.3\textwidth]{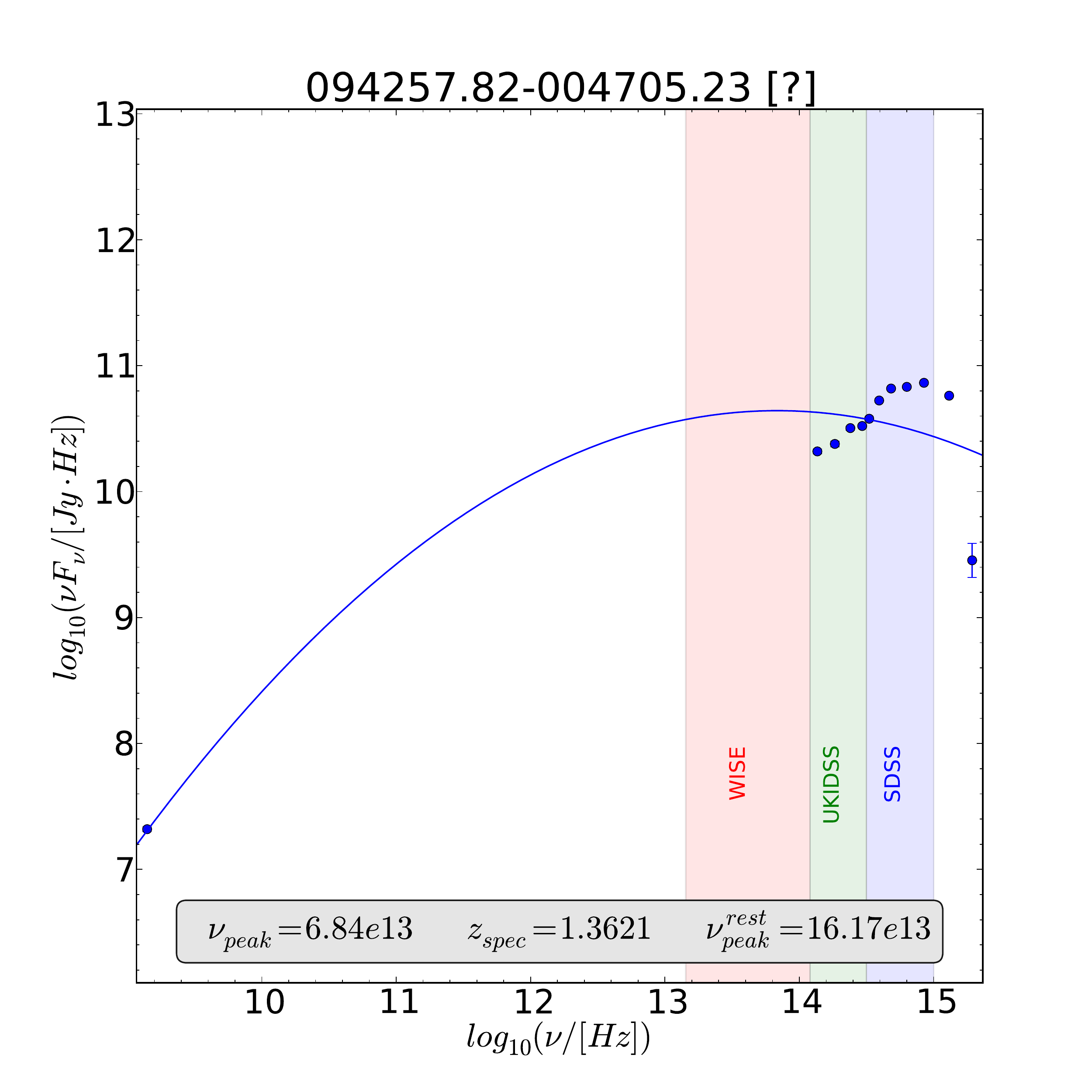}
\includegraphics[width=0.3\textwidth]{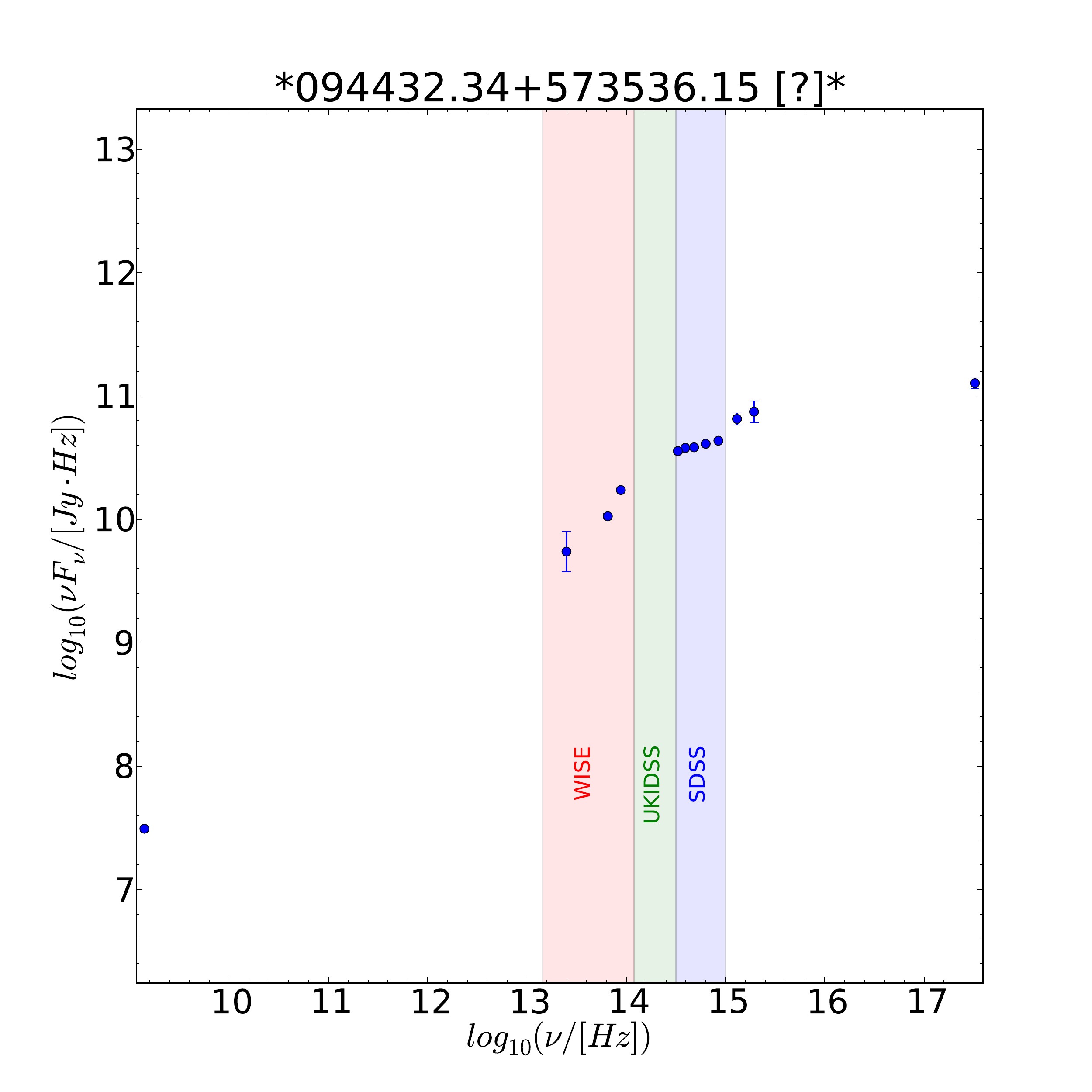}
\includegraphics[width=0.3\textwidth]{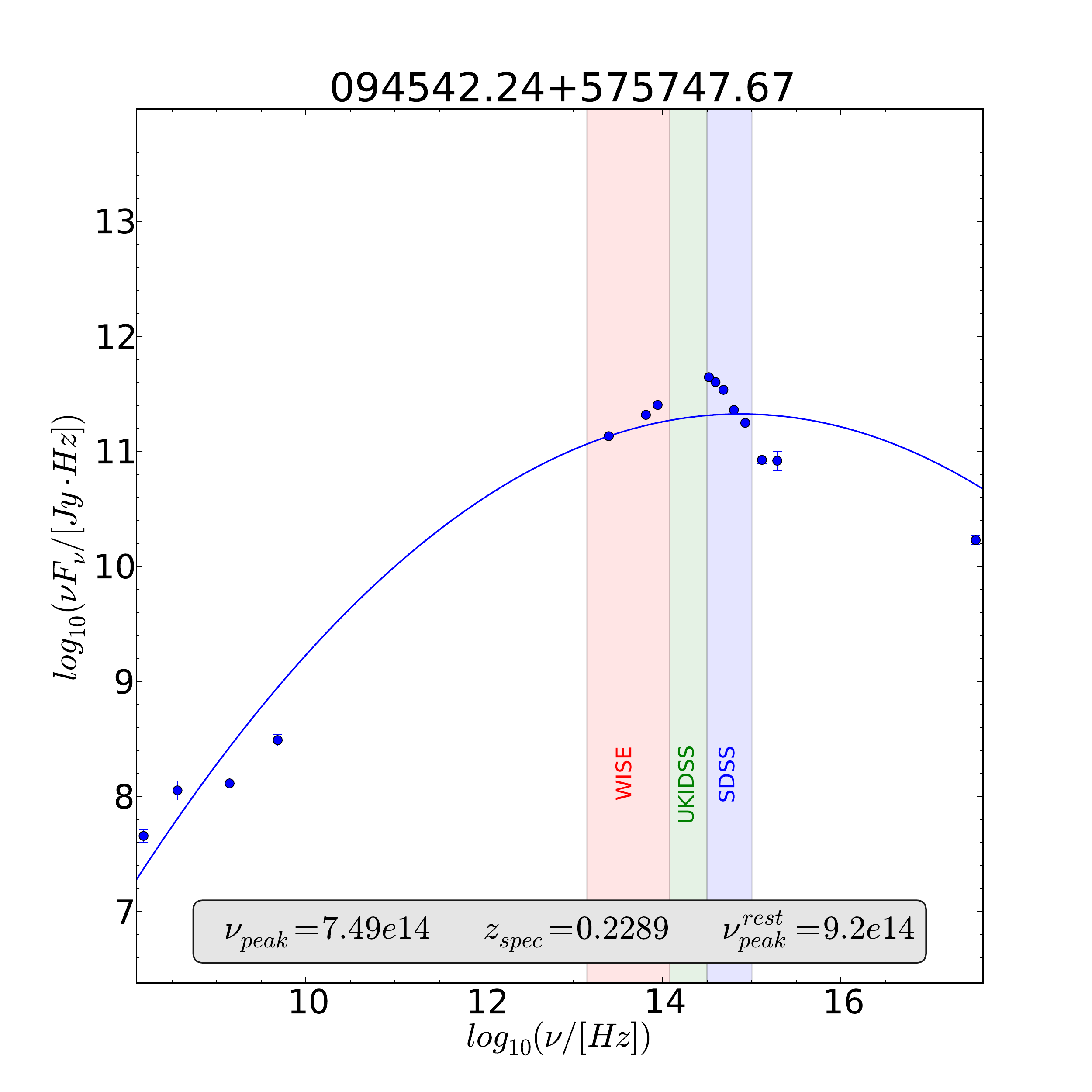}\\

\end{figure*}
\setcounter{figure}{0}
\begin{figure*}[htb!]
\caption{--Continued.}

\includegraphics[width=0.3\textwidth]{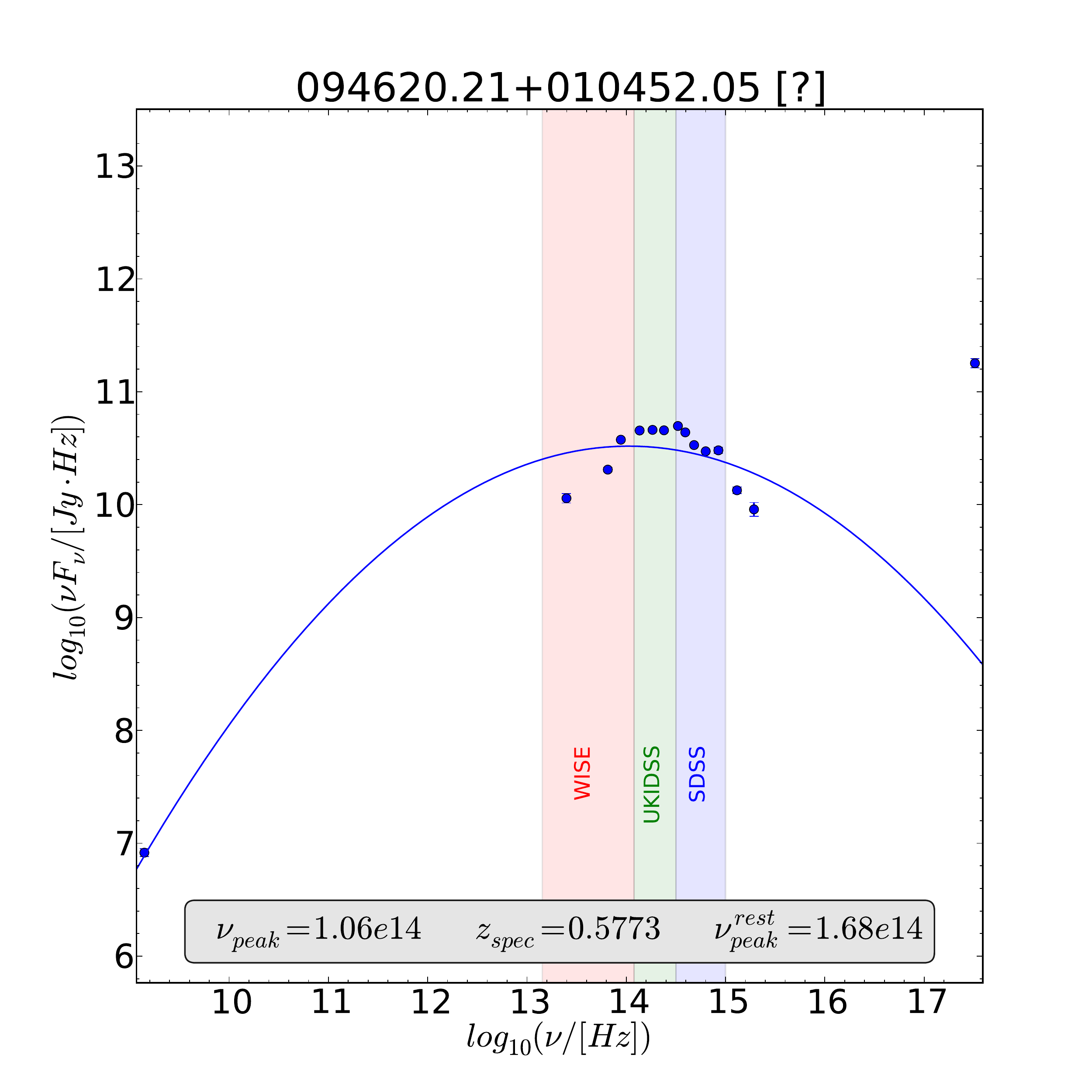}
\includegraphics[width=0.3\textwidth]{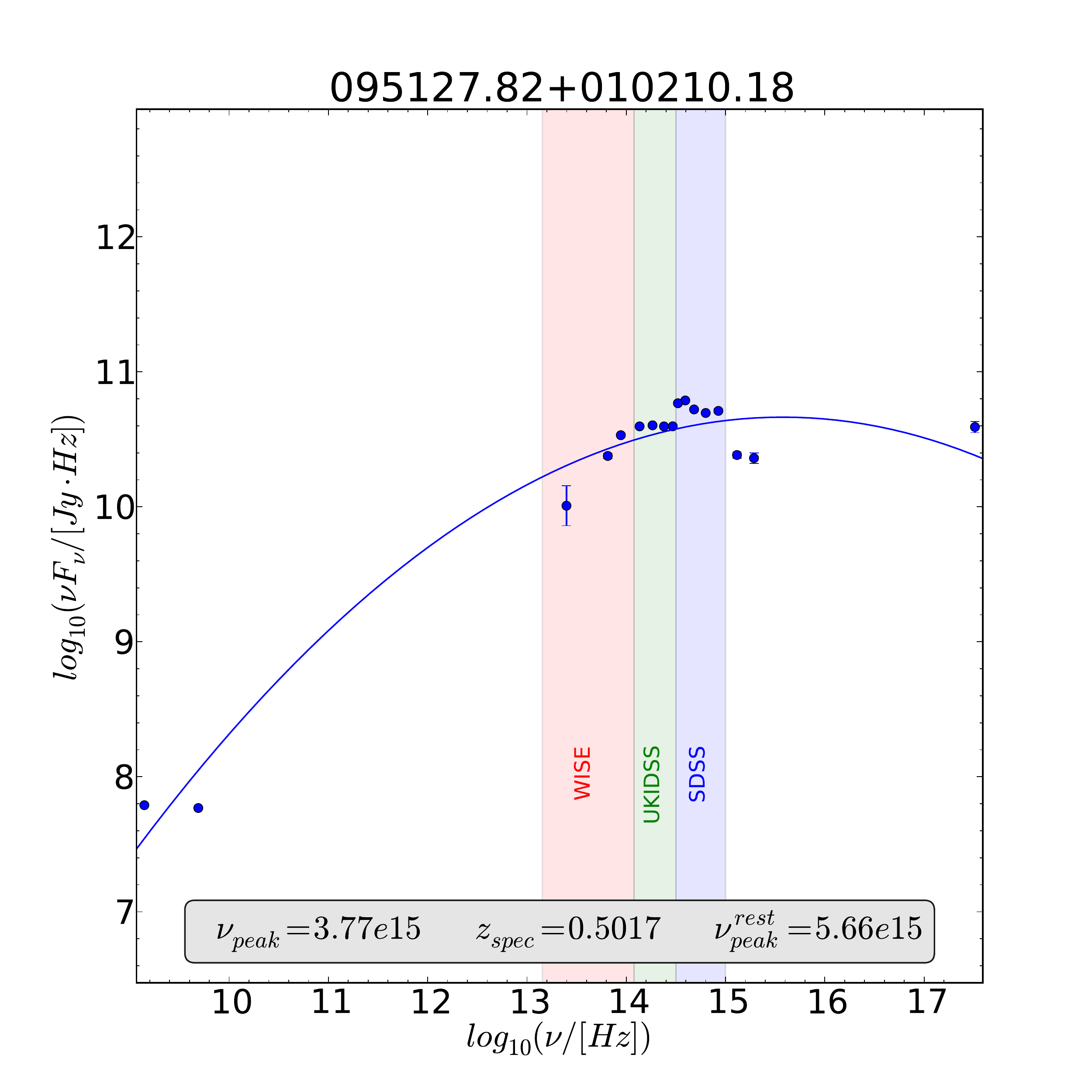}
\includegraphics[width=0.3\textwidth]{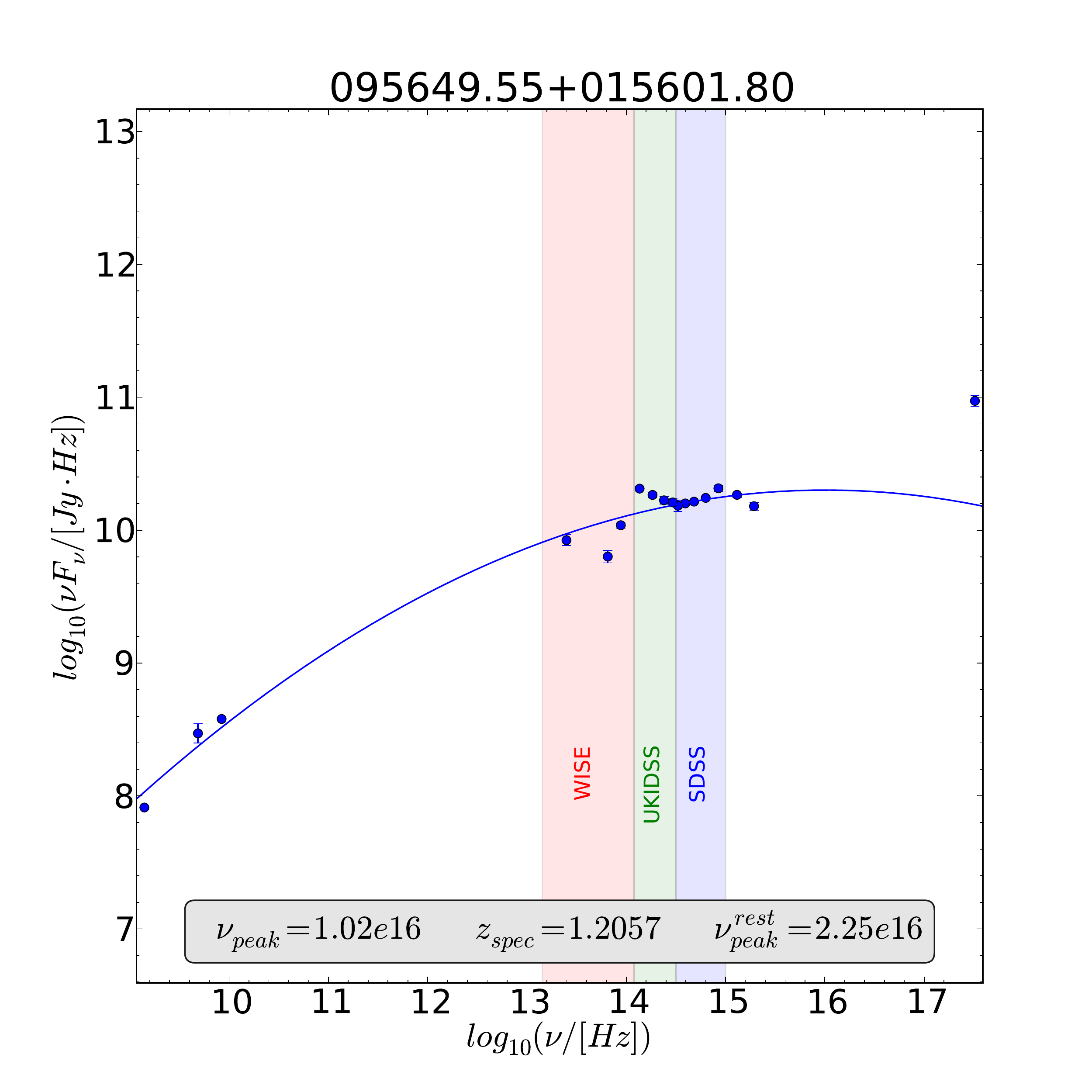}\\

\includegraphics[width=0.3\textwidth]{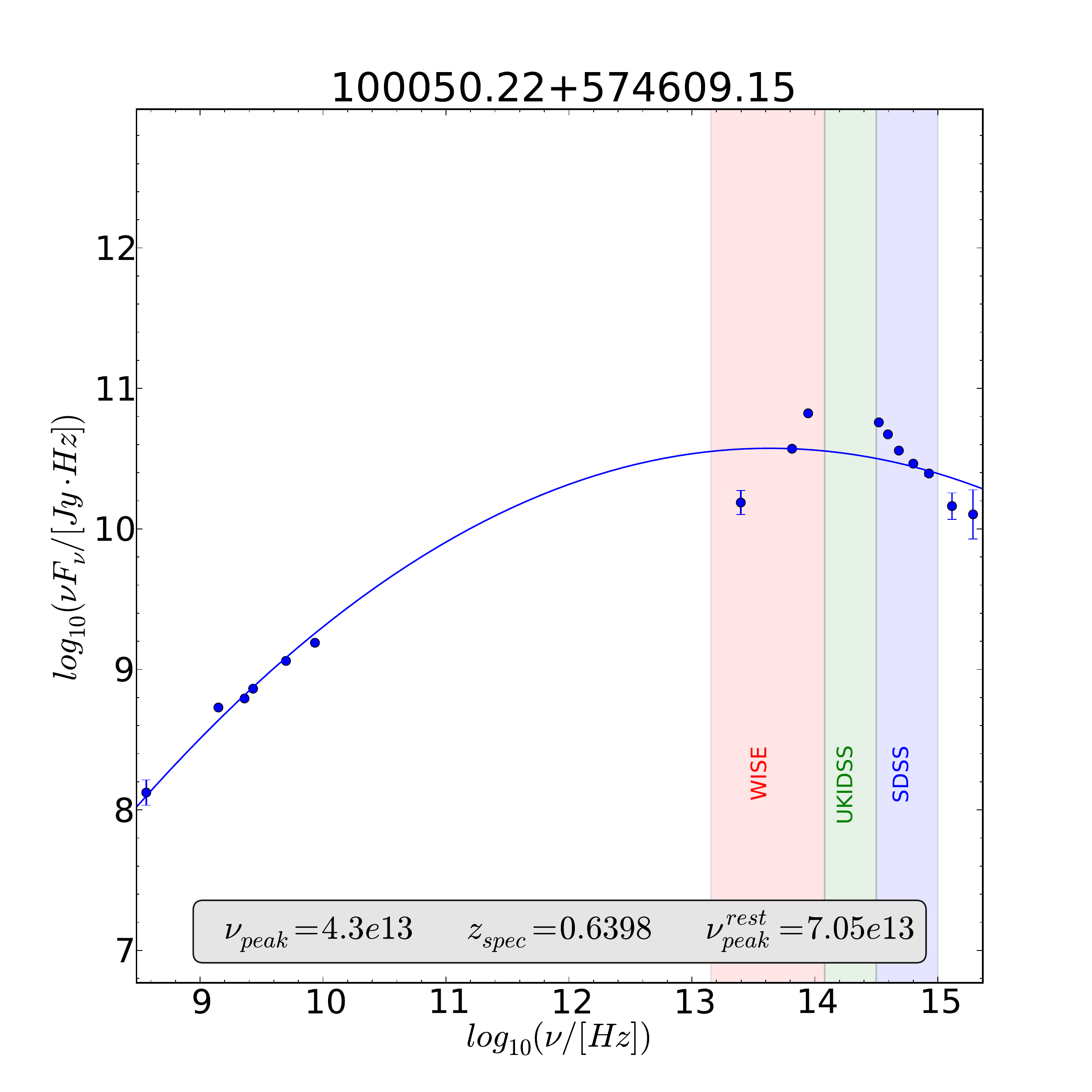}
\includegraphics[width=0.3\textwidth]{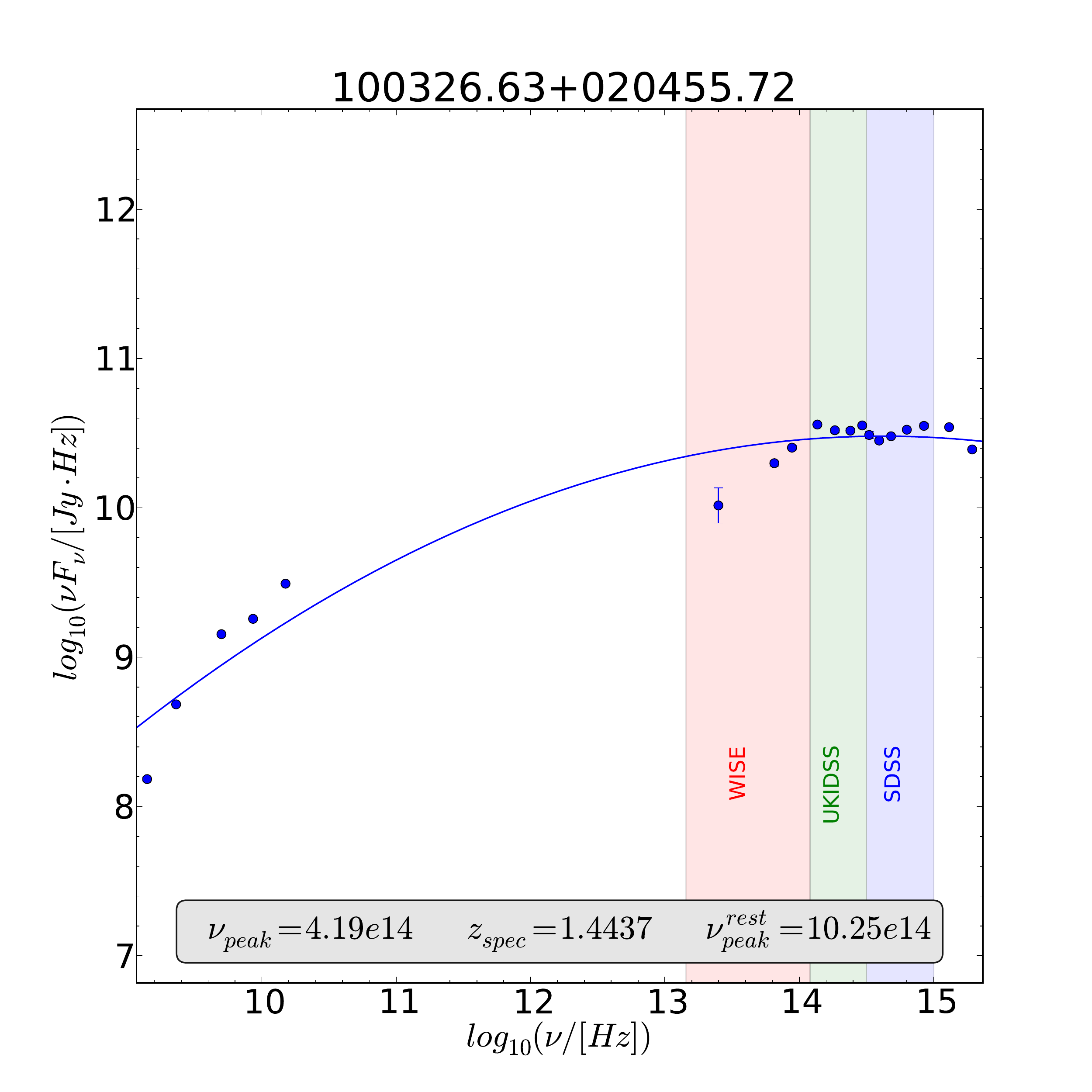}
\includegraphics[width=0.3\textwidth]{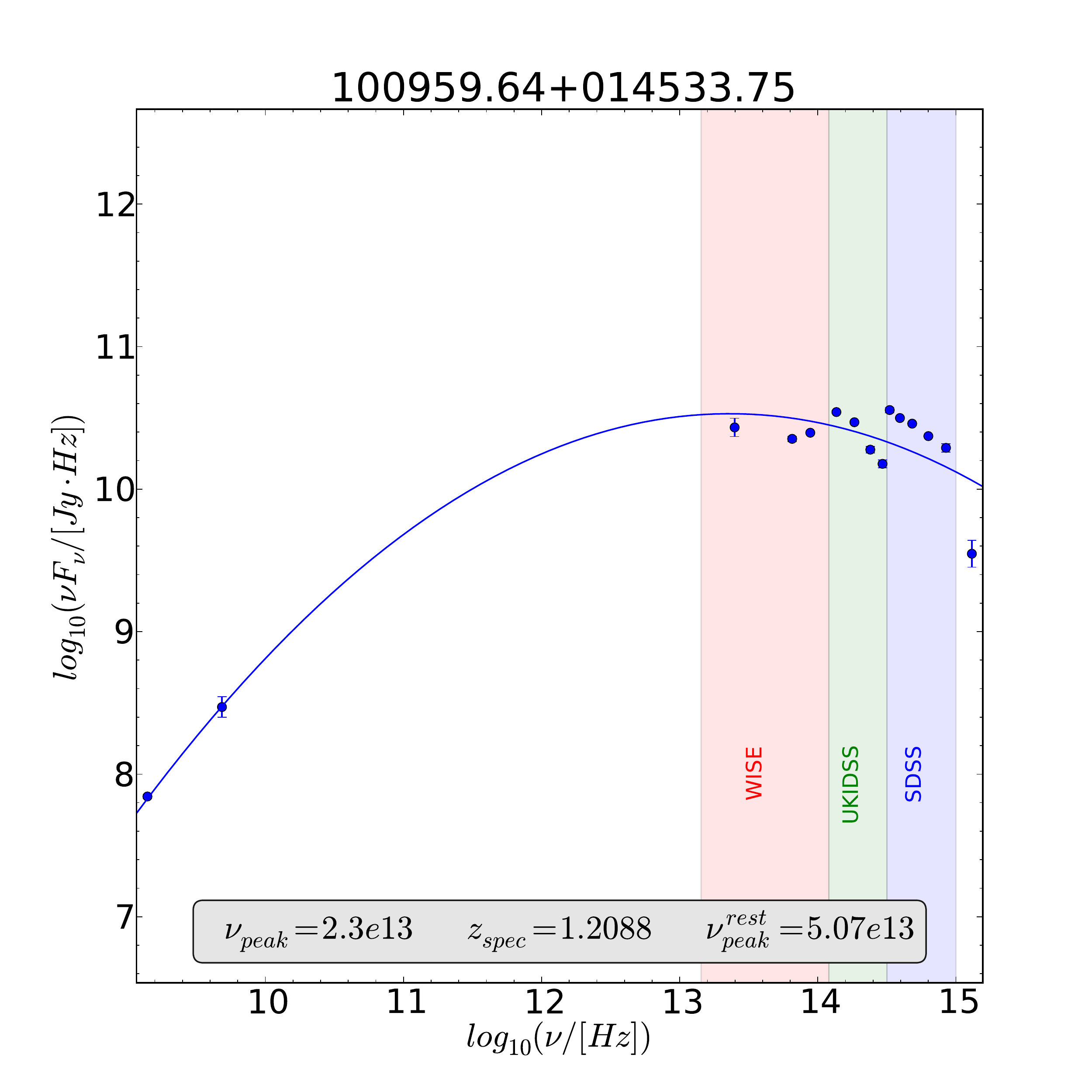}\\

\includegraphics[width=0.3\textwidth]{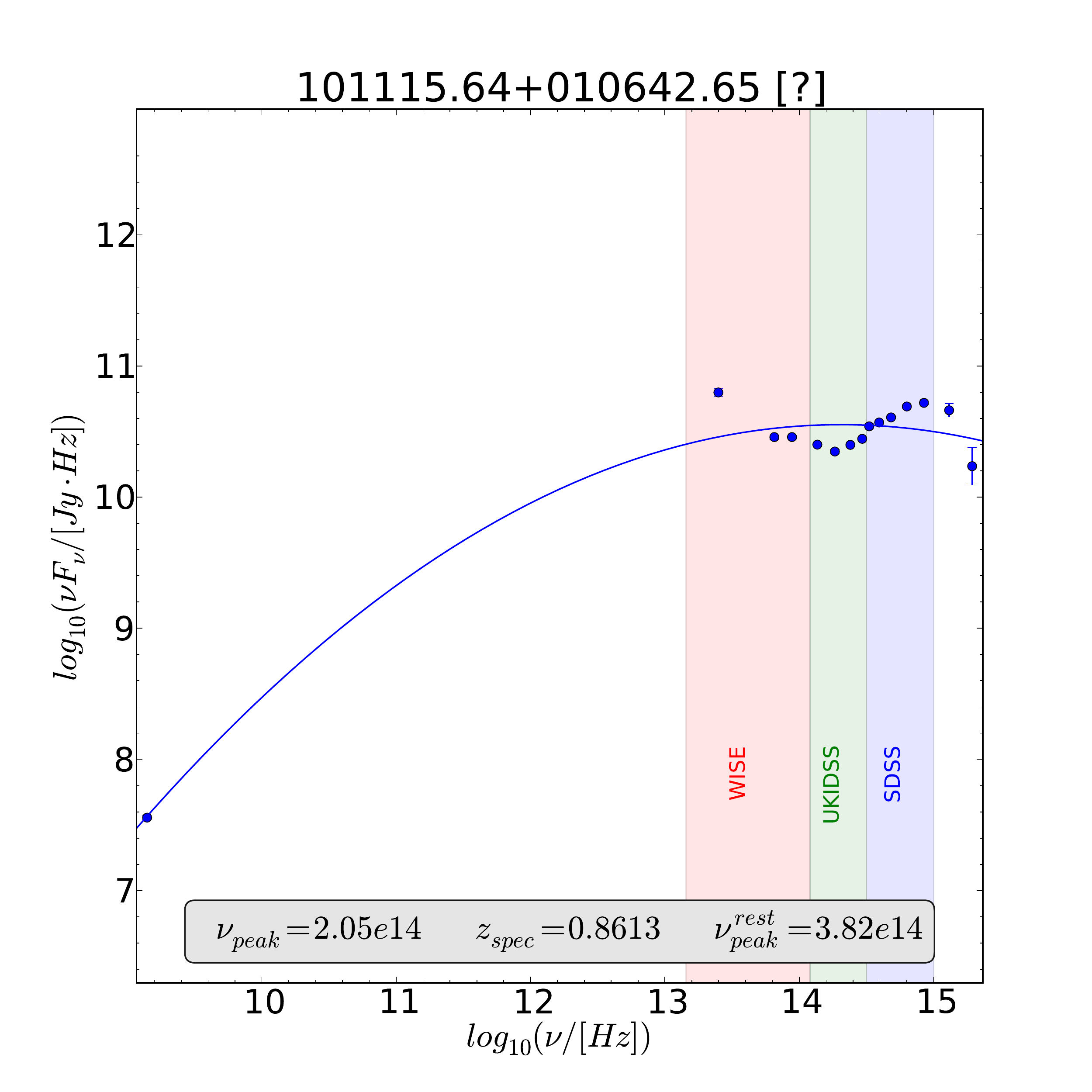}
\includegraphics[width=0.3\textwidth]{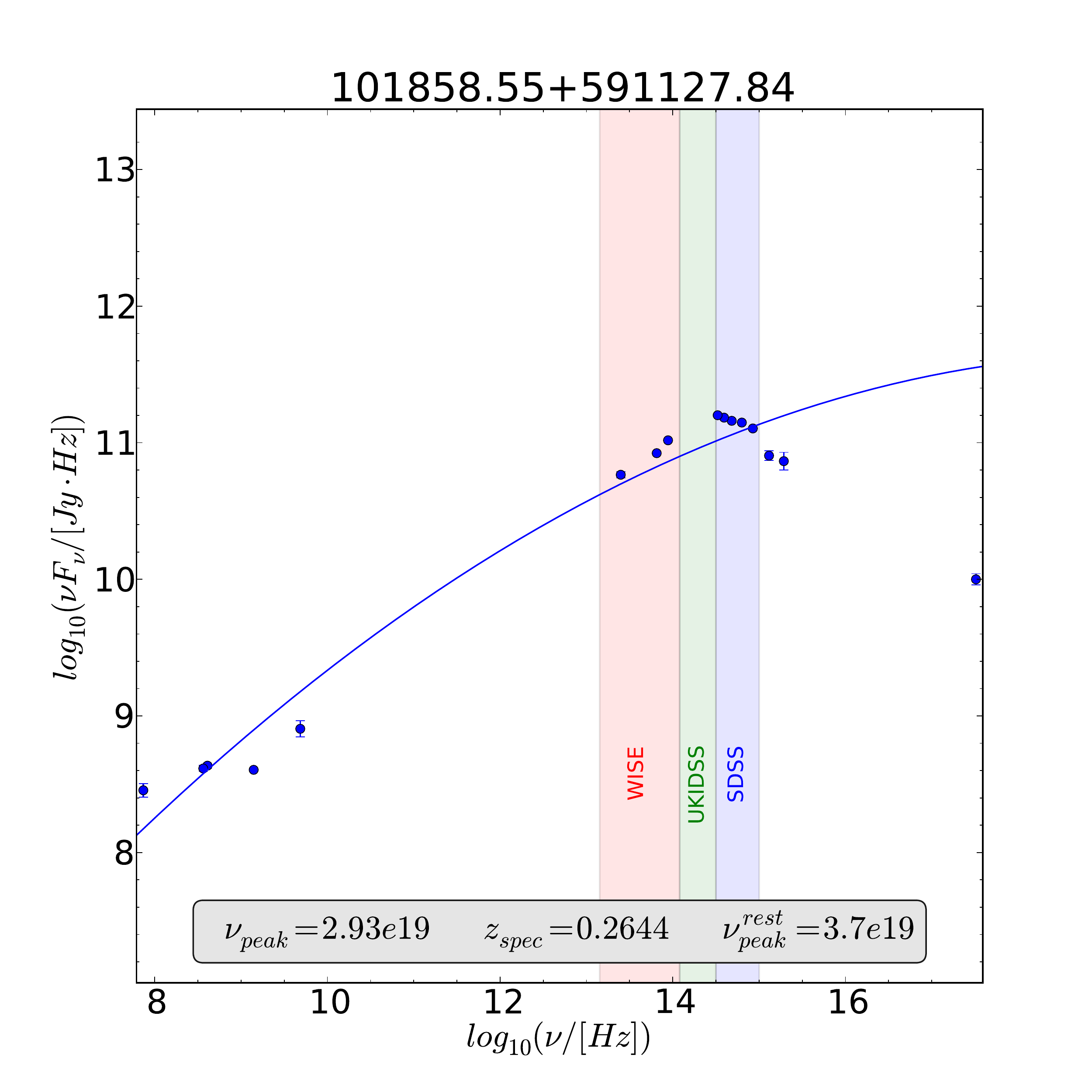}
\includegraphics[width=0.3\textwidth]{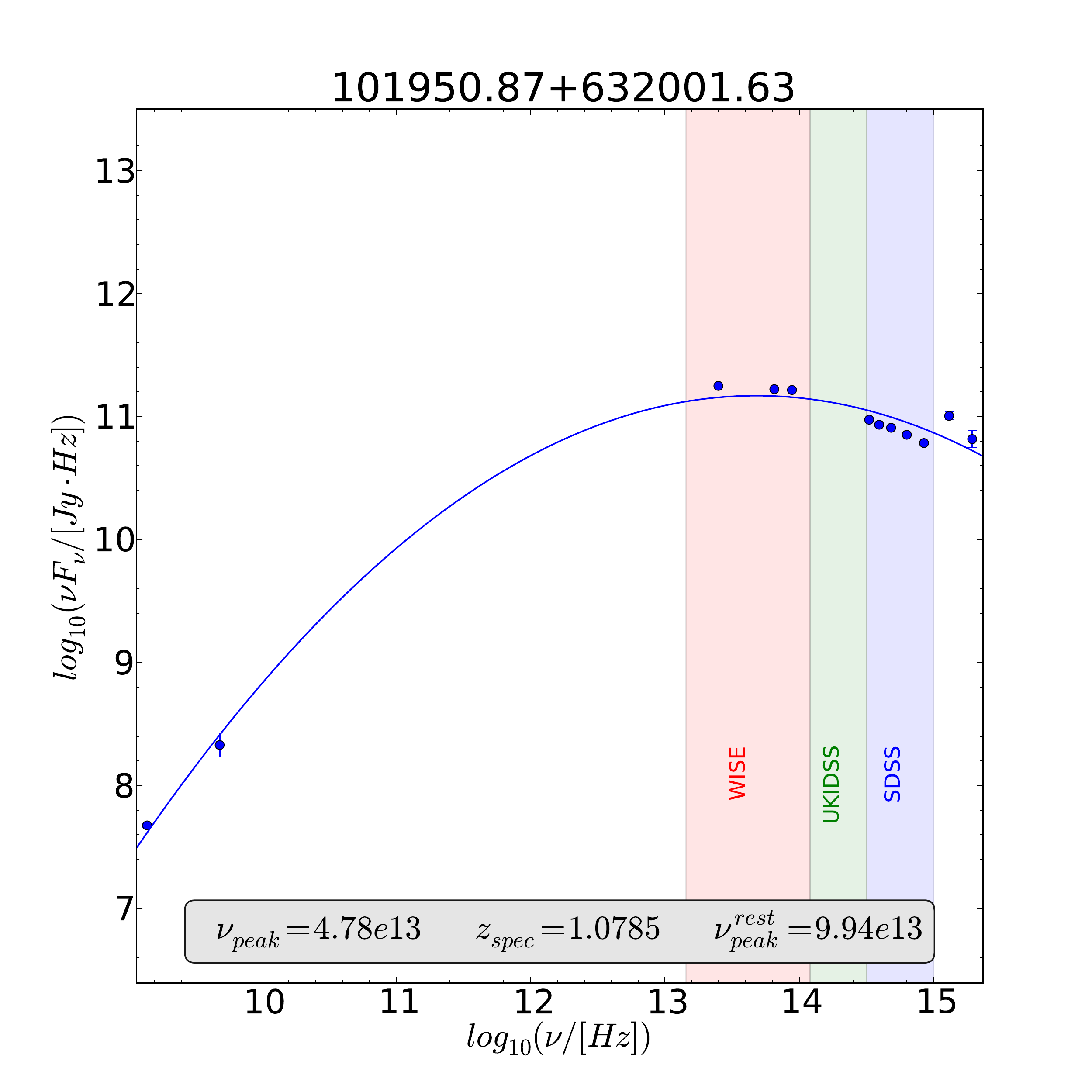}\\

\includegraphics[width=0.3\textwidth]{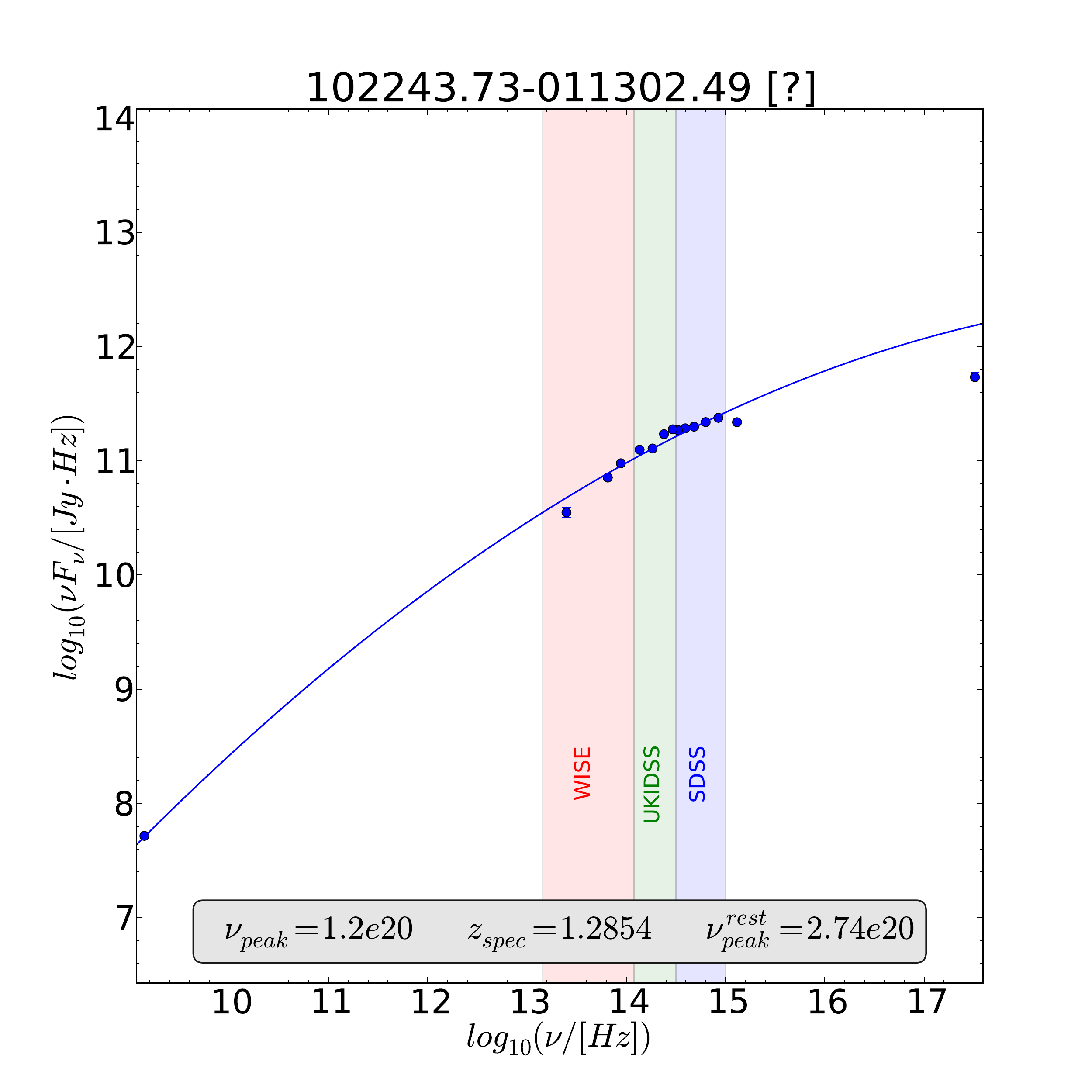}
\includegraphics[width=0.3\textwidth]{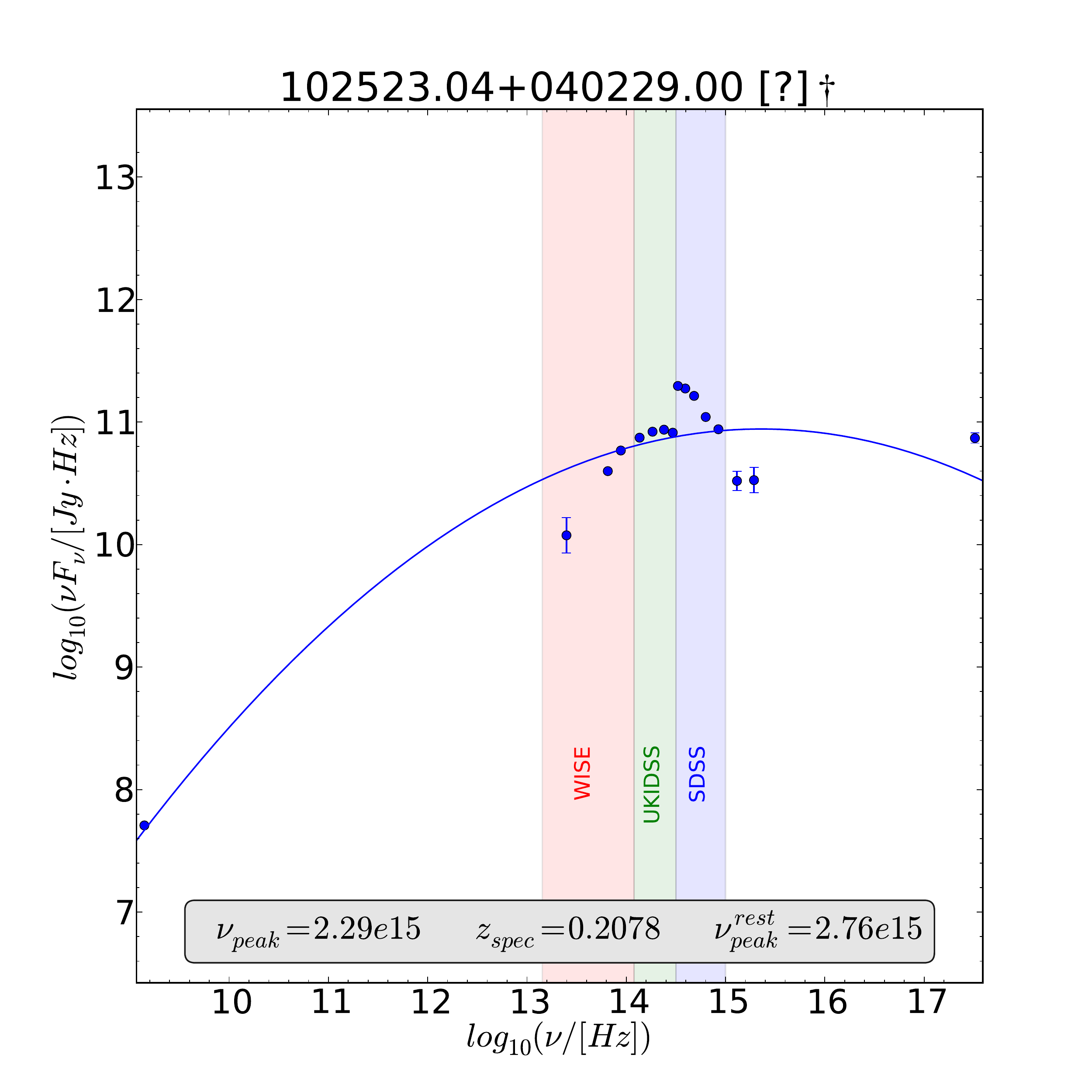}
\includegraphics[width=0.3\textwidth]{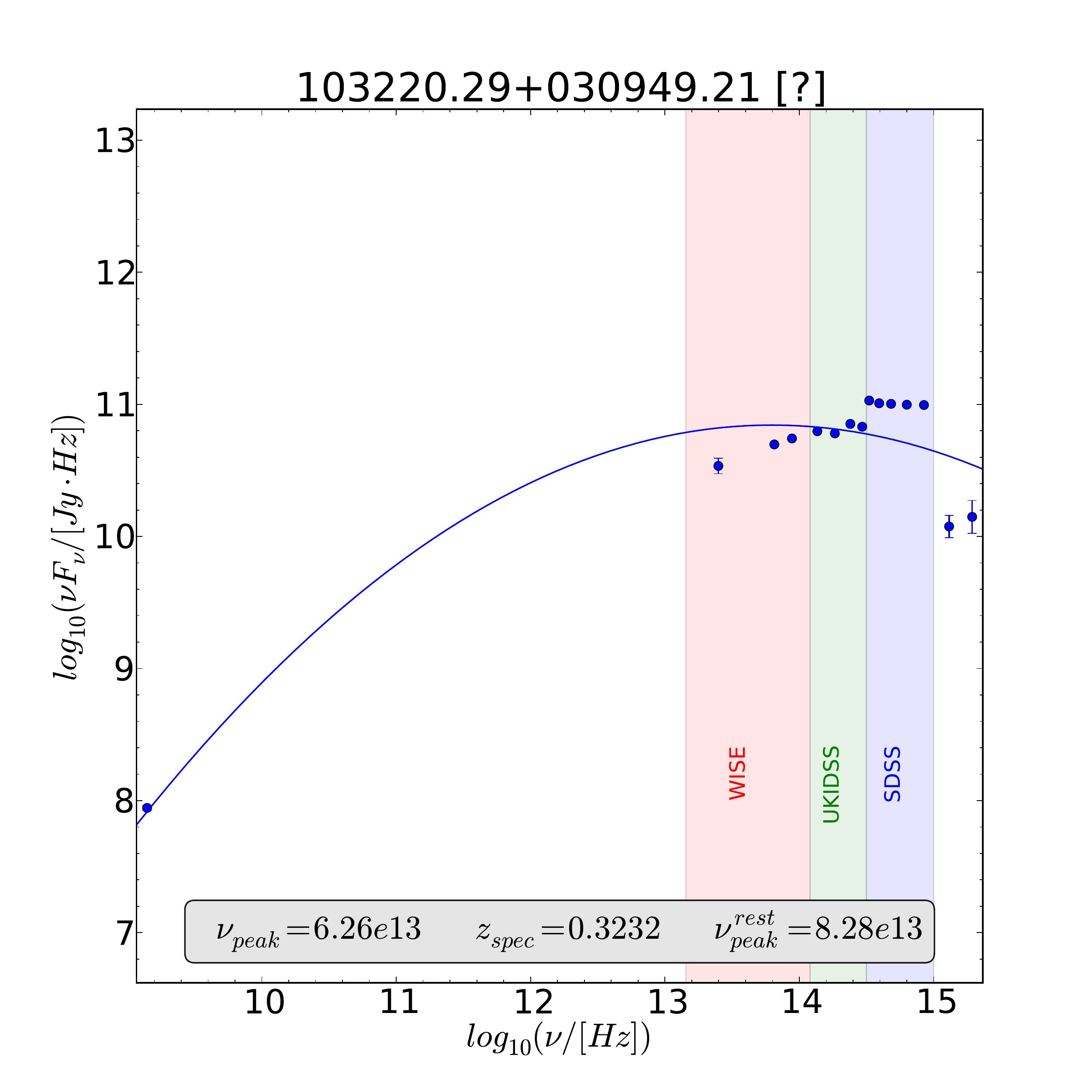}\\

\end{figure*}
\setcounter{figure}{0}
\begin{figure*}[htb!]
\caption{--Continued.}

\includegraphics[width=0.3\textwidth]{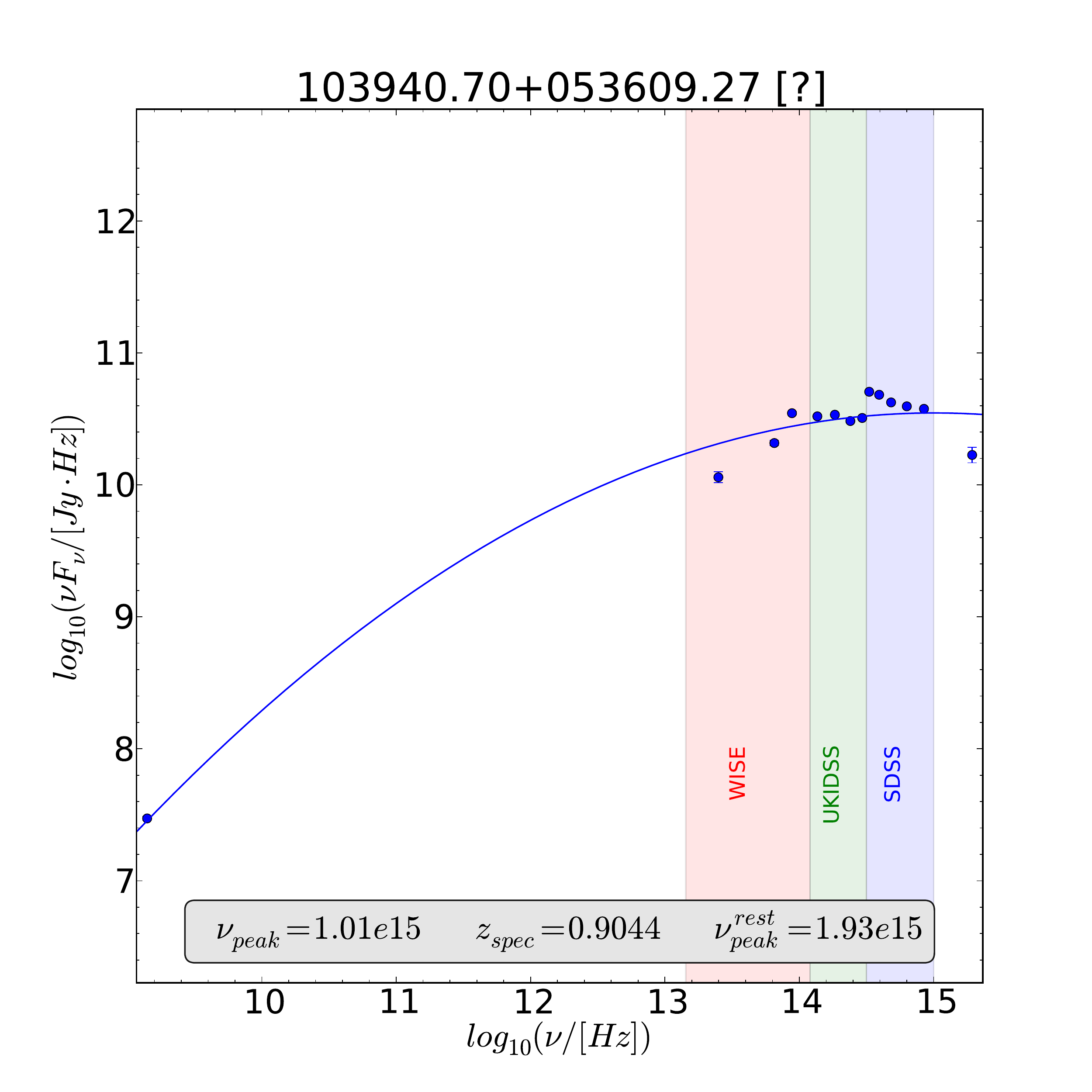}
\includegraphics[width=0.3\textwidth]{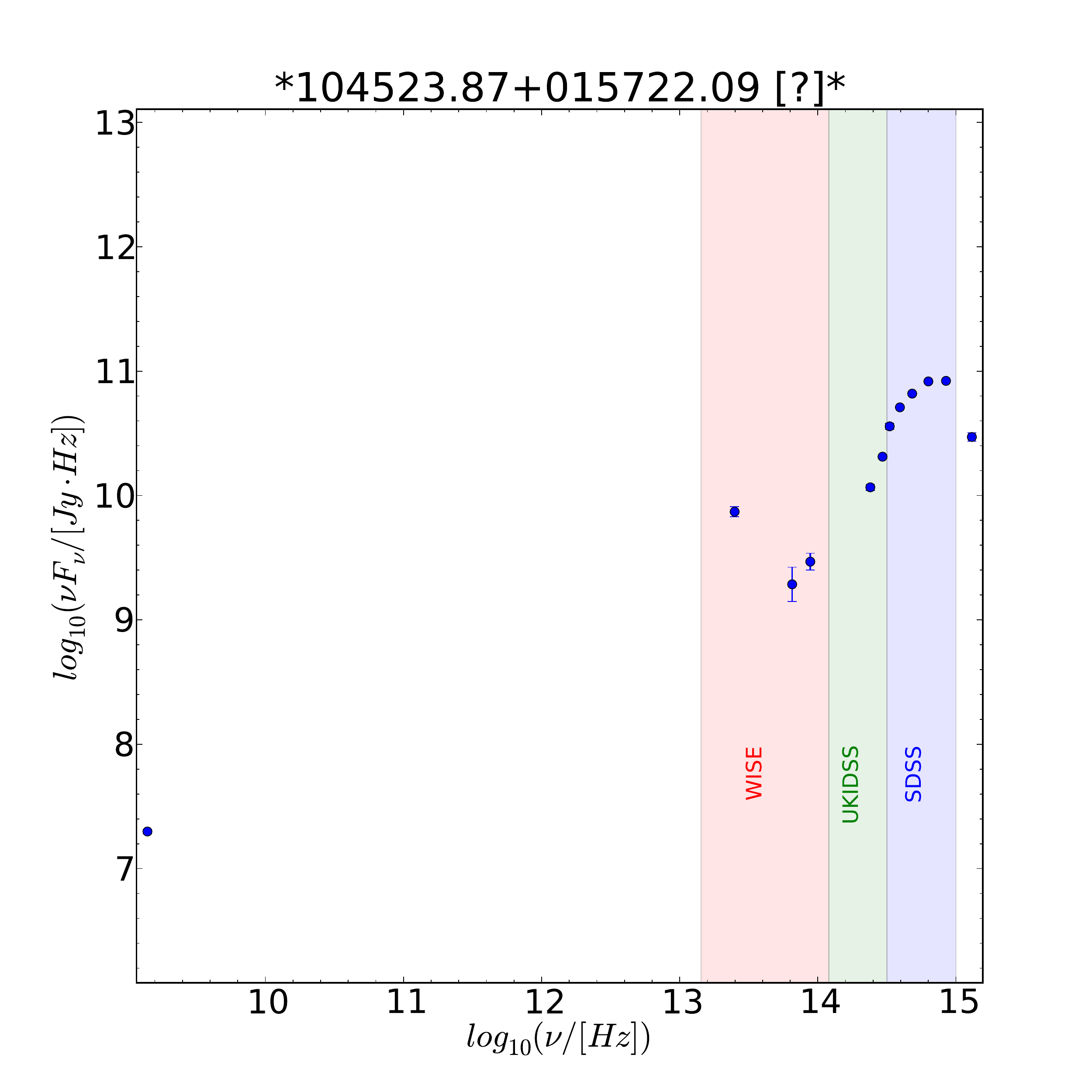}
\includegraphics[width=0.3\textwidth]{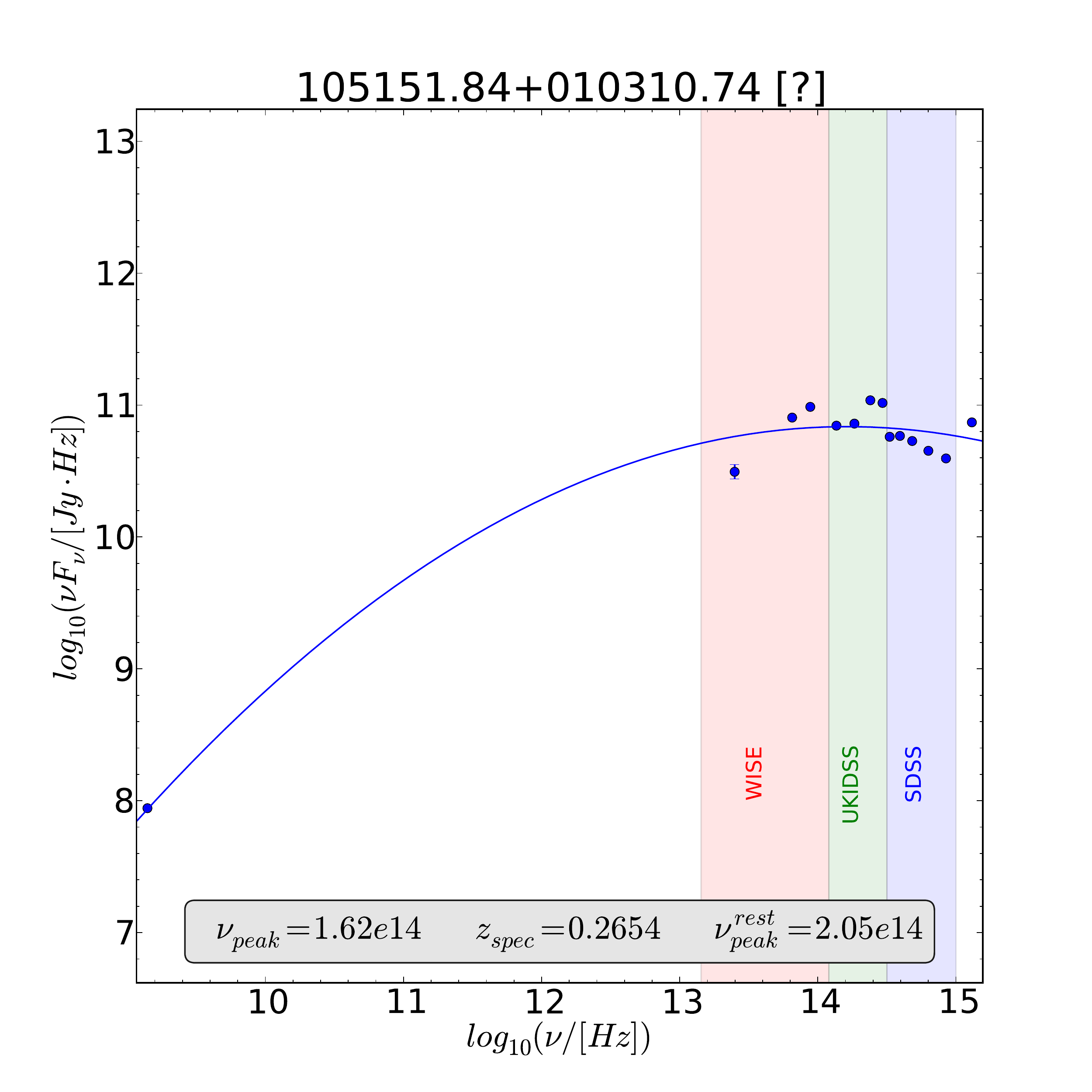}\\

\includegraphics[width=0.3\textwidth]{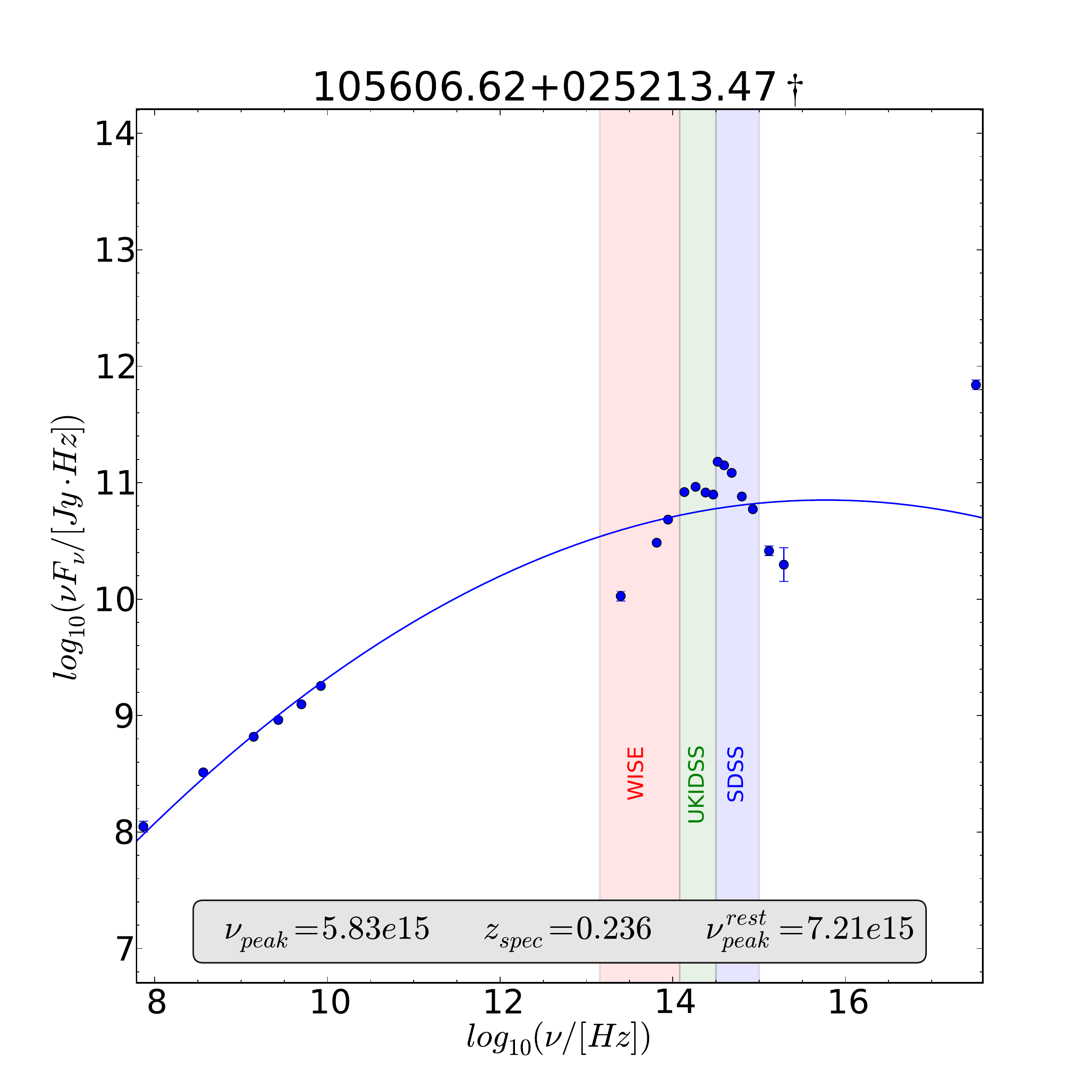}
\includegraphics[width=0.3\textwidth]{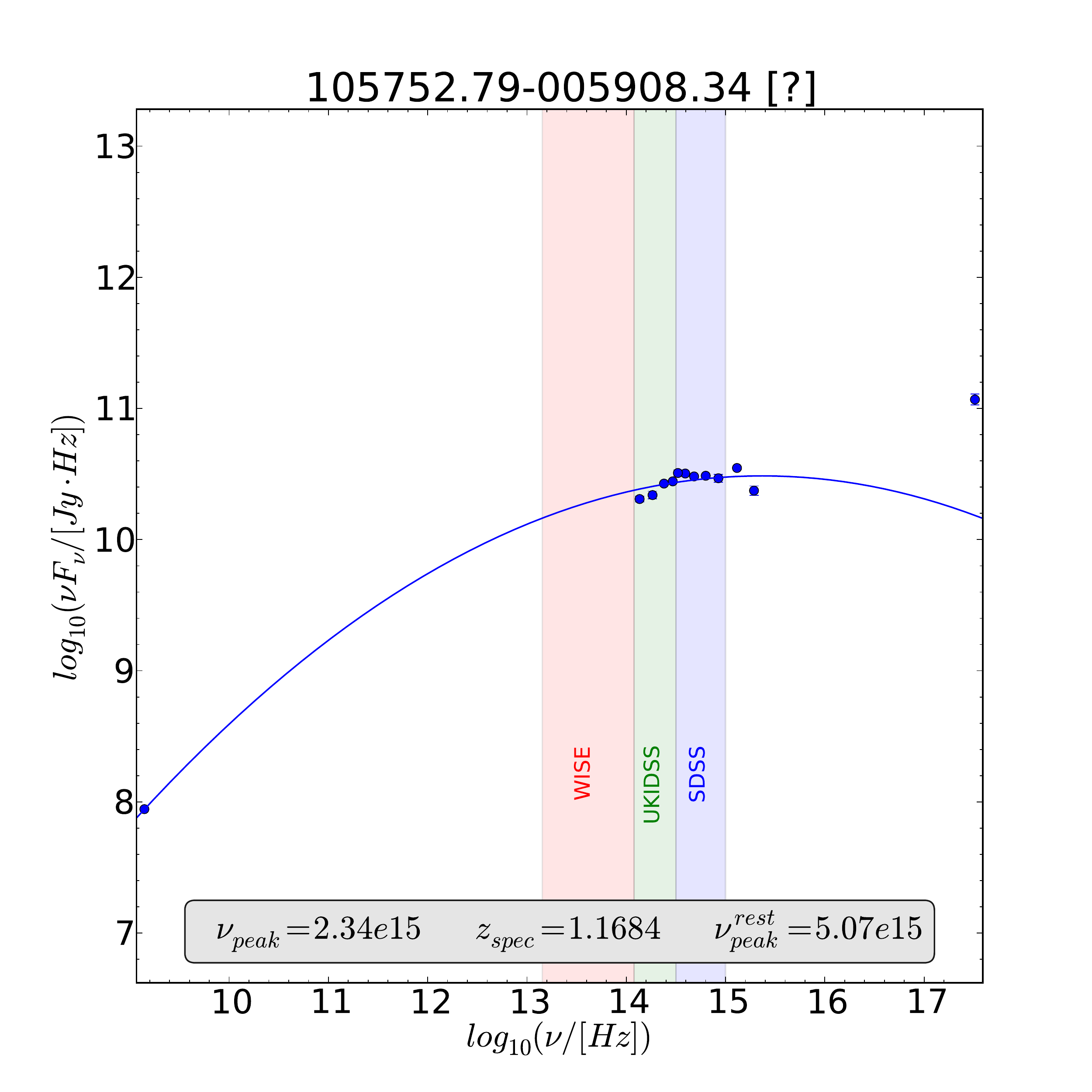}
\includegraphics[width=0.3\textwidth]{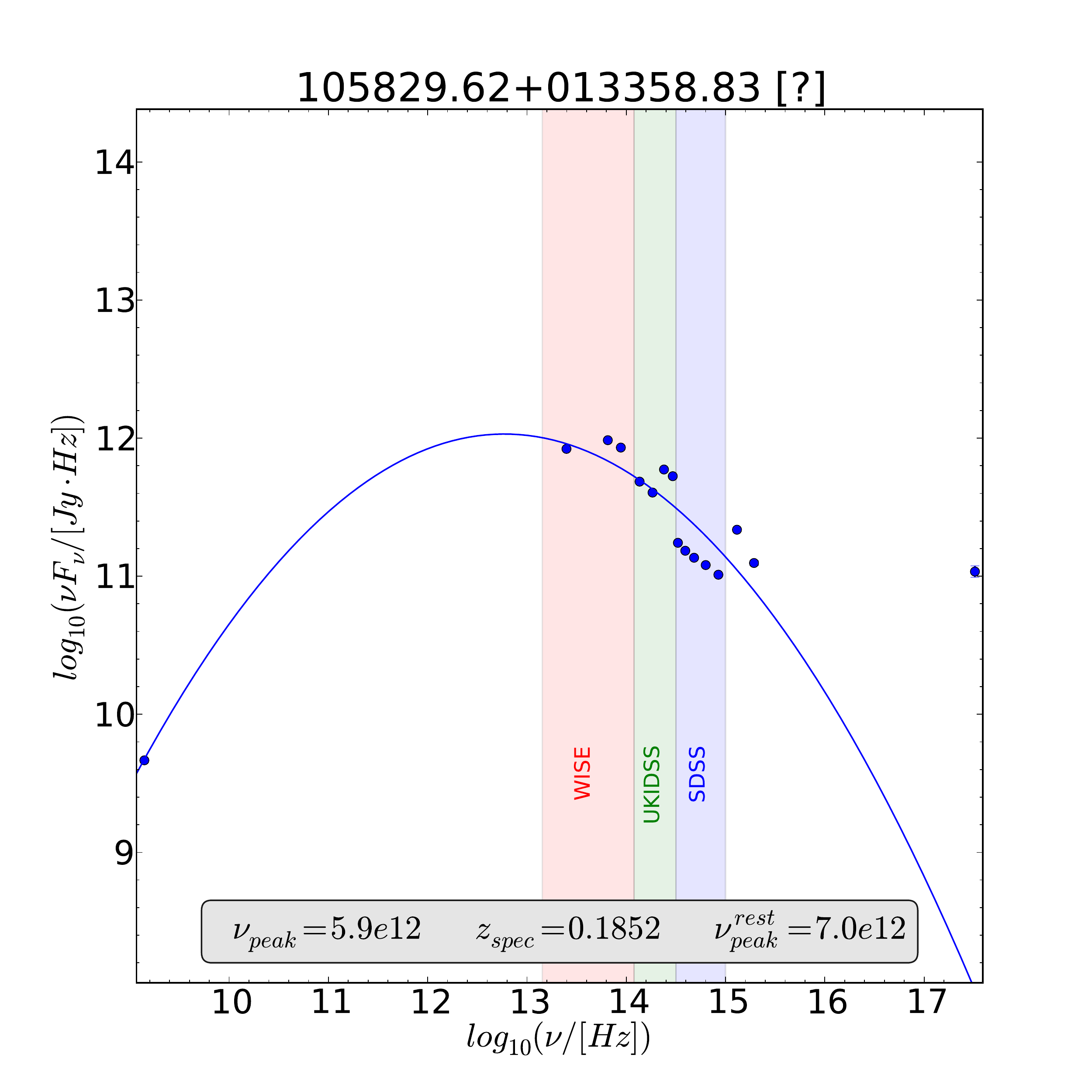}\\

\includegraphics[width=0.3\textwidth]{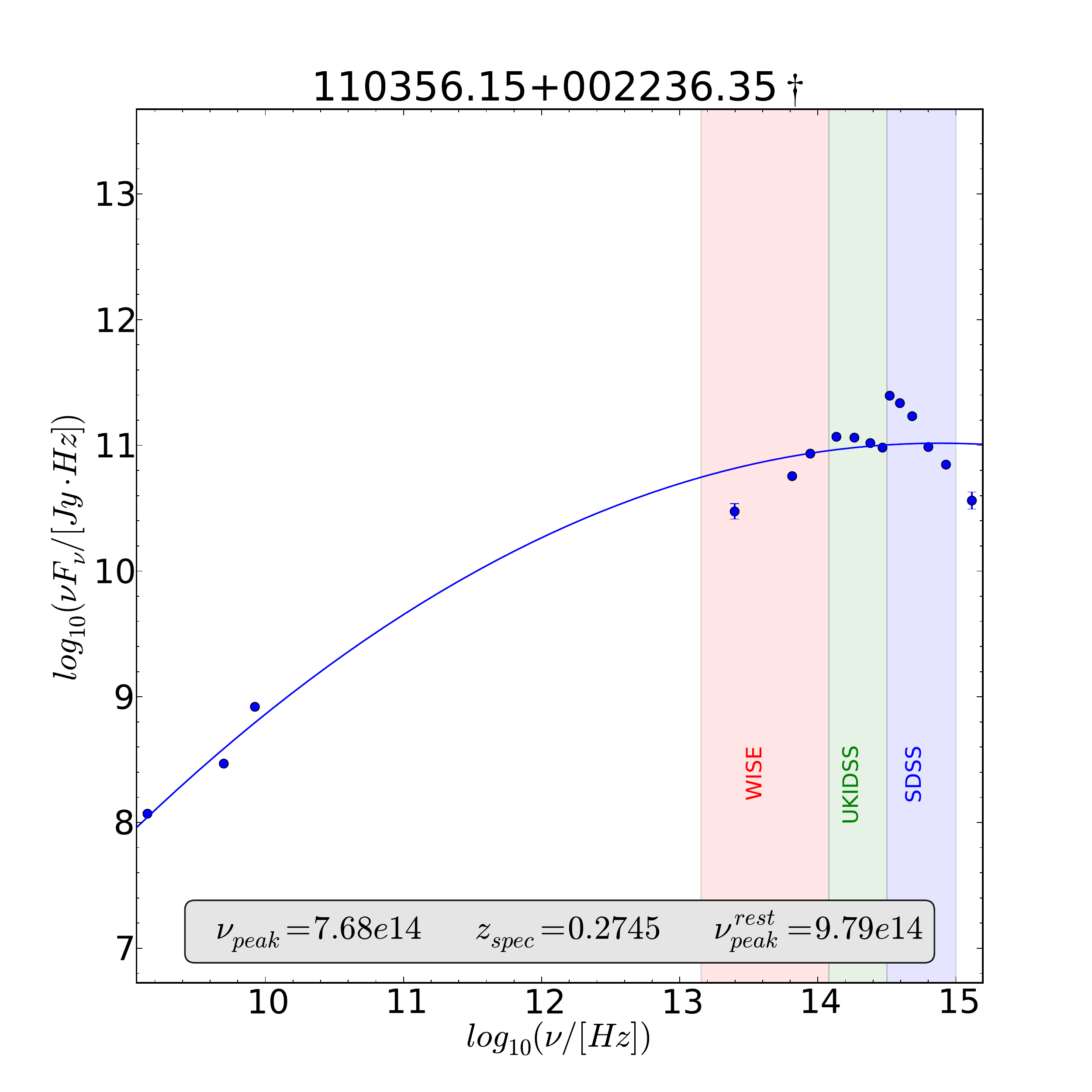}
\includegraphics[width=0.3\textwidth]{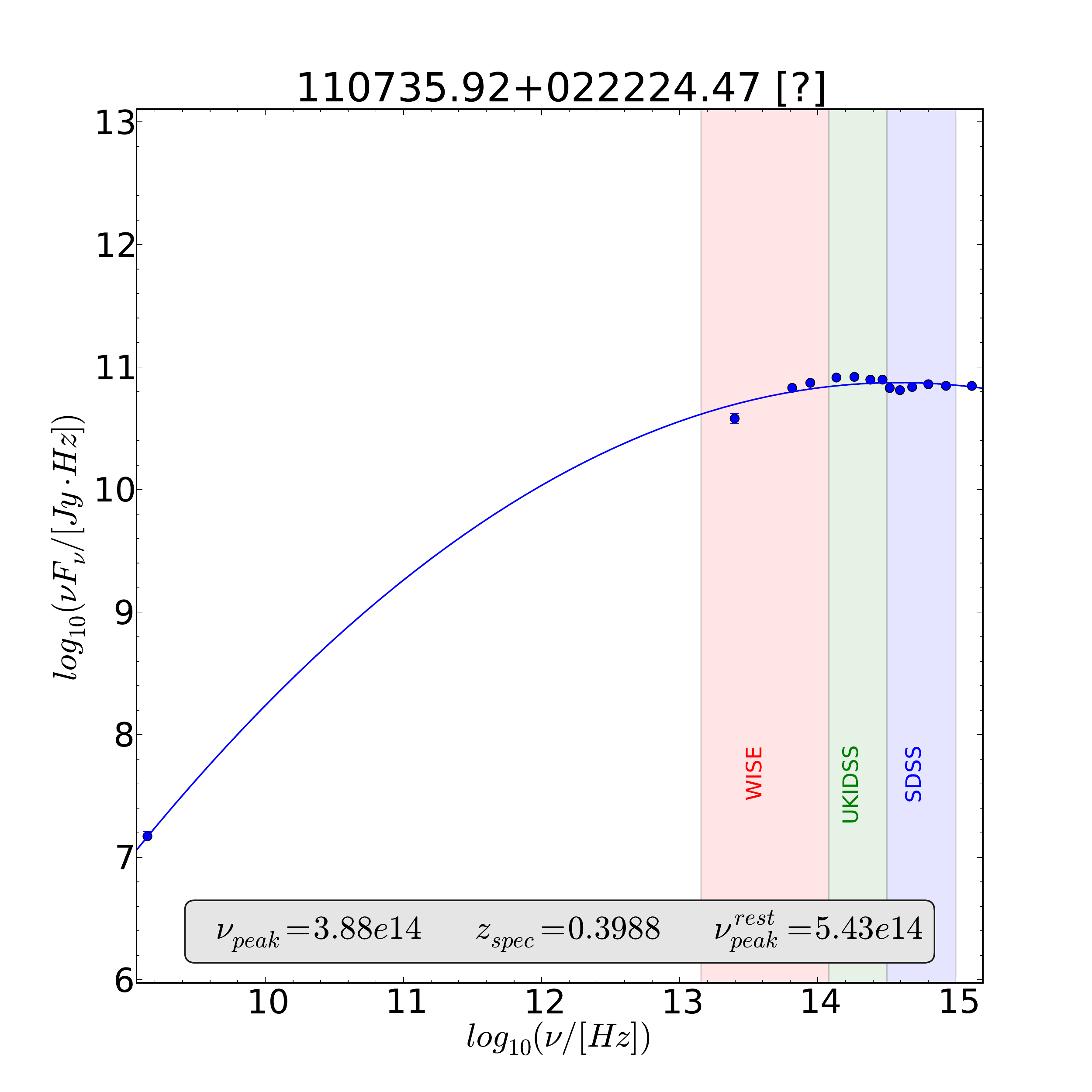}
\includegraphics[width=0.3\textwidth]{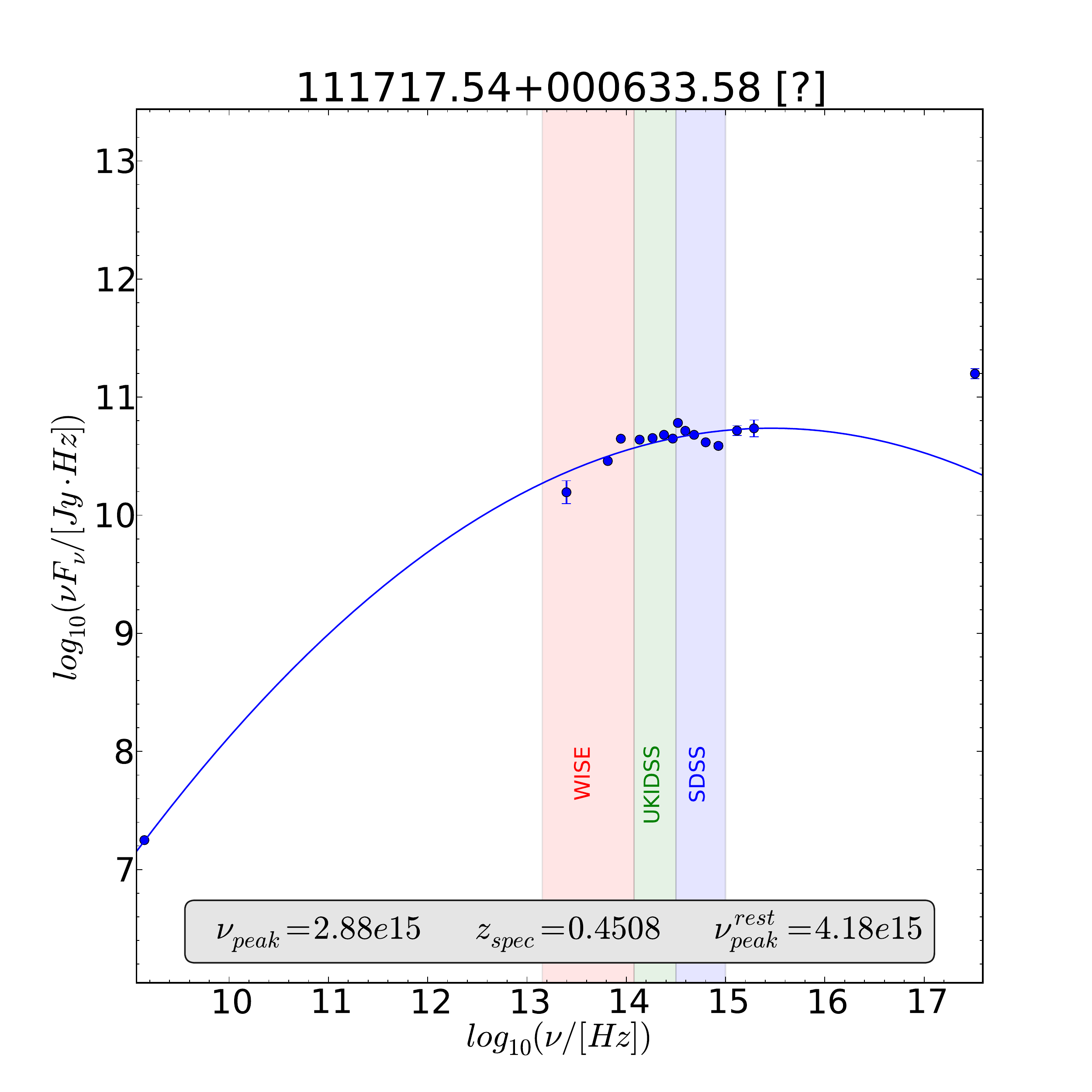}\\

\includegraphics[width=0.3\textwidth]{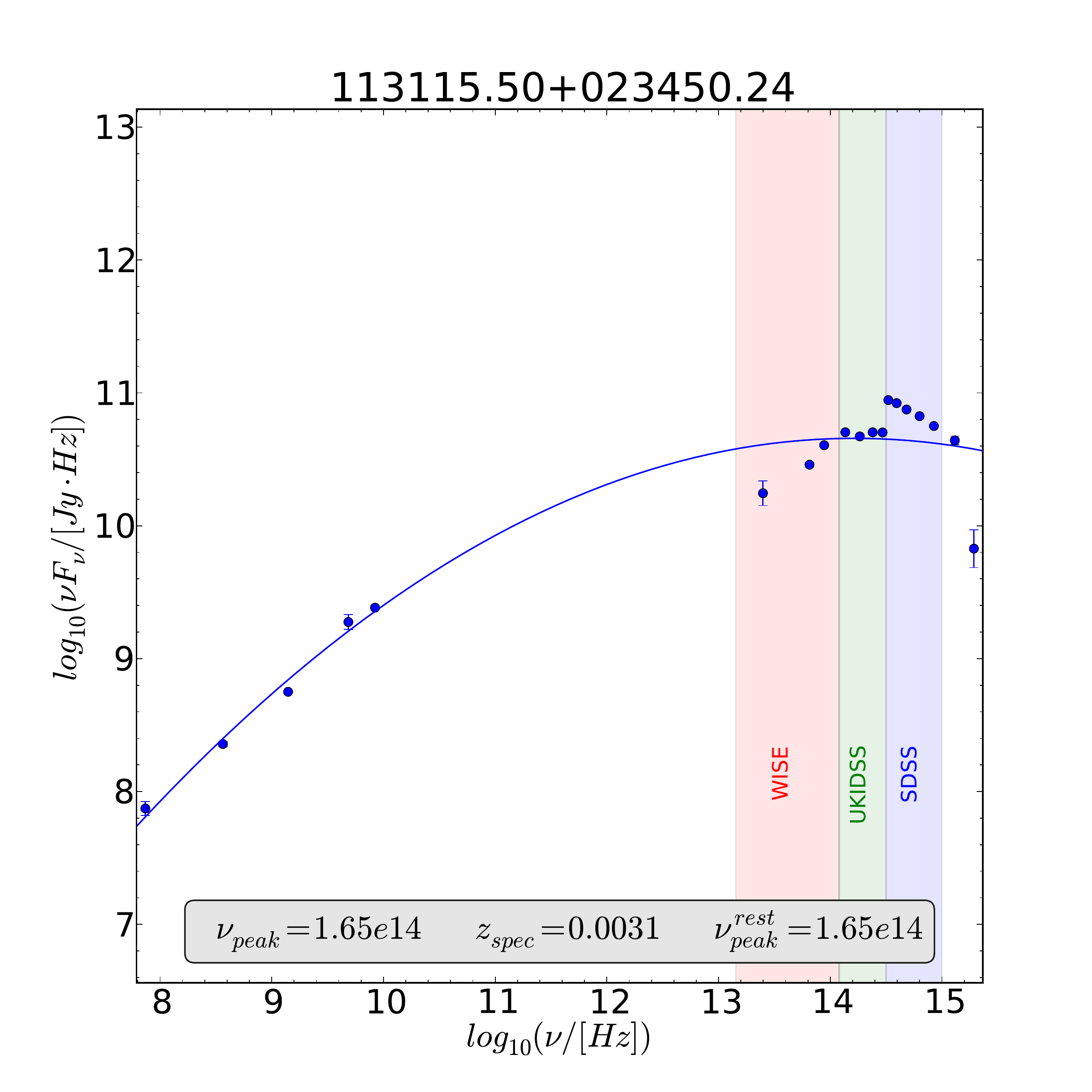}
\includegraphics[width=0.3\textwidth]{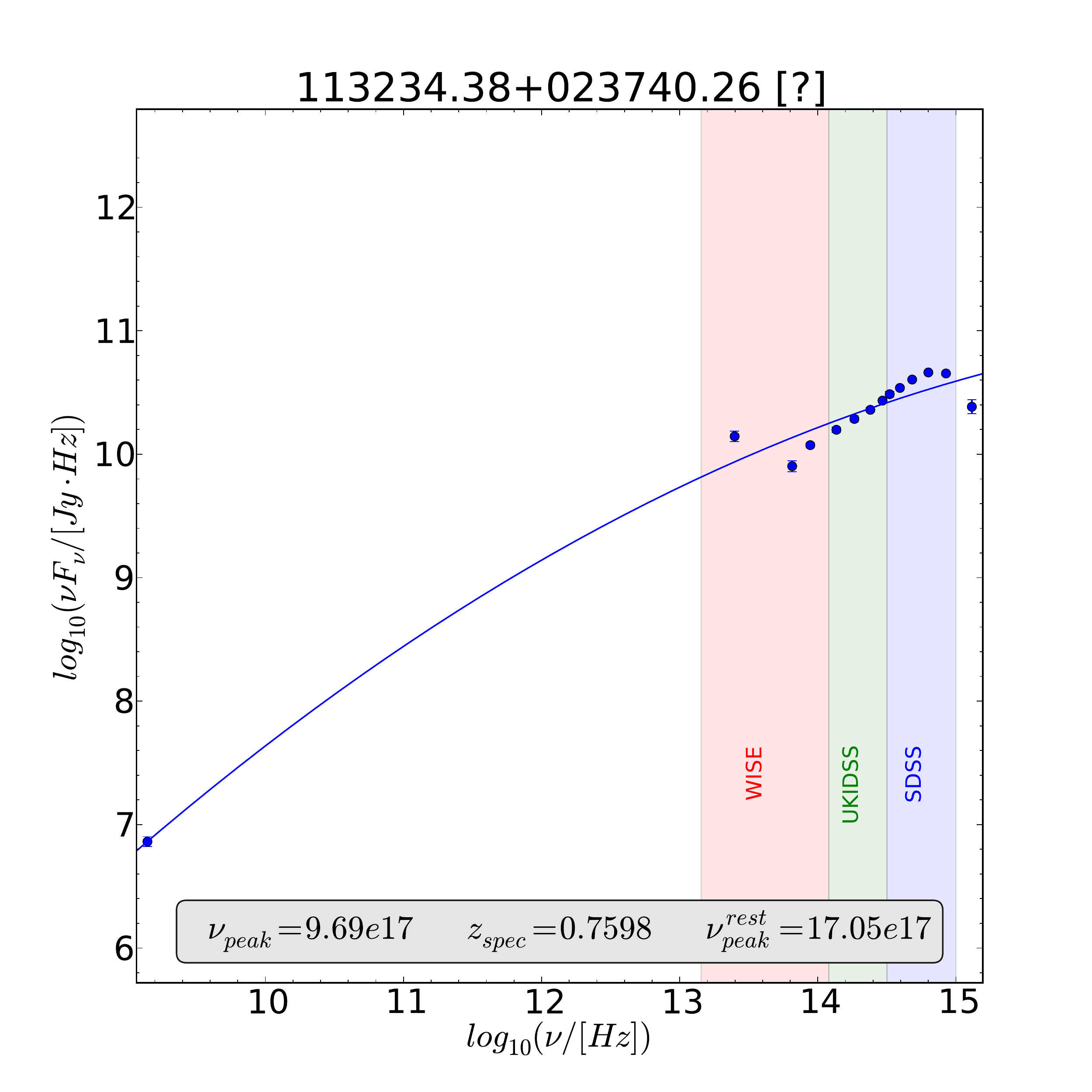}
\includegraphics[width=0.3\textwidth]{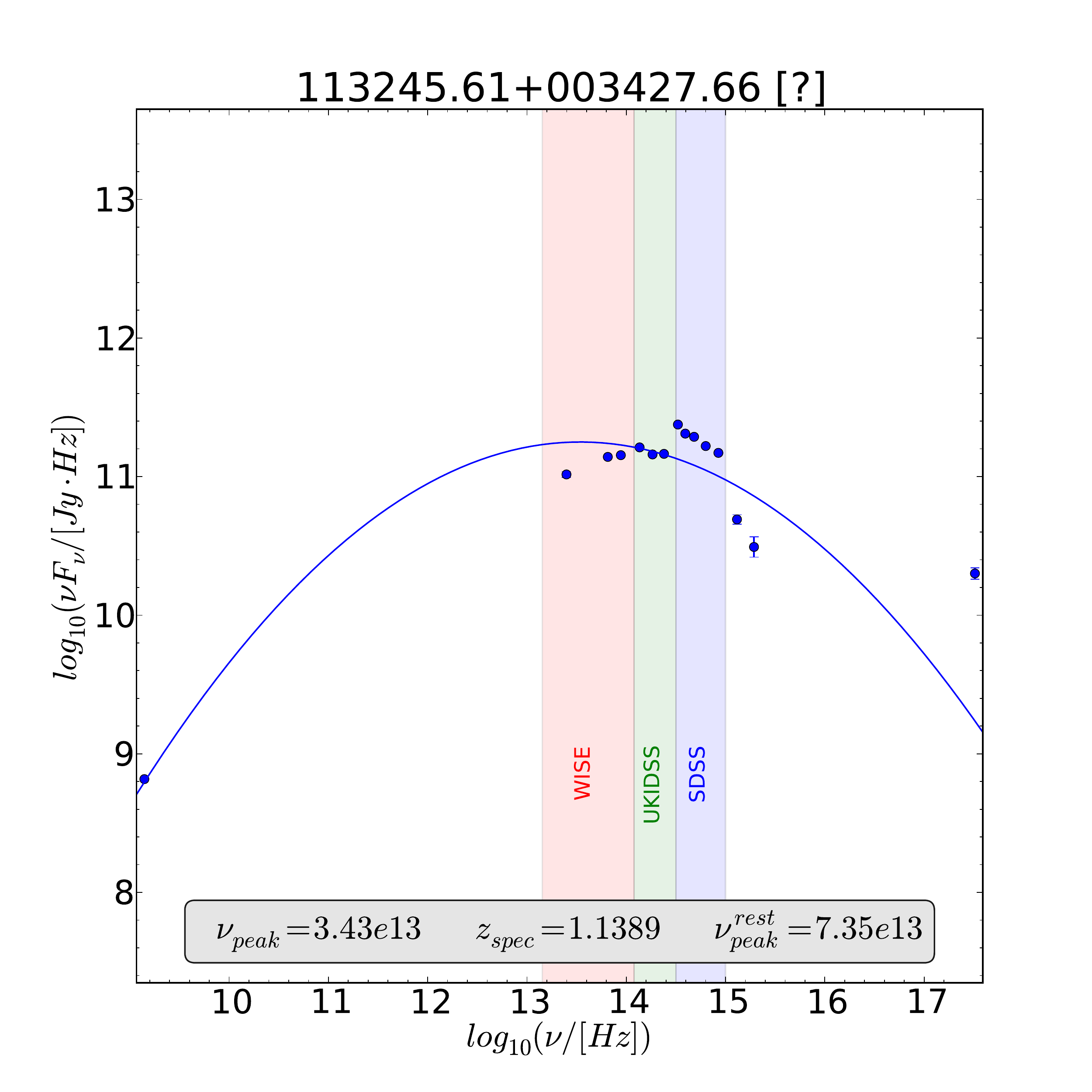}\\

\end{figure*}
\setcounter{figure}{0}
\begin{figure*}[htb!]
\caption{--Continued.}

\includegraphics[width=0.3\textwidth]{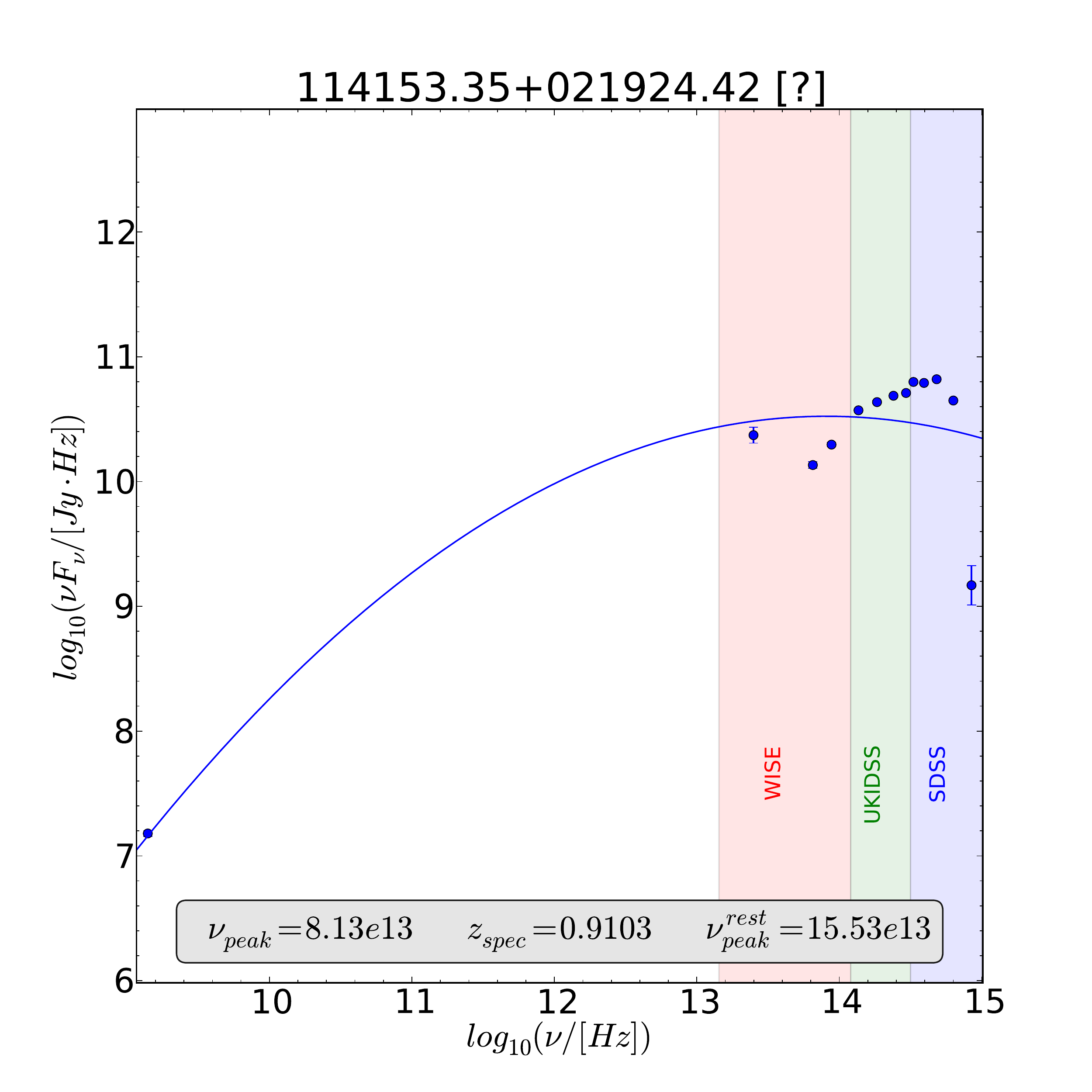}
\includegraphics[width=0.3\textwidth]{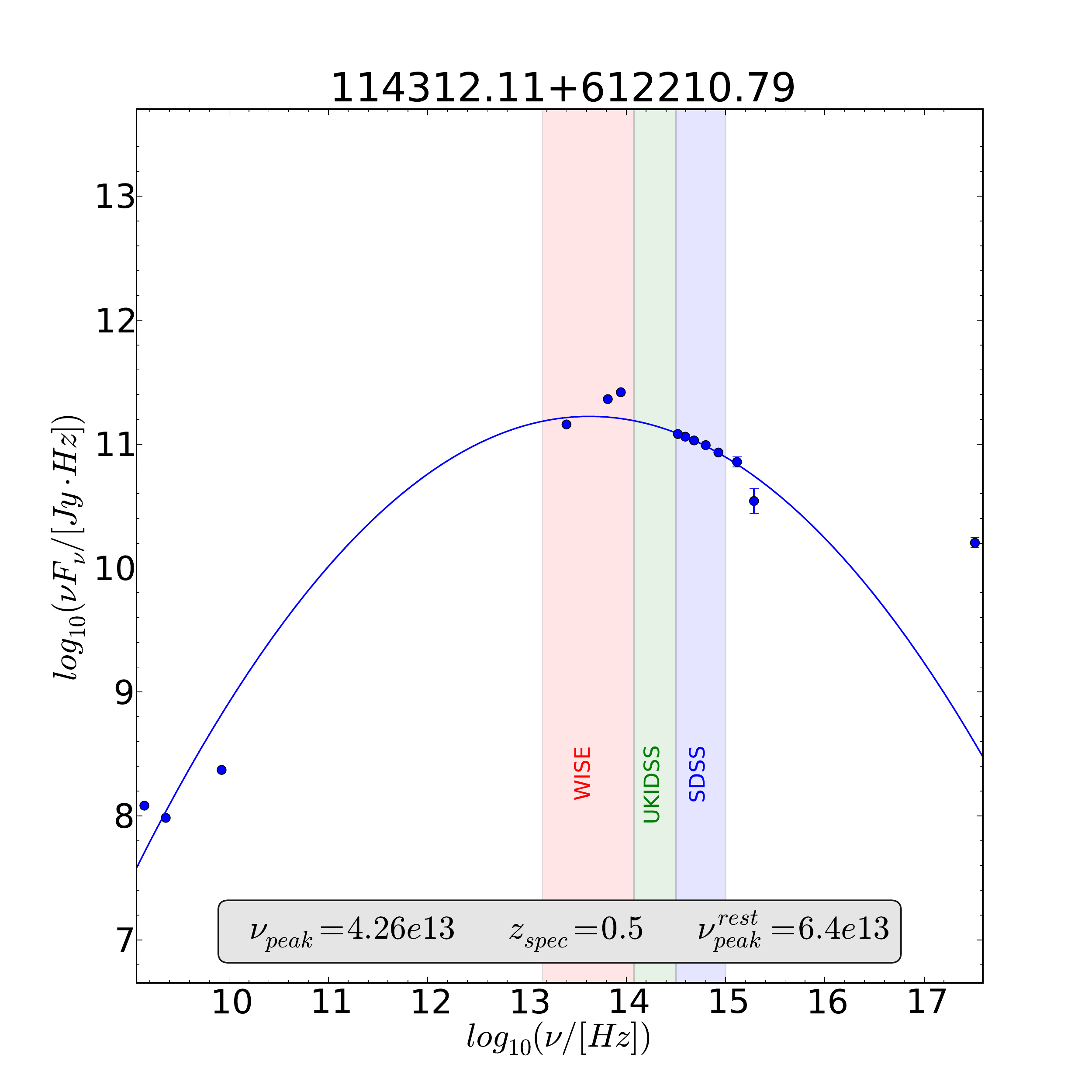}
\includegraphics[width=0.3\textwidth]{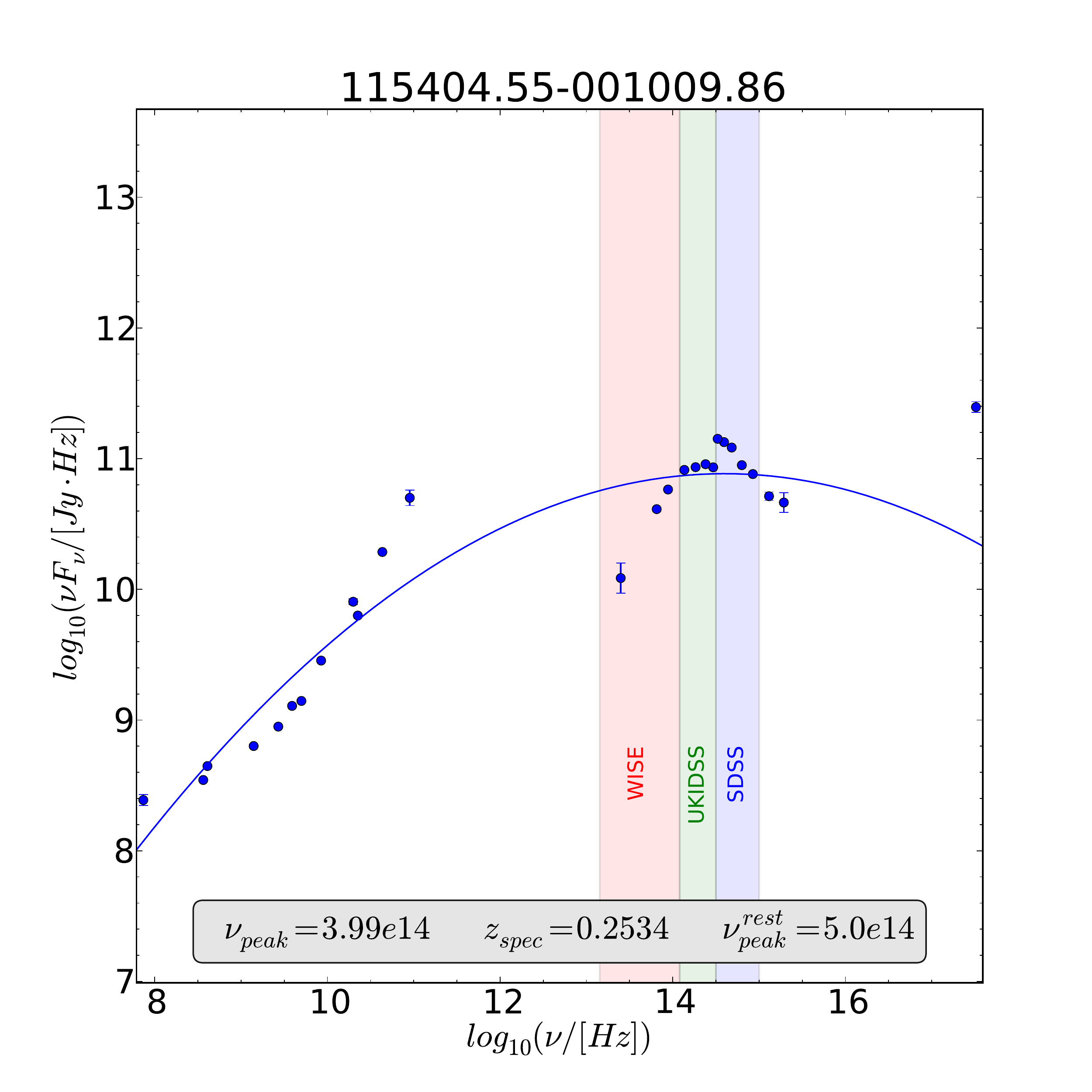}\\

\includegraphics[width=0.3\textwidth]{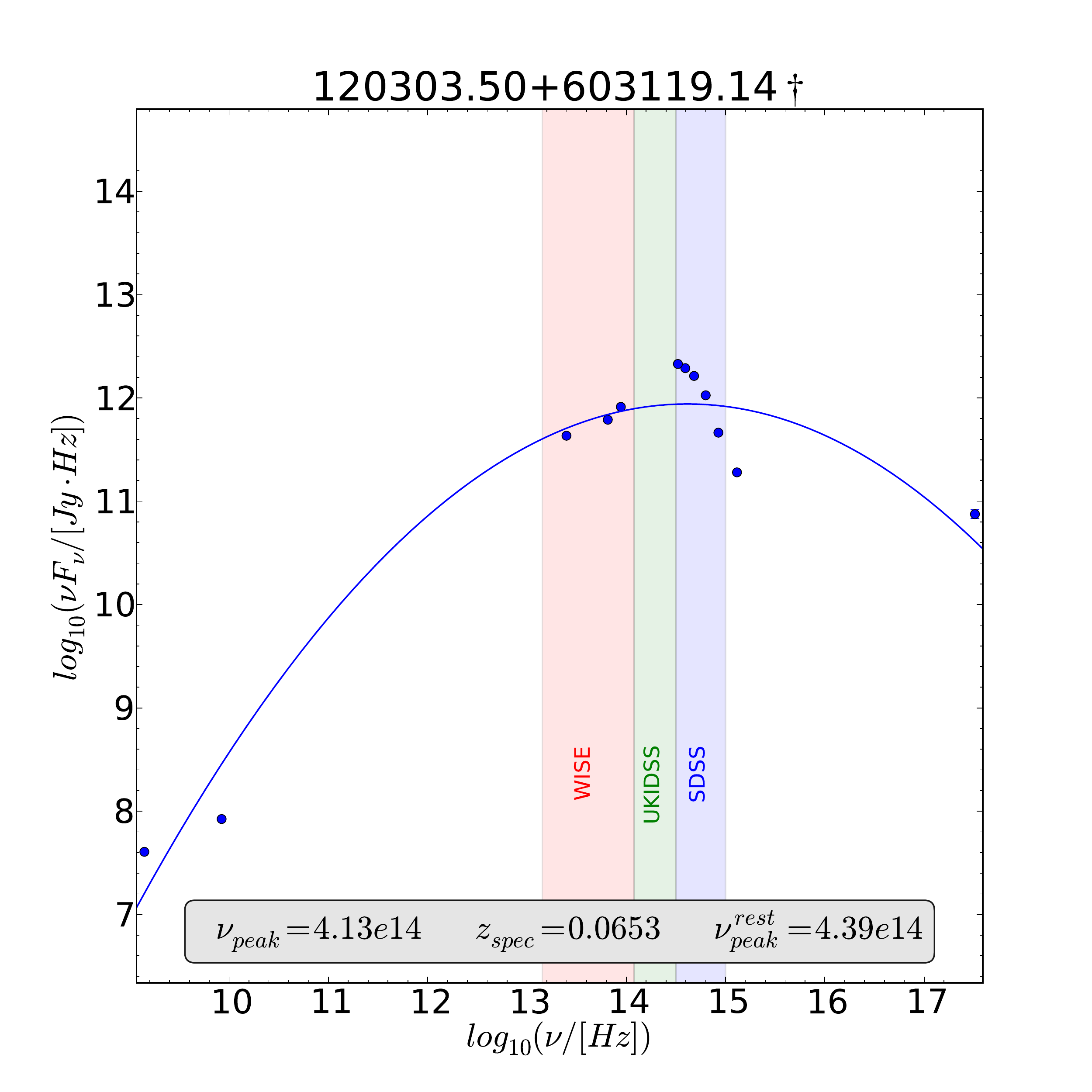}
\includegraphics[width=0.3\textwidth]{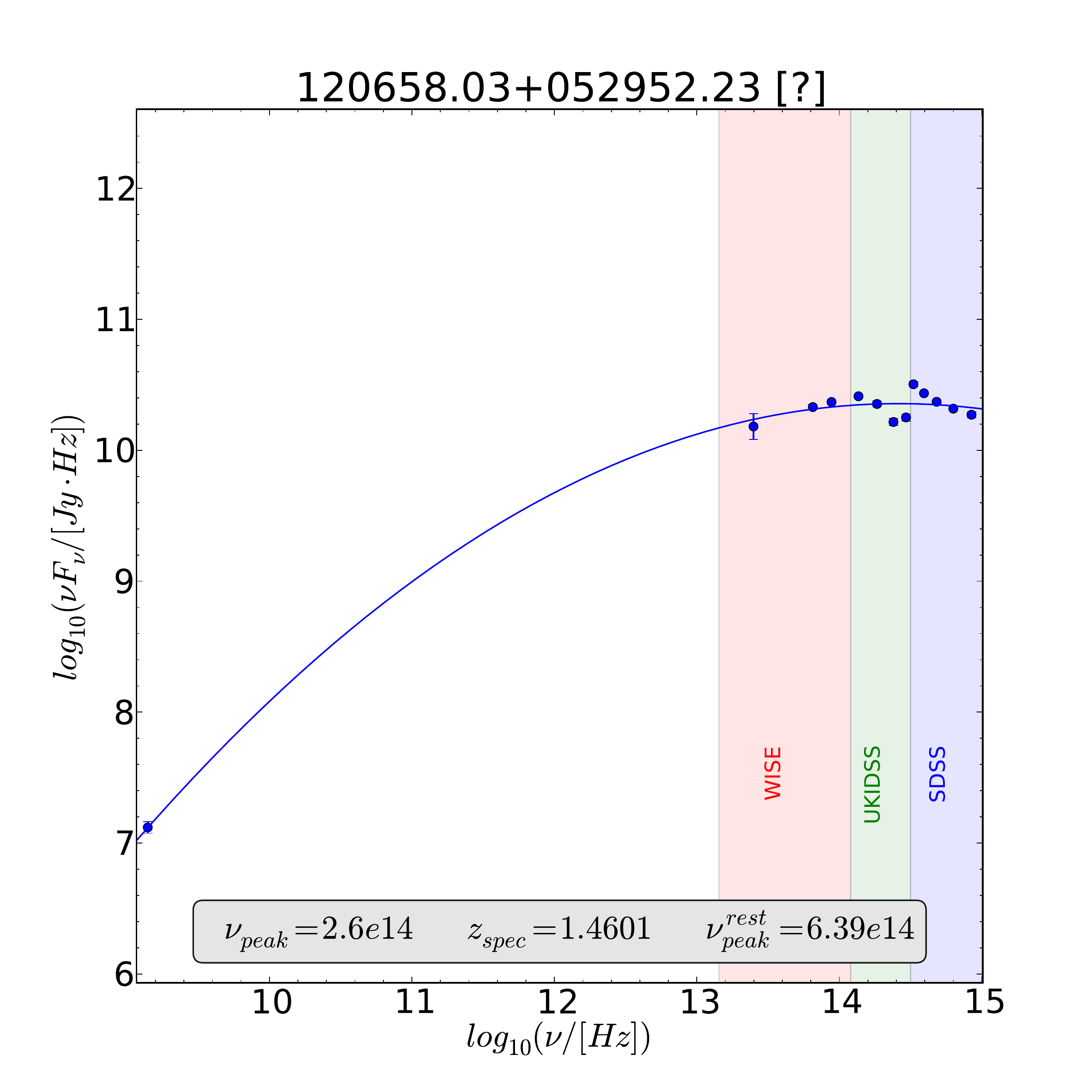}
\includegraphics[width=0.3\textwidth]{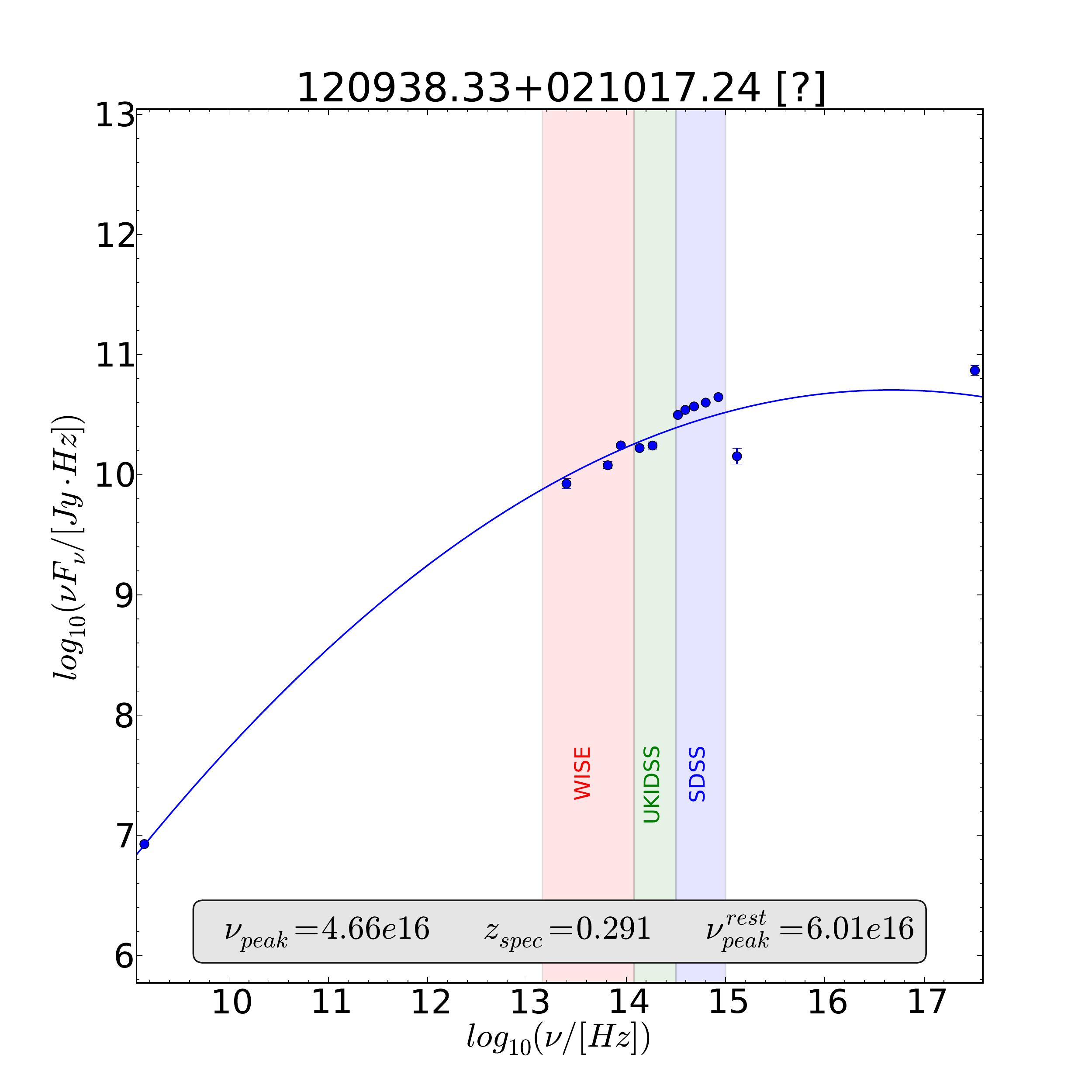}\\

\includegraphics[width=0.3\textwidth]{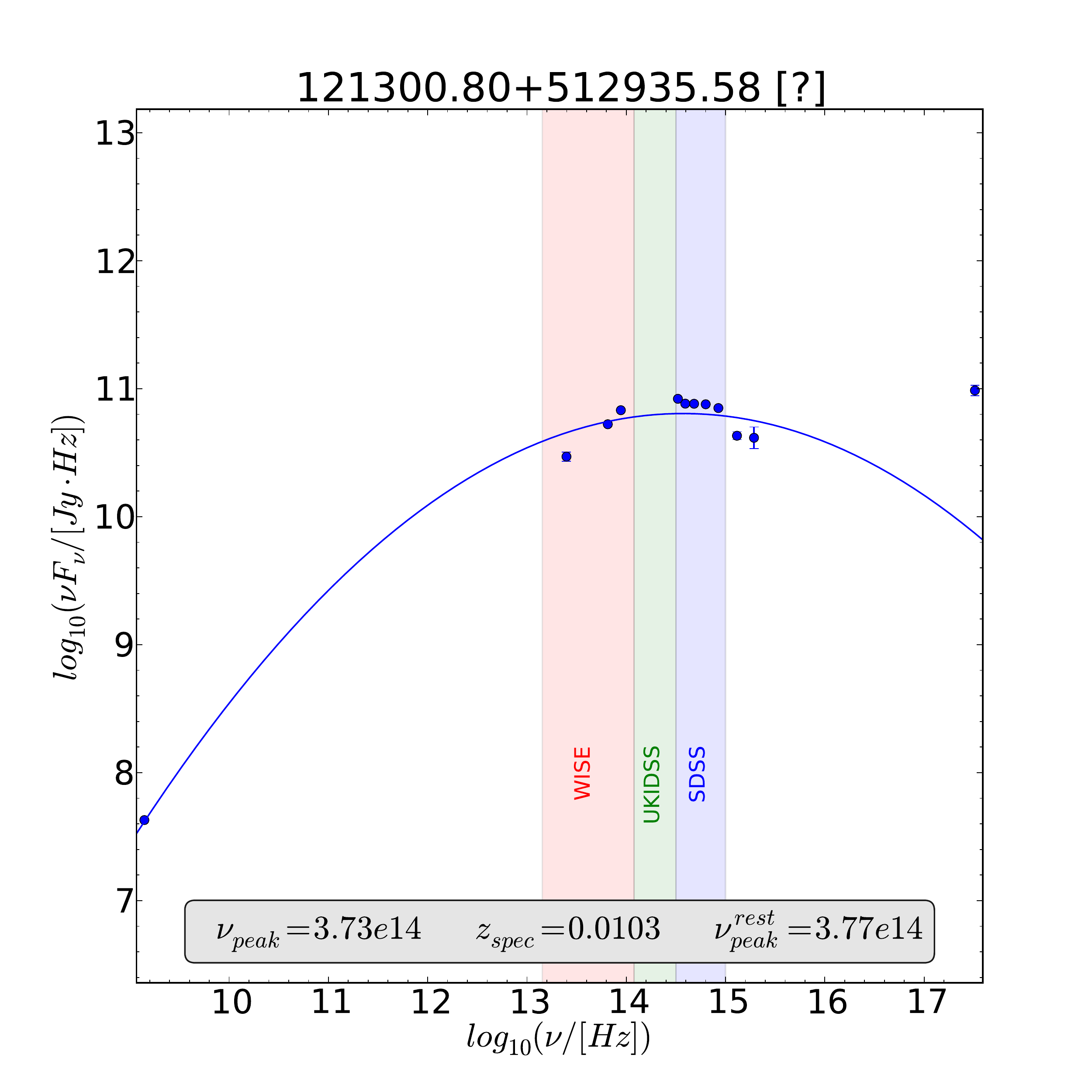}
\includegraphics[width=0.3\textwidth]{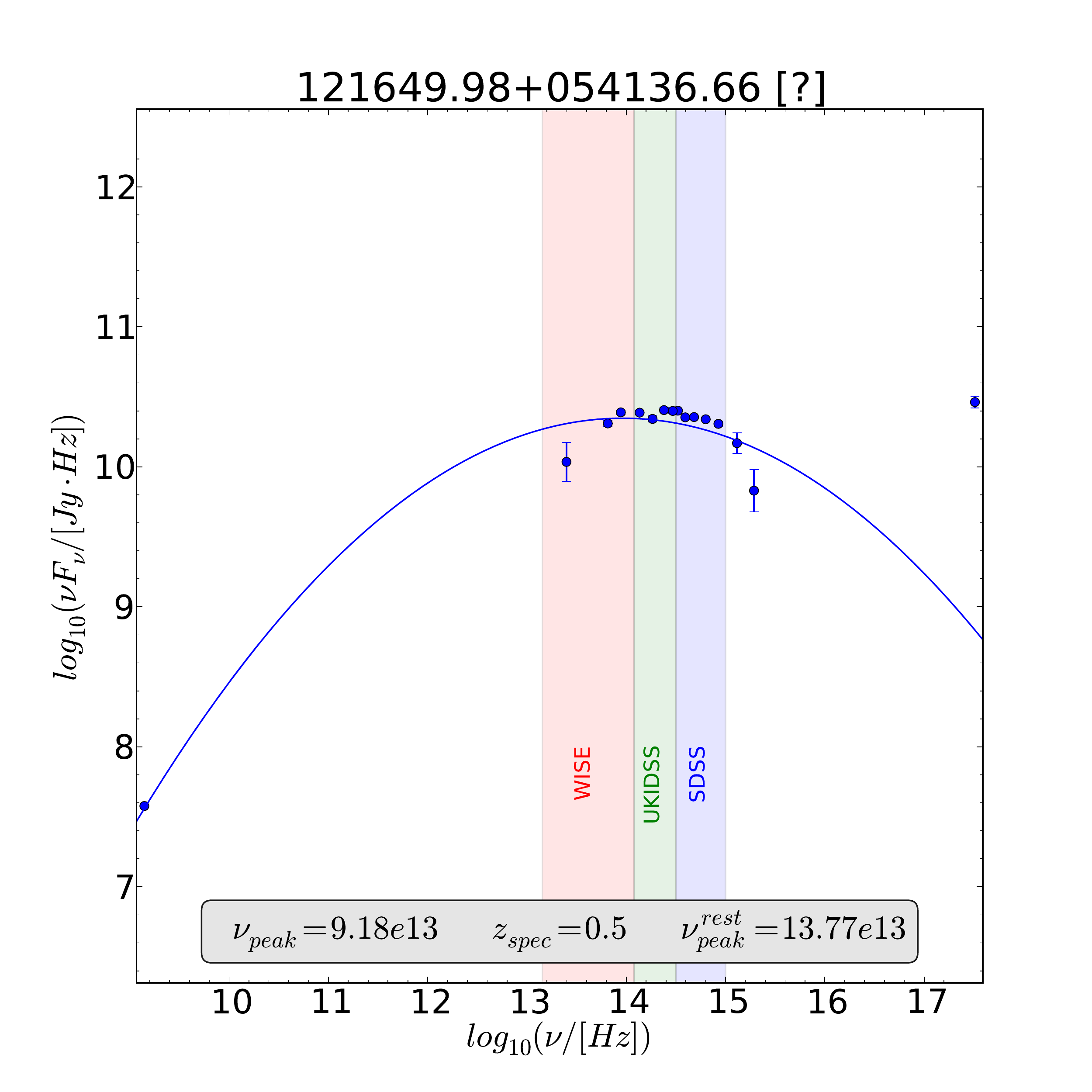}
\includegraphics[width=0.3\textwidth]{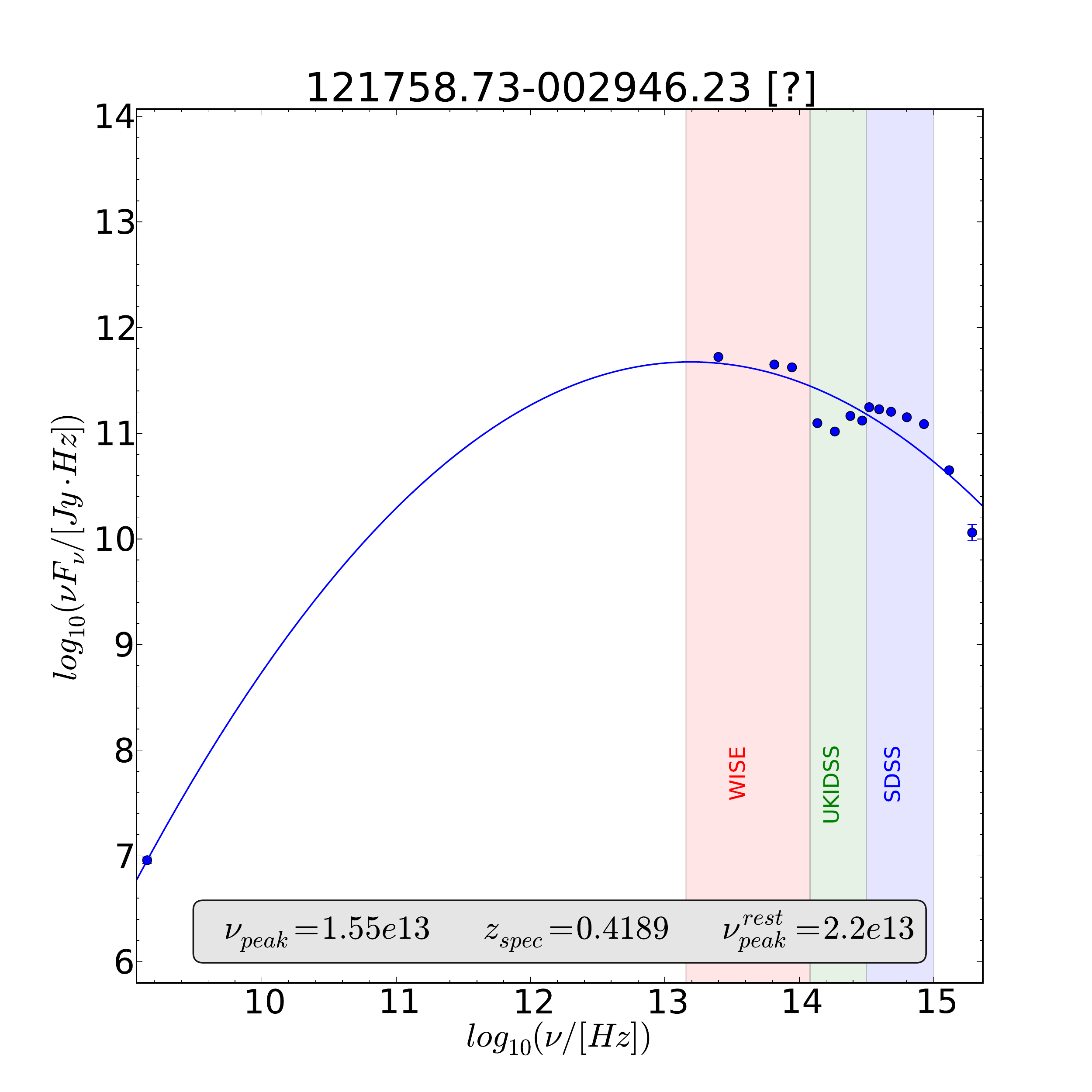}\\

\includegraphics[width=0.3\textwidth]{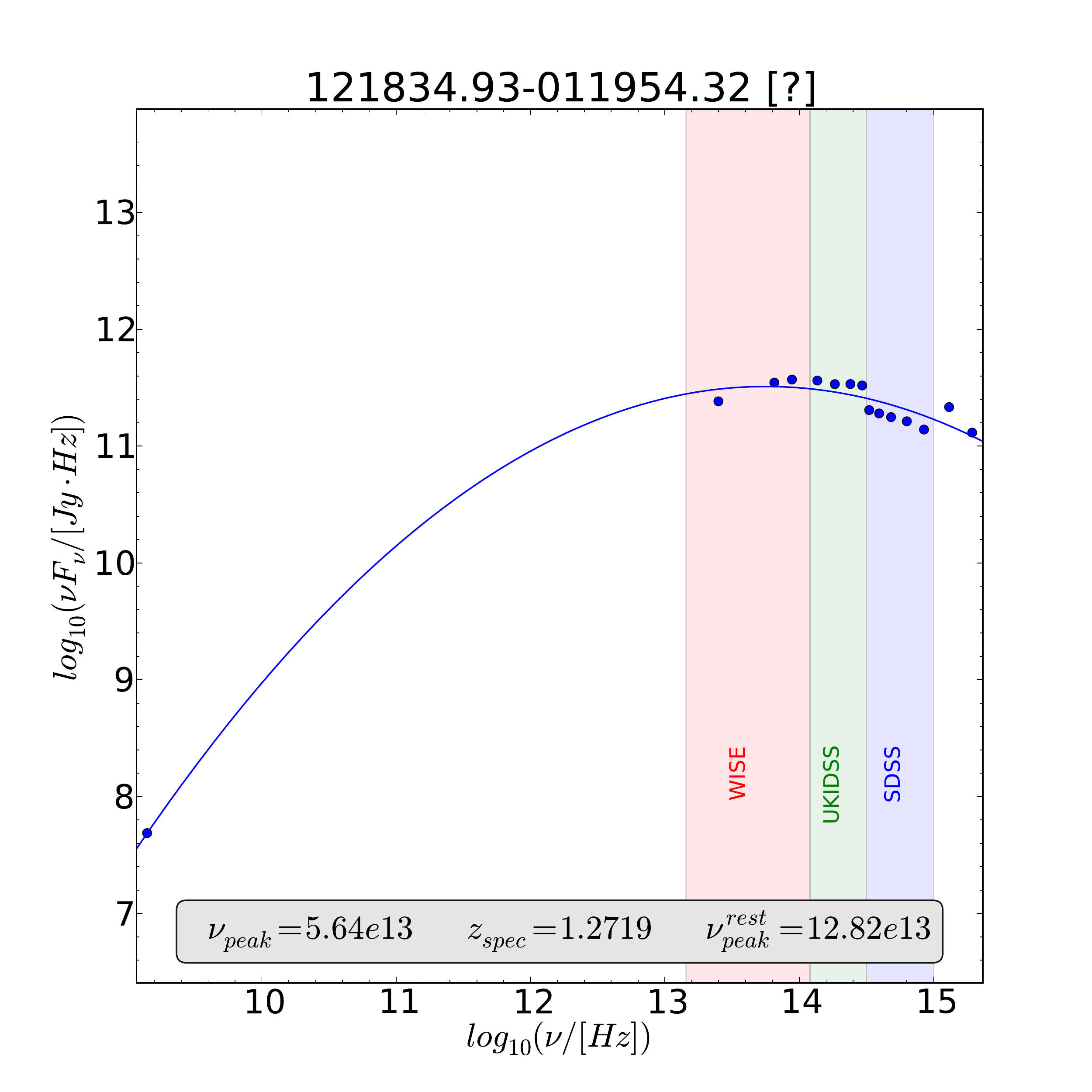}
\includegraphics[width=0.3\textwidth]{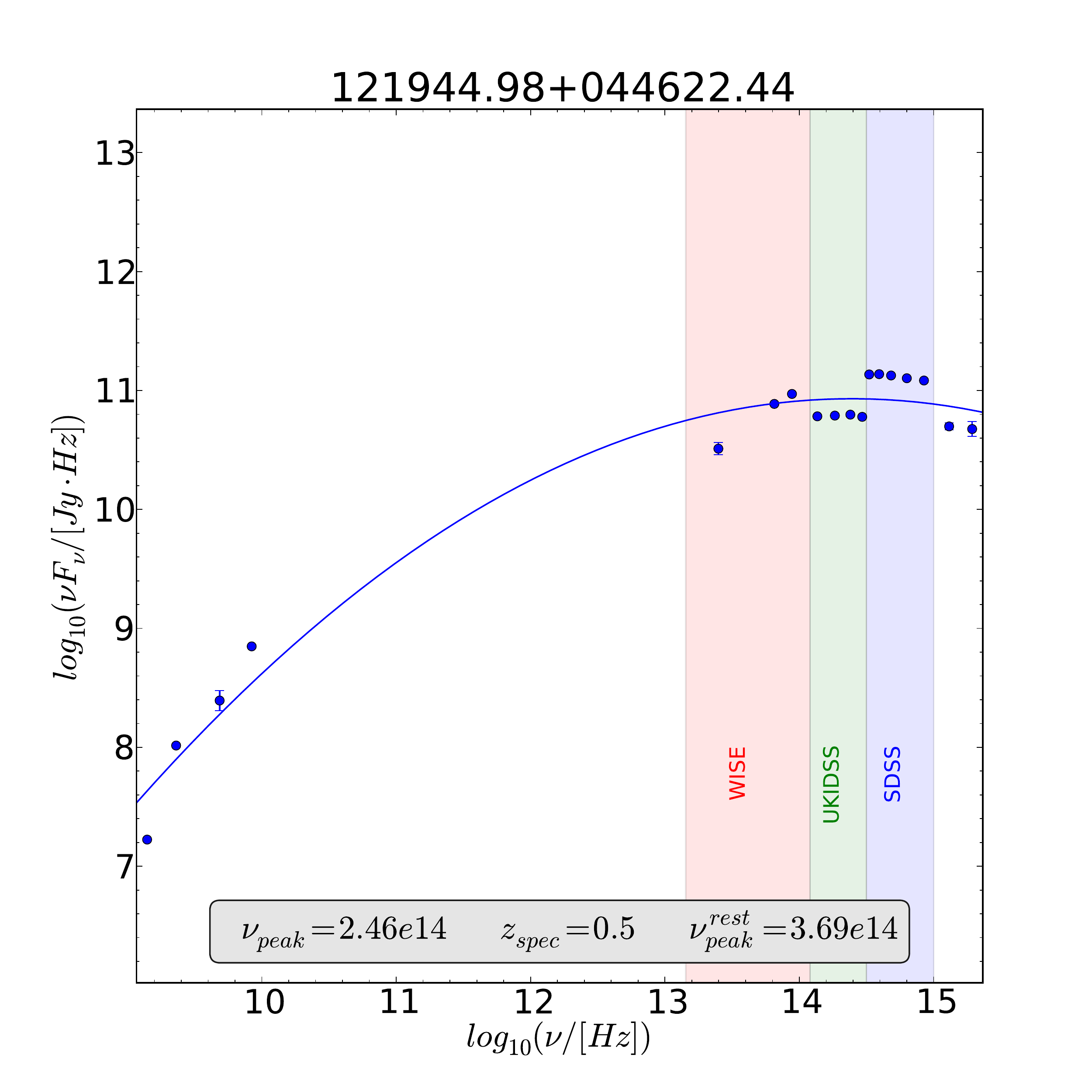}
\includegraphics[width=0.3\textwidth]{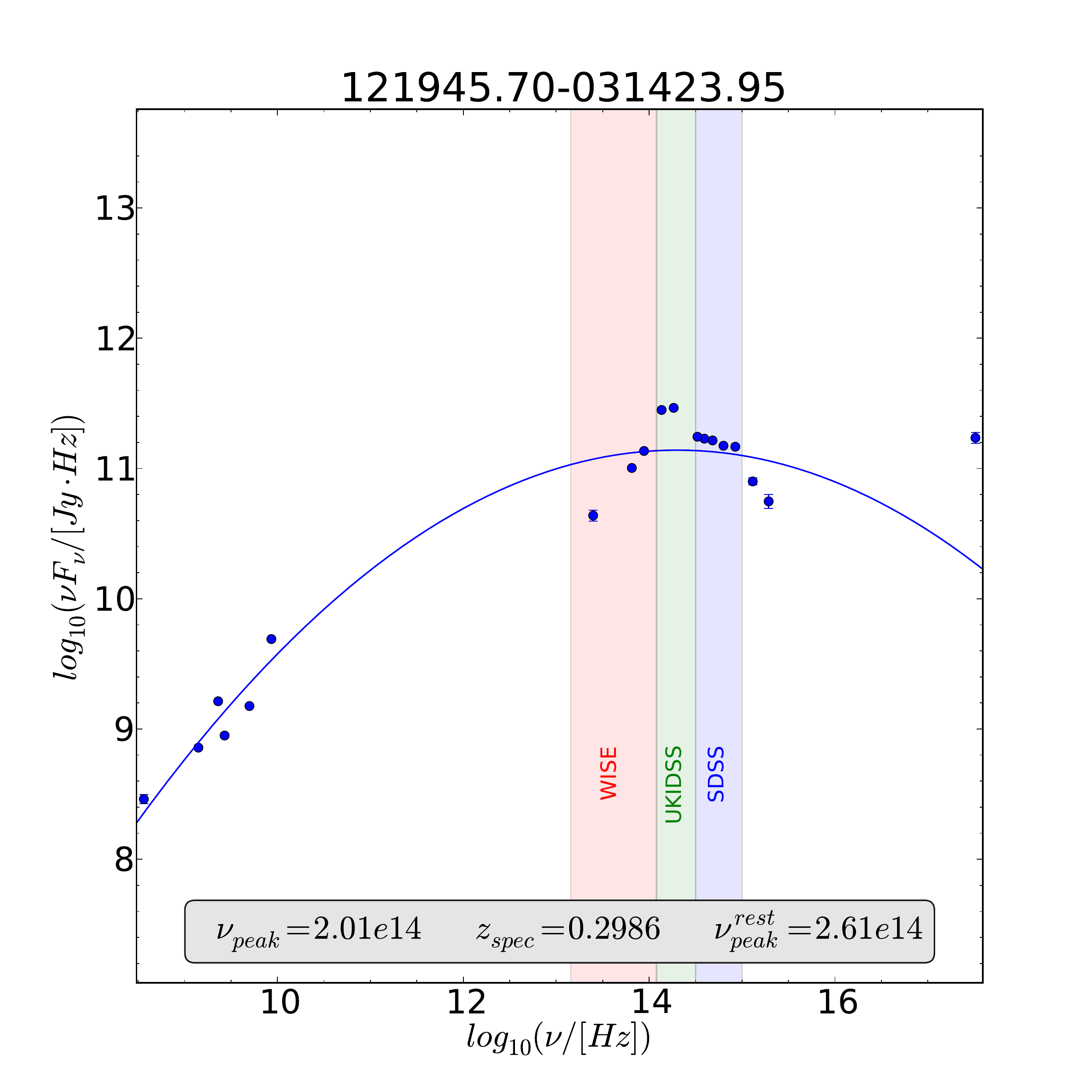}\\

\end{figure*}
\setcounter{figure}{0}
\begin{figure*}[htb!]
\caption{--Continued.}

\includegraphics[width=0.3\textwidth]{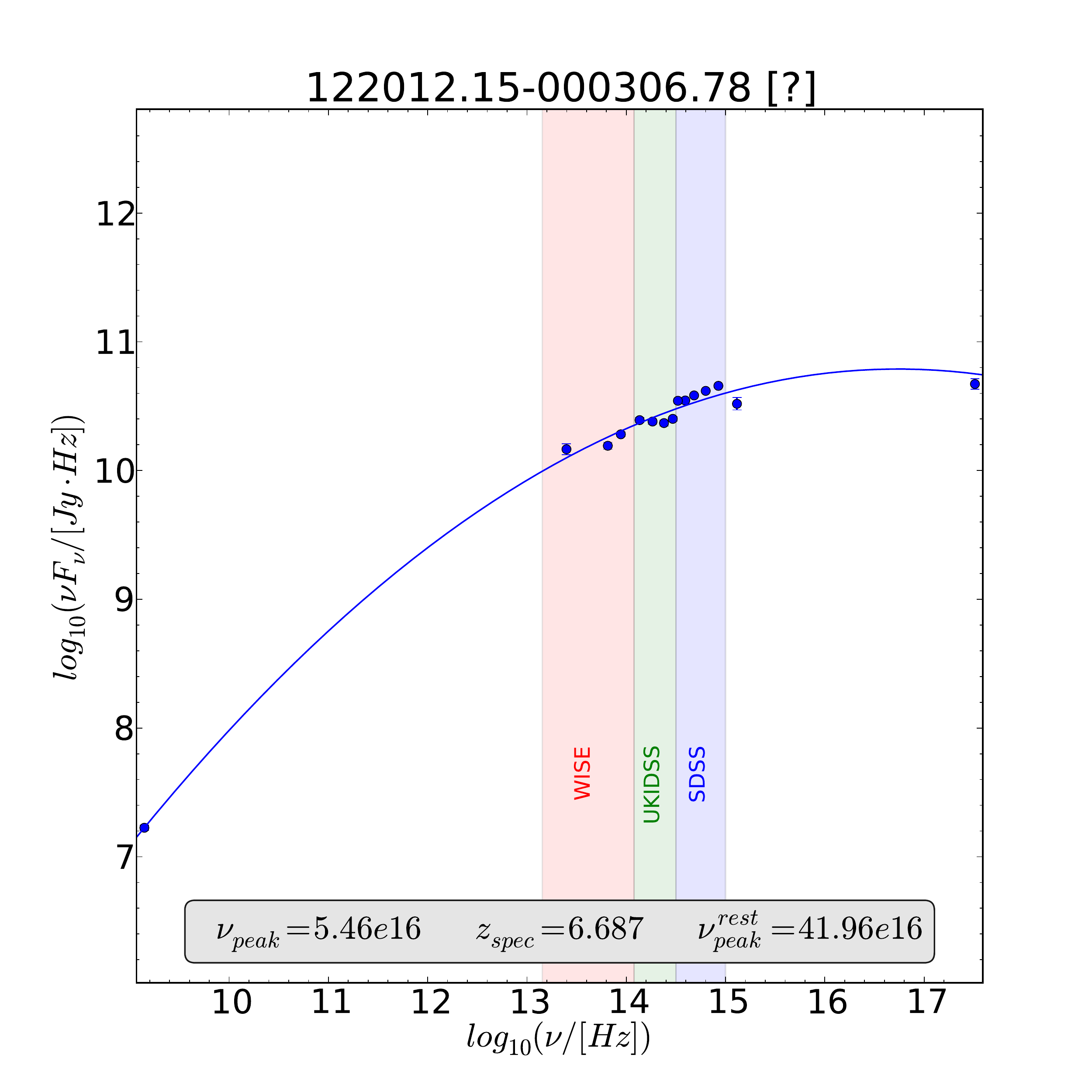}
\includegraphics[width=0.3\textwidth]{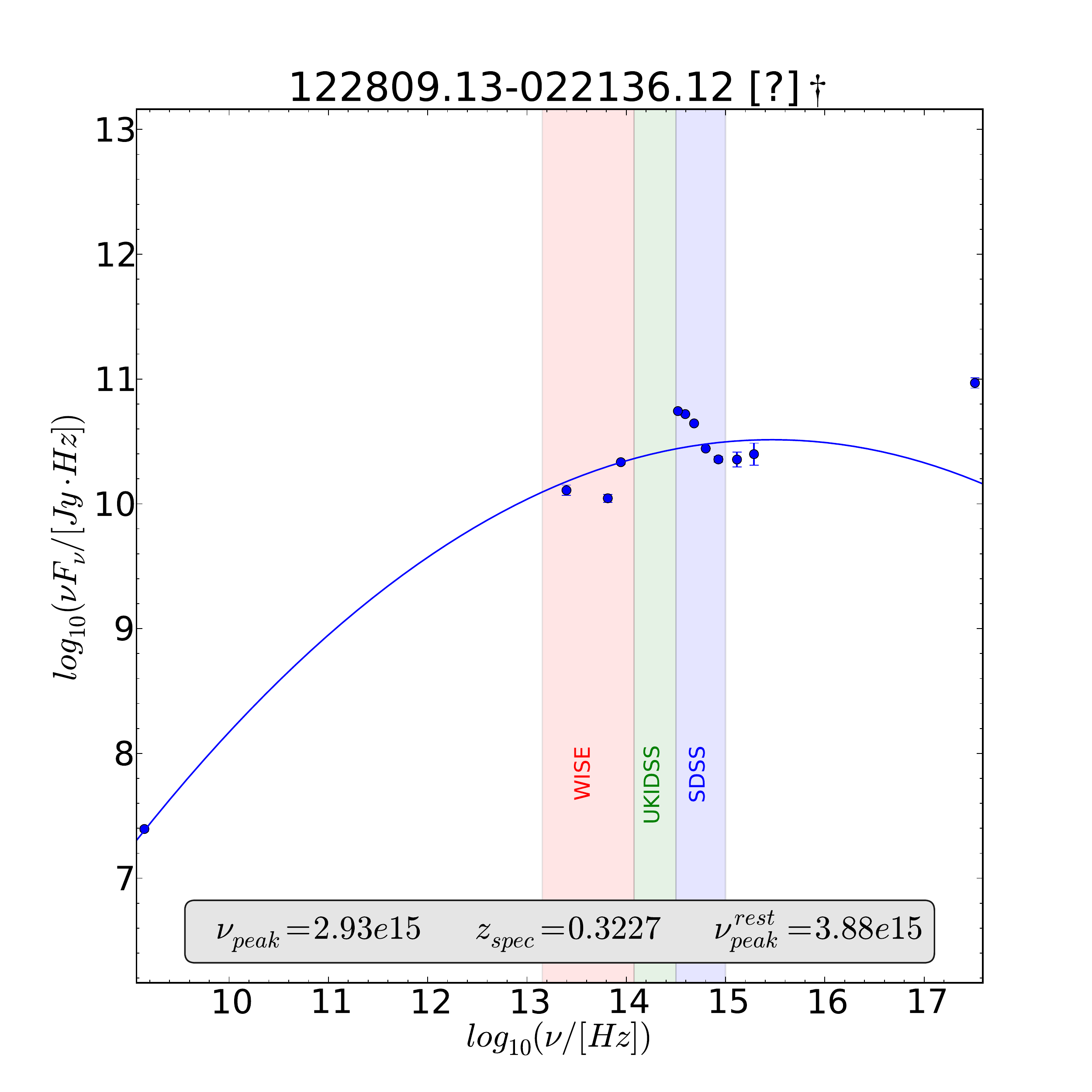}
\includegraphics[width=0.3\textwidth]{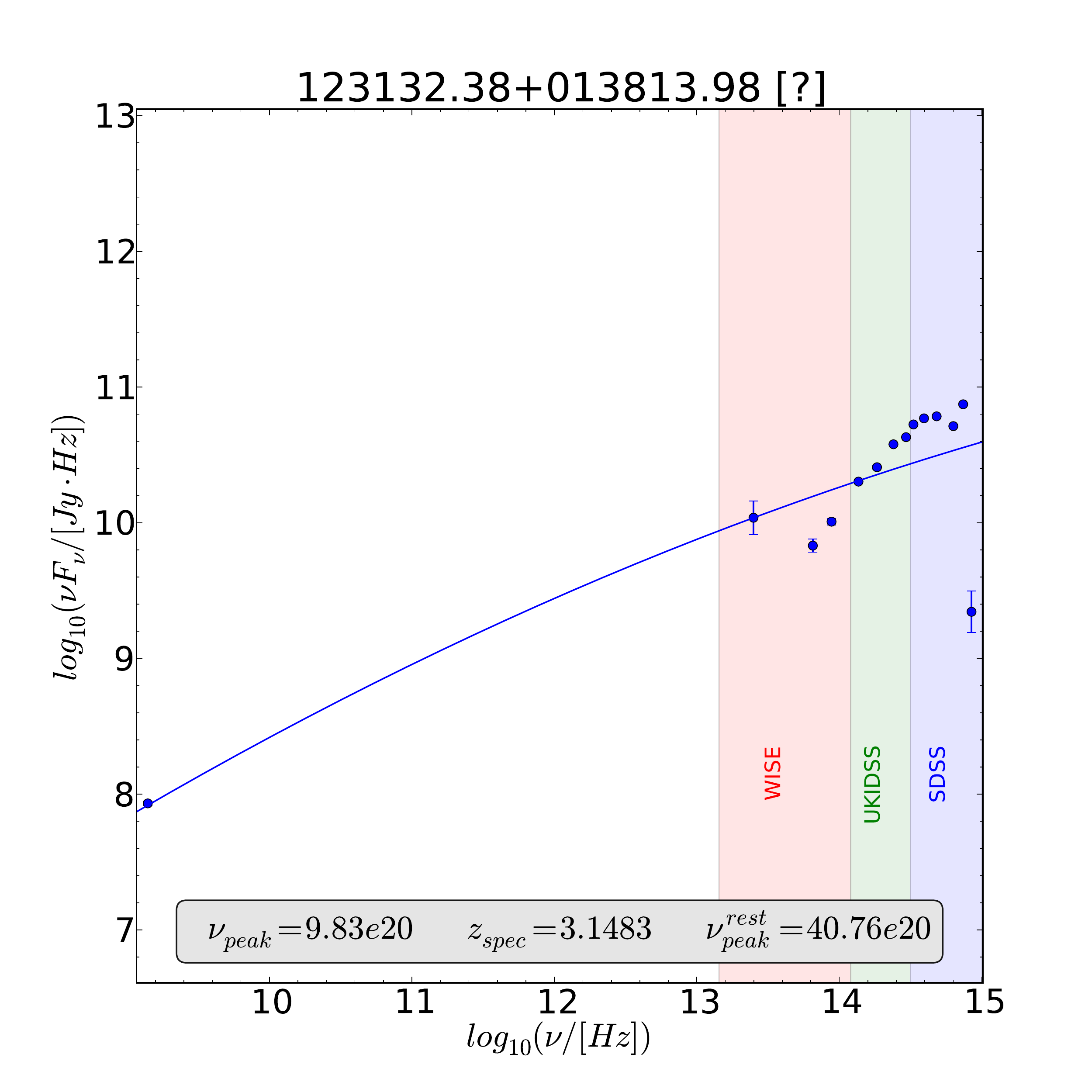}\\

\includegraphics[width=0.3\textwidth]{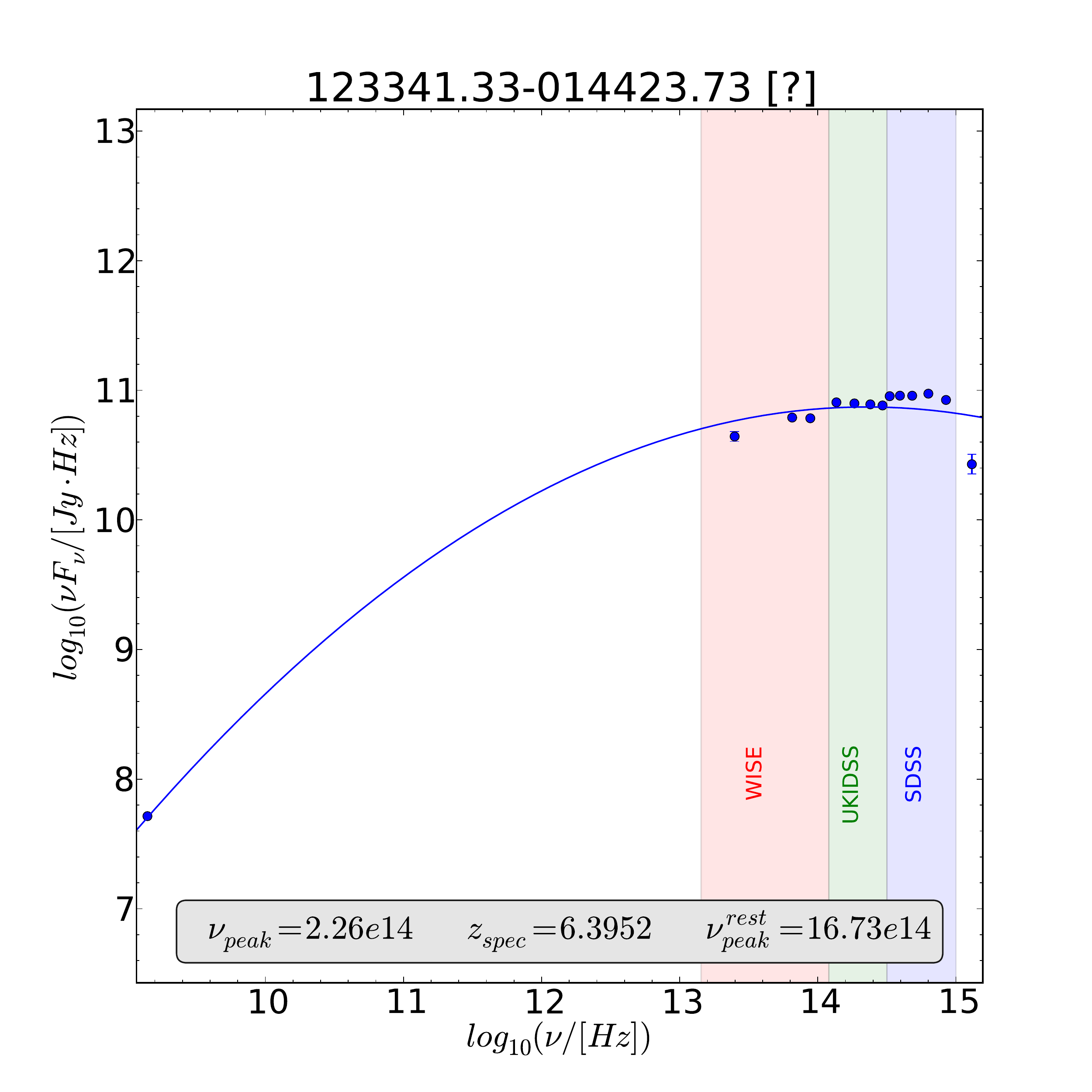}
\includegraphics[width=0.3\textwidth]{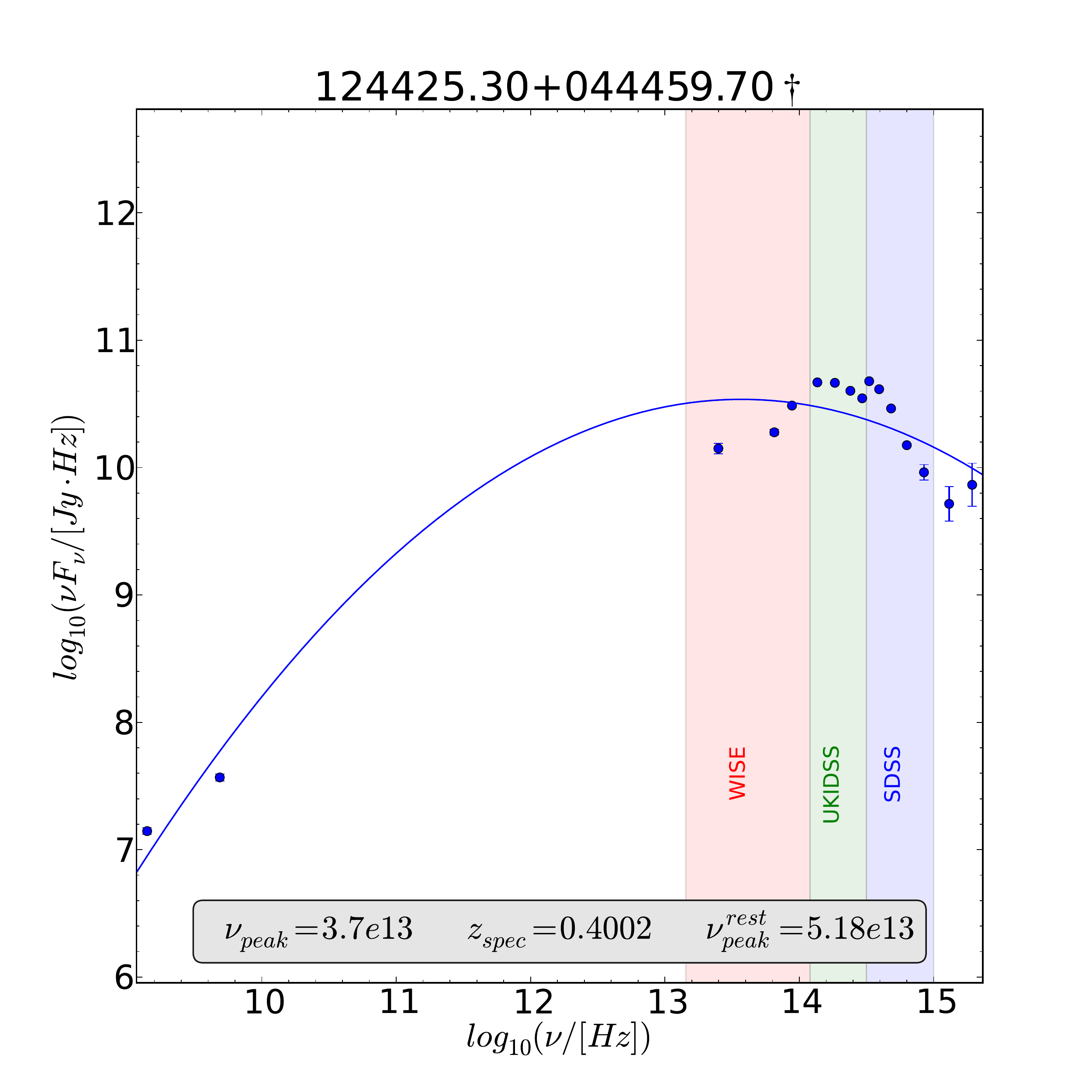}
\includegraphics[width=0.3\textwidth]{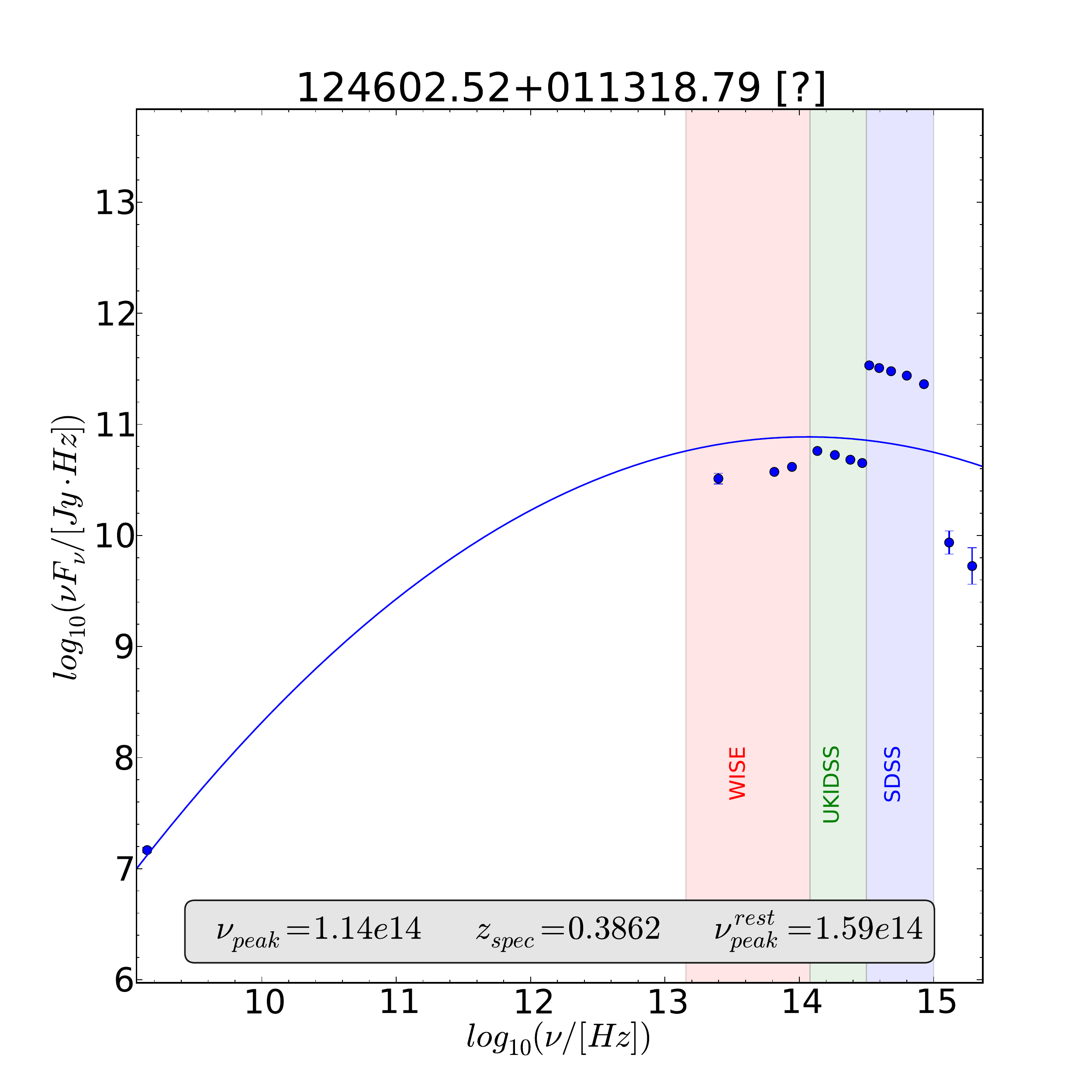}\\

\includegraphics[width=0.3\textwidth]{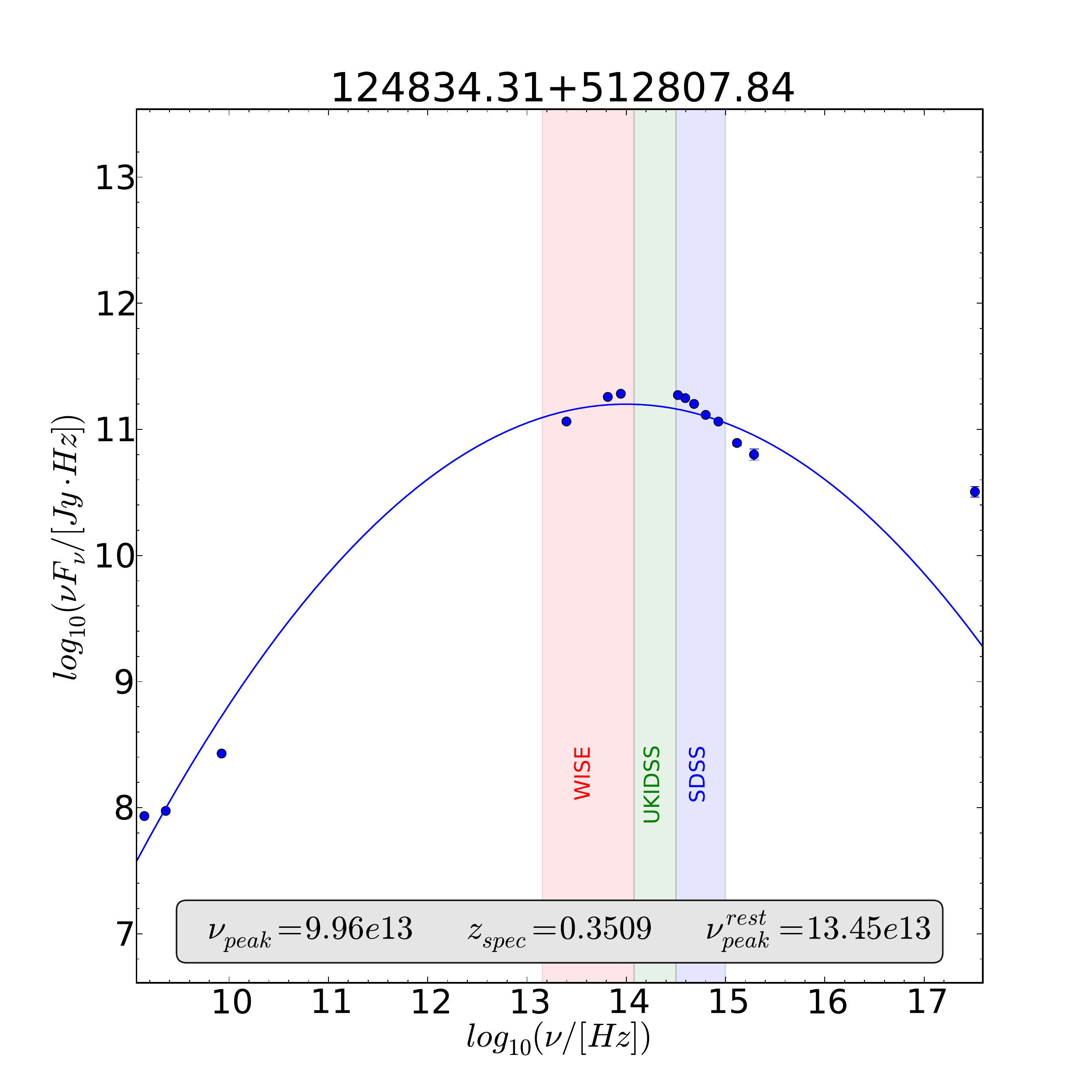}
\includegraphics[width=0.3\textwidth]{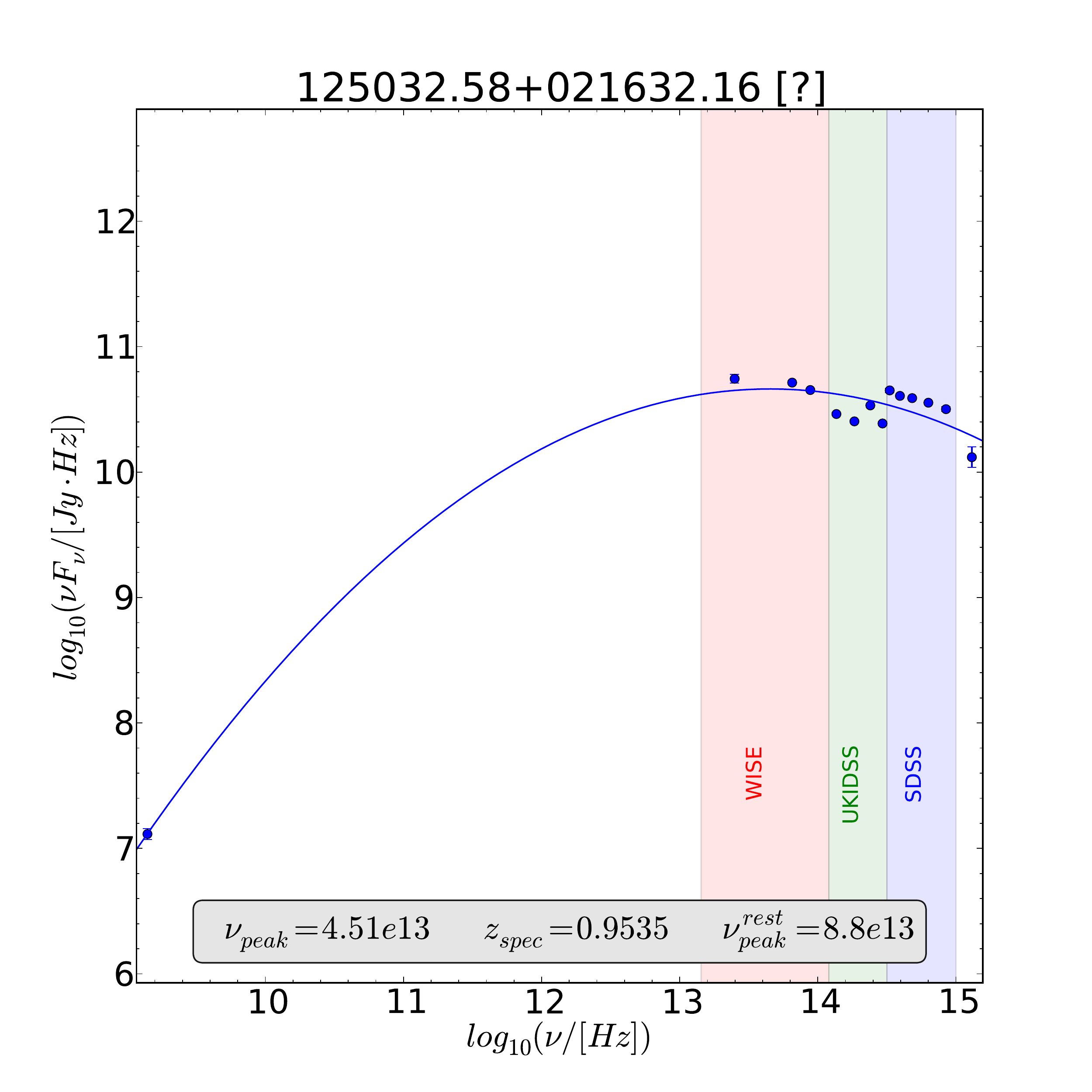}
\includegraphics[width=0.3\textwidth]{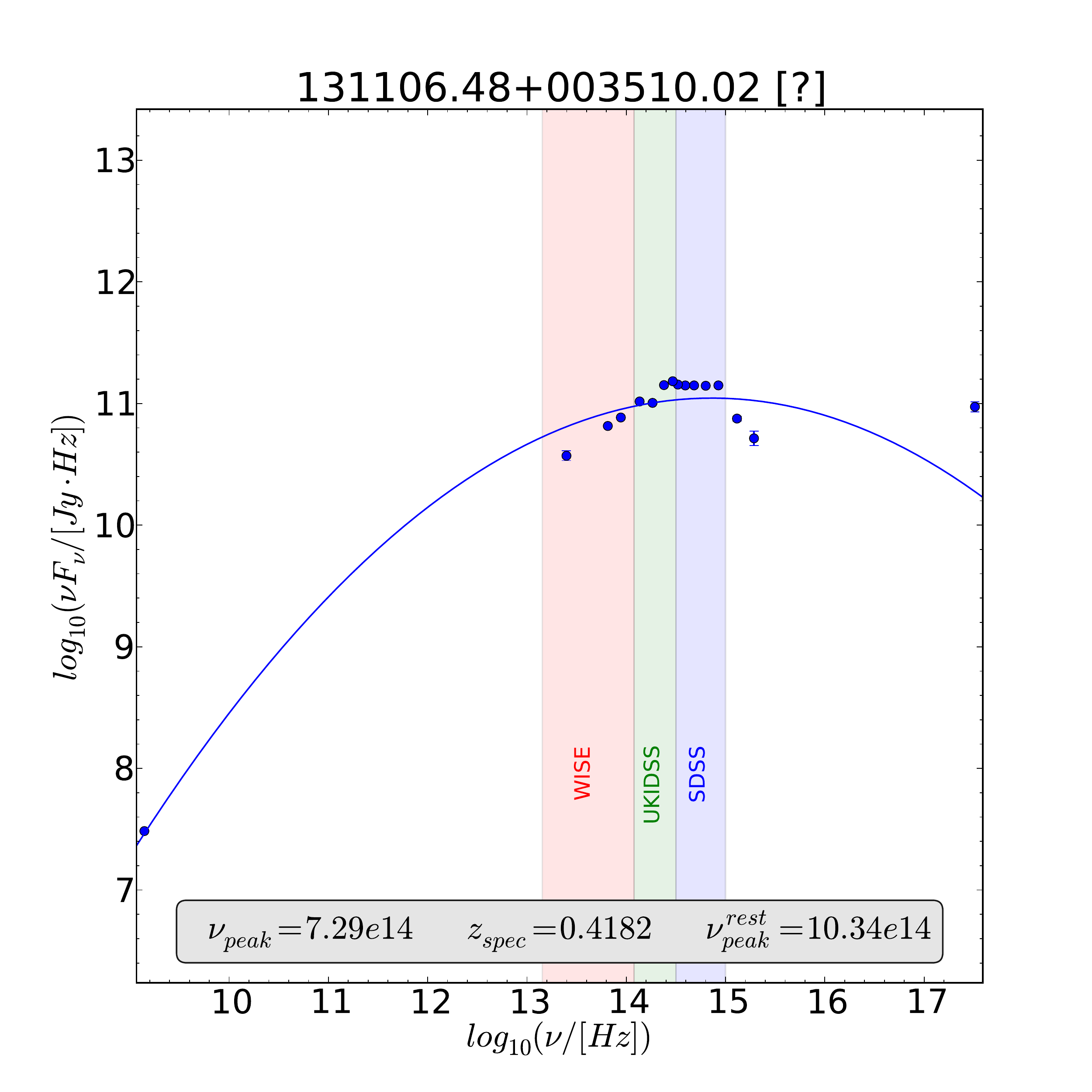}\\

\includegraphics[width=0.3\textwidth]{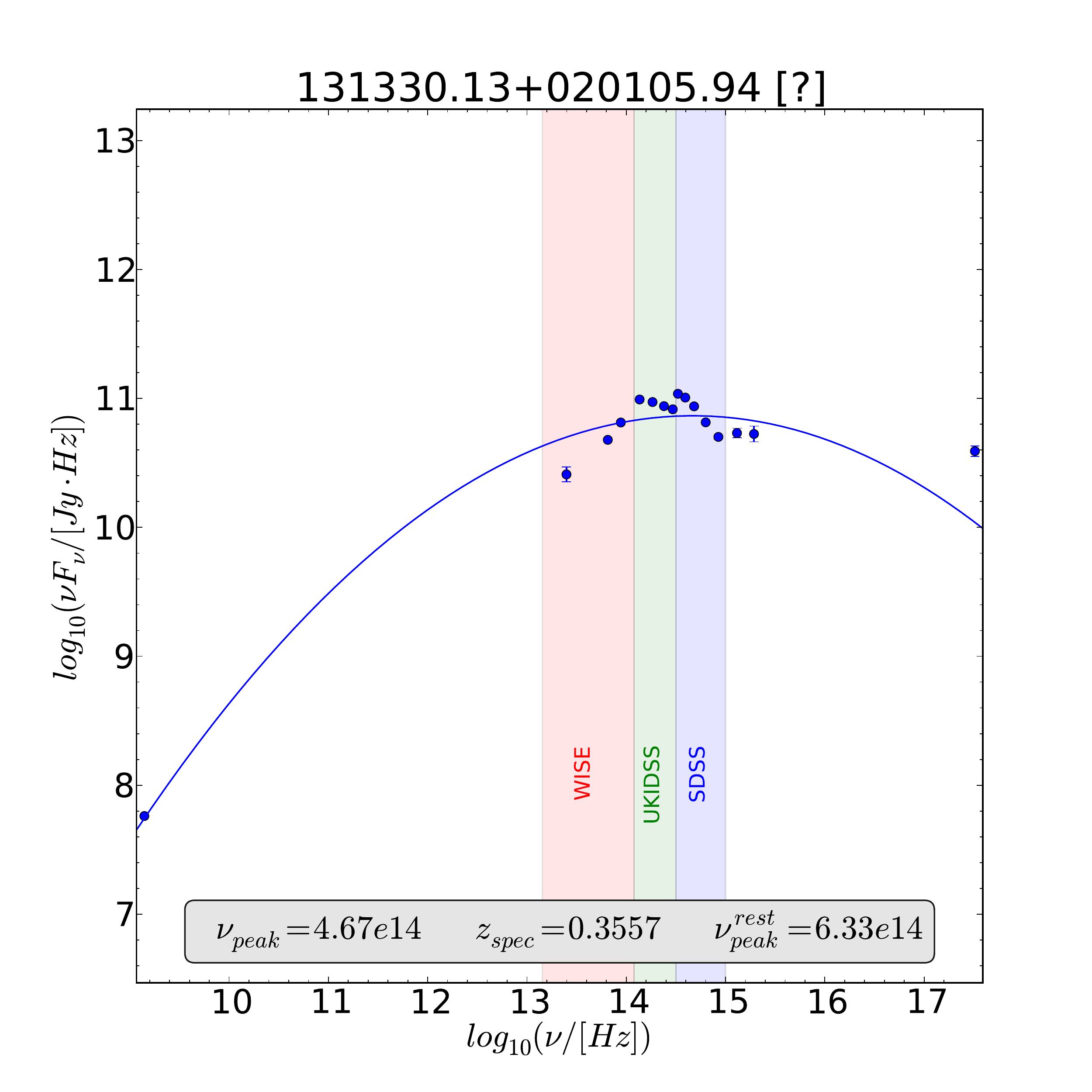}
\includegraphics[width=0.3\textwidth]{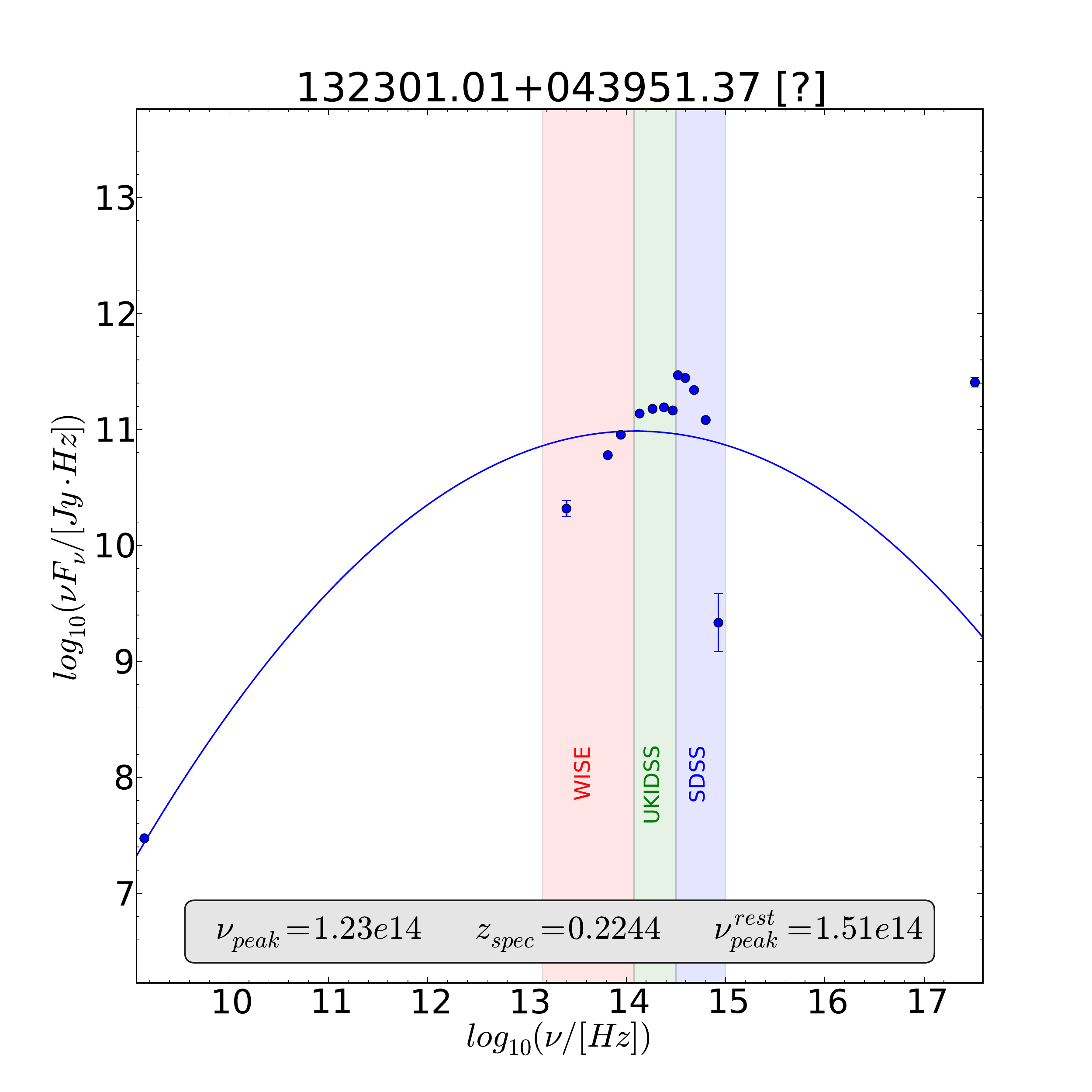}
\includegraphics[width=0.3\textwidth]{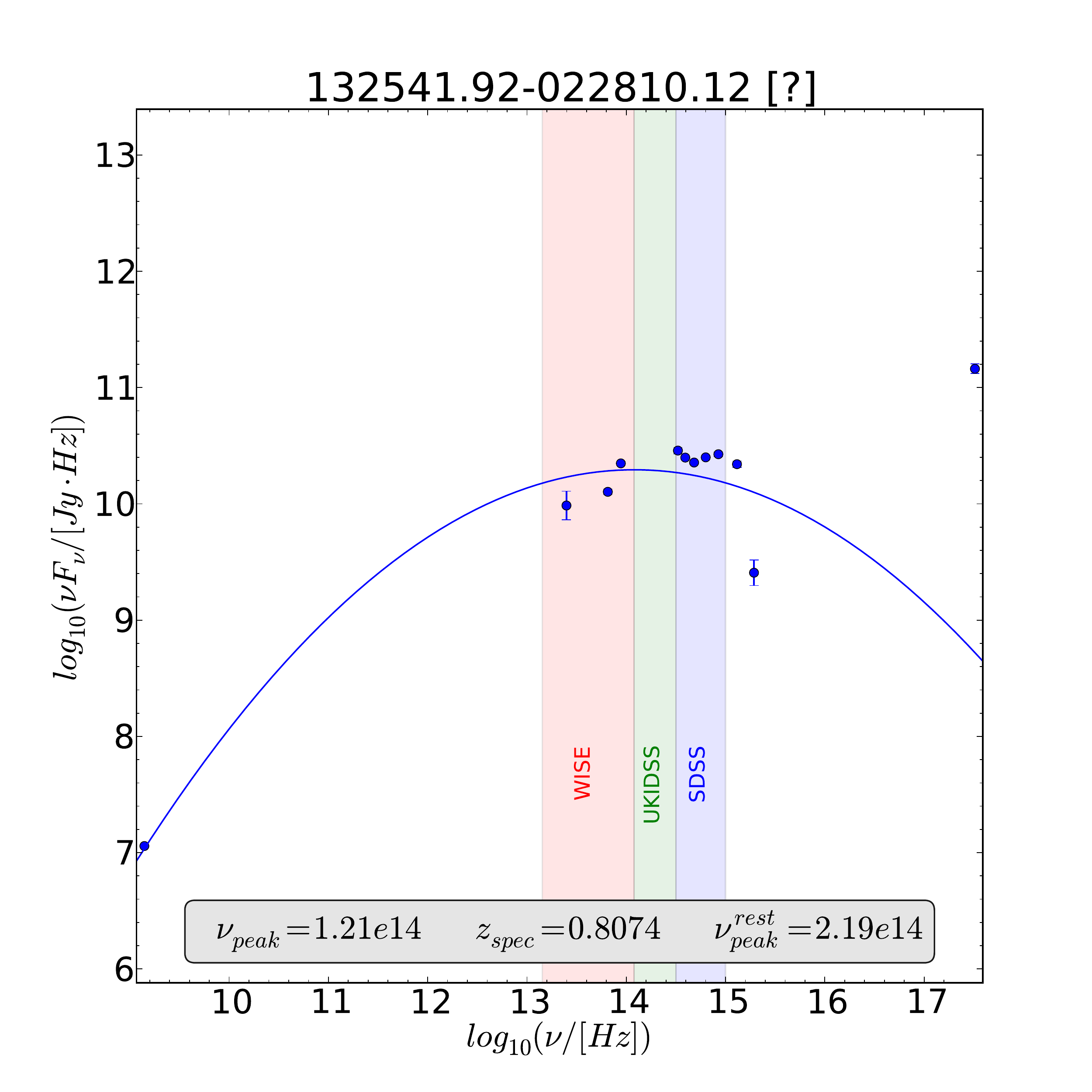}\\

\end{figure*}
\setcounter{figure}{0}
\begin{figure*}[htb!]
\caption{--Continued.}

\includegraphics[width=0.3\textwidth]{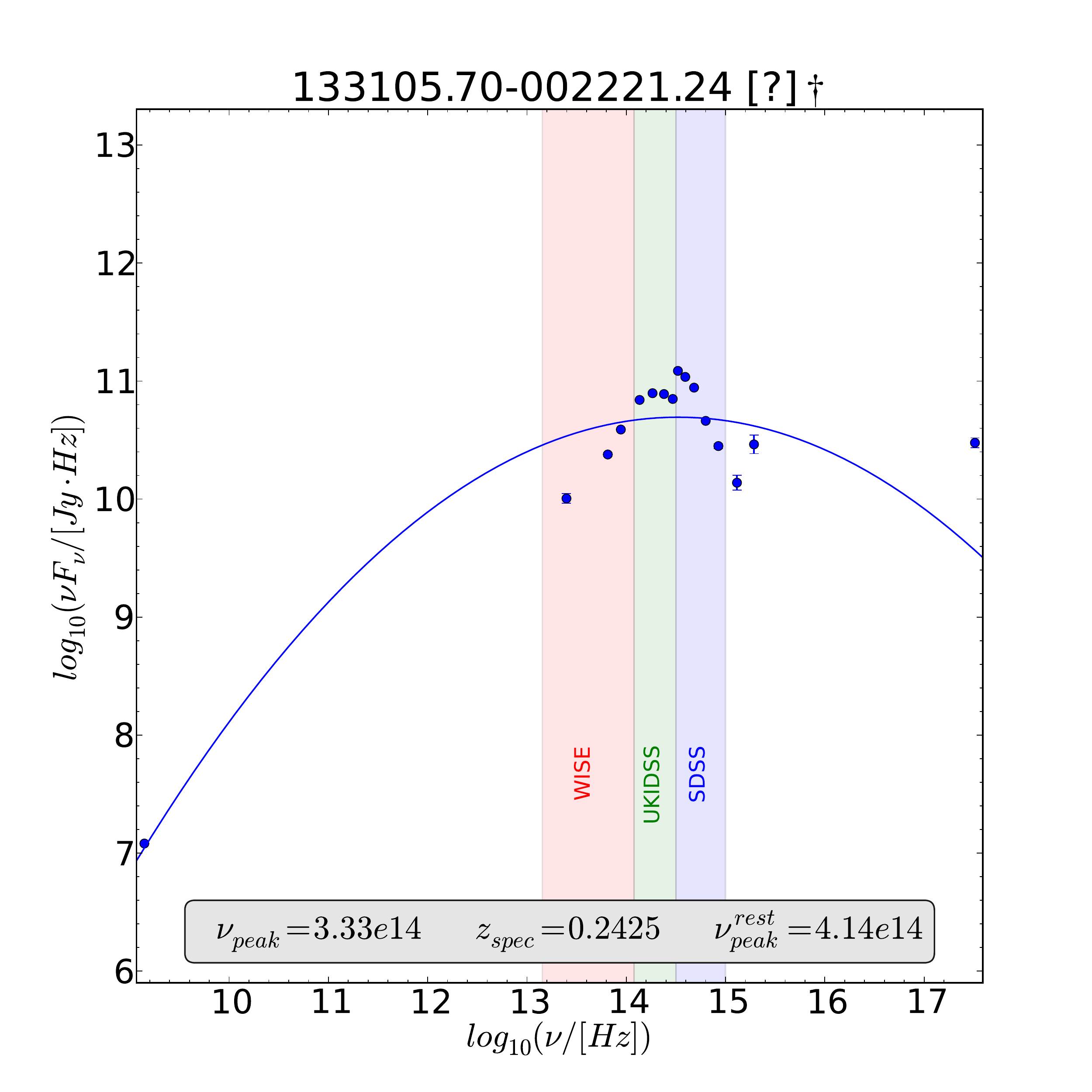}
\includegraphics[width=0.3\textwidth]{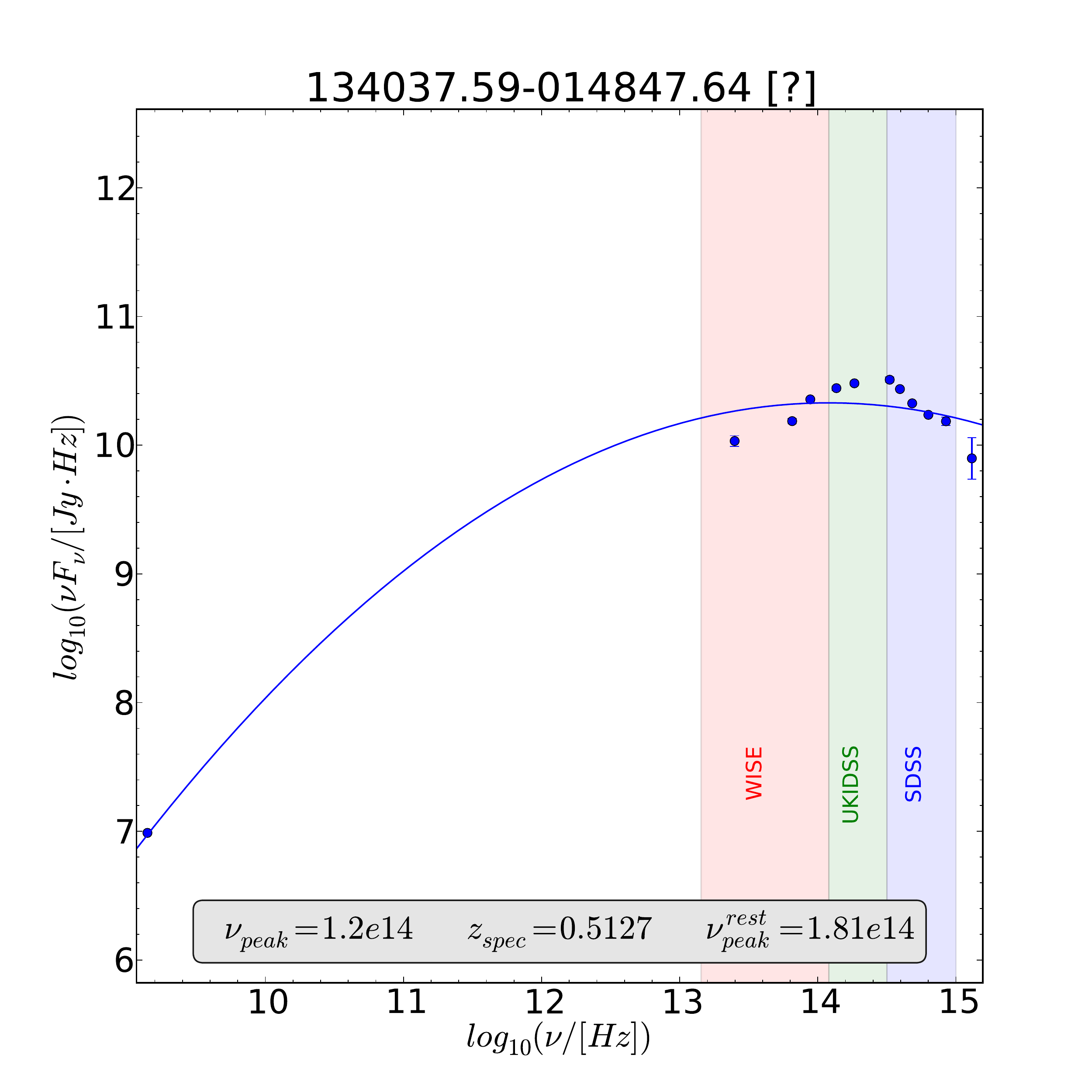}
\includegraphics[width=0.3\textwidth]{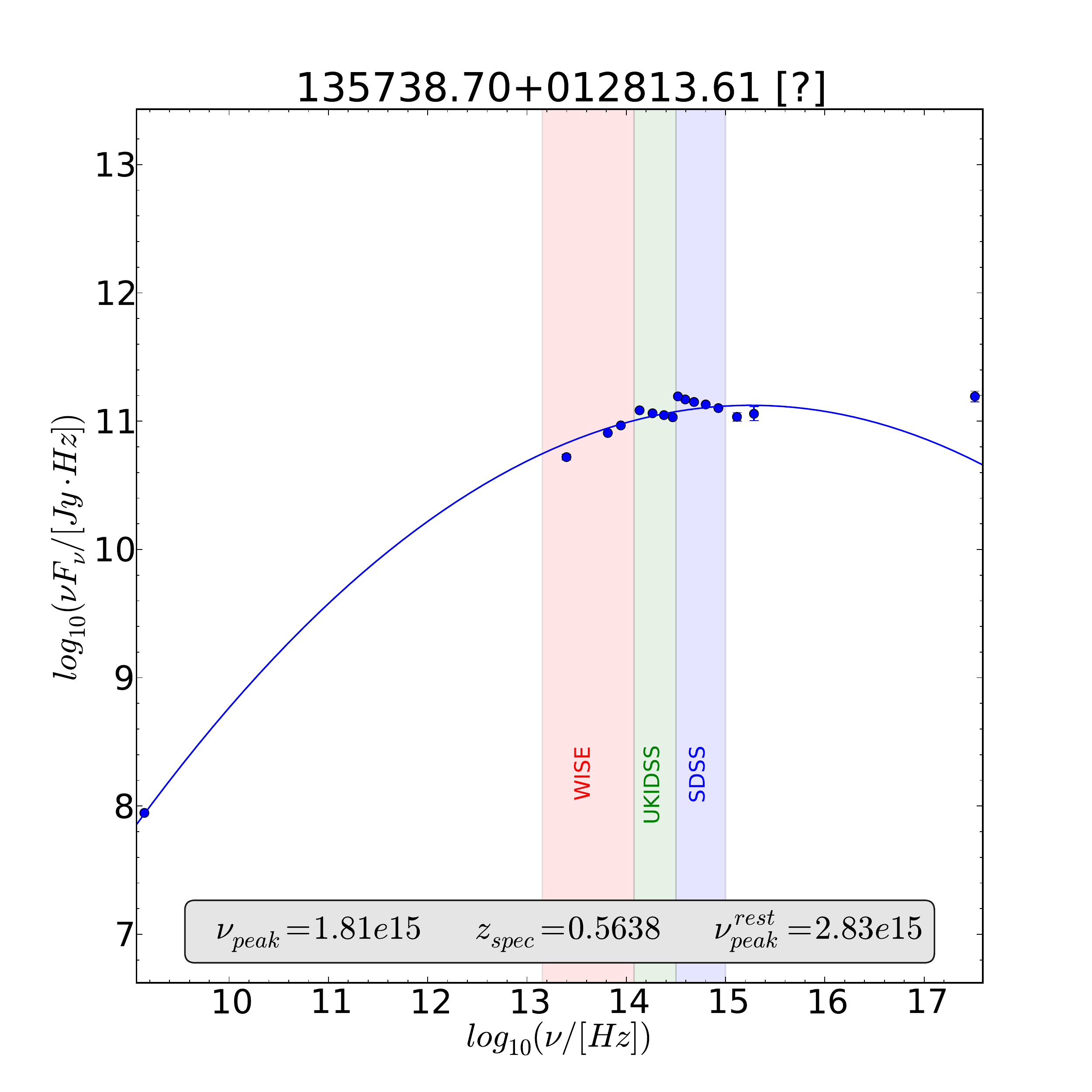}\\

\includegraphics[width=0.3\textwidth]{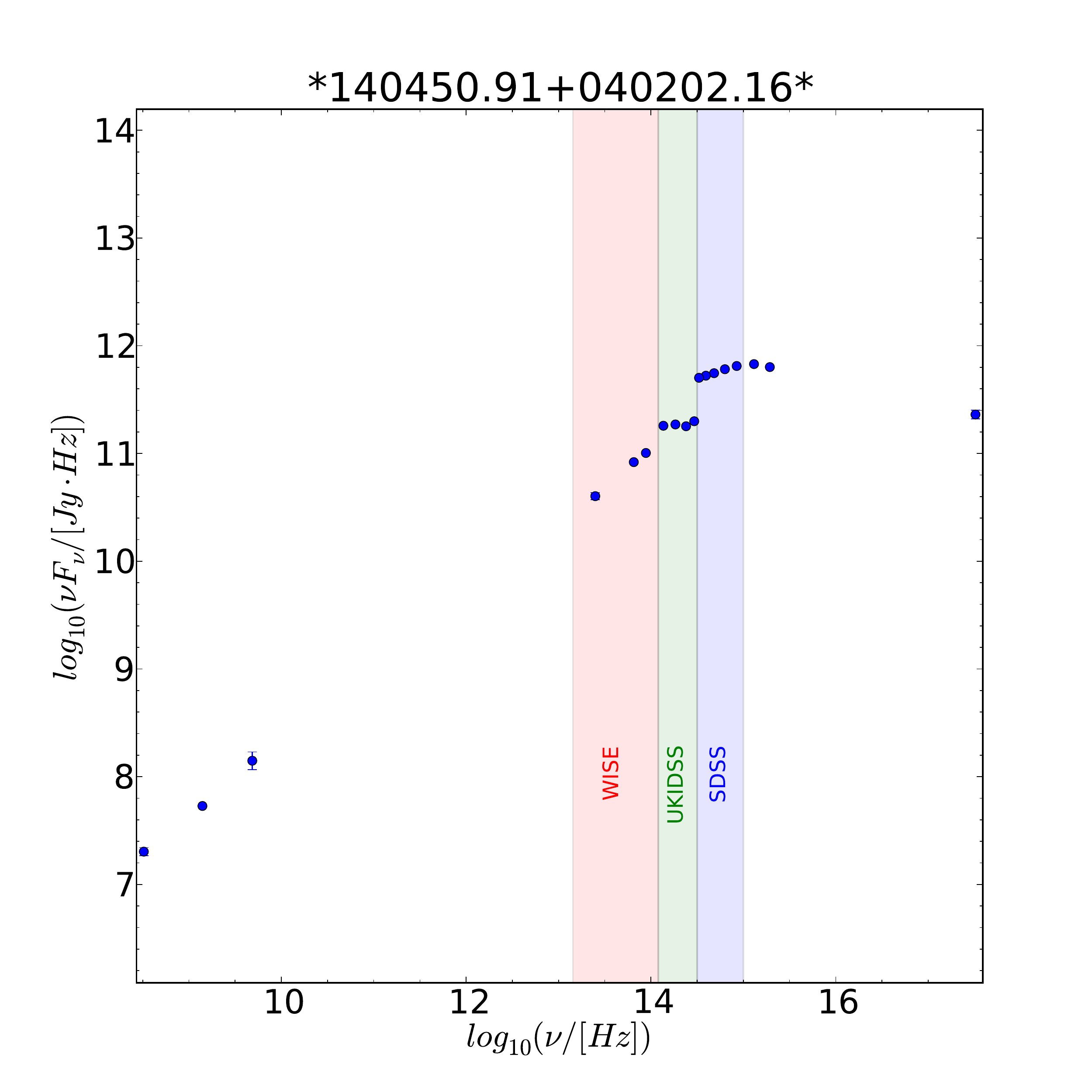}
\includegraphics[width=0.3\textwidth]{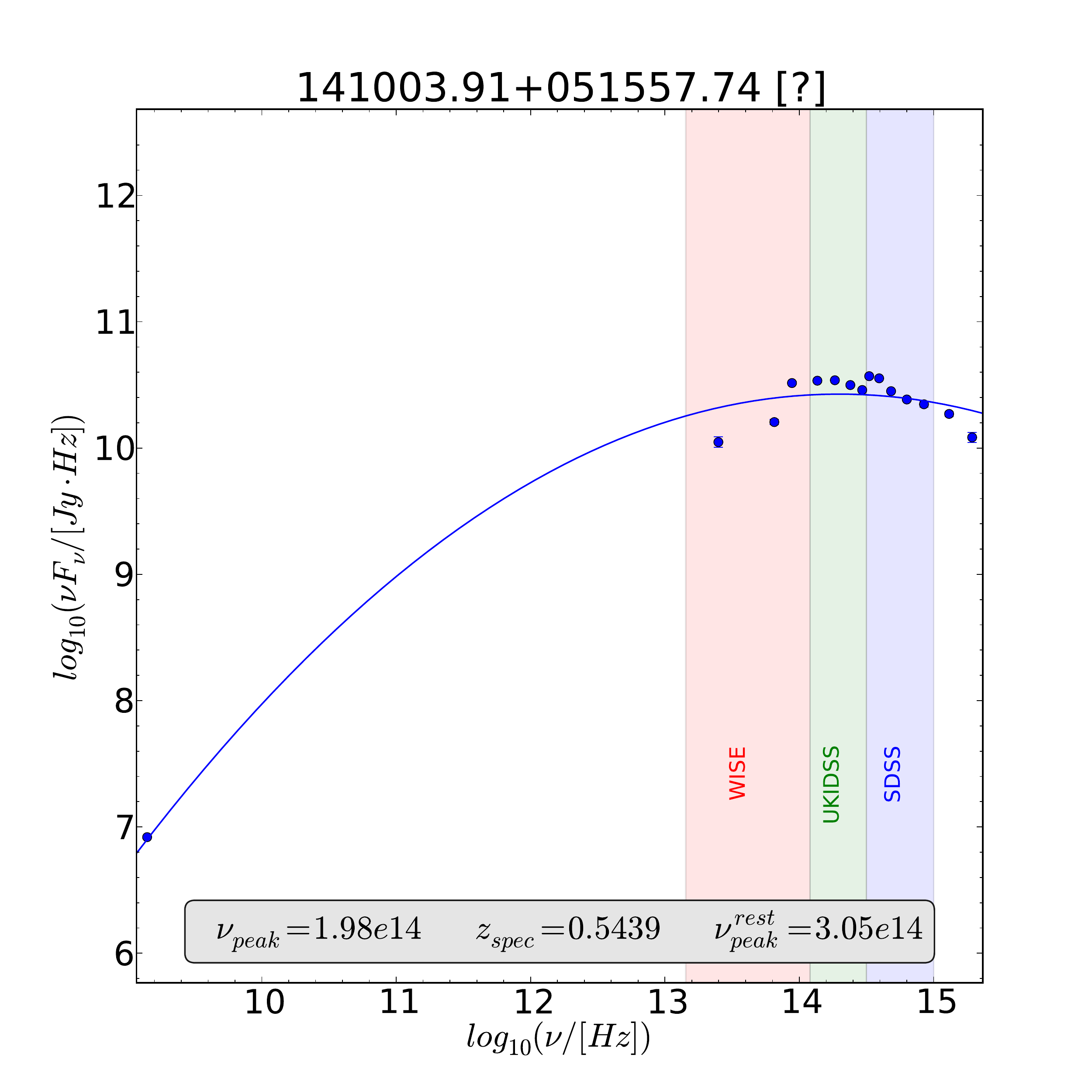}
\includegraphics[width=0.3\textwidth]{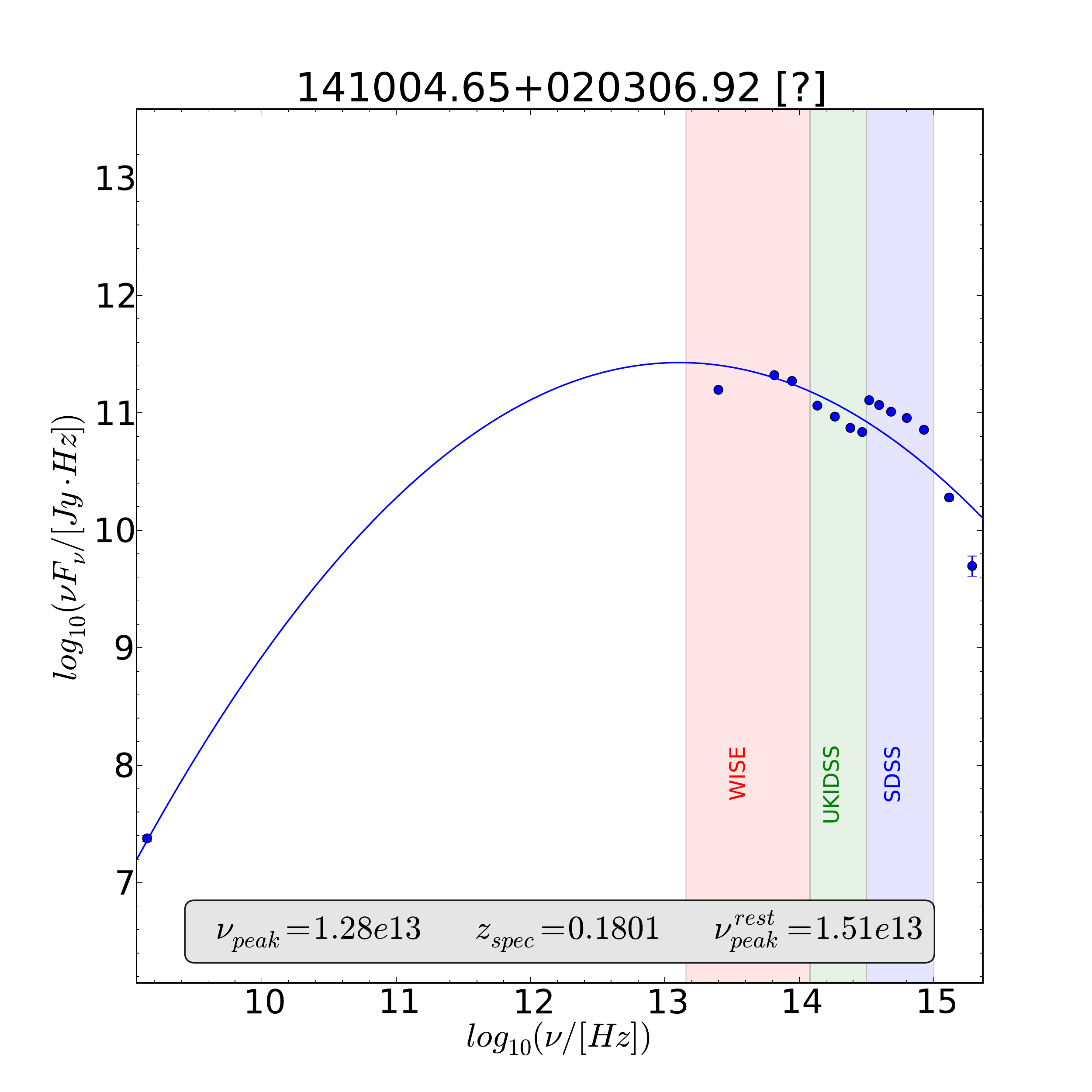}\\

\includegraphics[width=0.3\textwidth]{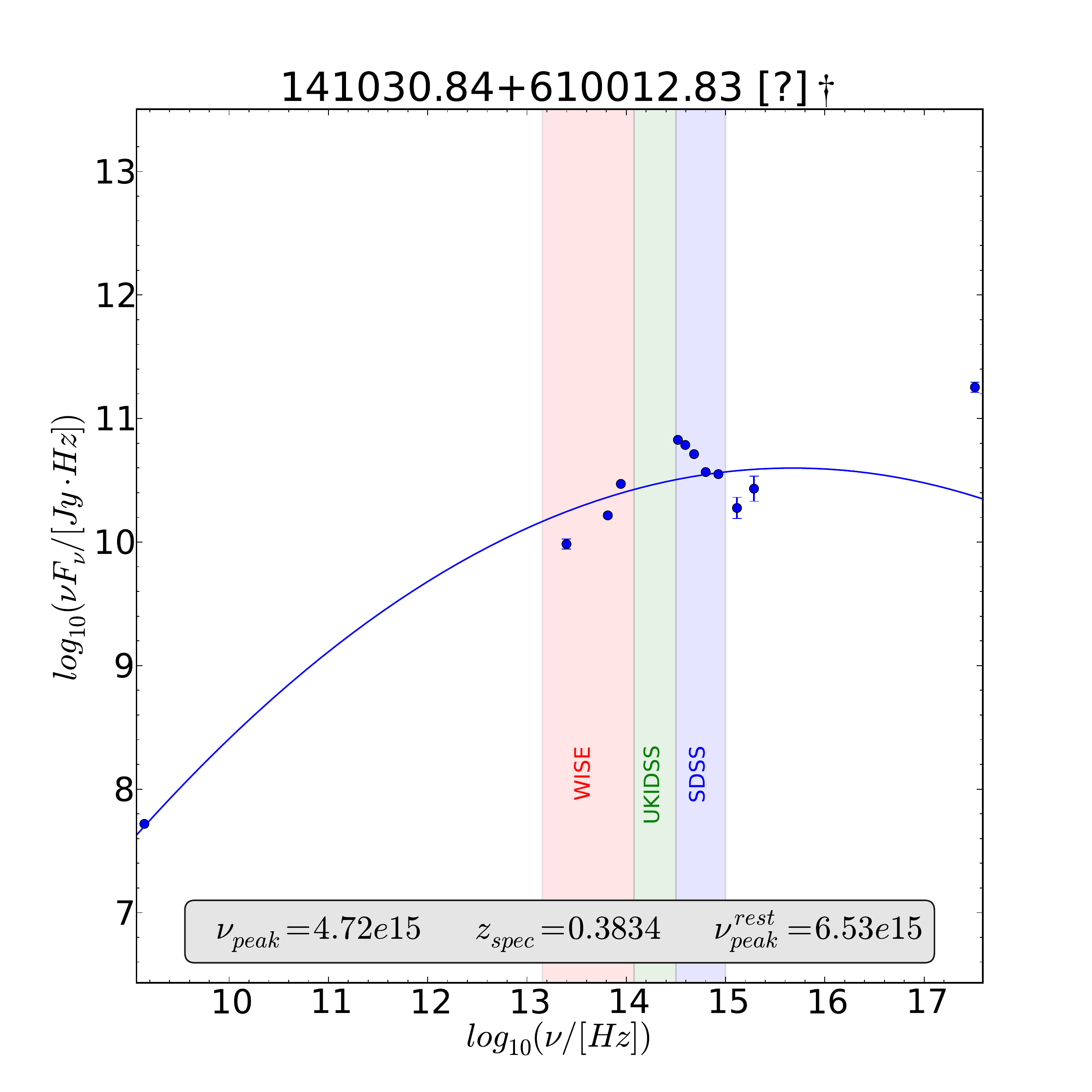}
\includegraphics[width=0.3\textwidth]{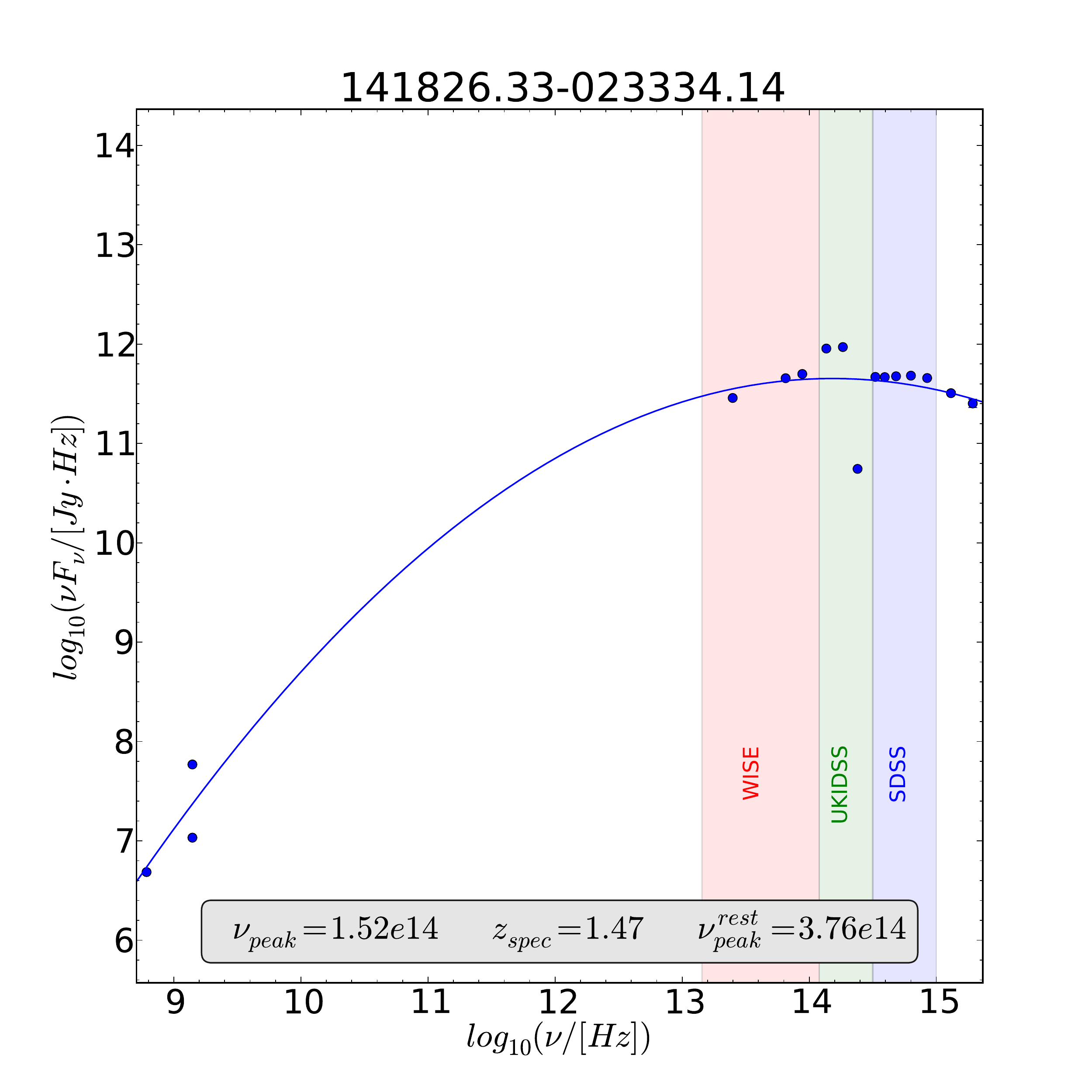}
\includegraphics[width=0.3\textwidth]{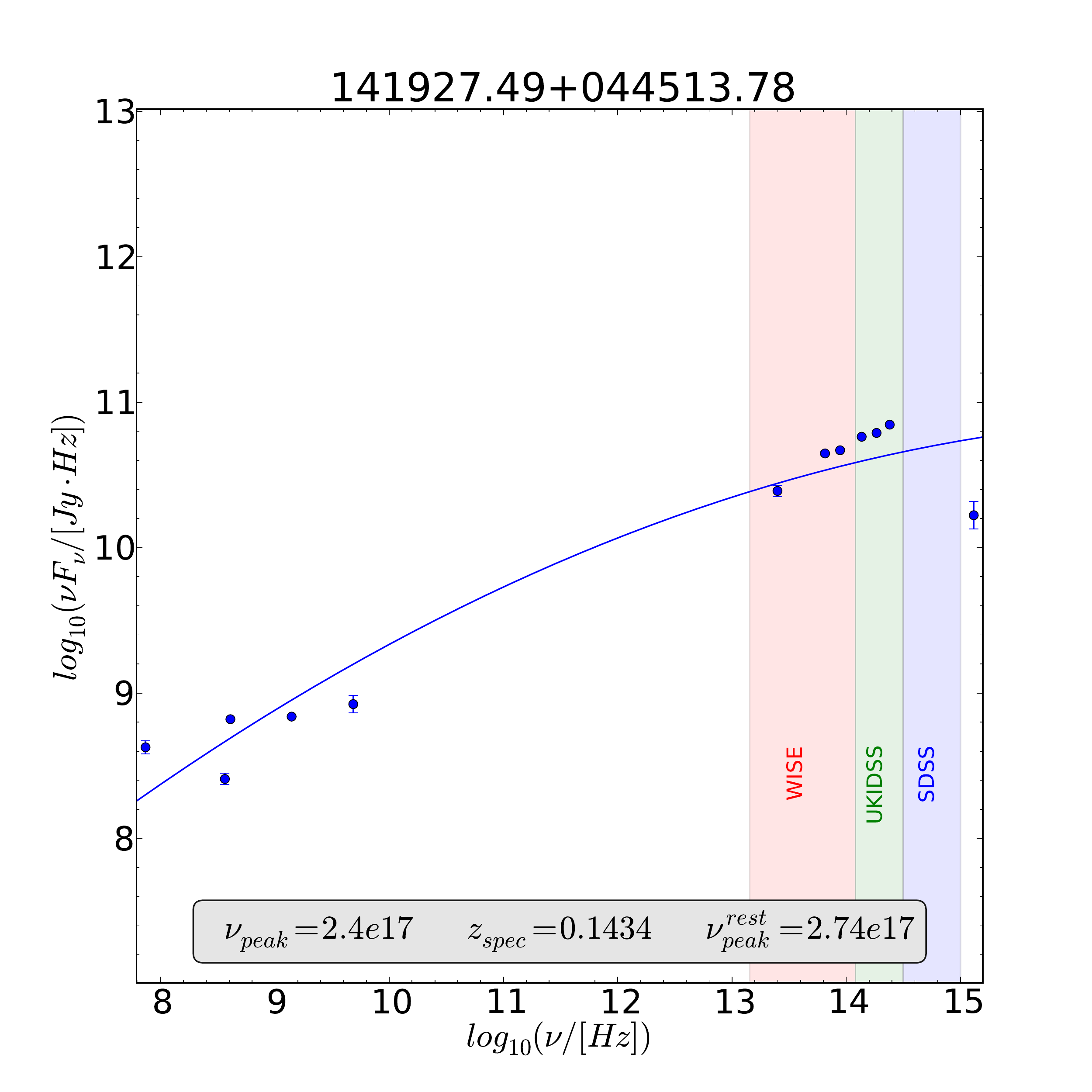}\\

\includegraphics[width=0.3\textwidth]{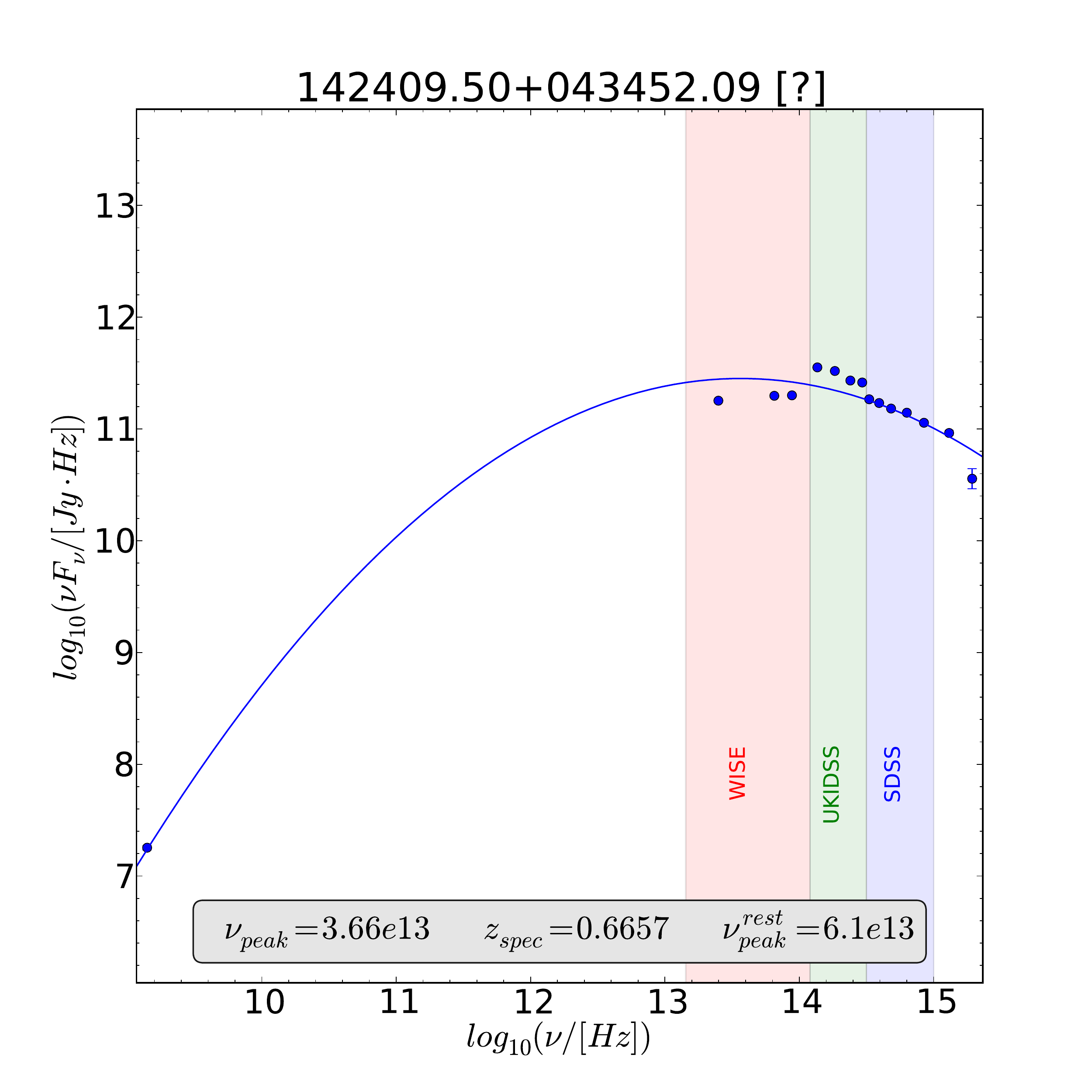}
\includegraphics[width=0.3\textwidth]{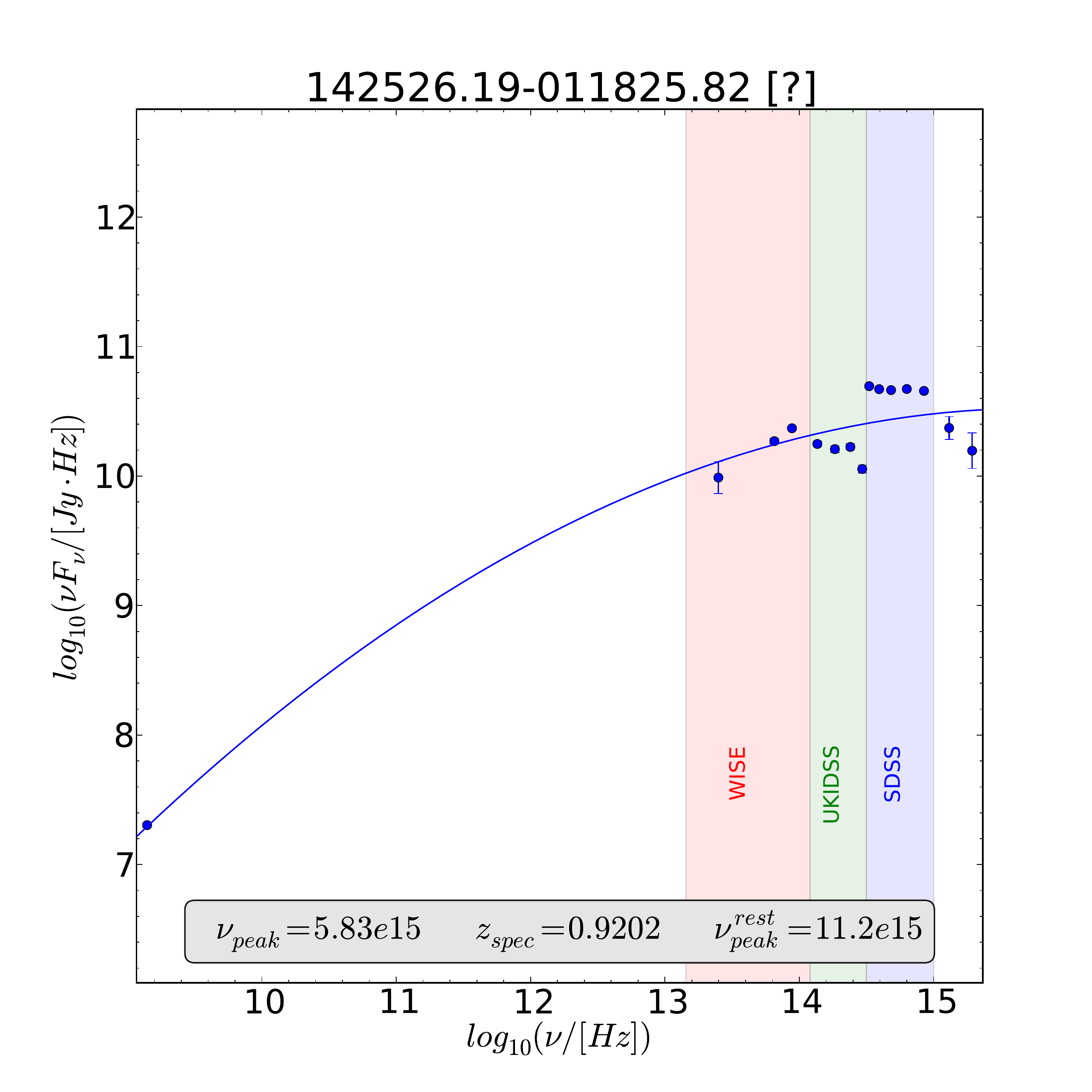}
\includegraphics[width=0.3\textwidth]{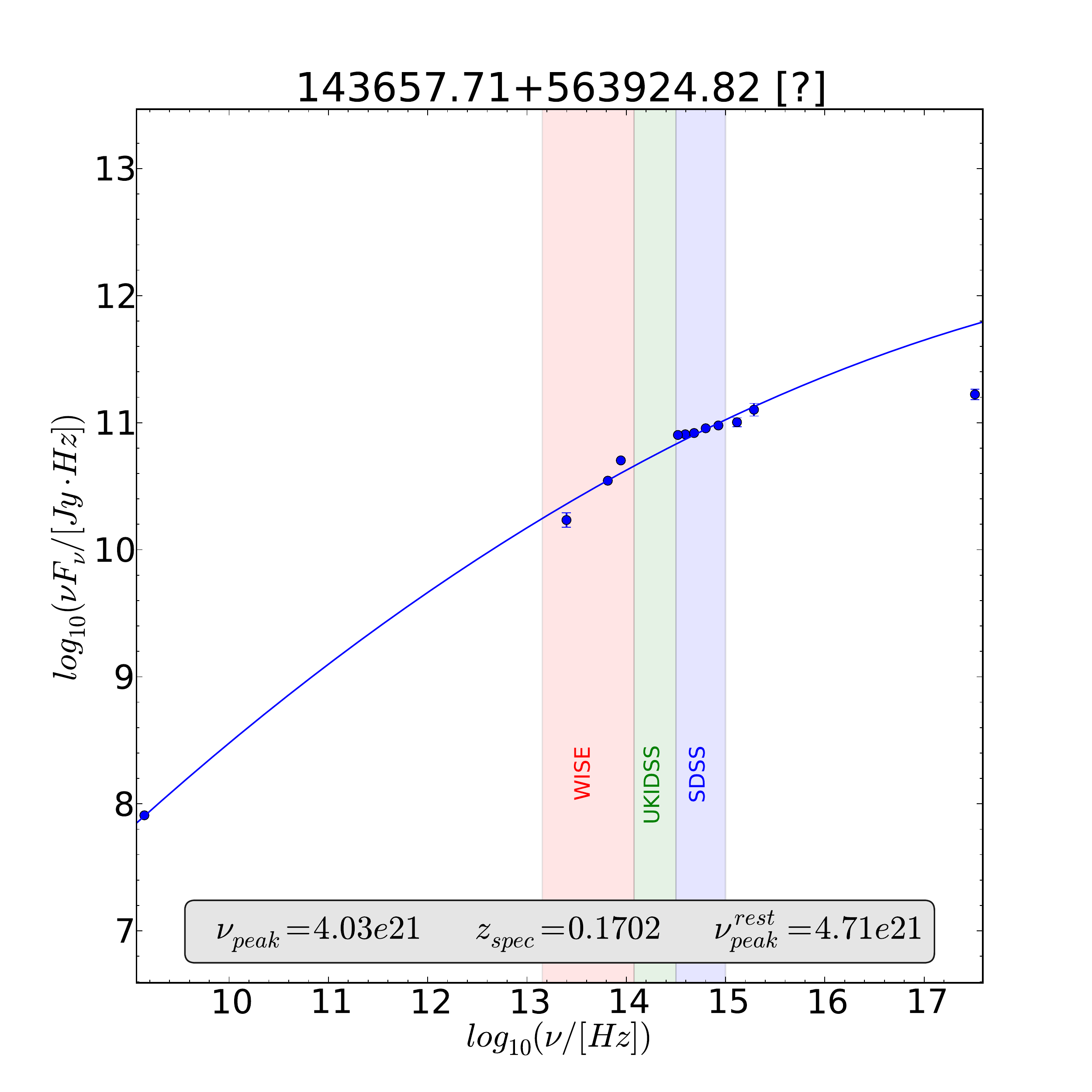}\\

\end{figure*}
\setcounter{figure}{0}
\begin{figure*}[htb!]
\caption{--Continued.}

\includegraphics[width=0.3\textwidth]{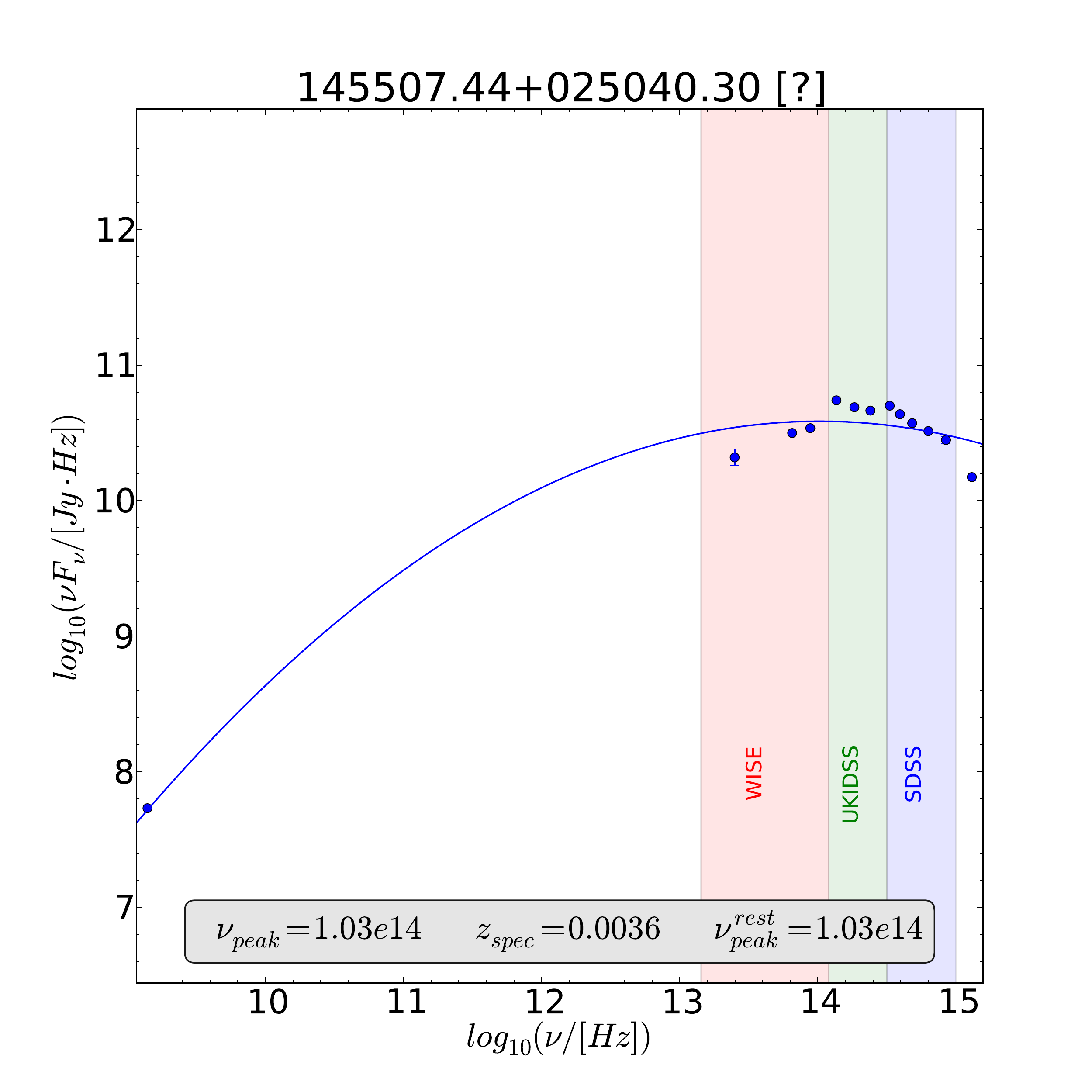}
\includegraphics[width=0.3\textwidth]{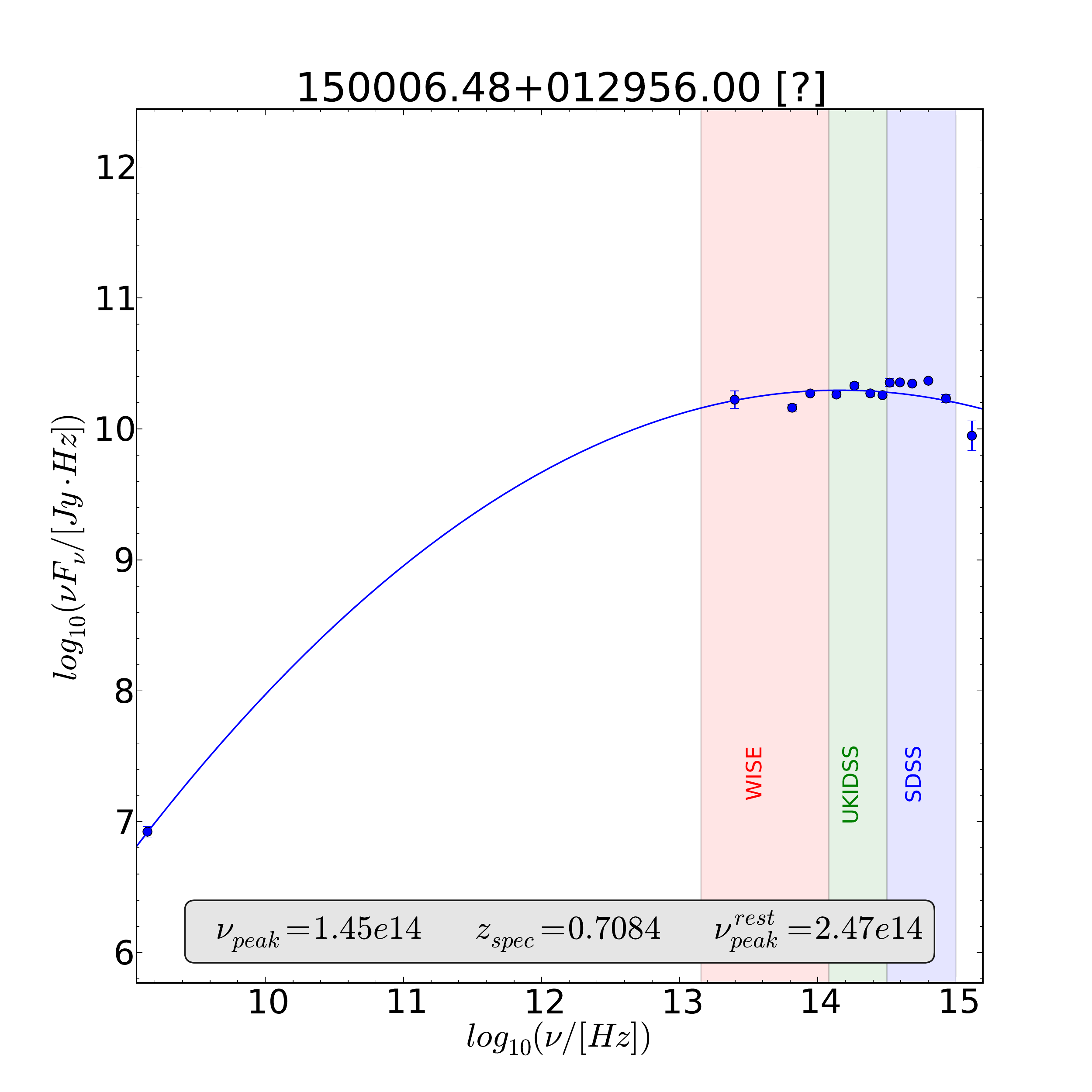}
\includegraphics[width=0.3\textwidth]{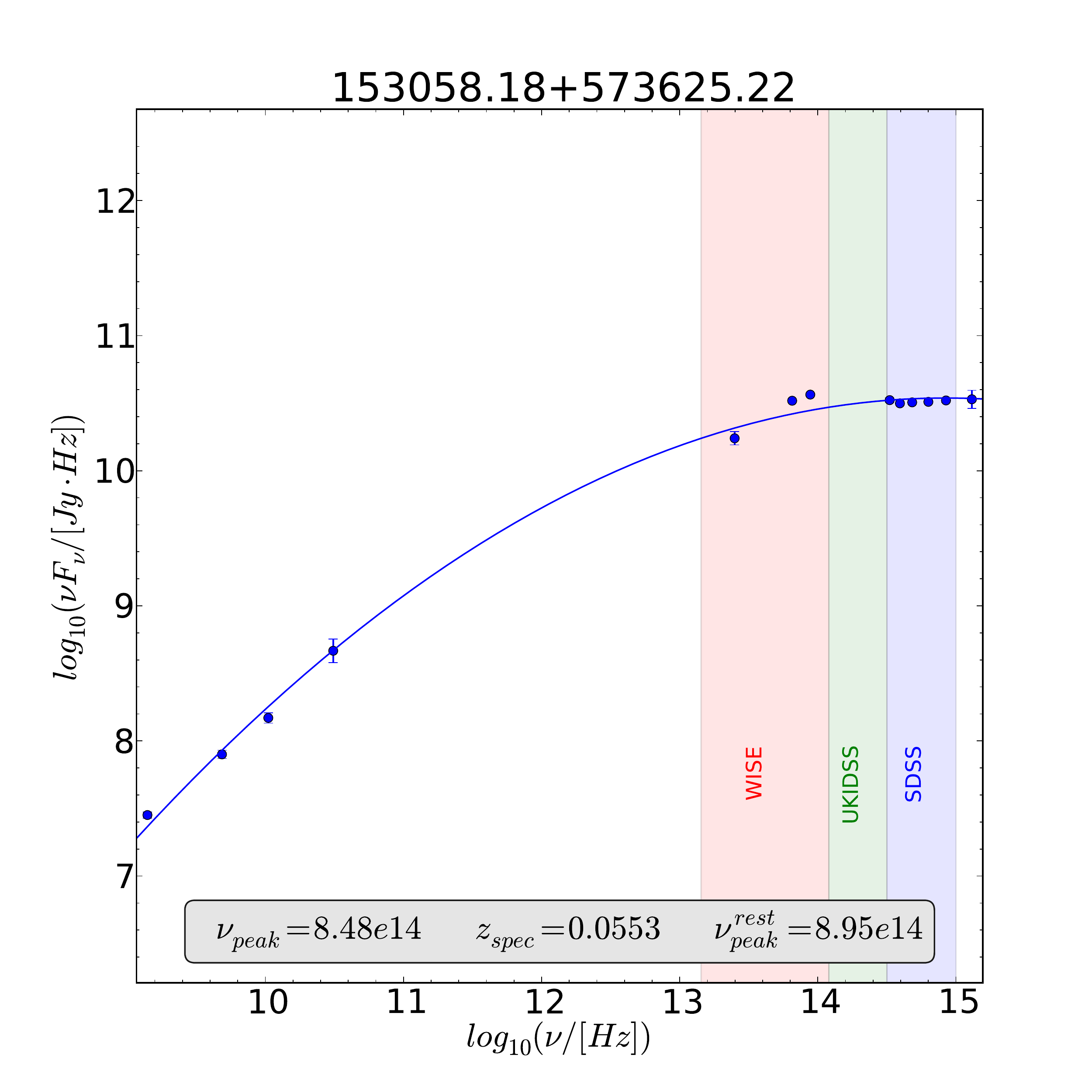}\\

\includegraphics[width=0.3\textwidth]{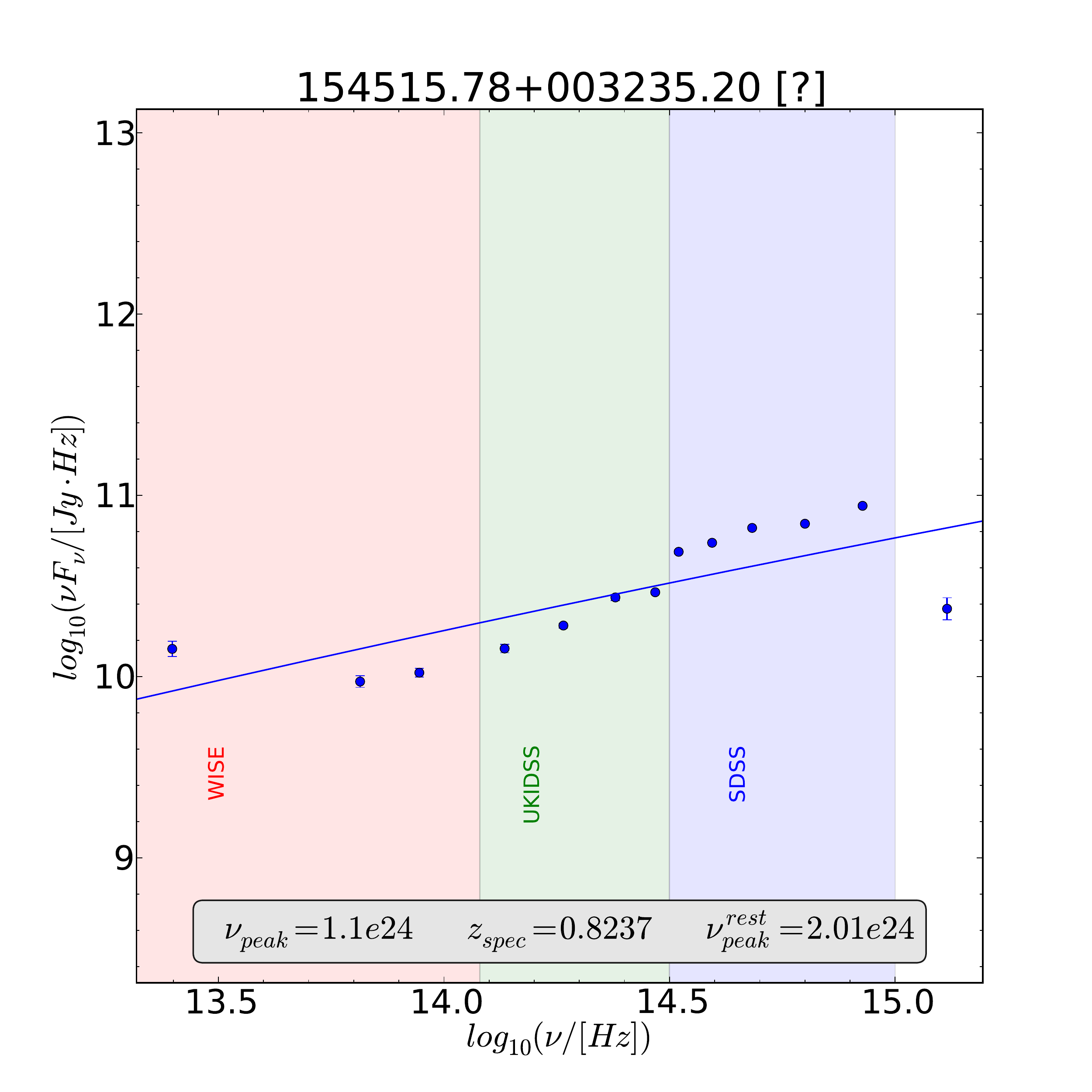}
\includegraphics[width=0.3\textwidth]{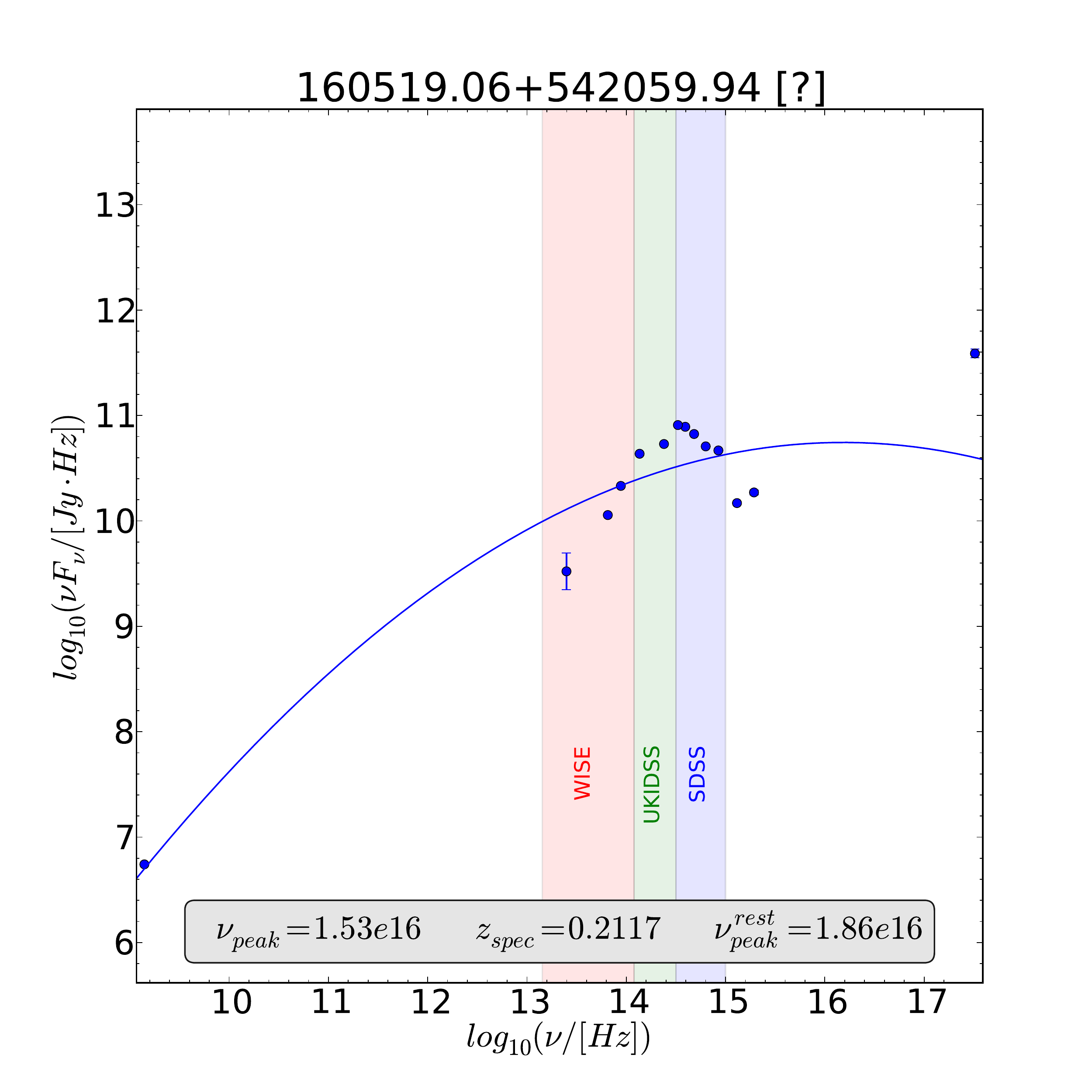}
\includegraphics[width=0.3\textwidth]{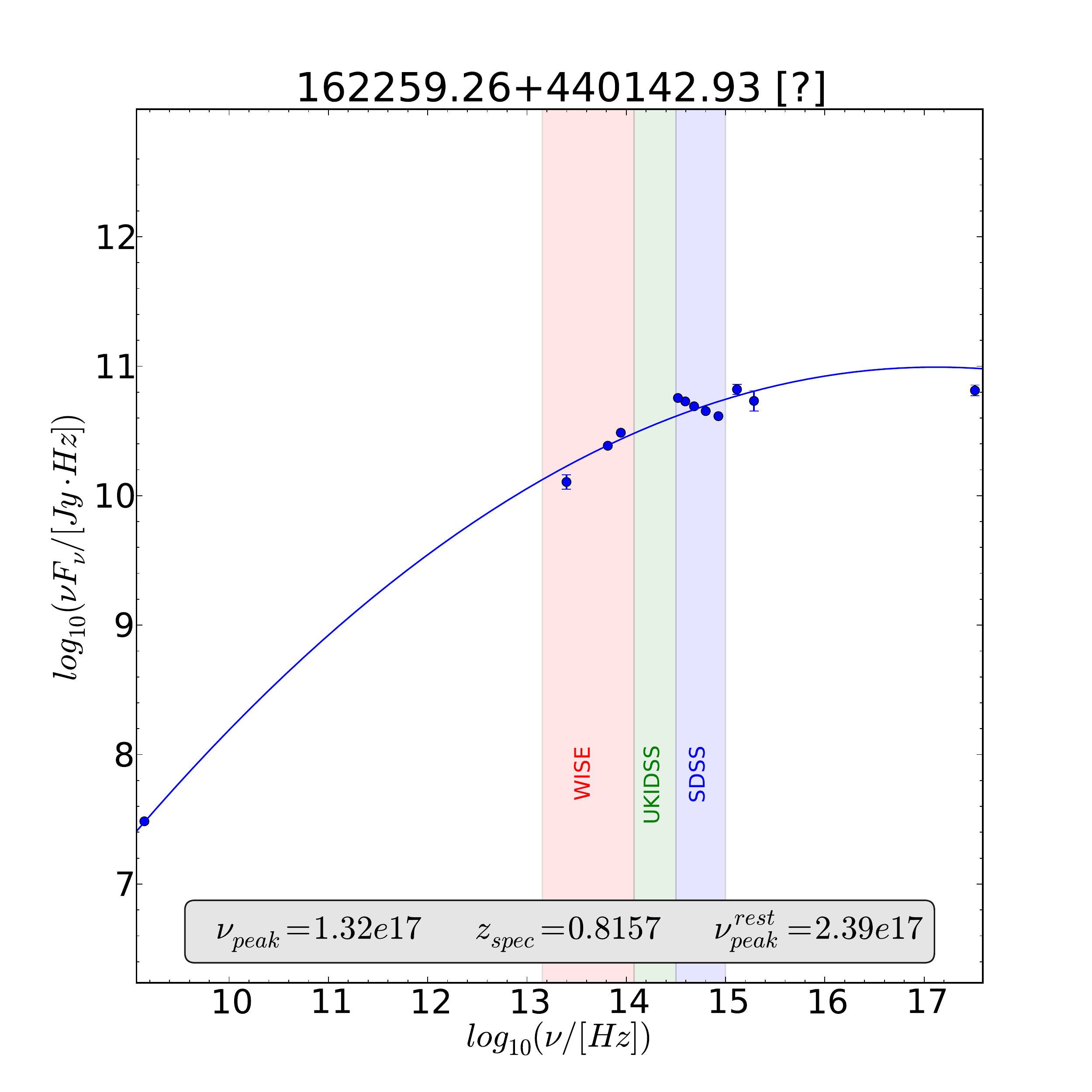}\\

\includegraphics[width=0.3\textwidth]{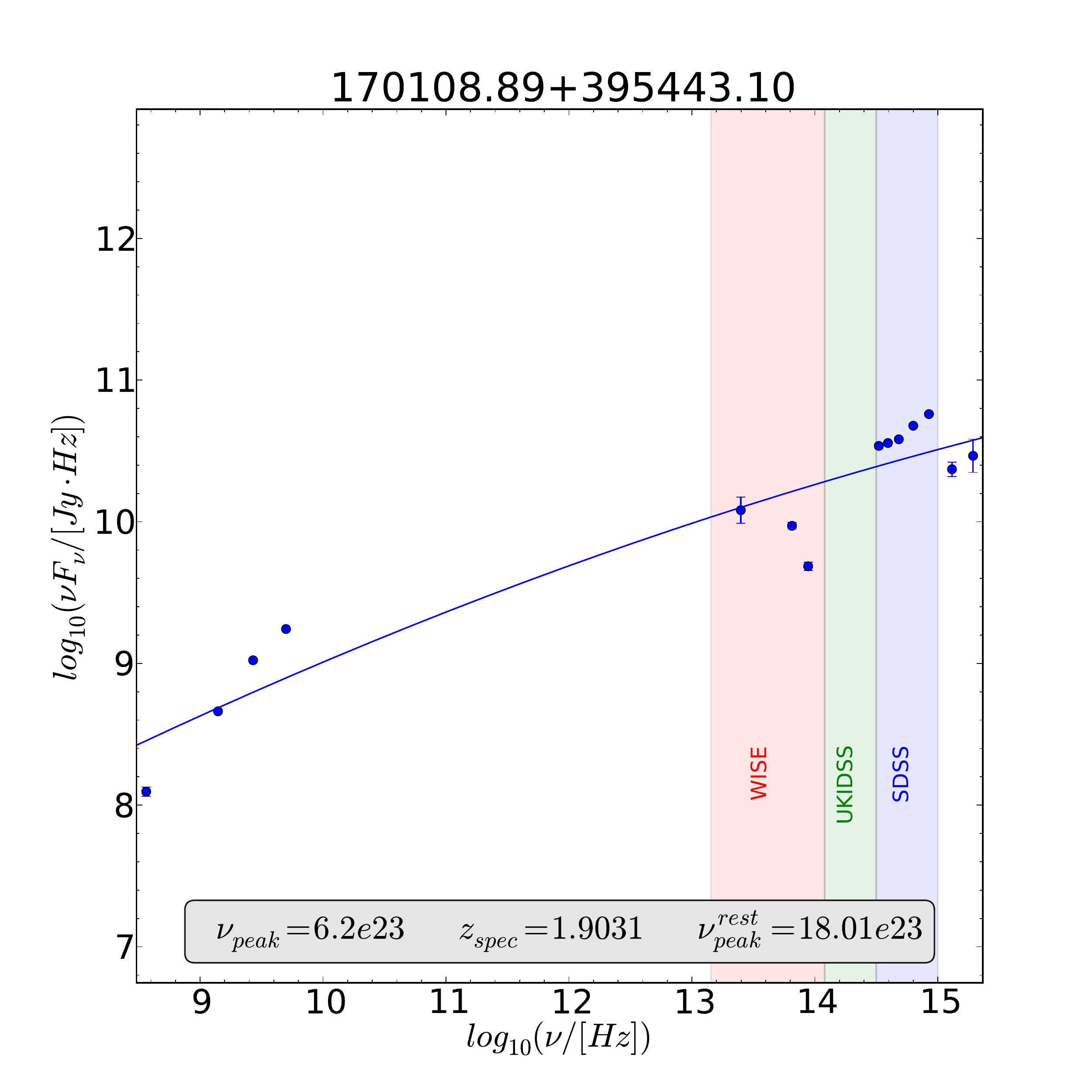}
\includegraphics[width=0.3\textwidth]{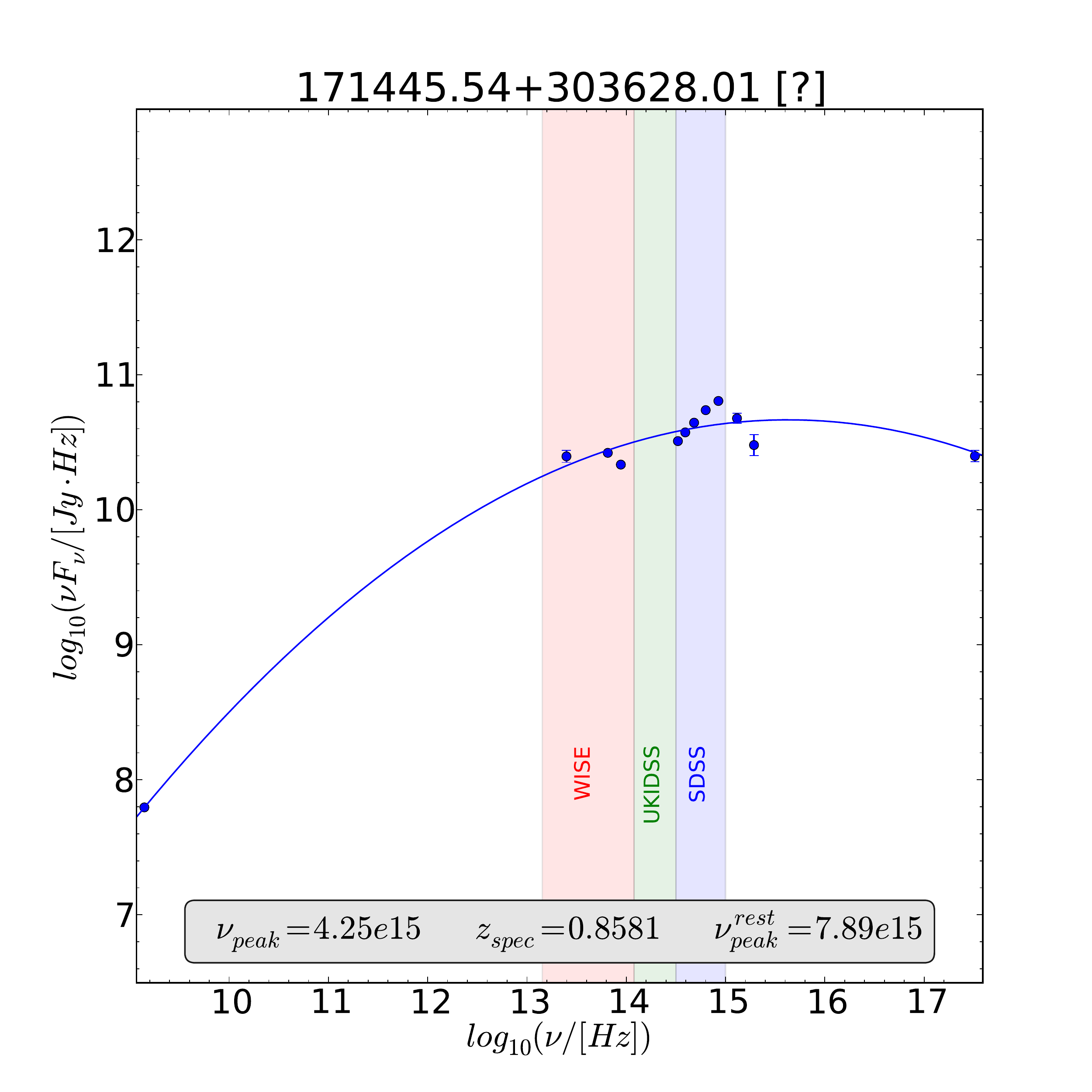}
\end{figure*}

\newpage
\begin{figure*}
\caption{Surface brightness (SB) profiles of the targets with detected
  host galaxies. Horizontal axis gives the distance from the center in
  arcsec and the vertical axis the surface brightness in
  mag/sq. arcsec. In addition to the observed SB, we show the SB of
  the total core + host galaxy model (solid line), the core model
  (dashed line) and the host galaxy model (dotted line).}
\includegraphics[width=5.5cm]{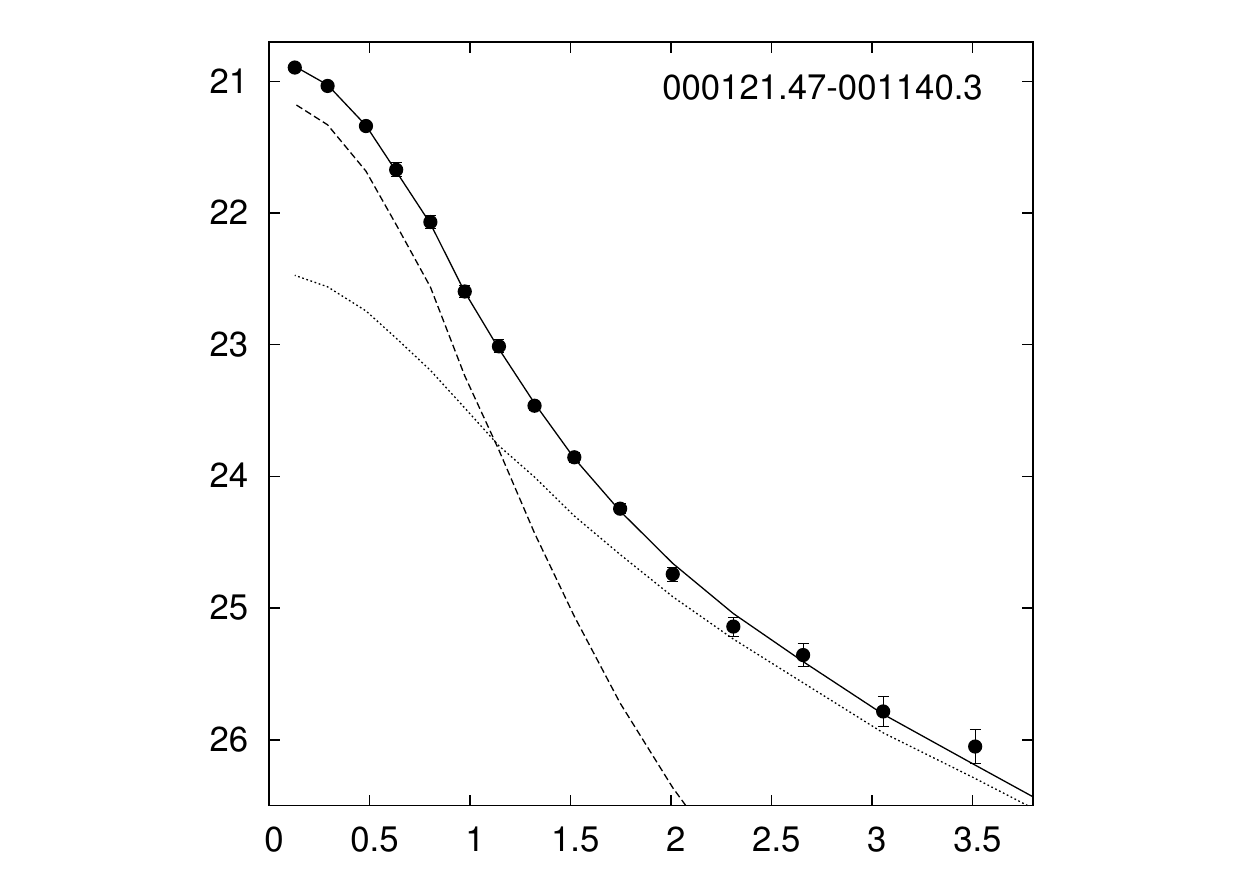}
\hspace*{-1.5cm}
\includegraphics[width=5.5cm]{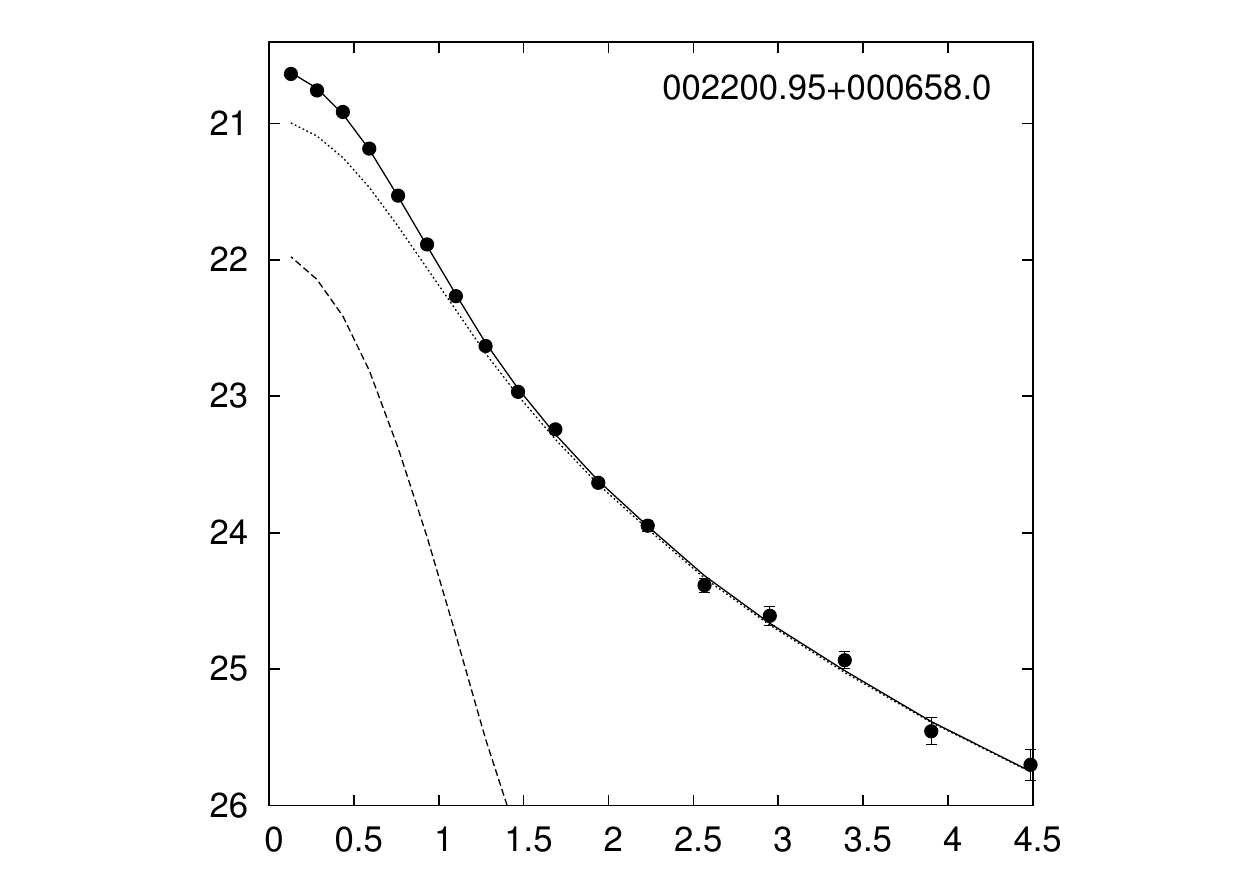}
\hspace*{-1.5cm}
\includegraphics[width=5.5cm]{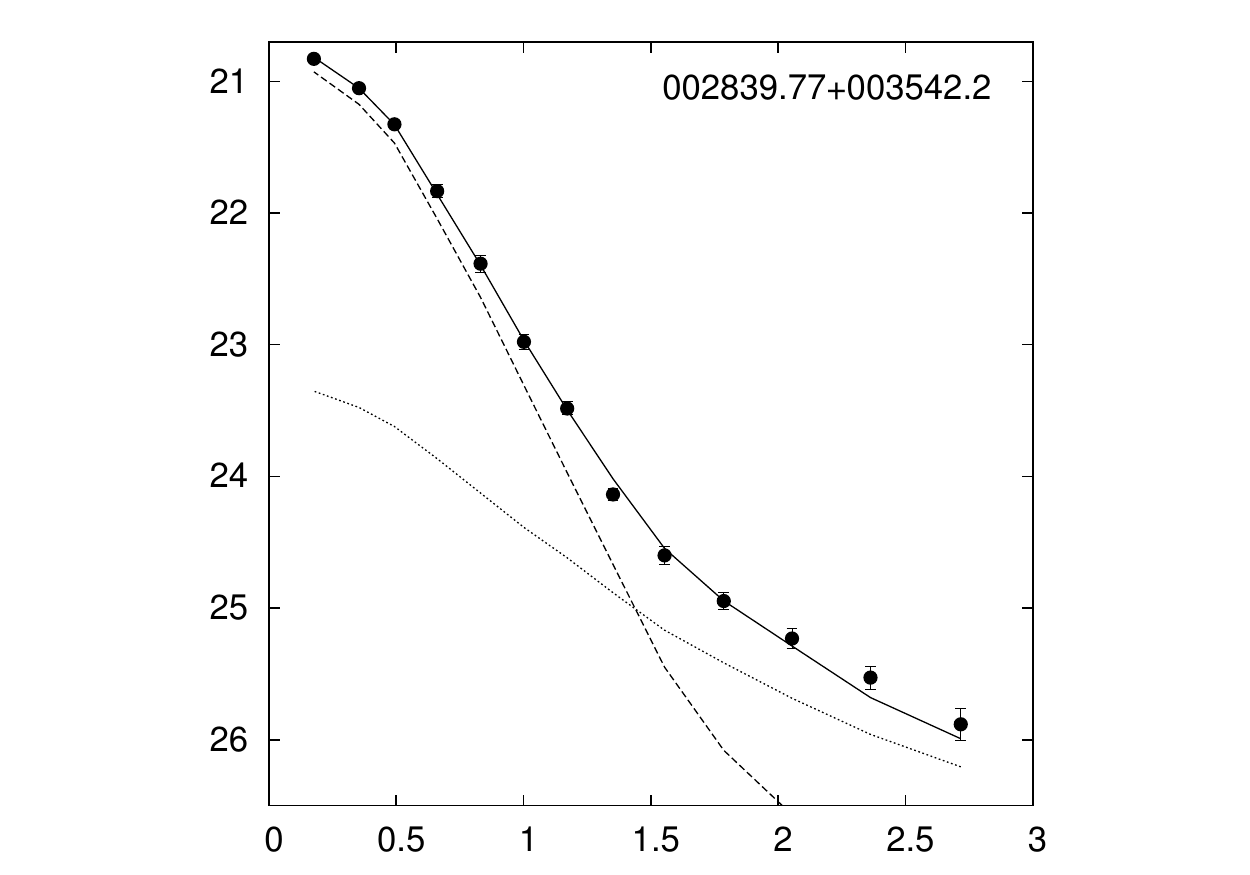}
\hspace*{-1.5cm}
\includegraphics[width=5.5cm]{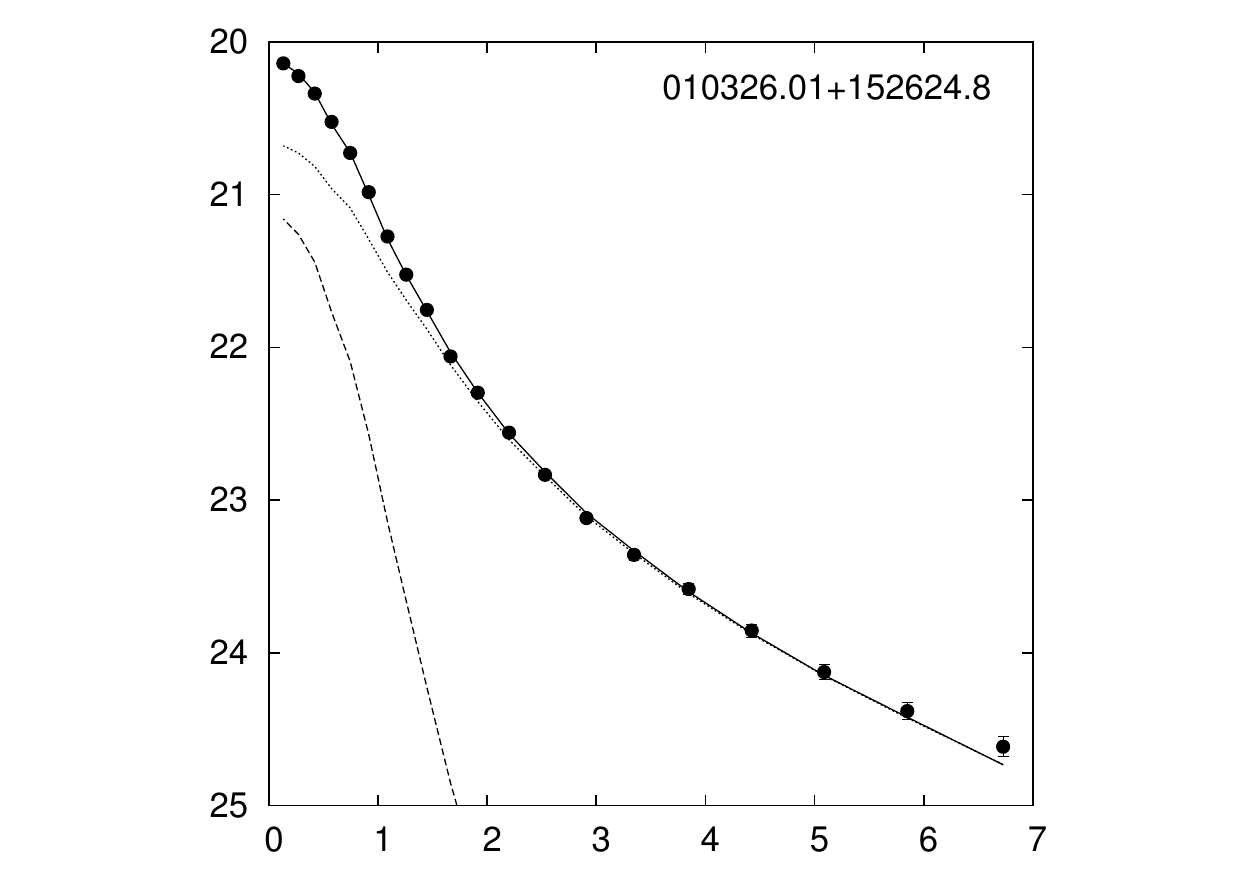}\\

\includegraphics[width=5.5cm]{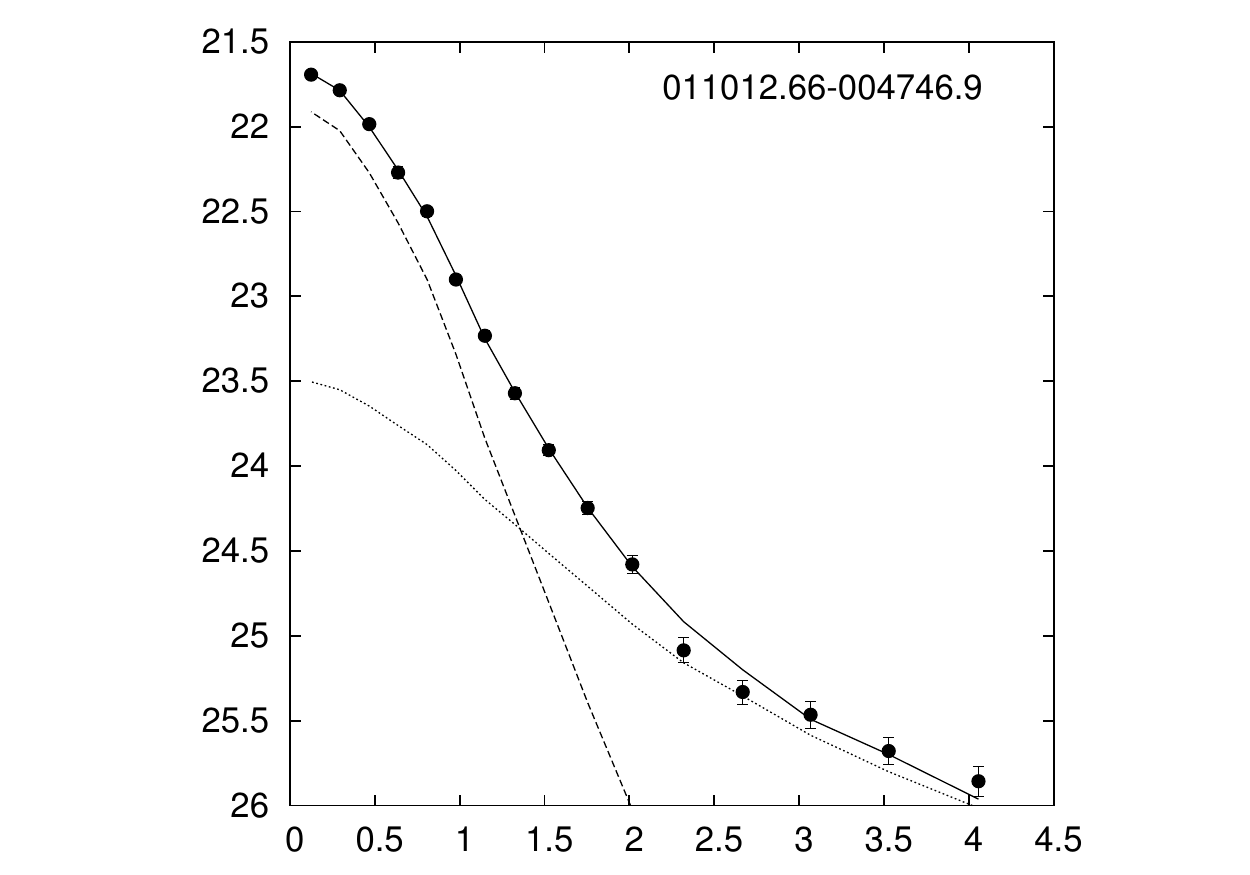}
\hspace*{-1.5cm}
\includegraphics[width=5.5cm]{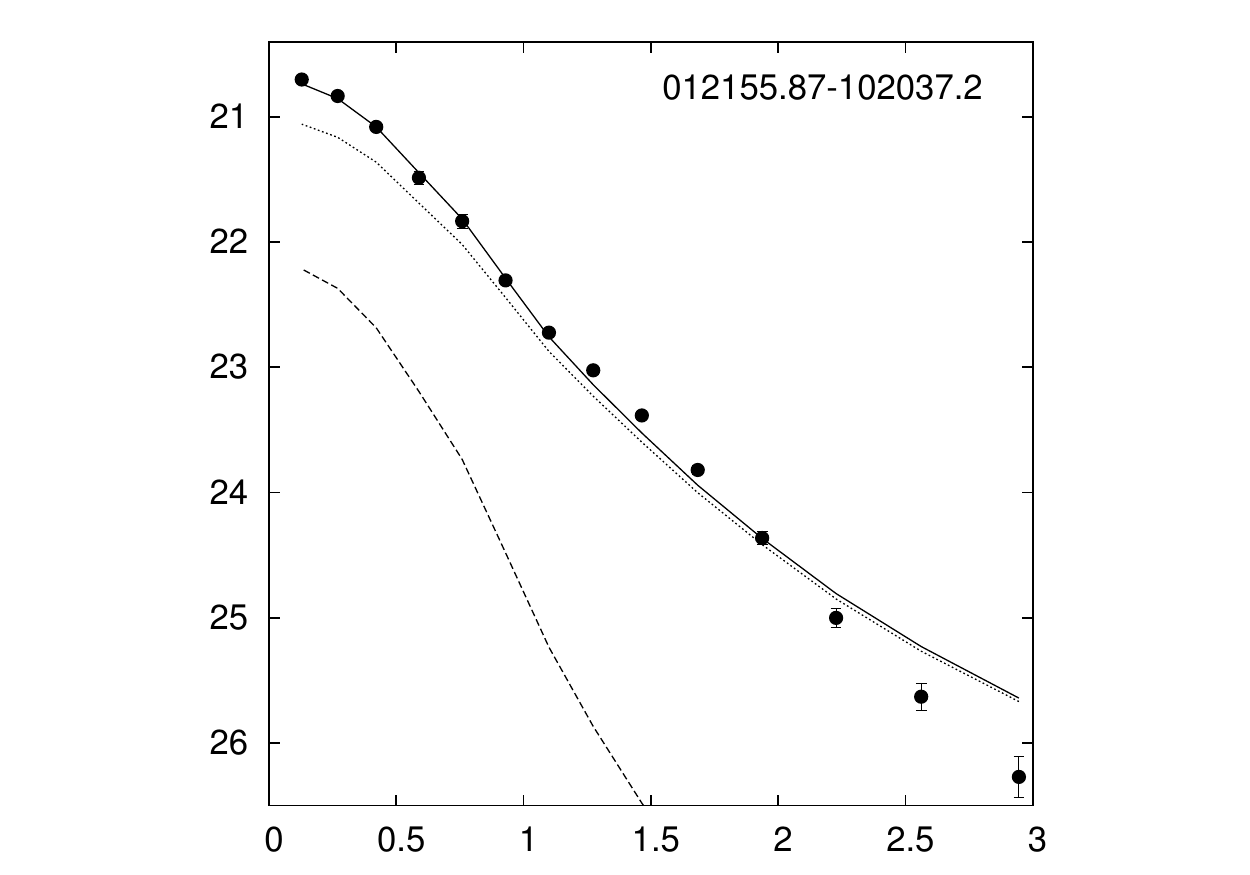}
\hspace*{-1.5cm}
\includegraphics[width=5.5cm]{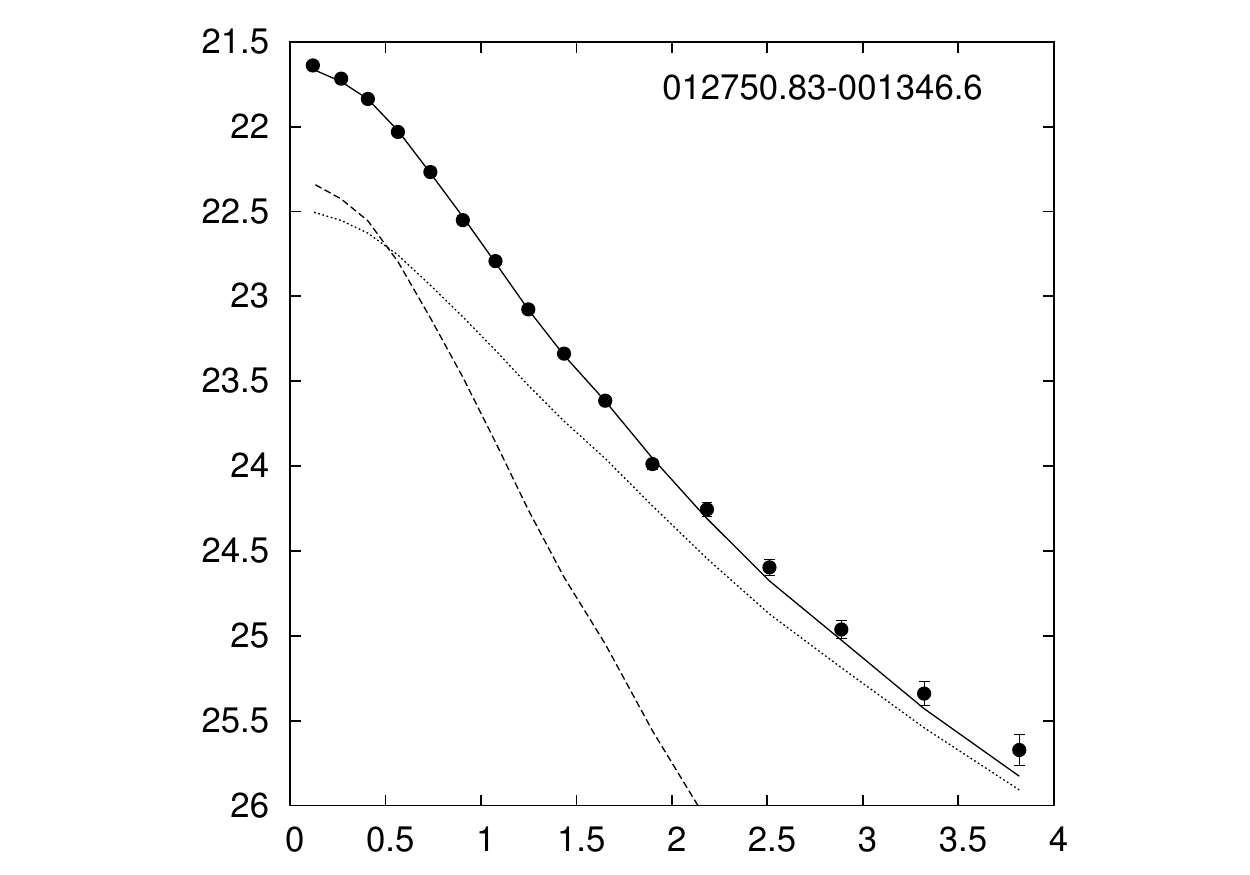}
\hspace*{-1.5cm}
\includegraphics[width=5.5cm]{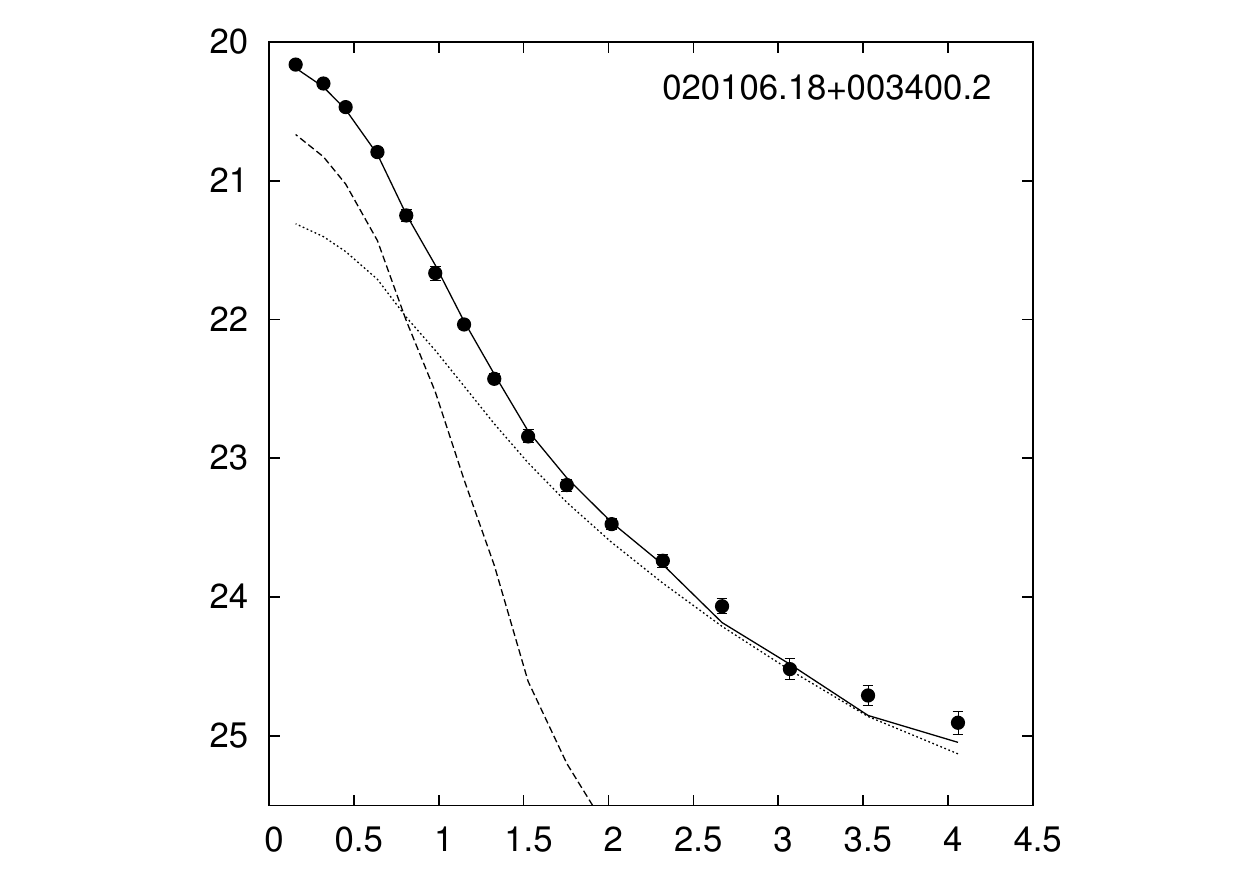}\\

\includegraphics[width=5.5cm]{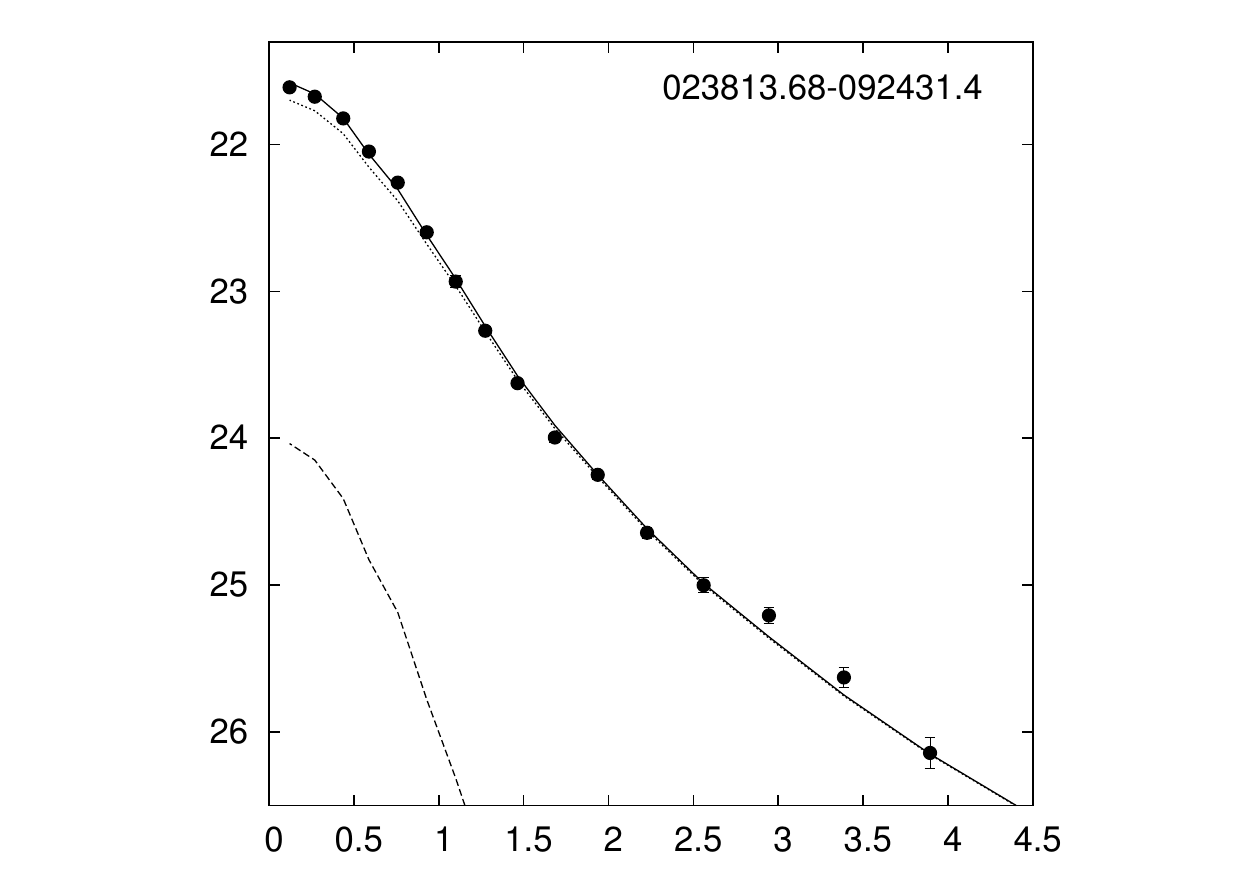}
\hspace*{-1.5cm}
\includegraphics[width=5.5cm]{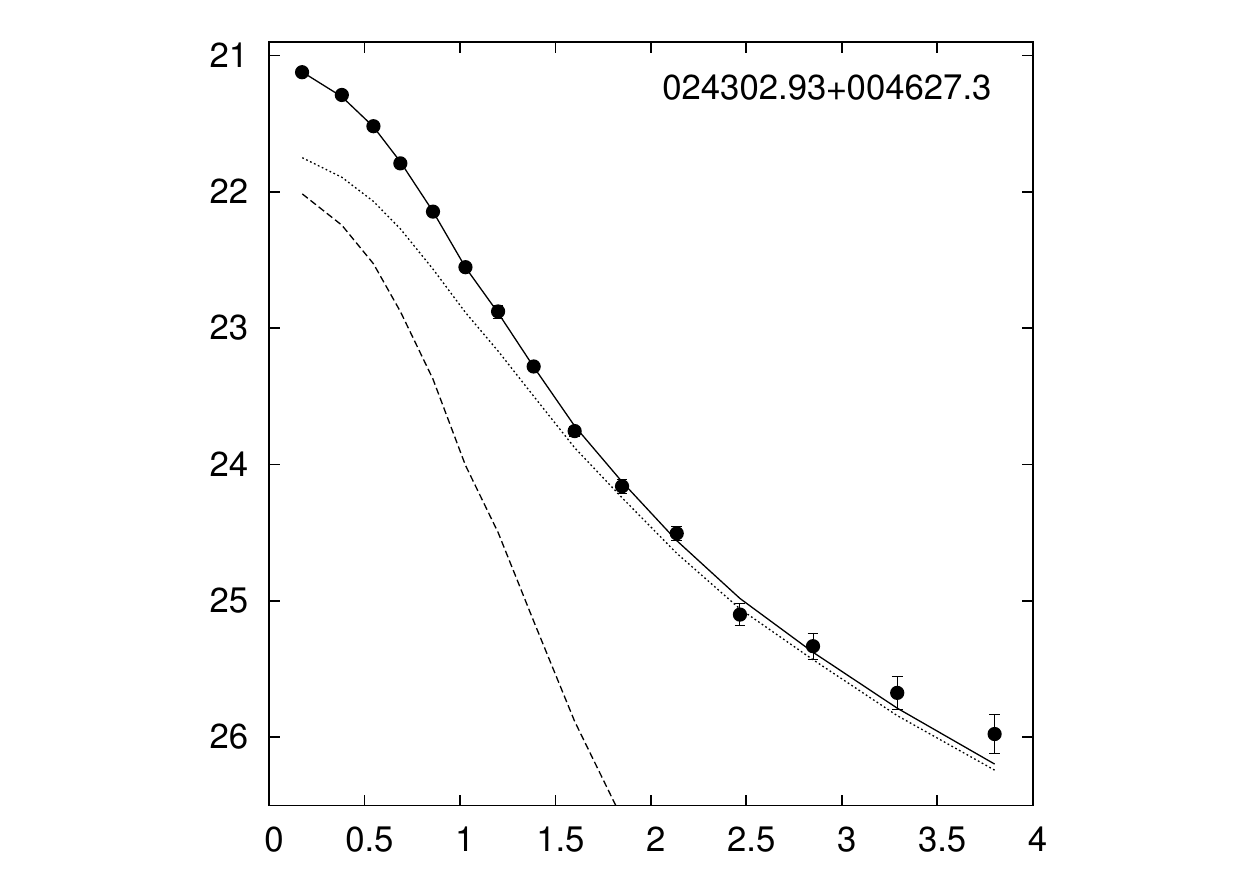}
\hspace*{-1.5cm}
\includegraphics[width=5.5cm]{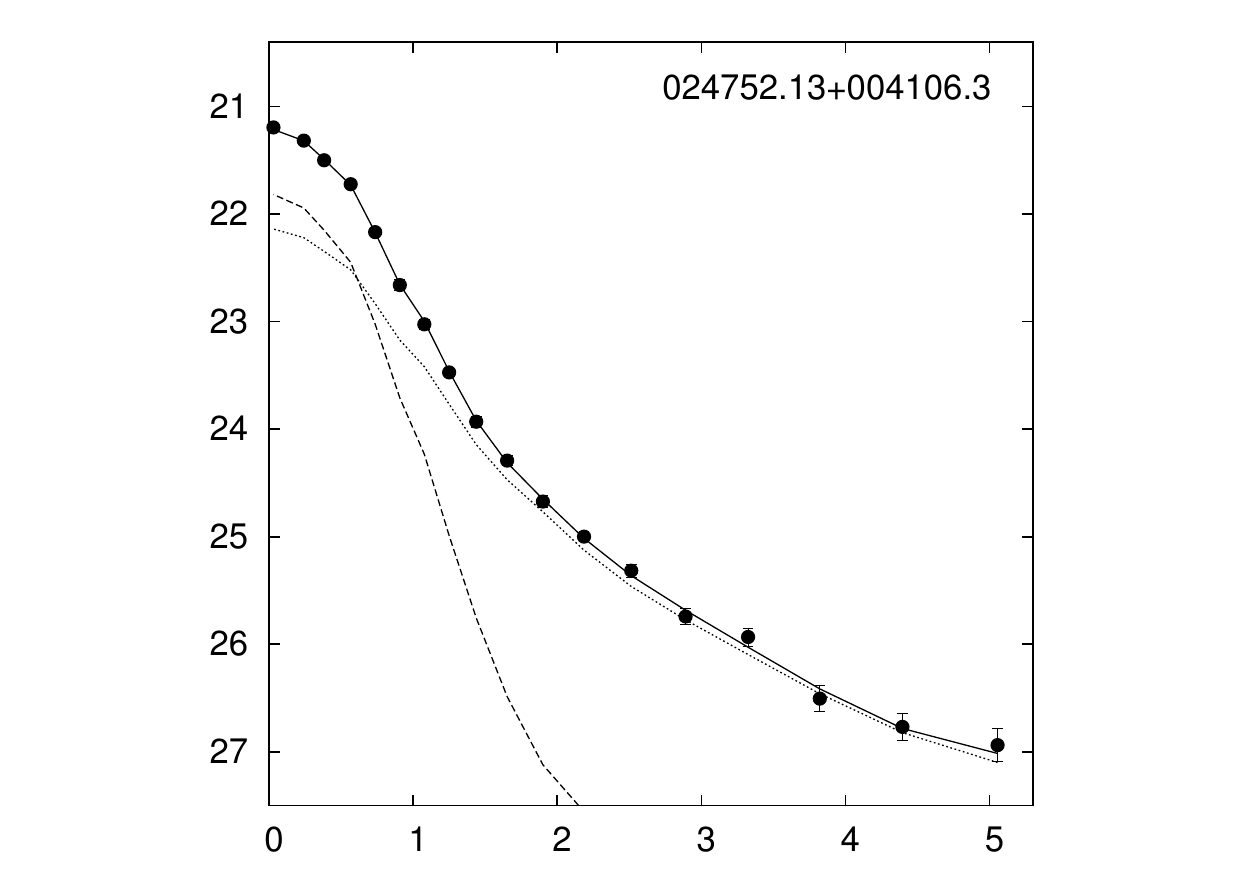}
\hspace*{-1.5cm}
\includegraphics[width=5.5cm]{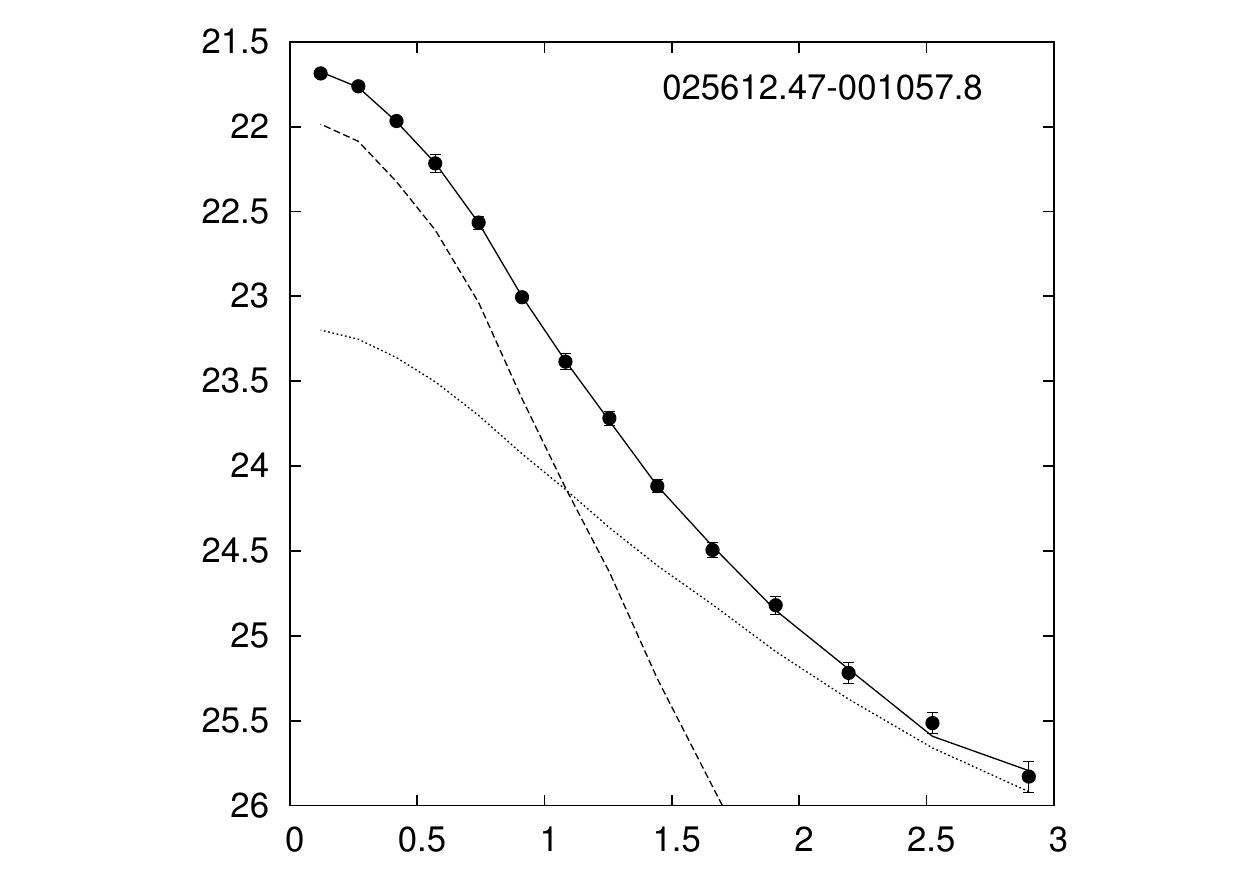}\\

\includegraphics[width=5.5cm]{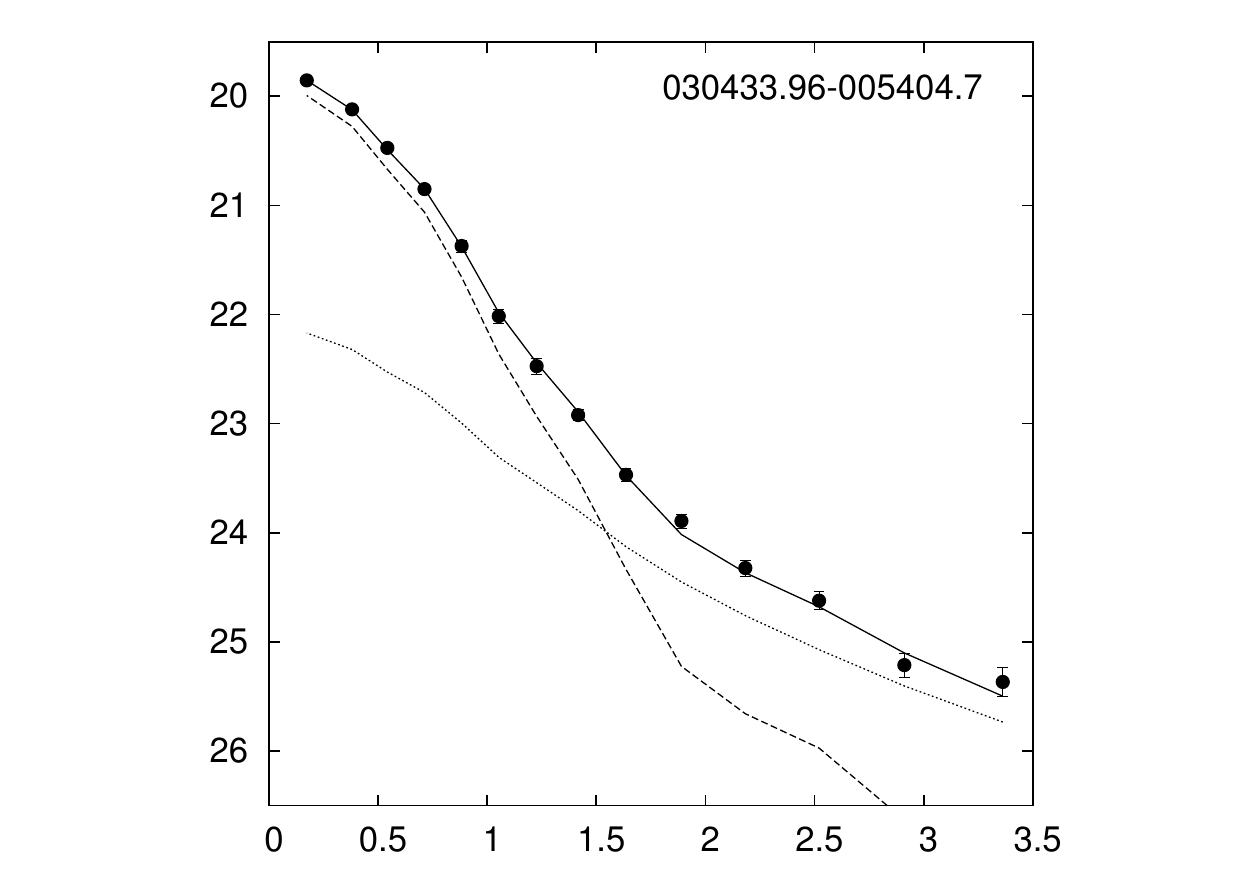}
\hspace*{-1.5cm}
\includegraphics[width=5.5cm]{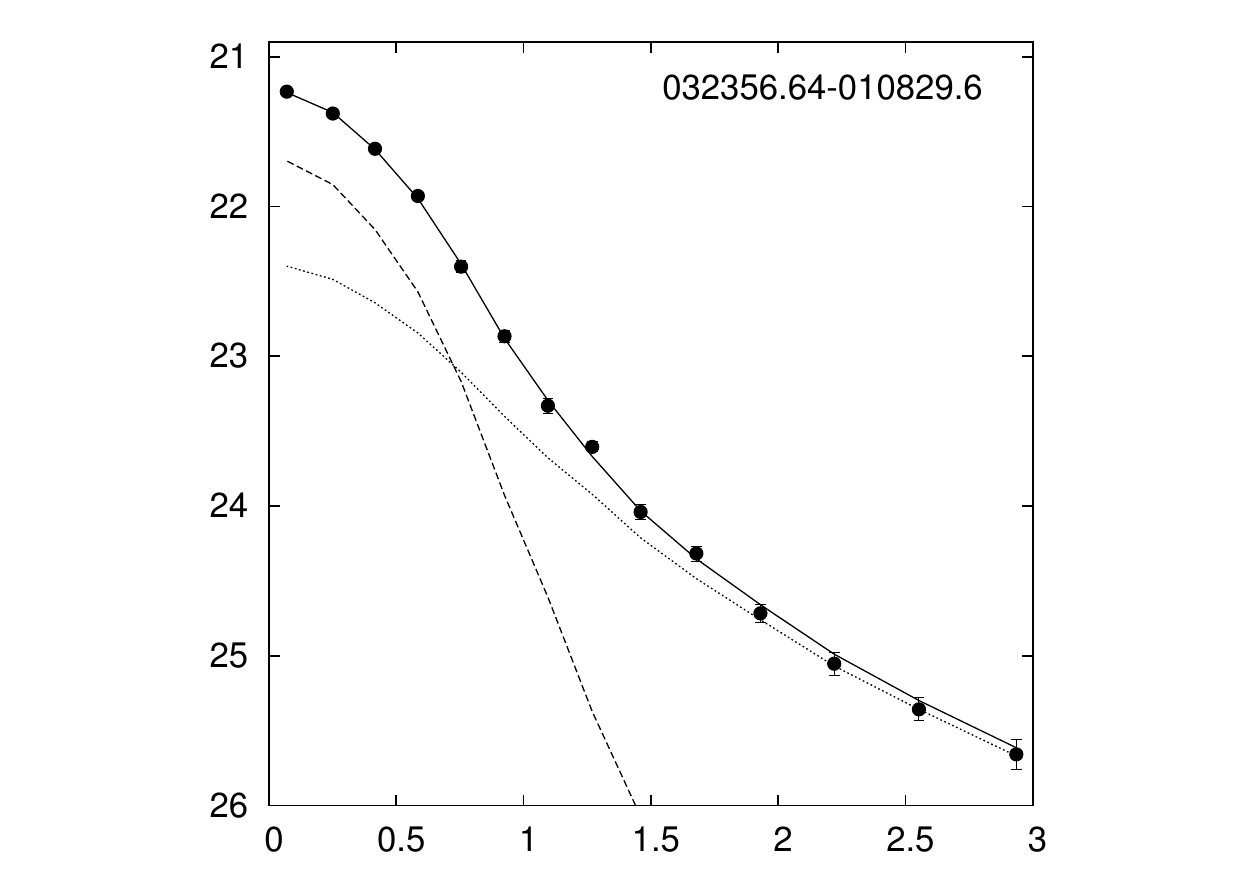}
\hspace*{-1.5cm}
\includegraphics[width=5.5cm]{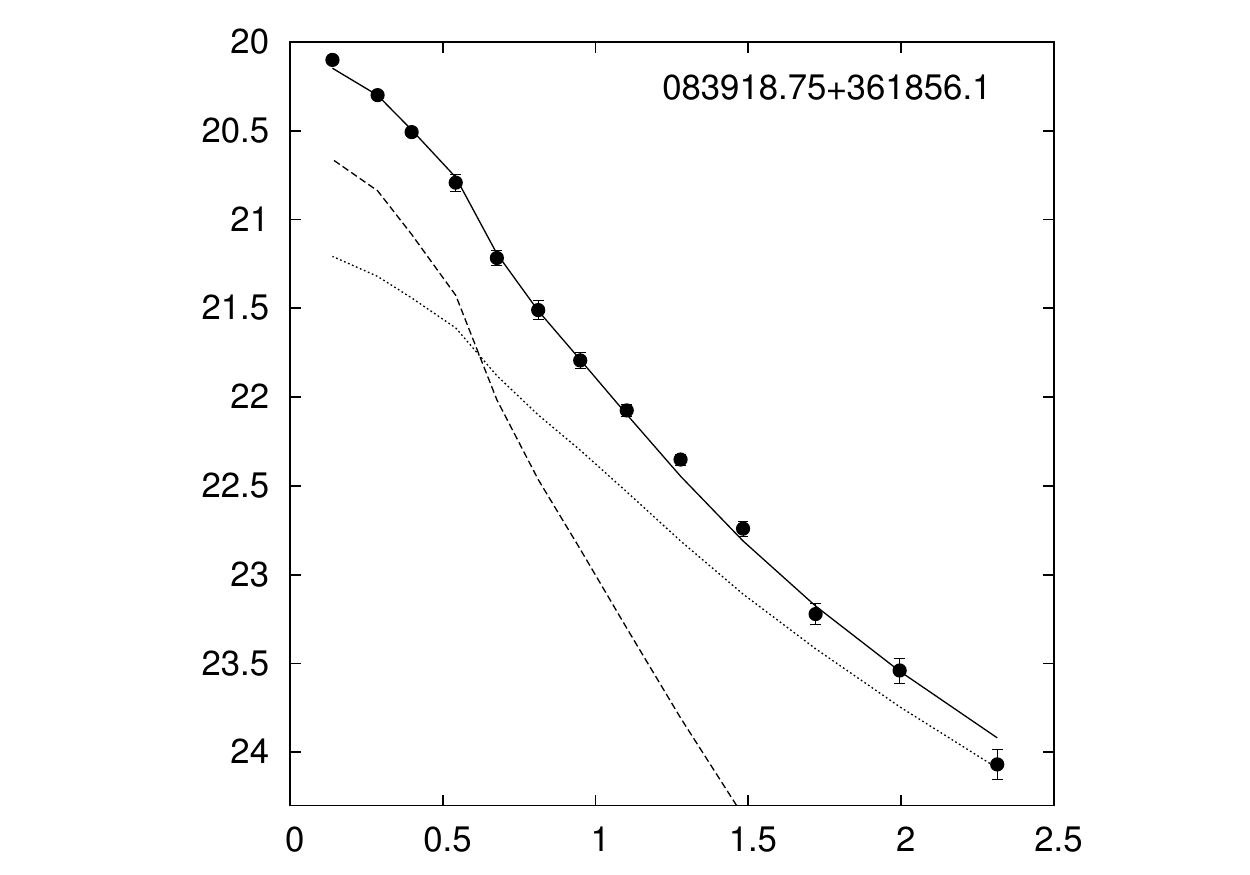}
\hspace*{-1.5cm}
\includegraphics[width=5.5cm]{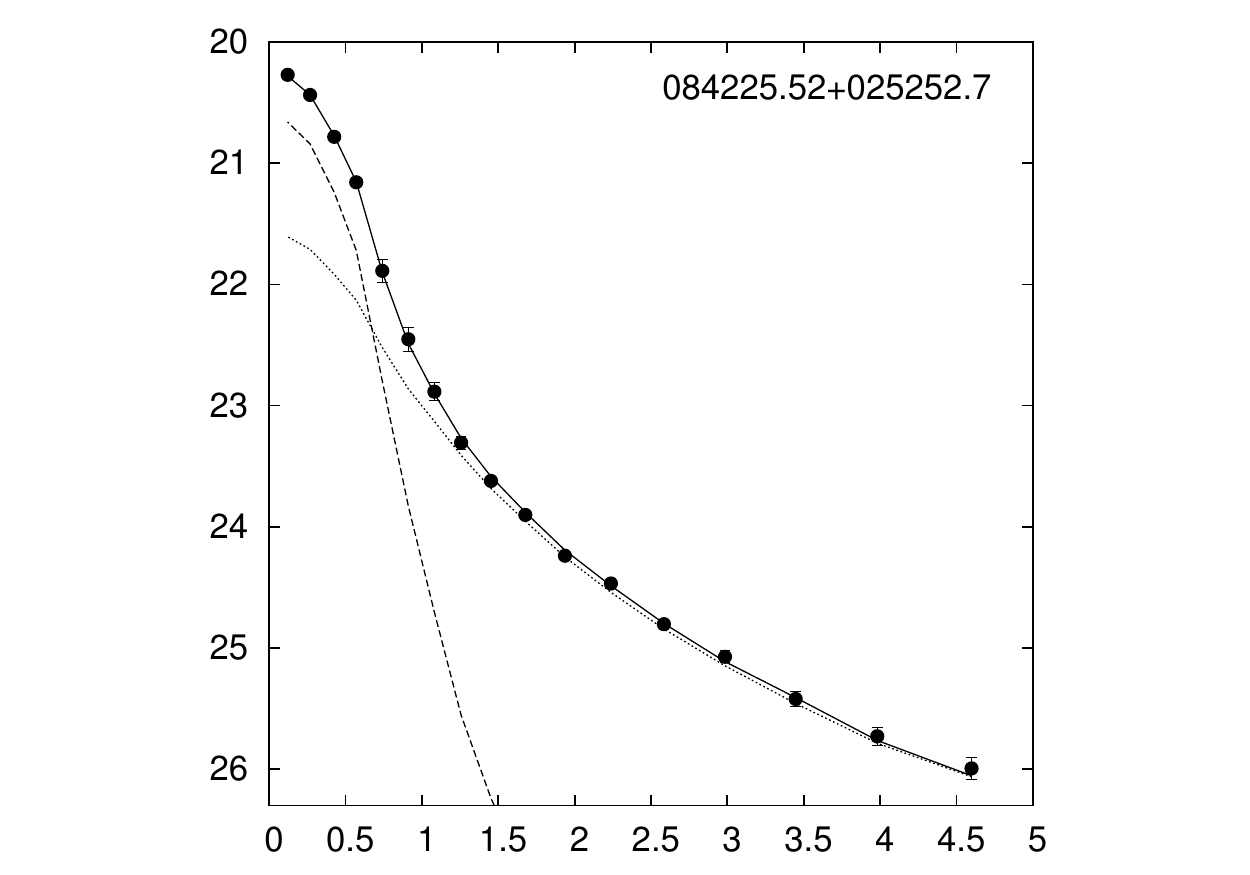}\\

\includegraphics[width=5.5cm]{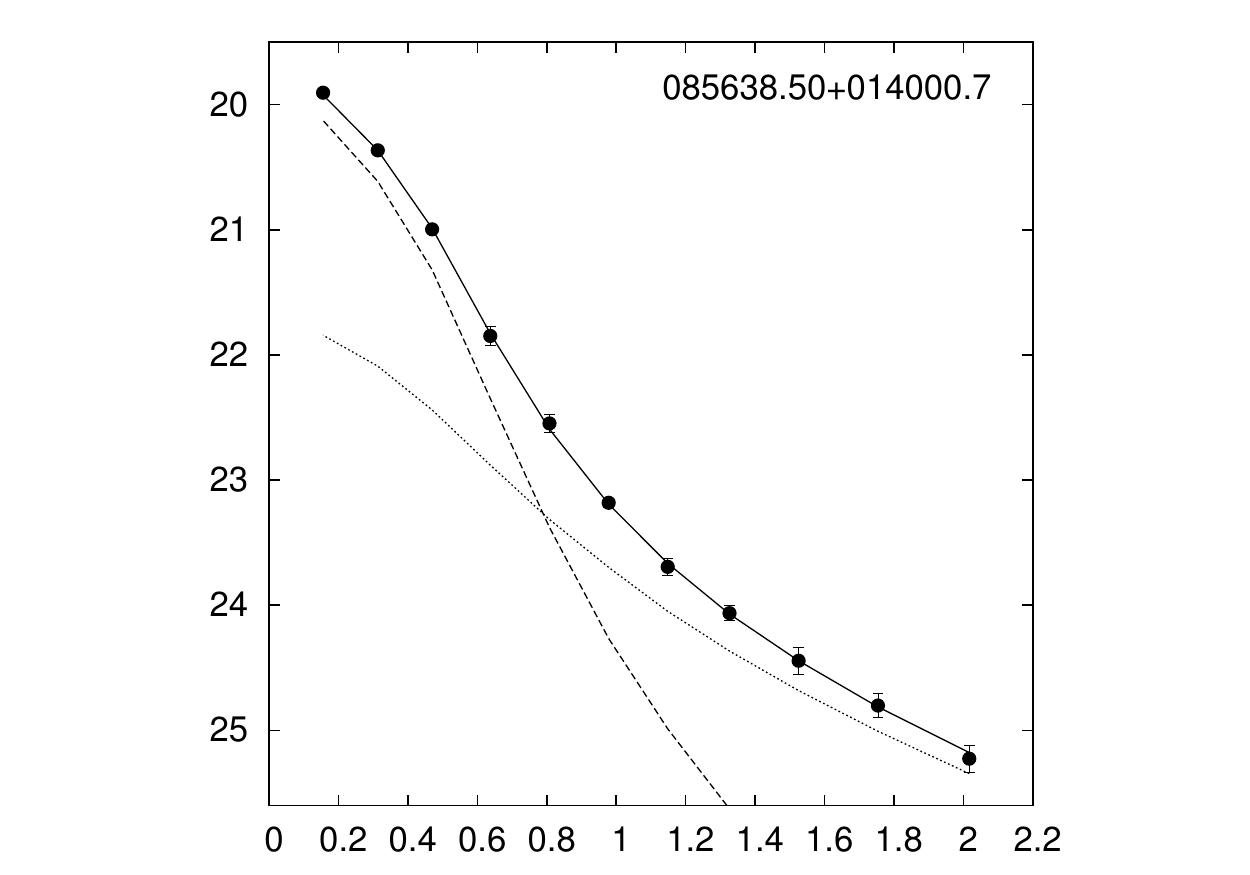}
\hspace*{-1.5cm}
\includegraphics[width=5.5cm]{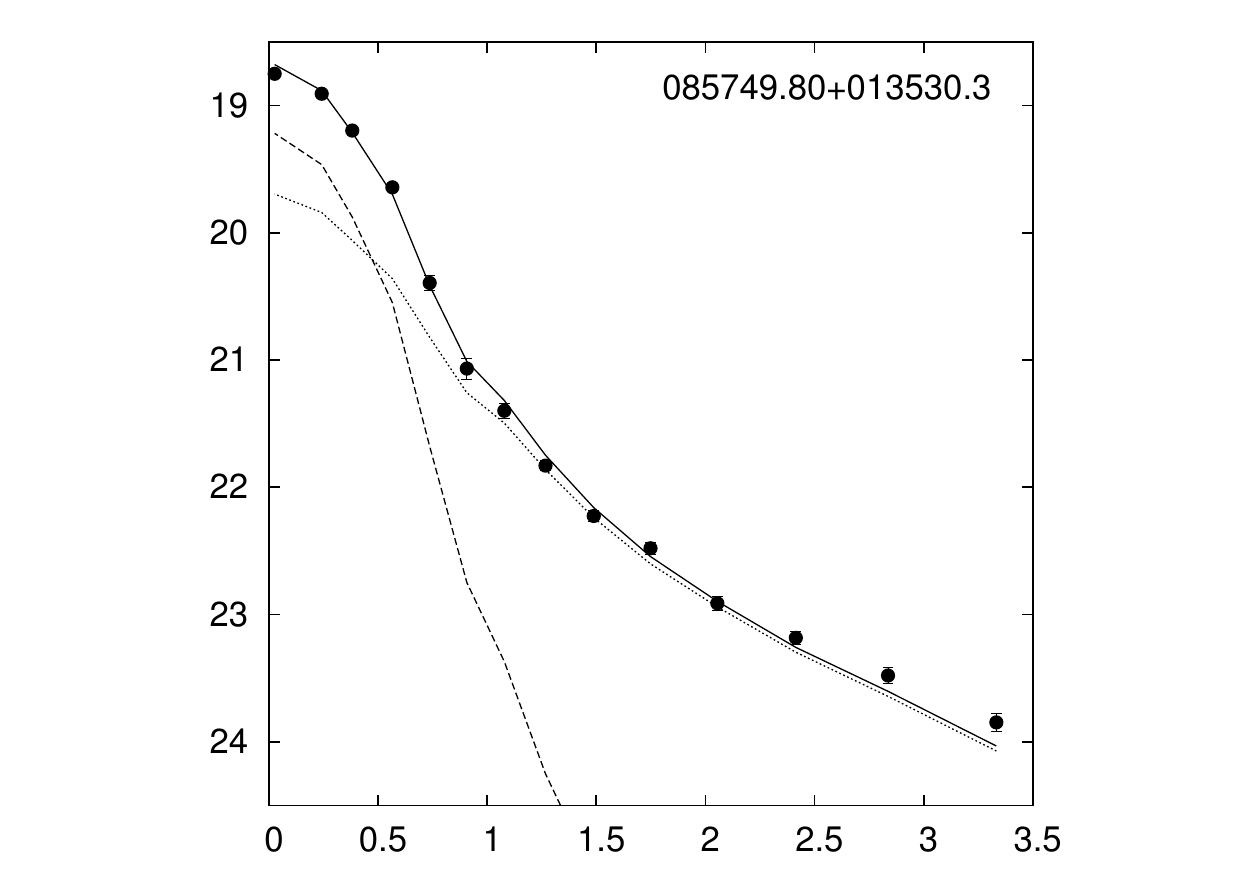}
\hspace*{-1.5cm}
\includegraphics[width=5.5cm]{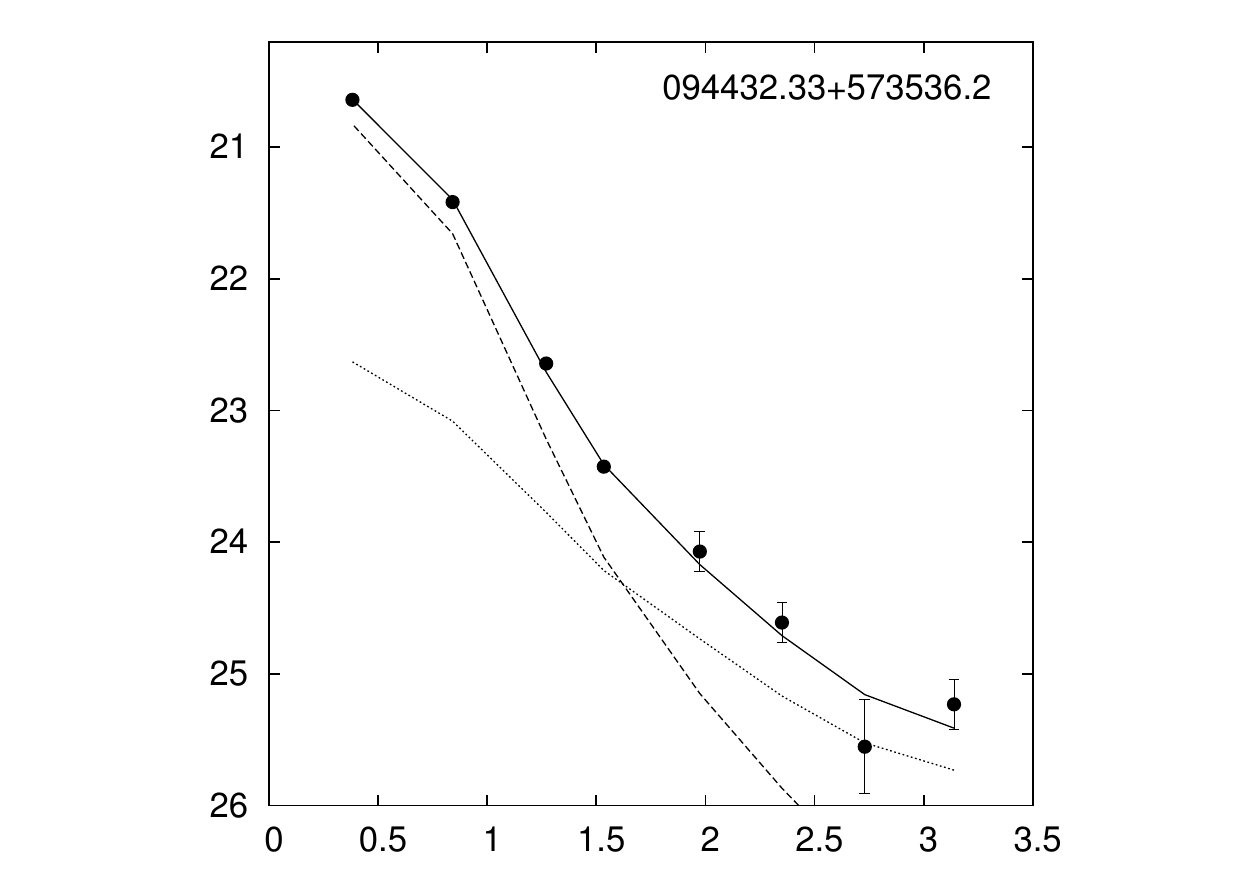}
\hspace*{-1.5cm}
\includegraphics[width=5.5cm]{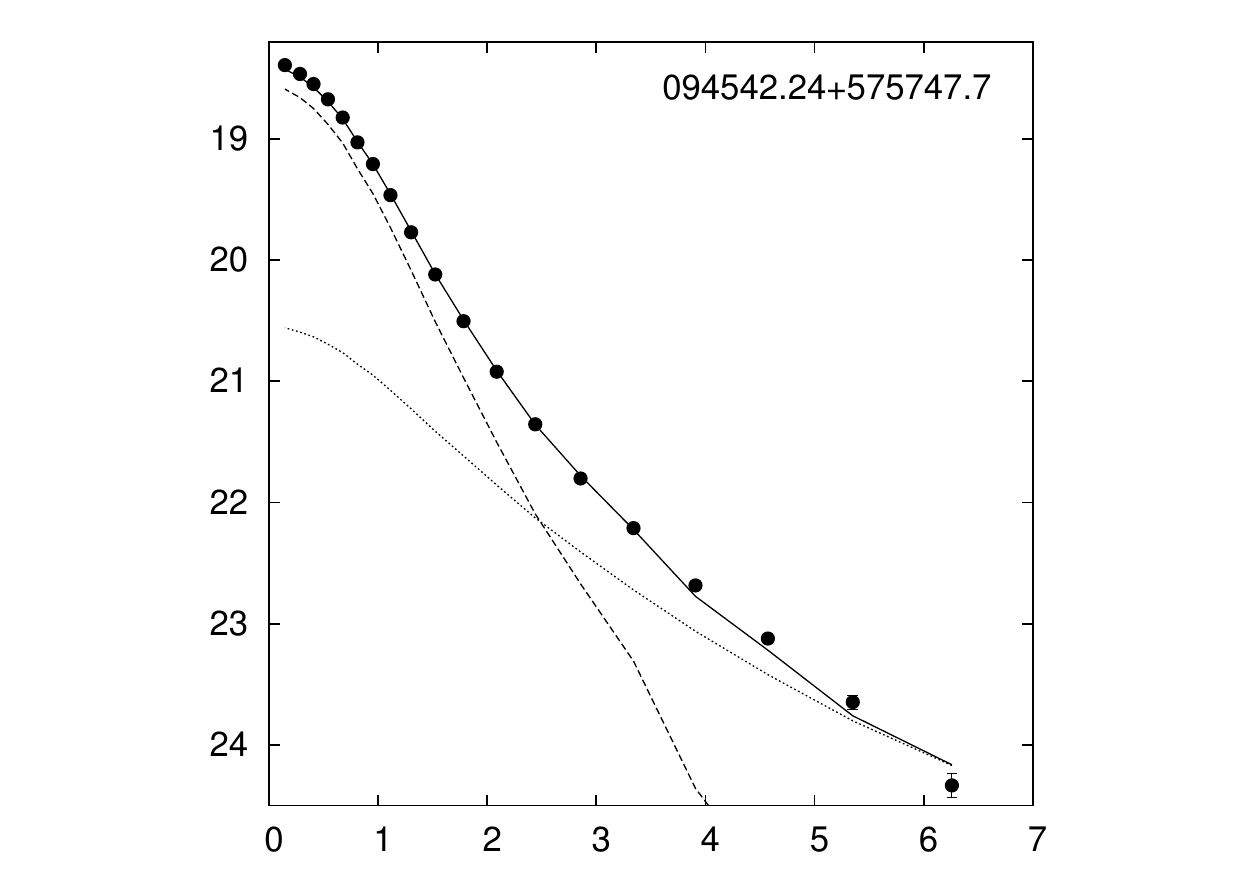}\\

\end{figure*}

\setcounter{figure}{0}

\begin{figure*}

\caption{--Continued.
}

\includegraphics[width=5.5cm]{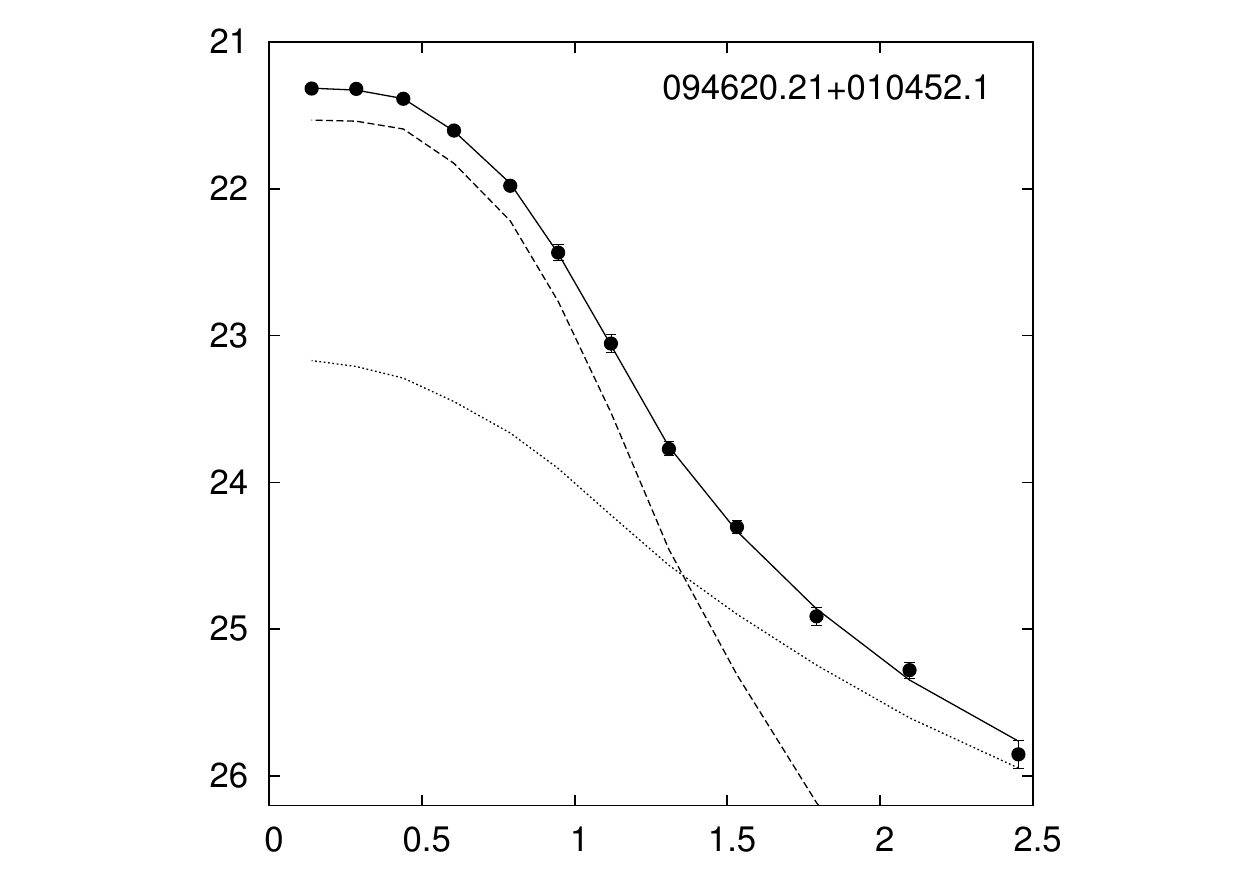}
\hspace*{-1.5cm}
\includegraphics[width=5.5cm]{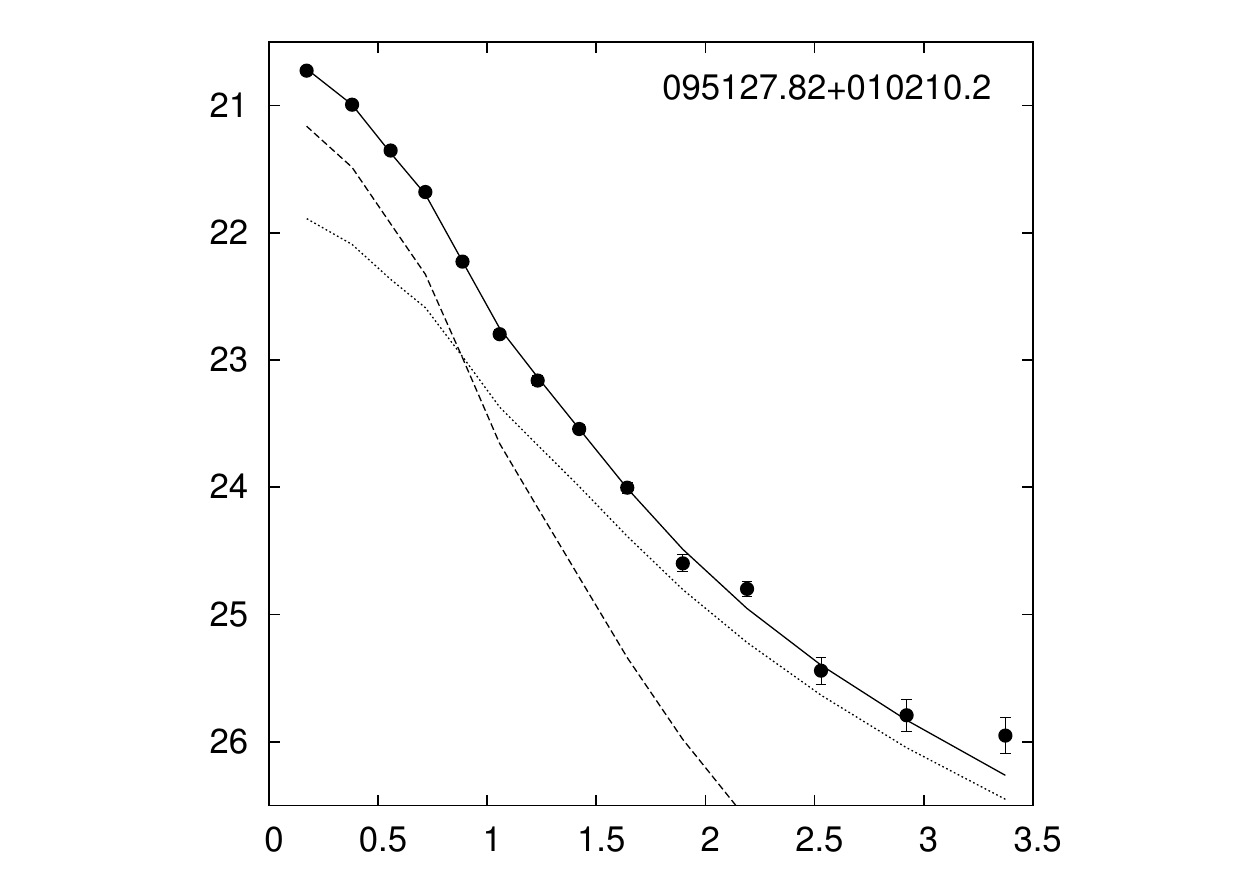}
\hspace*{-1.5cm}
\includegraphics[width=5.5cm]{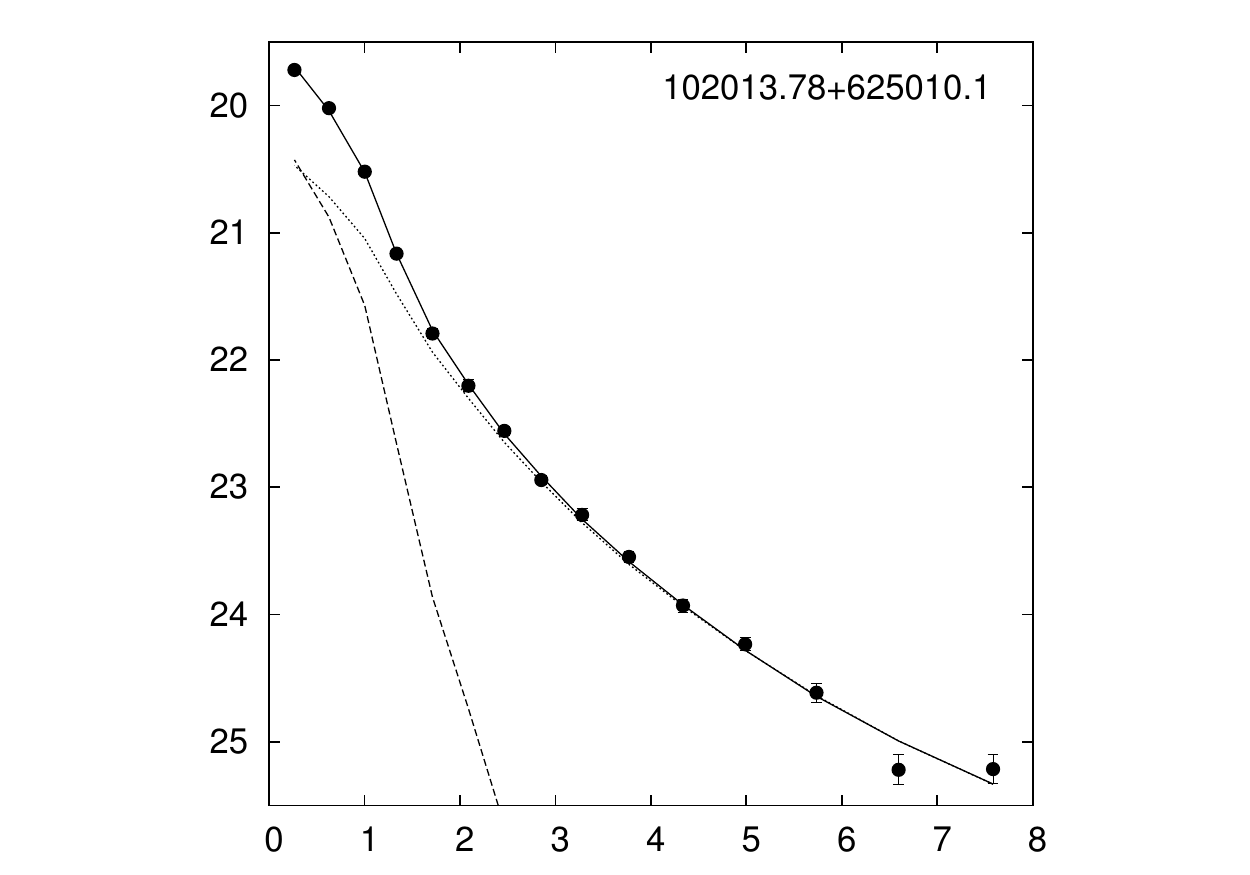}
\hspace*{-1.5cm}
\includegraphics[width=5.5cm]{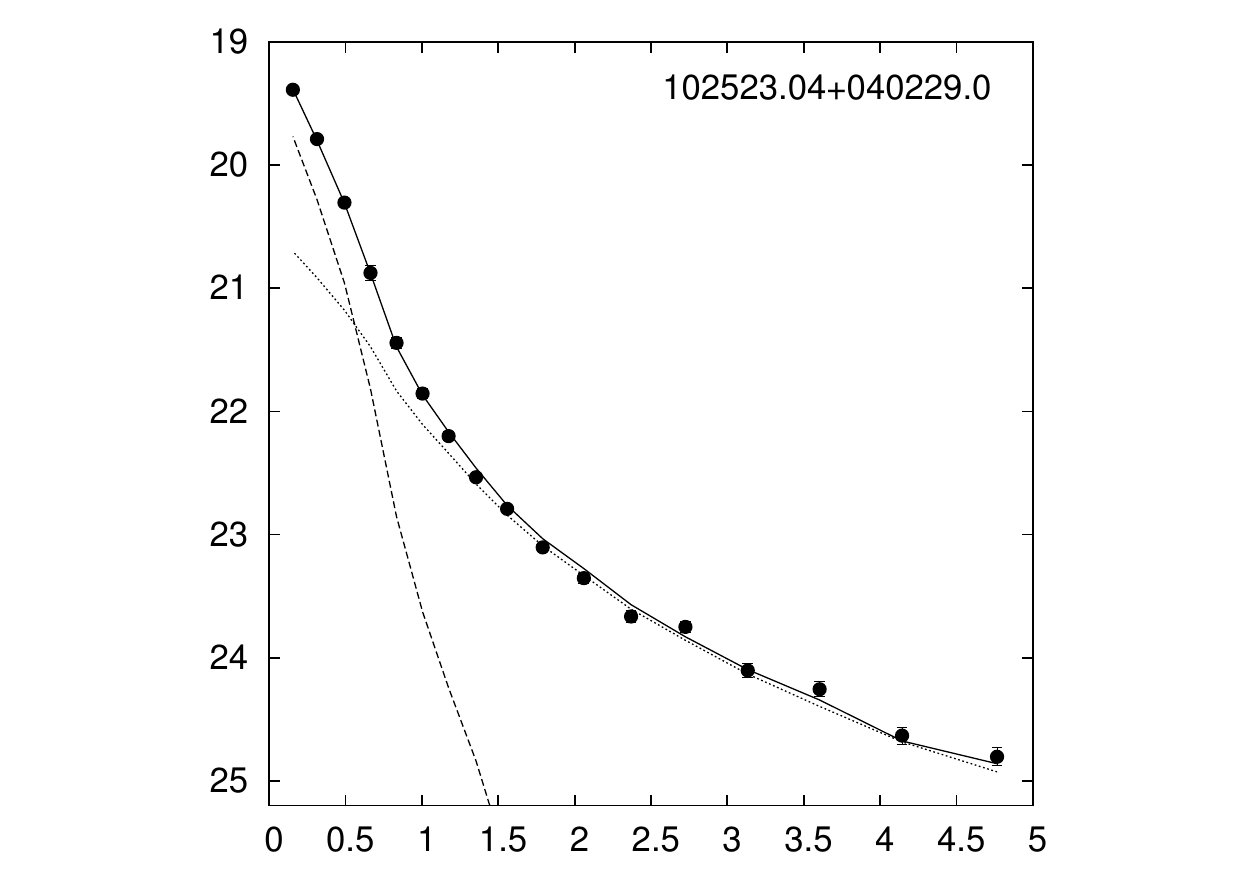}

\includegraphics[width=5.5cm]{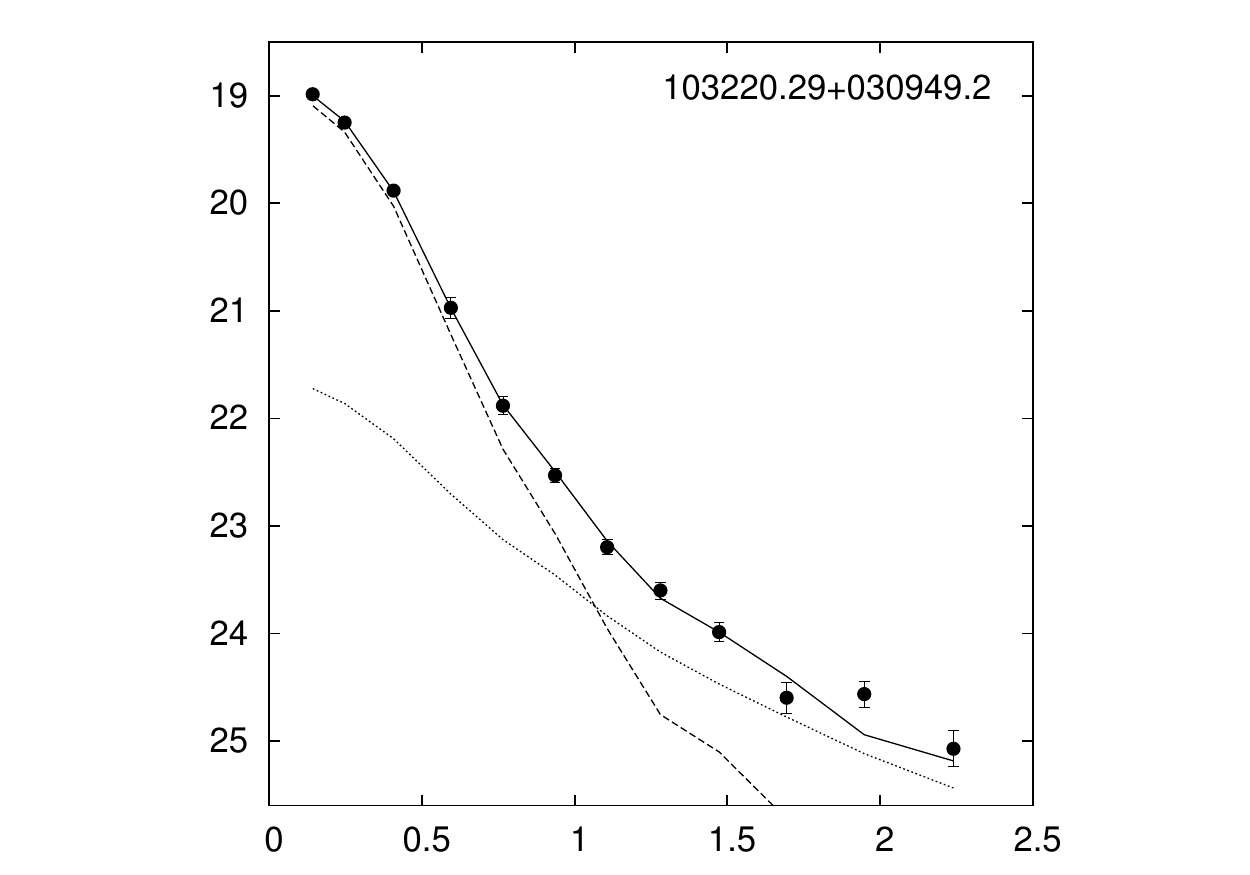}
\hspace*{-1.5cm}
\includegraphics[width=5.5cm]{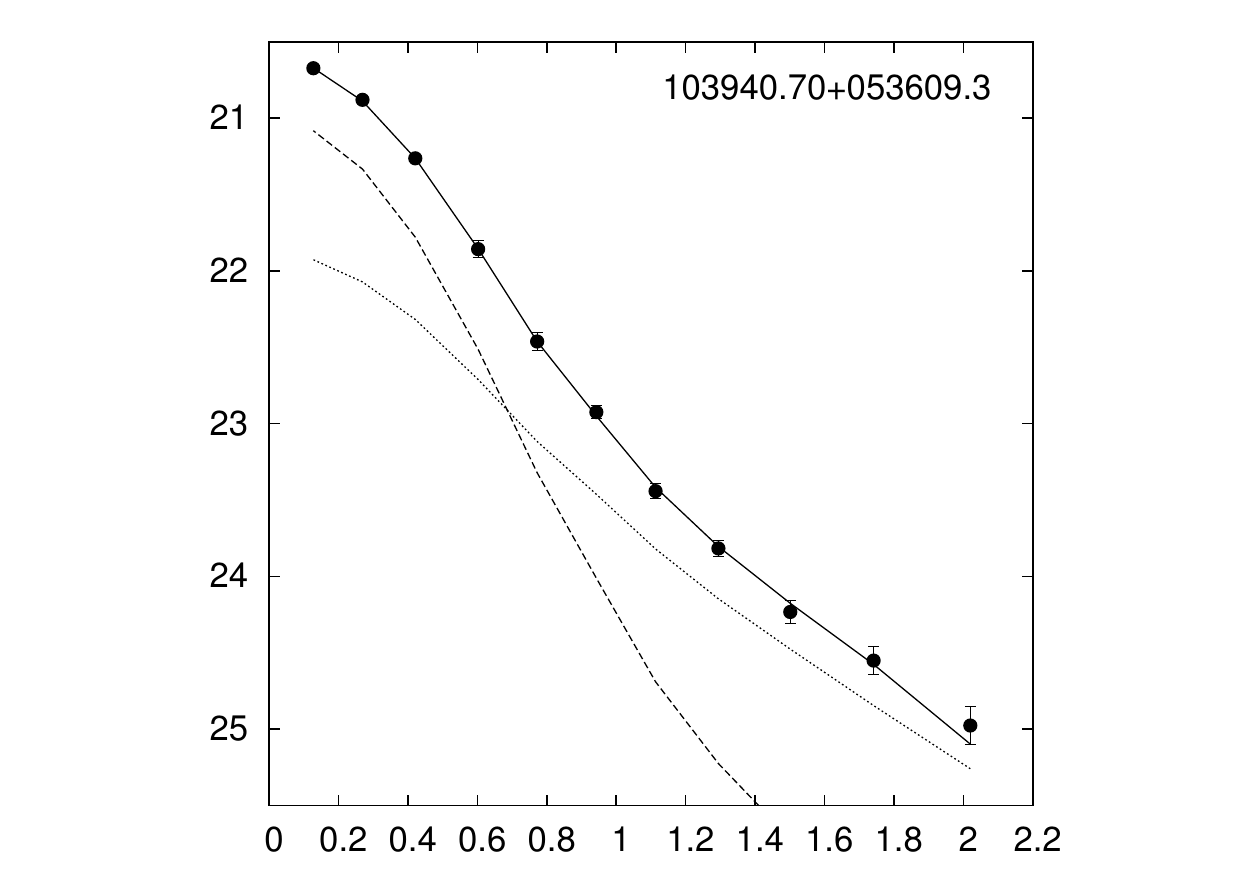}
\hspace*{-1.5cm}
\includegraphics[width=5.5cm]{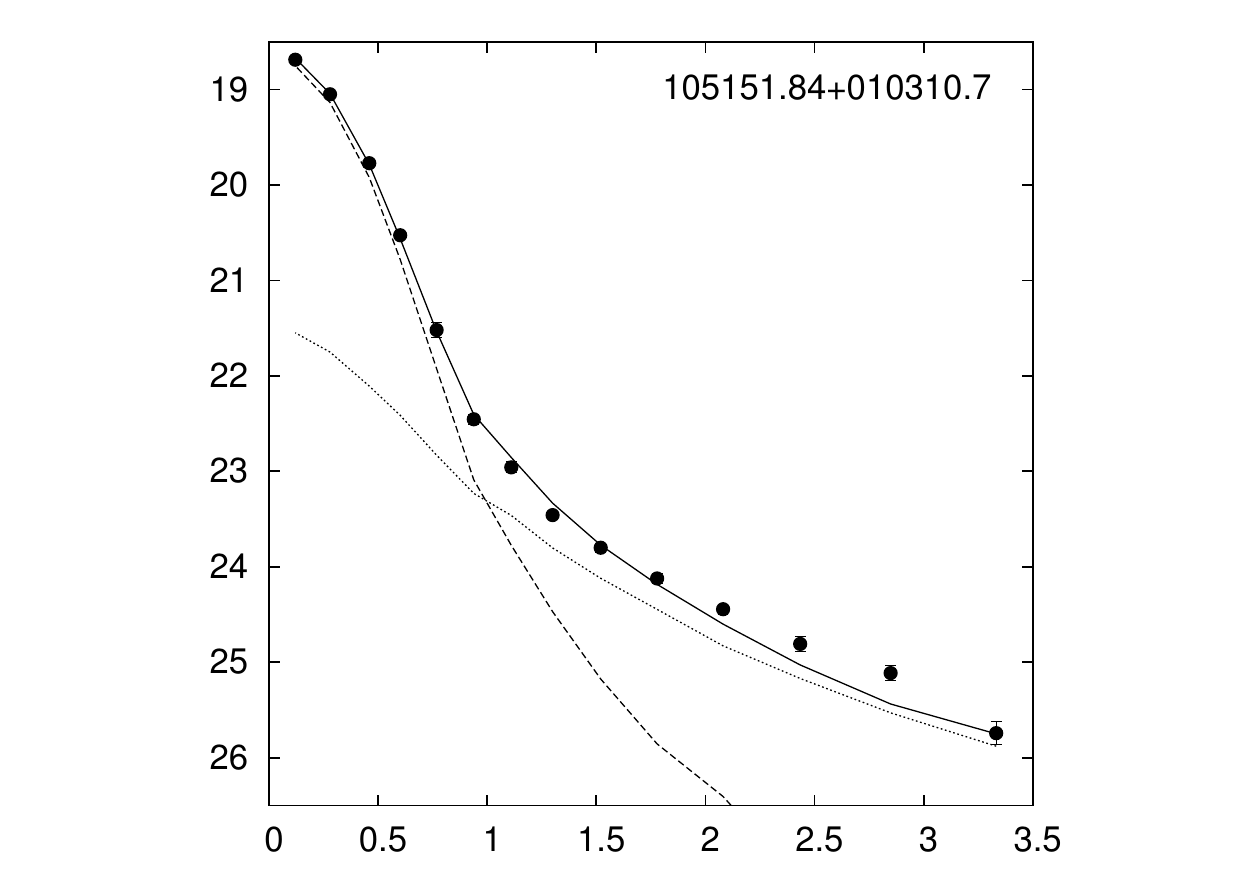}
\hspace*{-1.5cm}
\includegraphics[width=5.5cm]{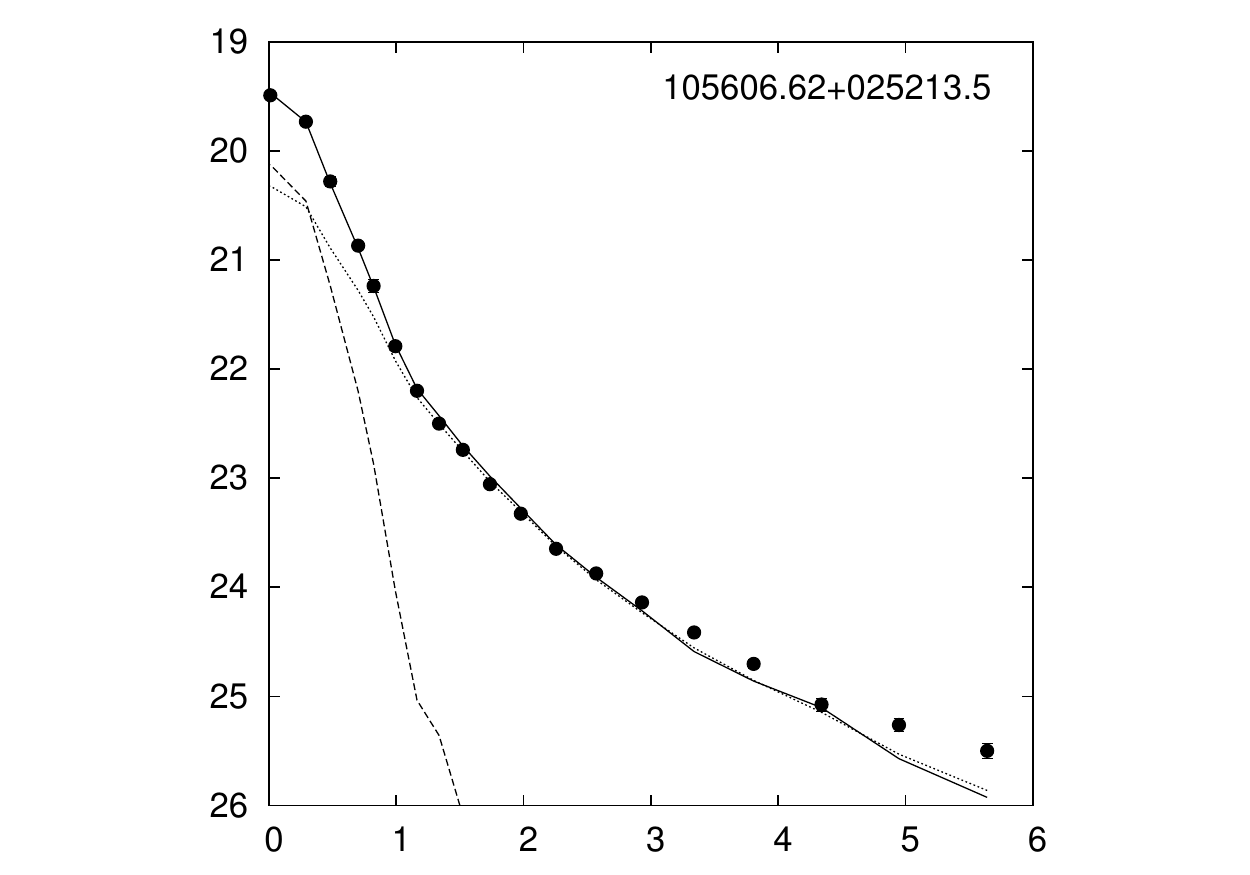}\\

\includegraphics[width=5.5cm]{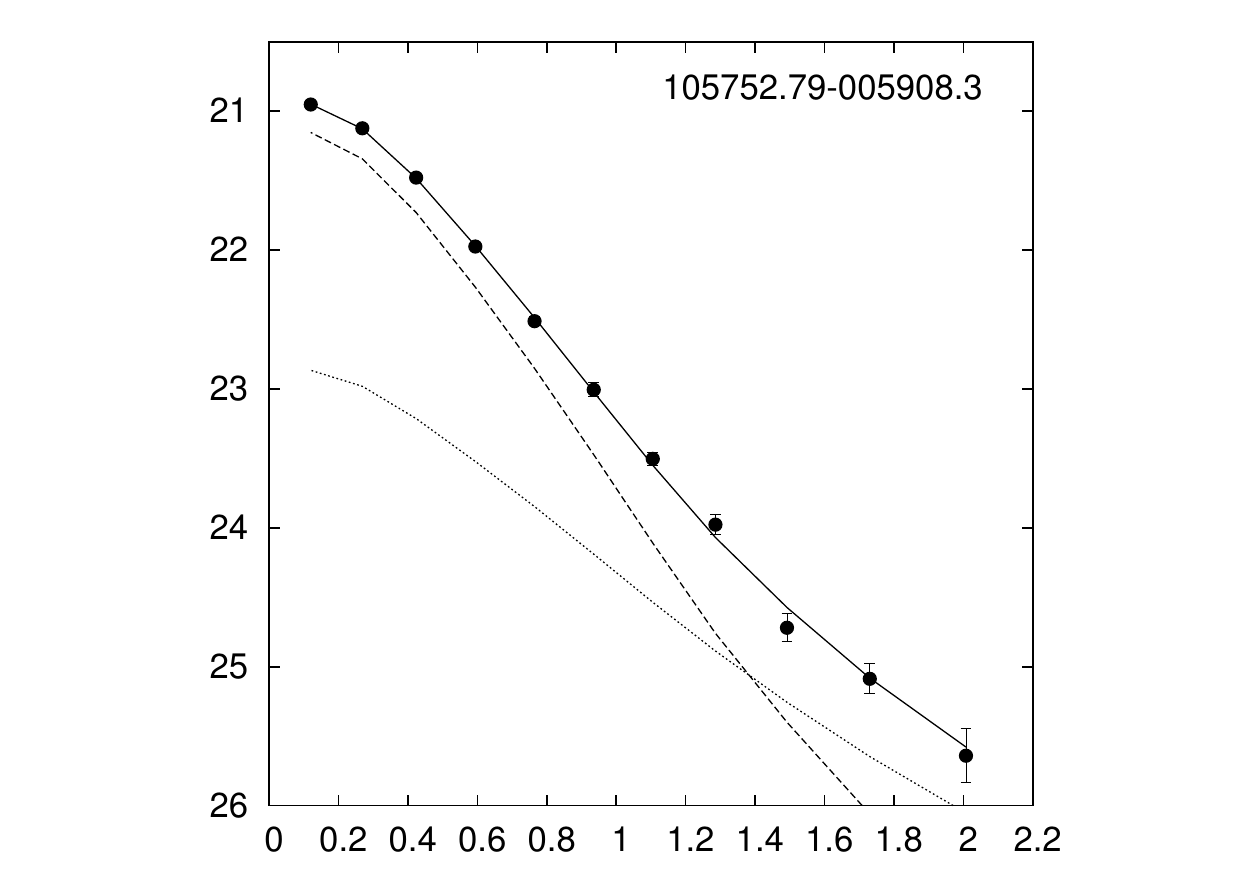}
\hspace*{-1.5cm}
\includegraphics[width=5.5cm]{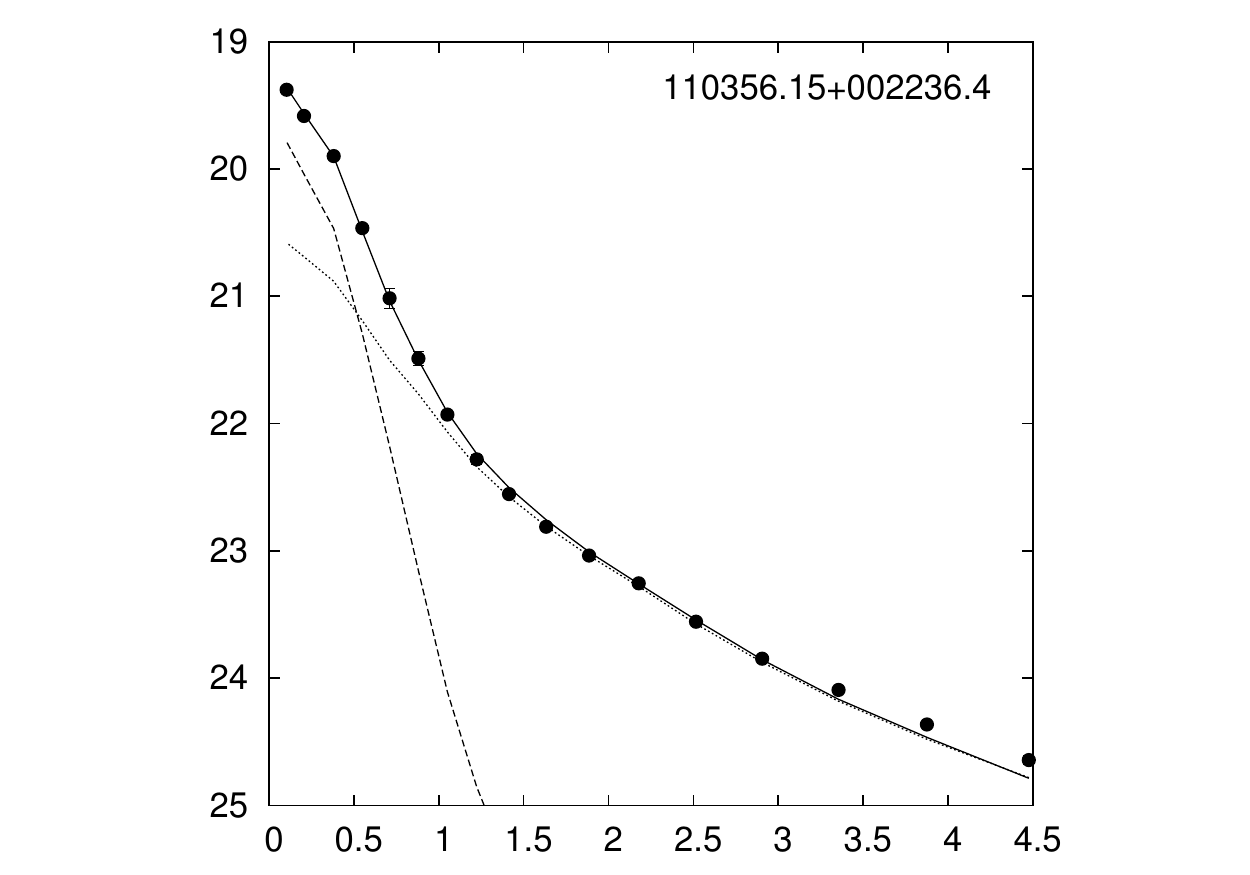}
\hspace*{-1.5cm}
\includegraphics[width=5.5cm]{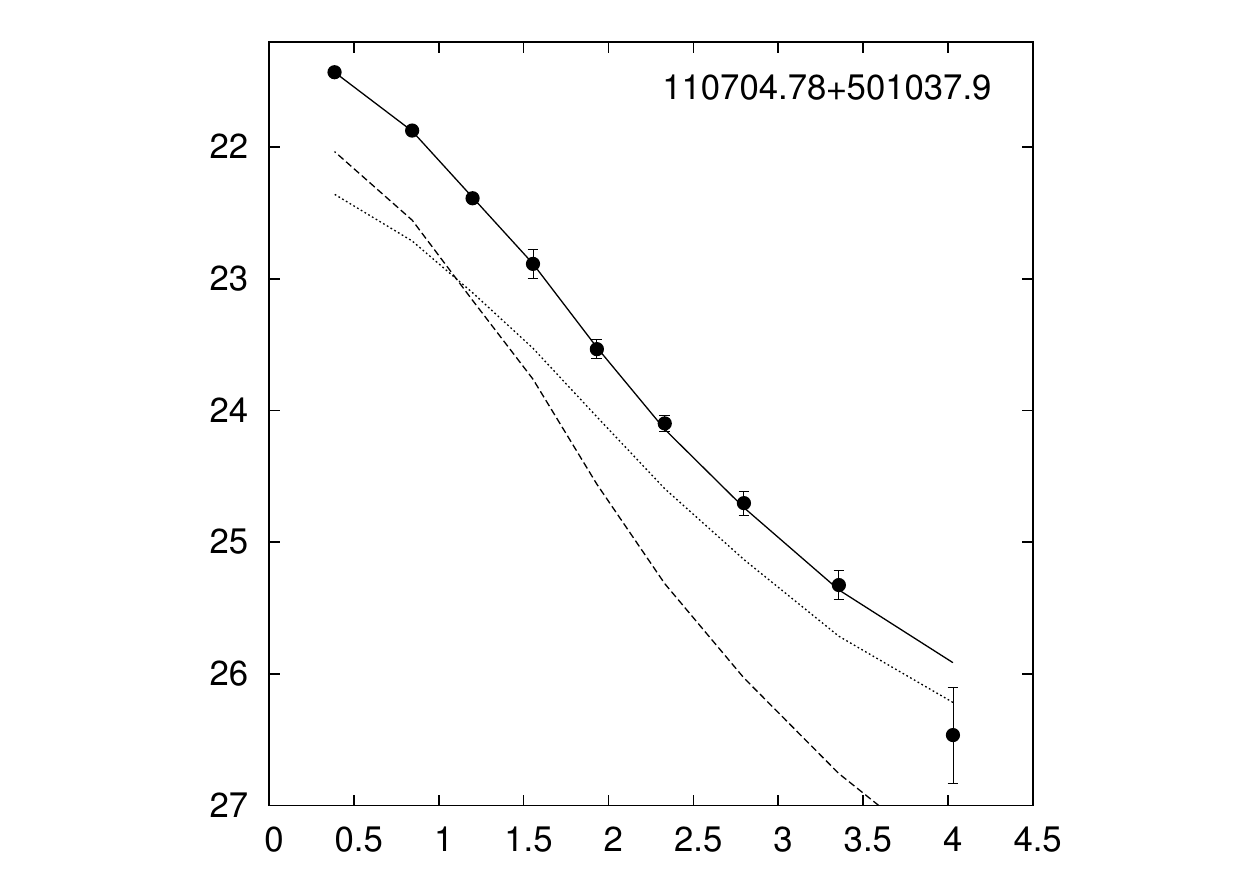}
\hspace*{-1.5cm}
\includegraphics[width=5.5cm]{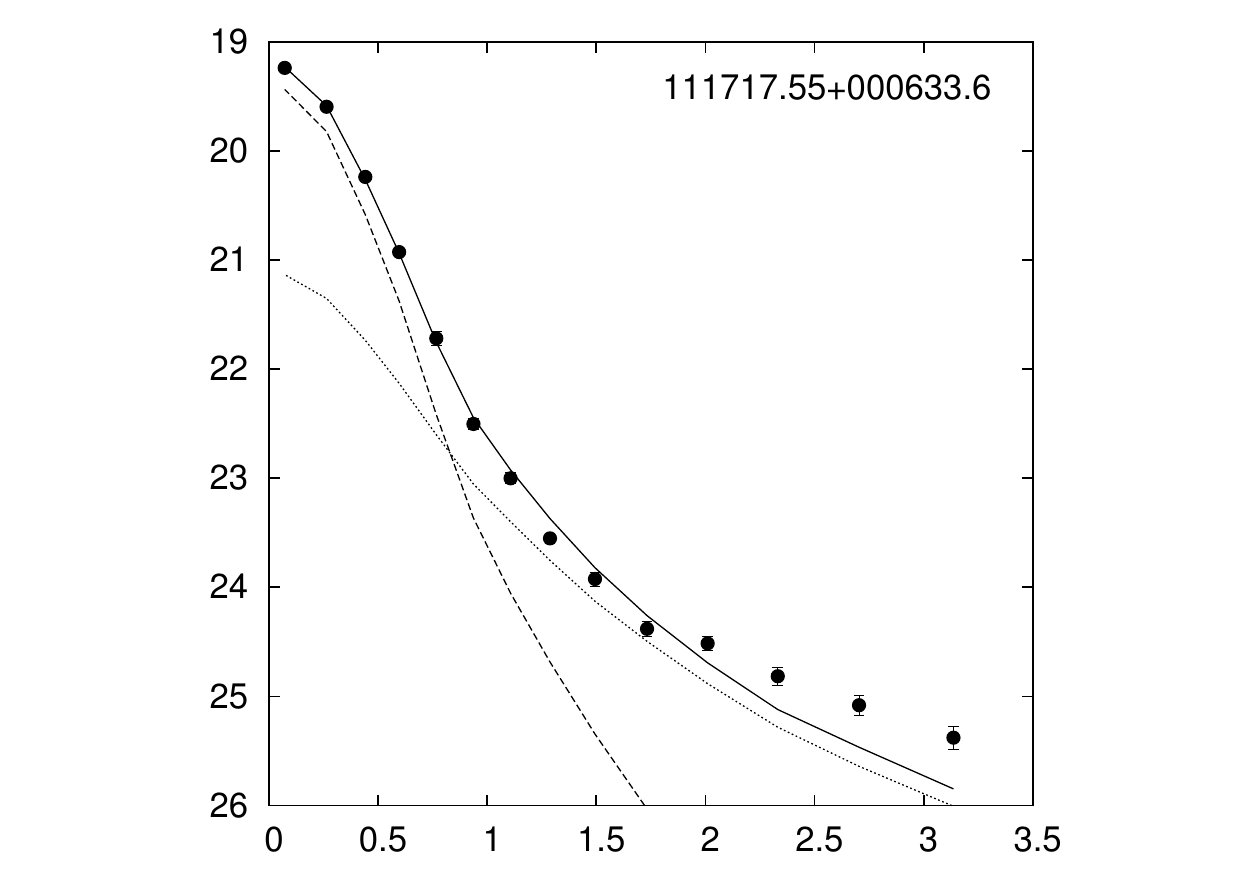}\\

\includegraphics[width=5.5cm]{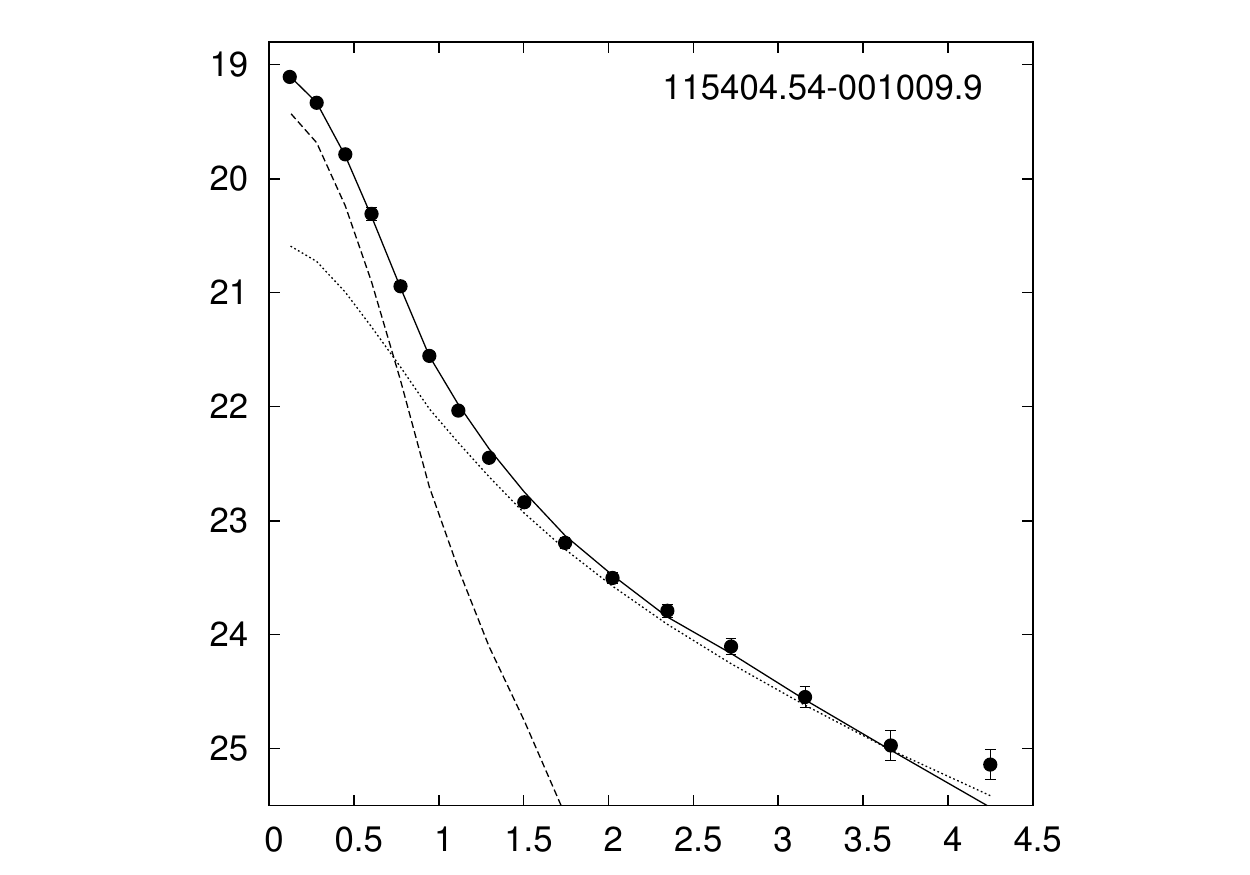}
\hspace*{-1.5cm}
\includegraphics[width=5.5cm]{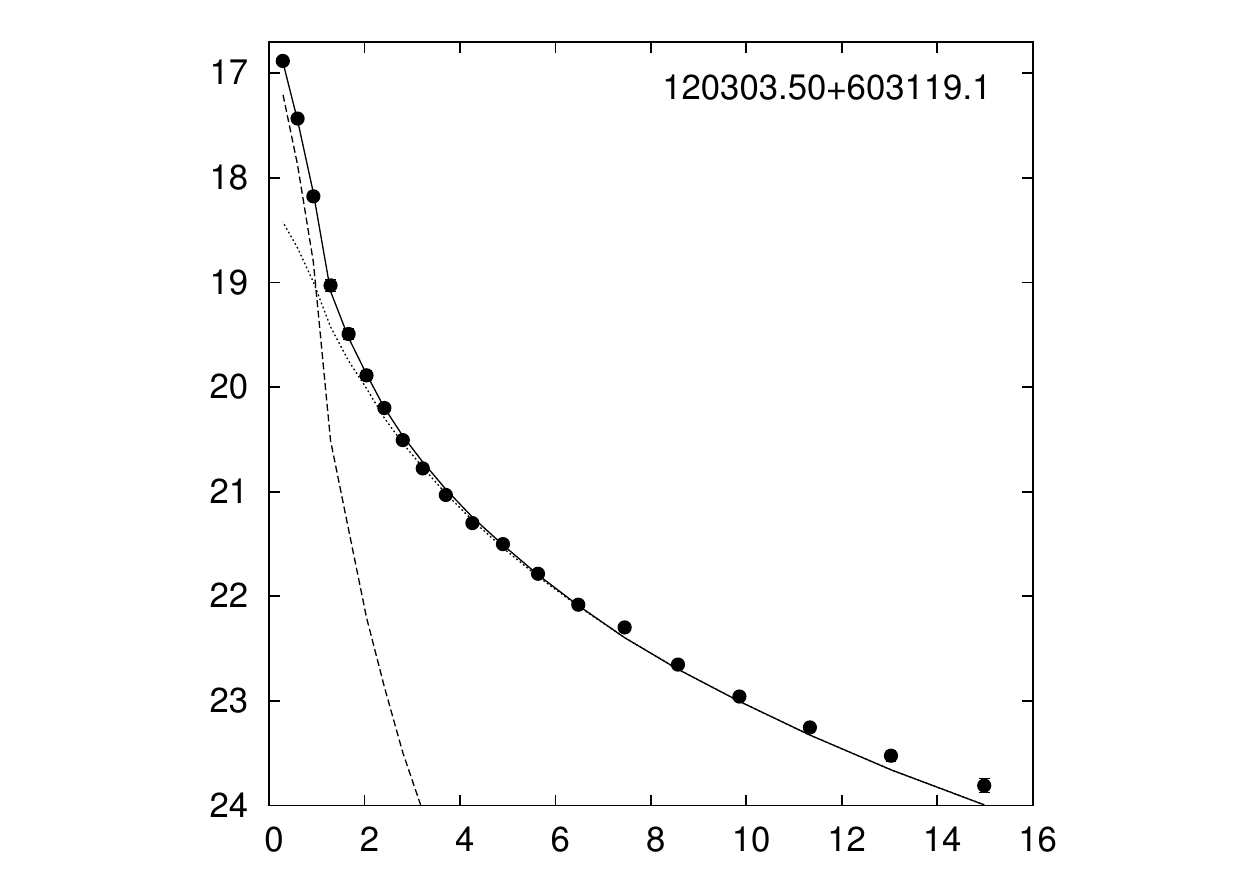}
\hspace*{-1.5cm}
\includegraphics[width=5.5cm]{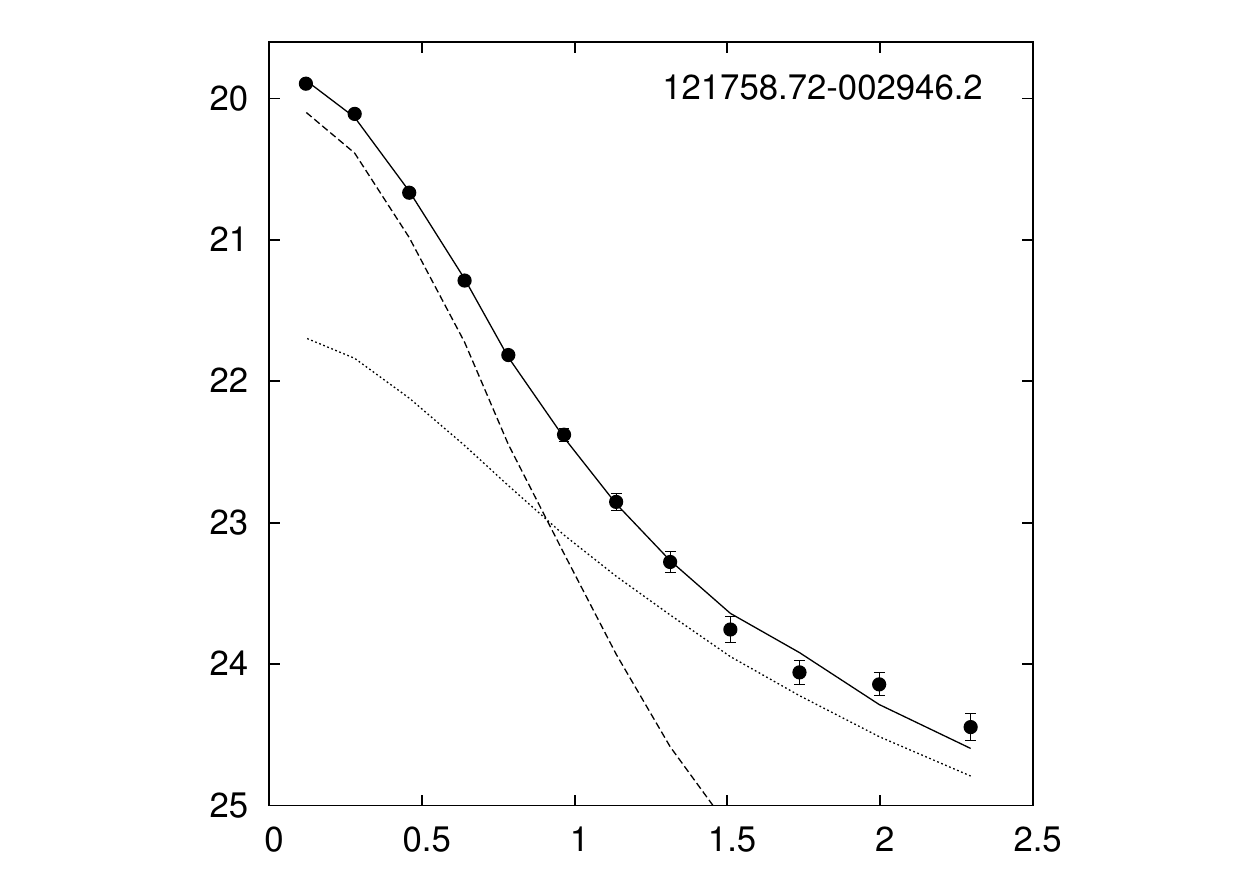}
\hspace*{-1.5cm}
\includegraphics[width=5.5cm]{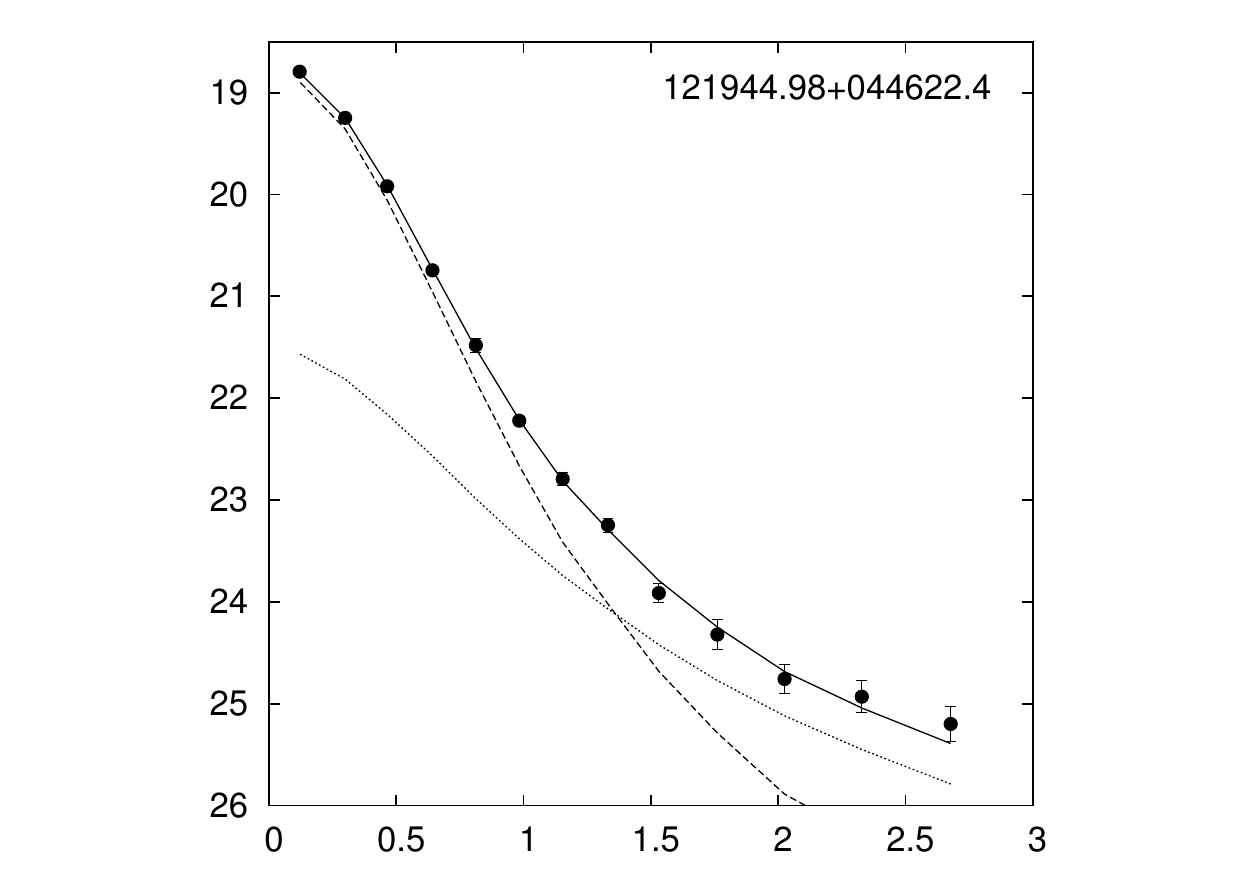}

\includegraphics[width=5.5cm]{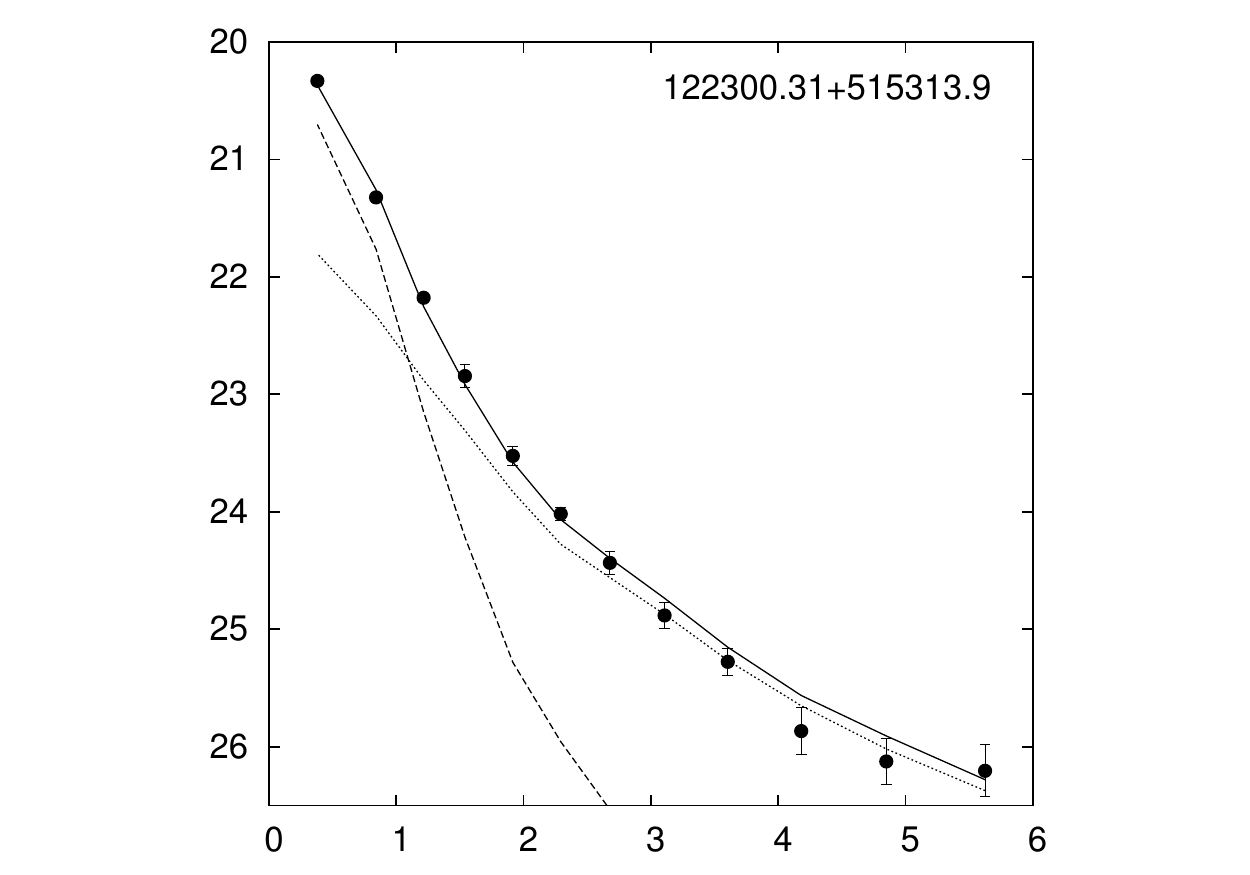}
\hspace*{-1.5cm}
\includegraphics[width=5.5cm]{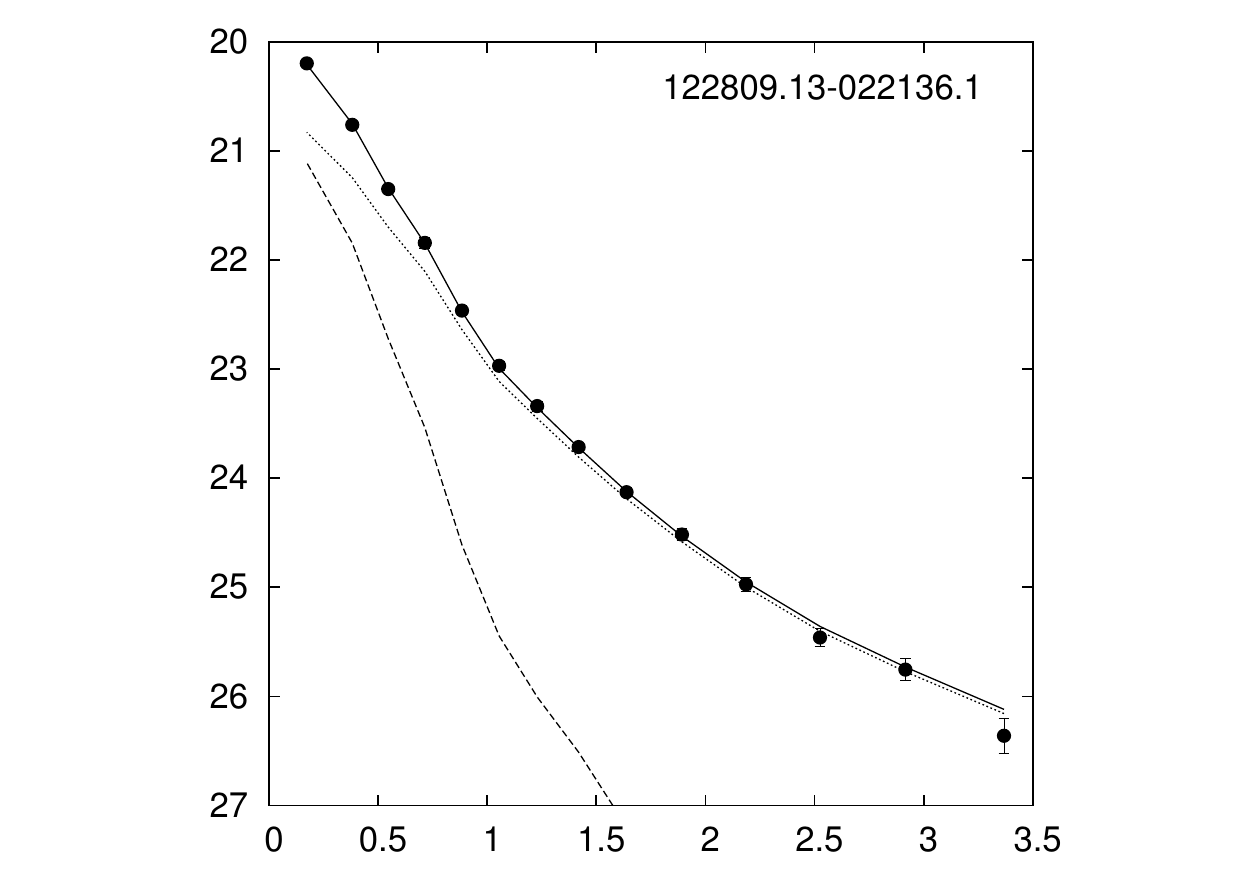}
\hspace*{-1.5cm}
\includegraphics[width=5.5cm]{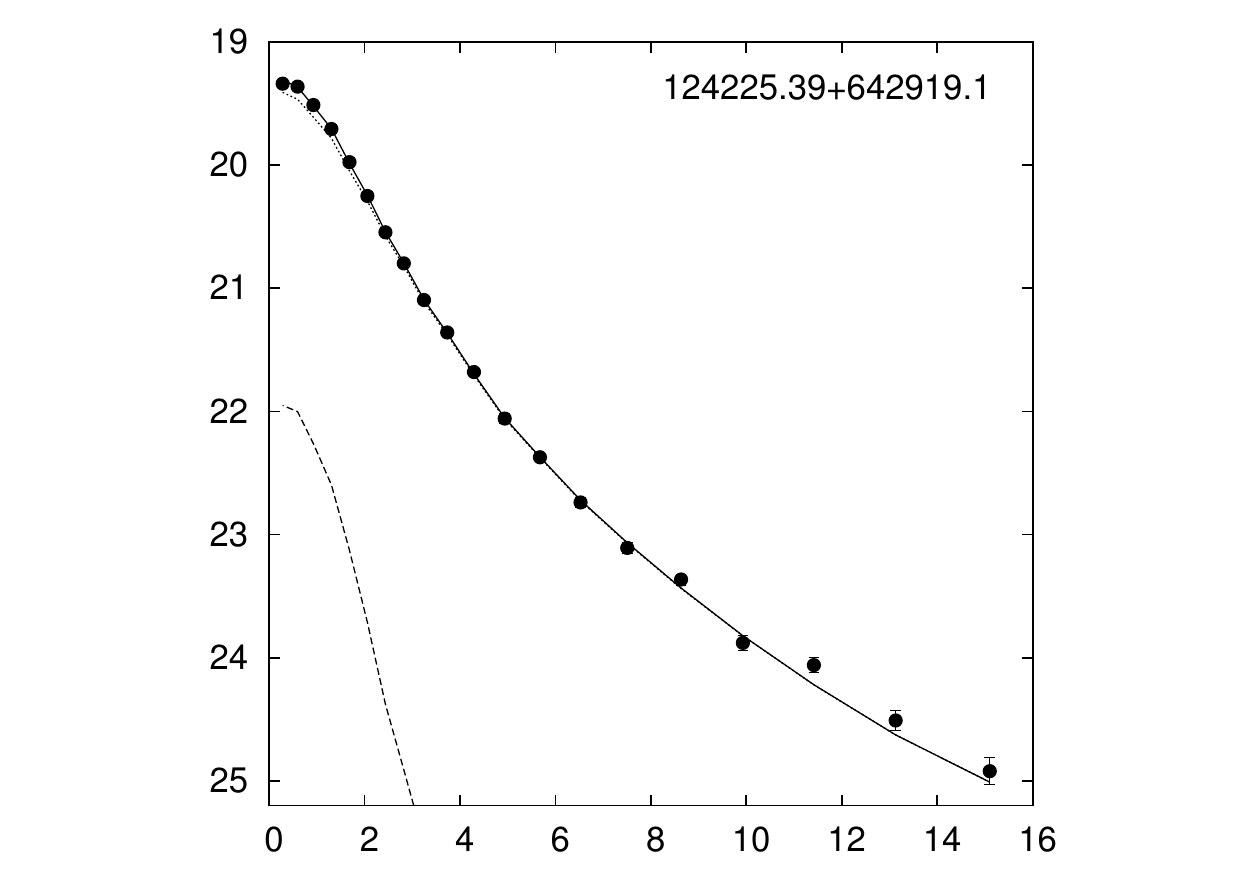}
\hspace*{-1.5cm}
\includegraphics[width=5.5cm]{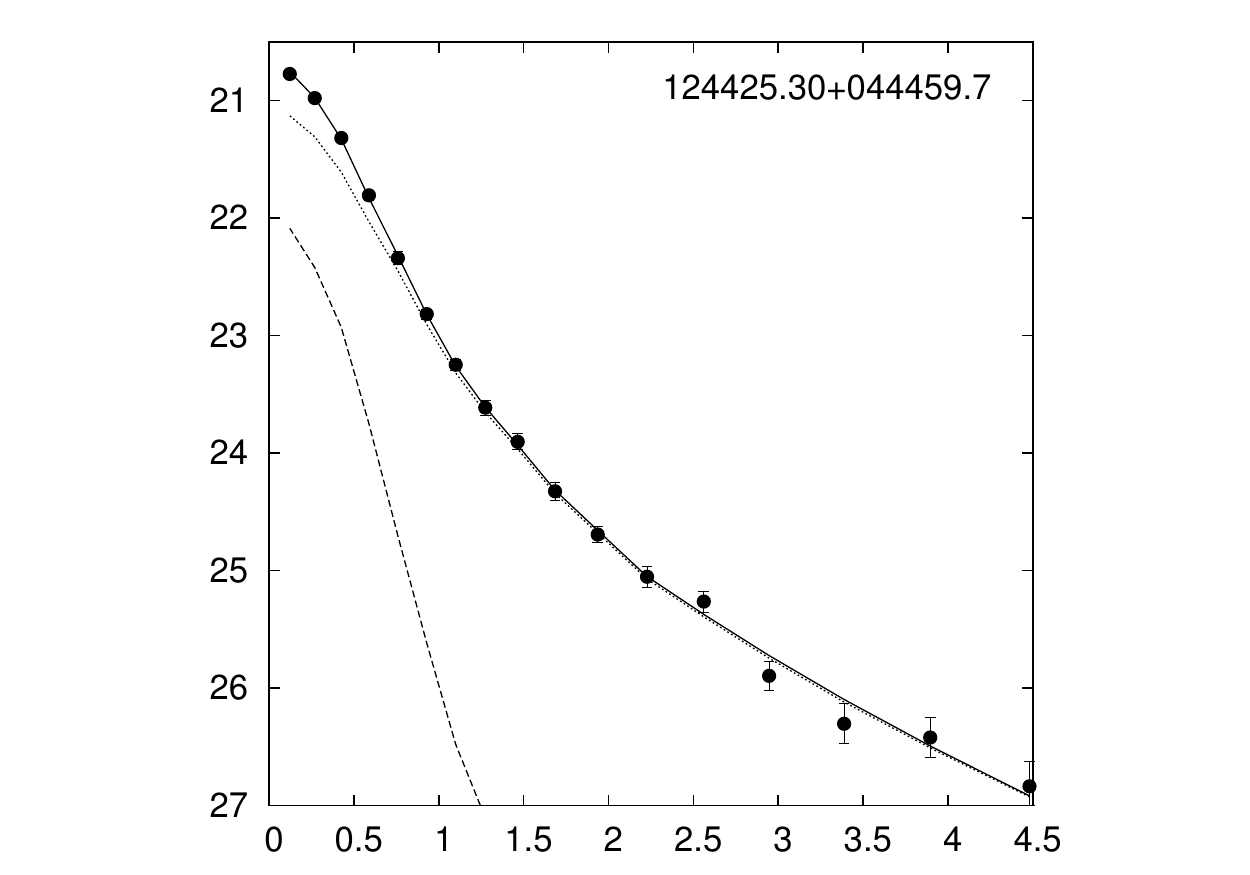}\\

\end{figure*}

\setcounter{figure}{0}

\begin{figure*}
\caption{--Continued.}
\includegraphics[width=5.5cm]{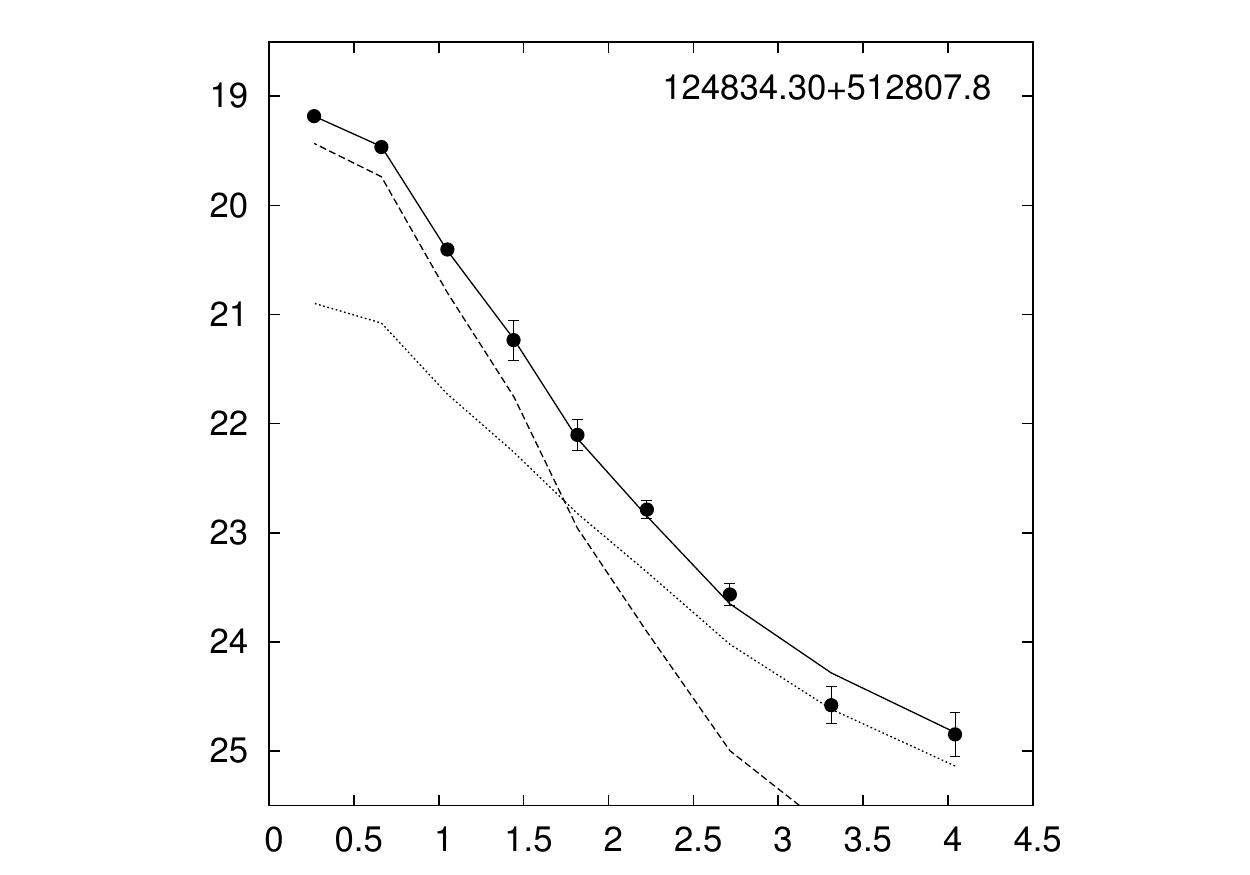}
\hspace*{-1.5cm}
\includegraphics[width=5.5cm]{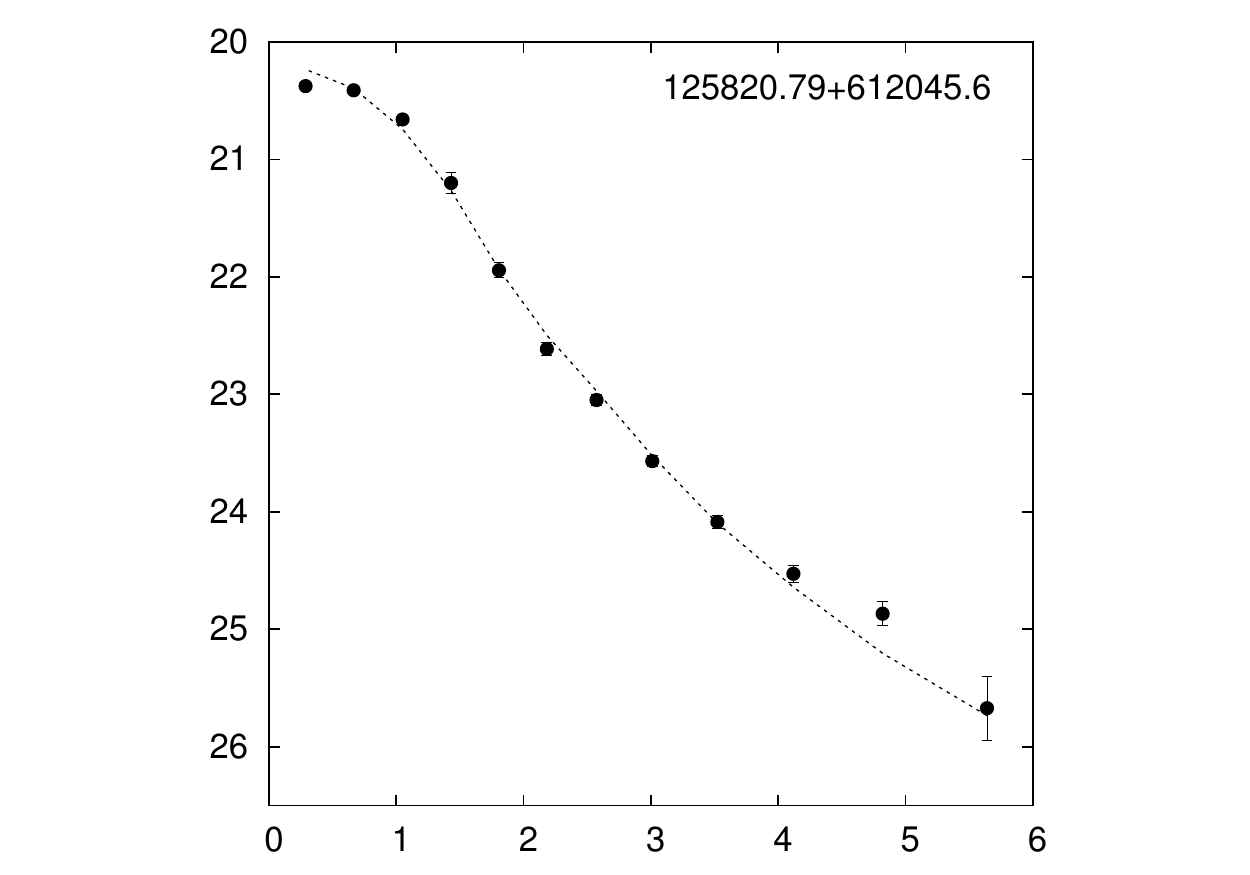}
\hspace*{-1.5cm}
\includegraphics[width=5.5cm]{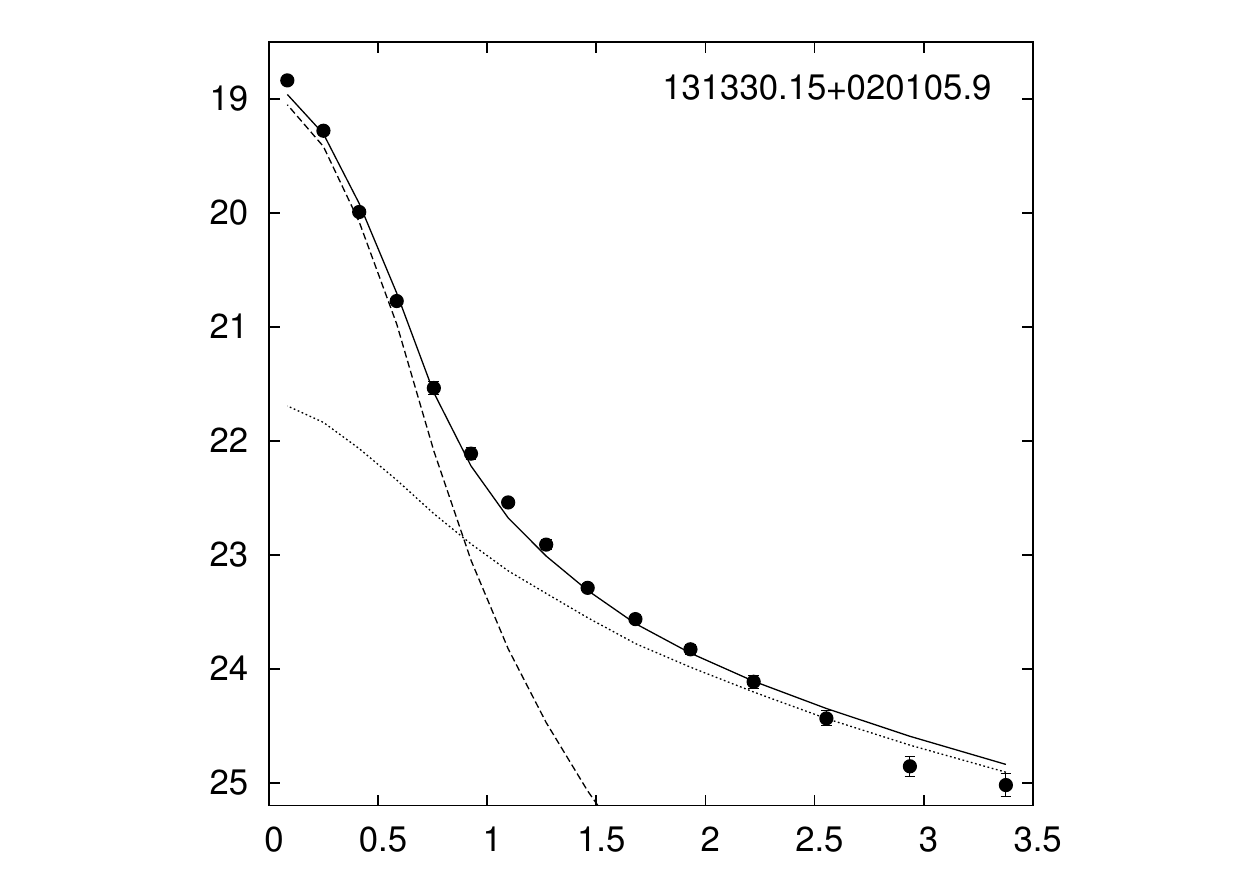}
\hspace*{-1.5cm}
\includegraphics[width=5.5cm]{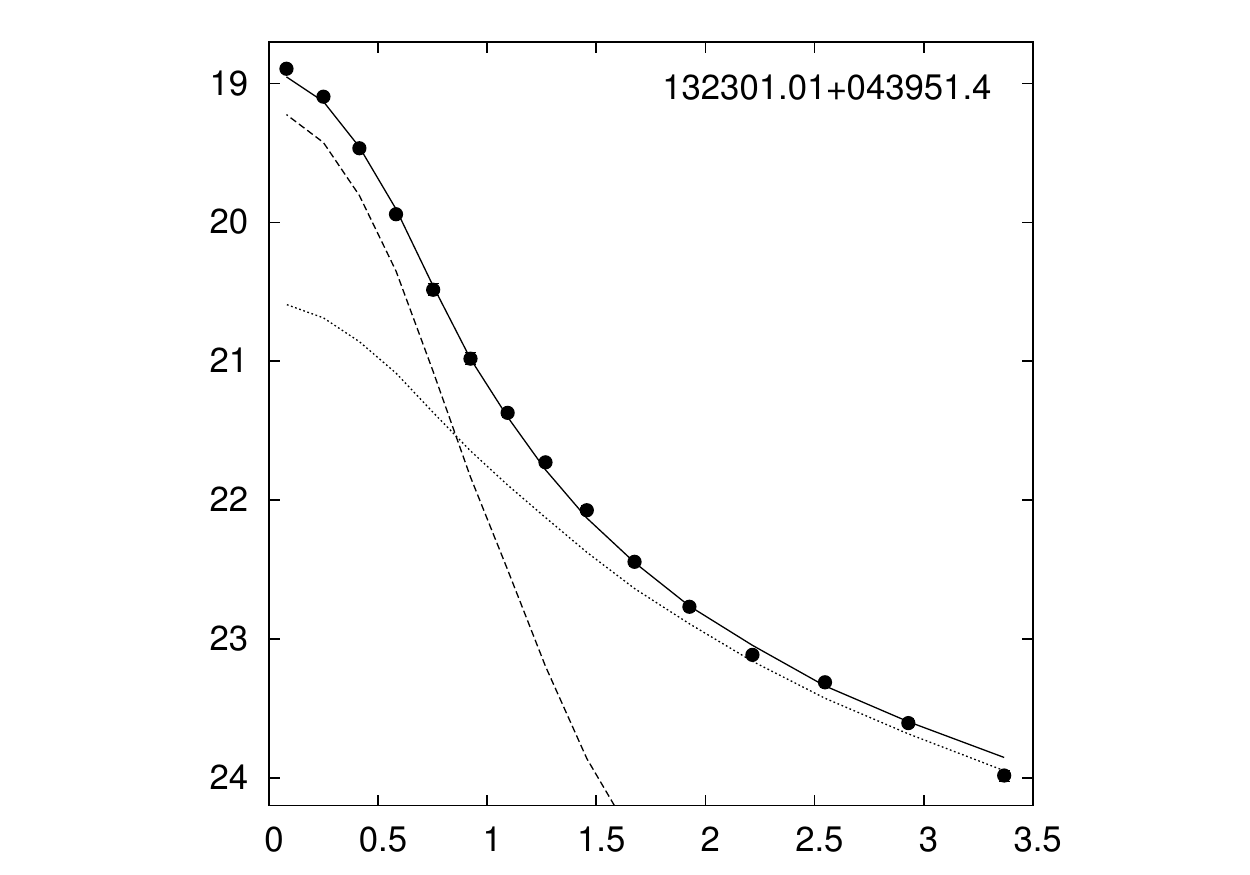}\\

\includegraphics[width=5.5cm]{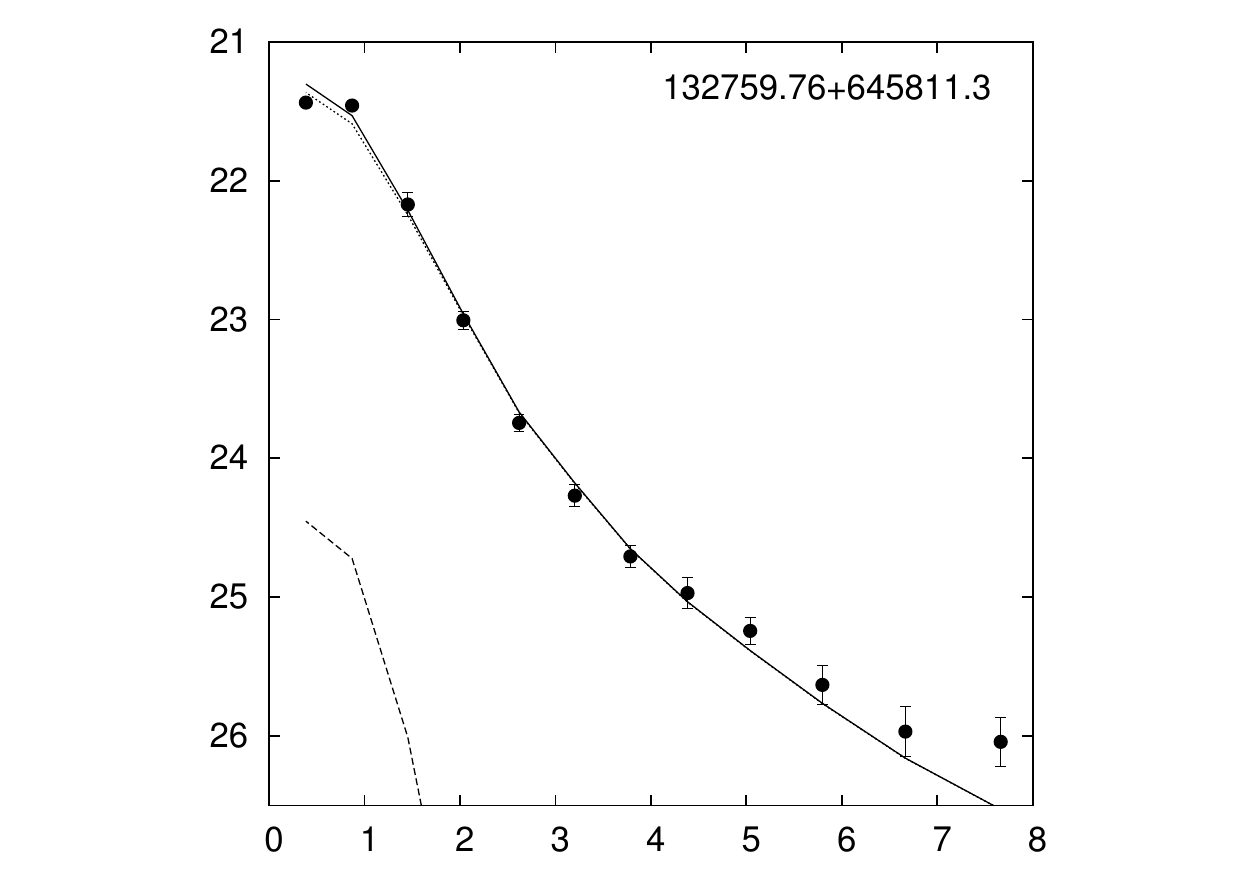}
\hspace*{-1.5cm}
\includegraphics[width=5.5cm]{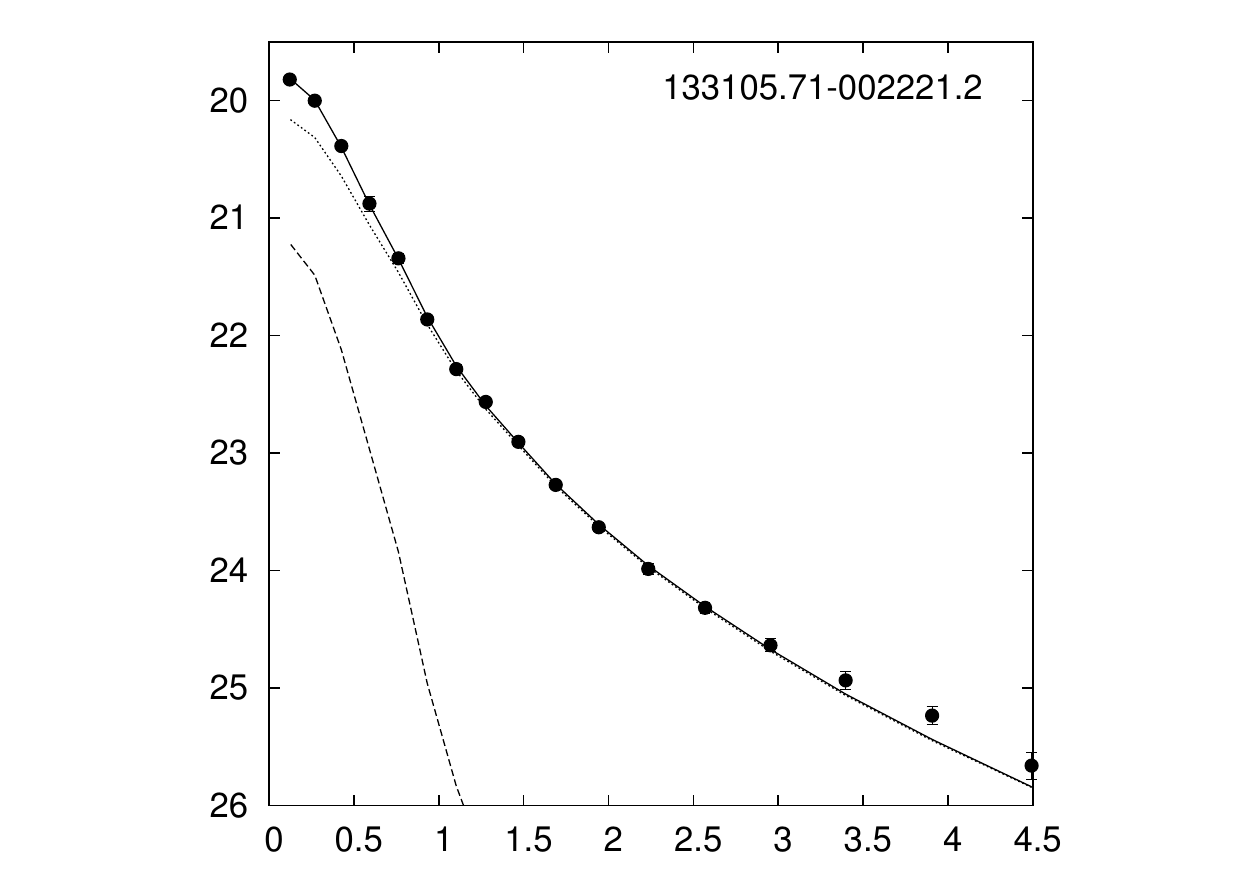}
\hspace*{-1.5cm}
\includegraphics[width=5.5cm]{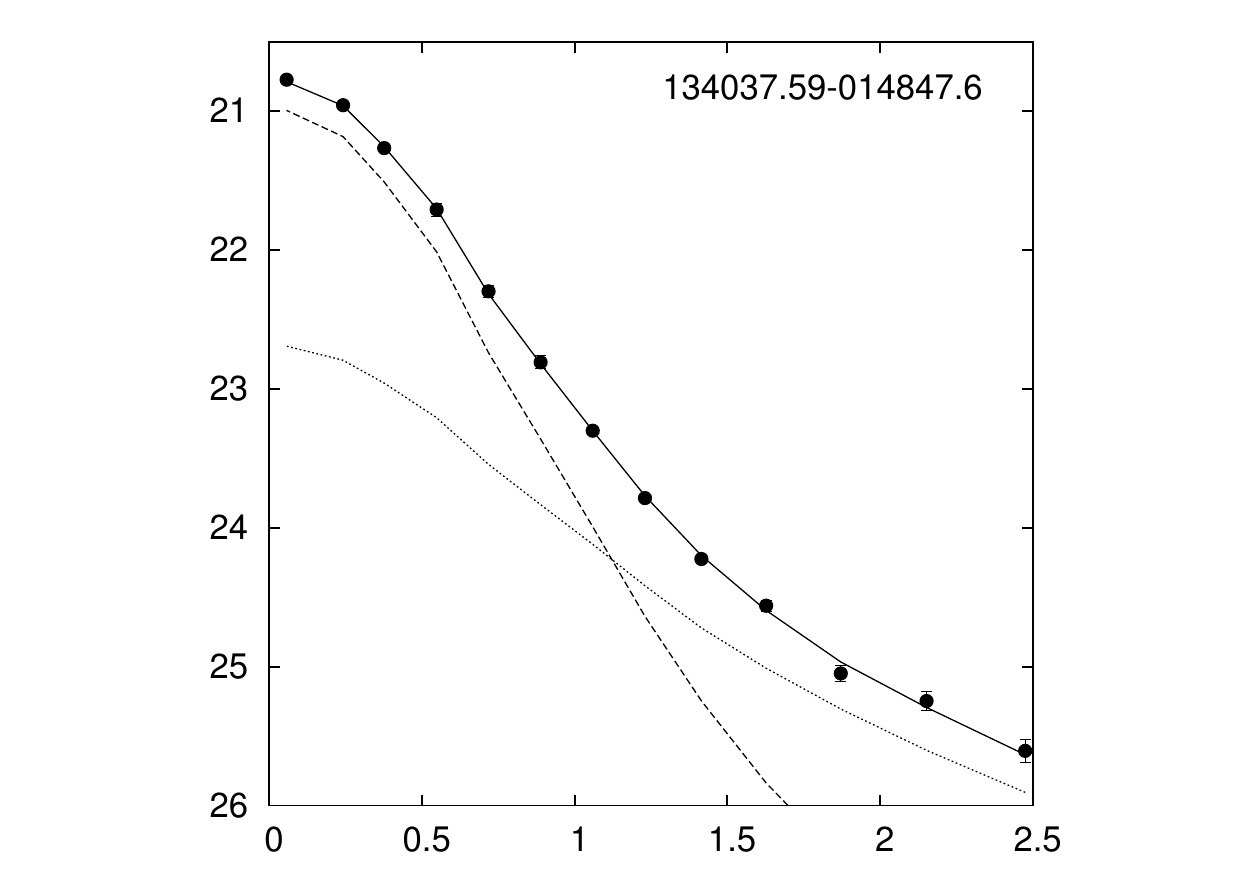}
\hspace*{-1.5cm}
\includegraphics[width=5.5cm]{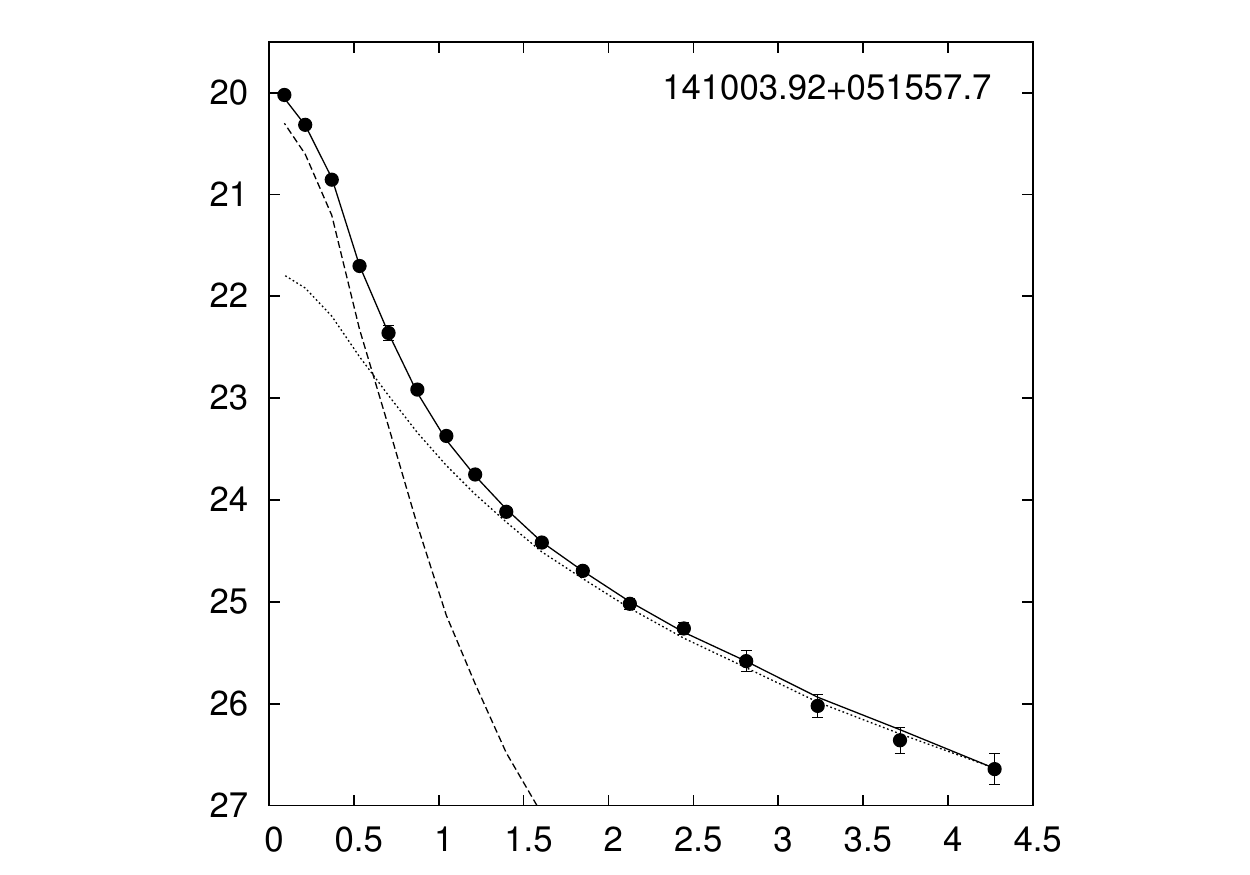}\\

\includegraphics[width=5.5cm]{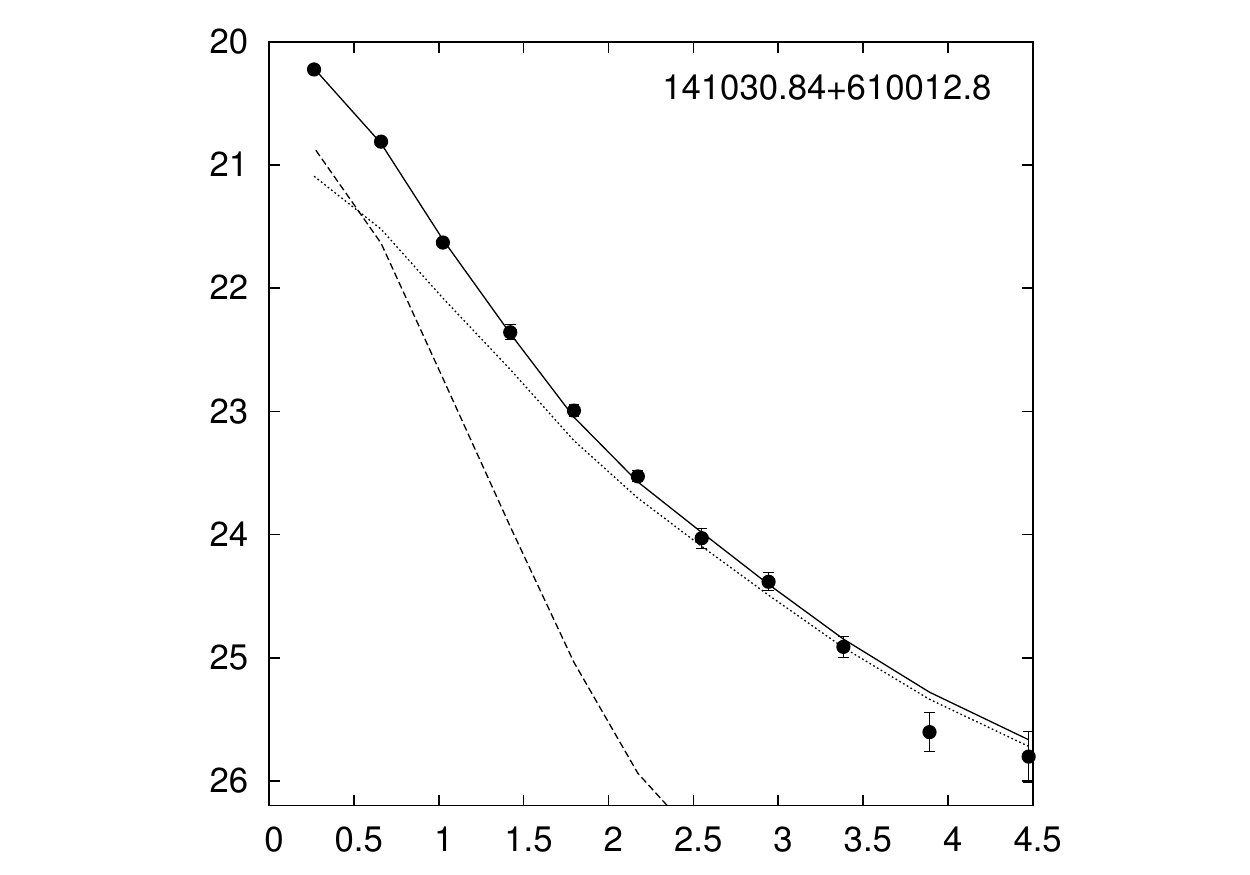}
\hspace*{-1.5cm}
\includegraphics[width=5.5cm]{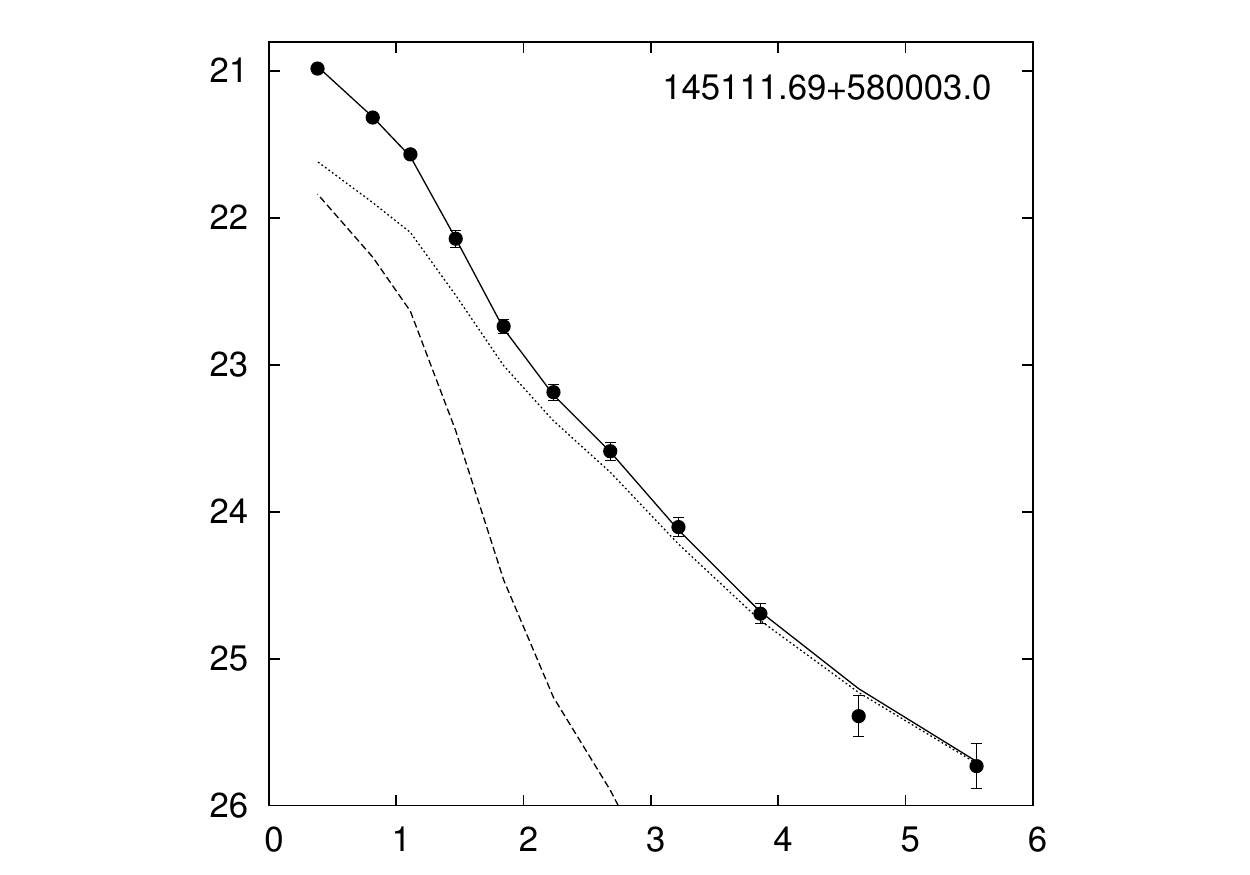}
\hspace*{-1.5cm}
\includegraphics[width=5.5cm]{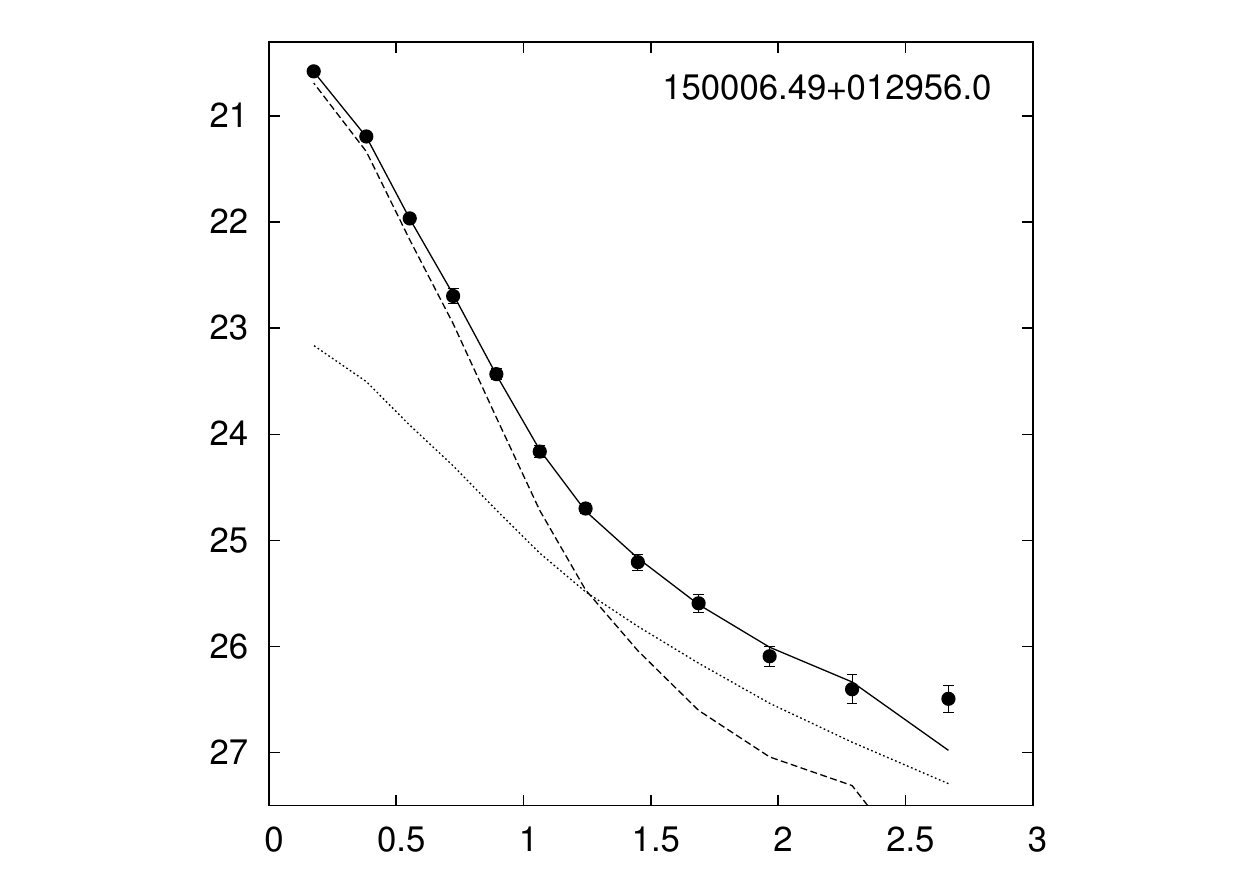}
\hspace*{-1.5cm}
\includegraphics[width=5.5cm]{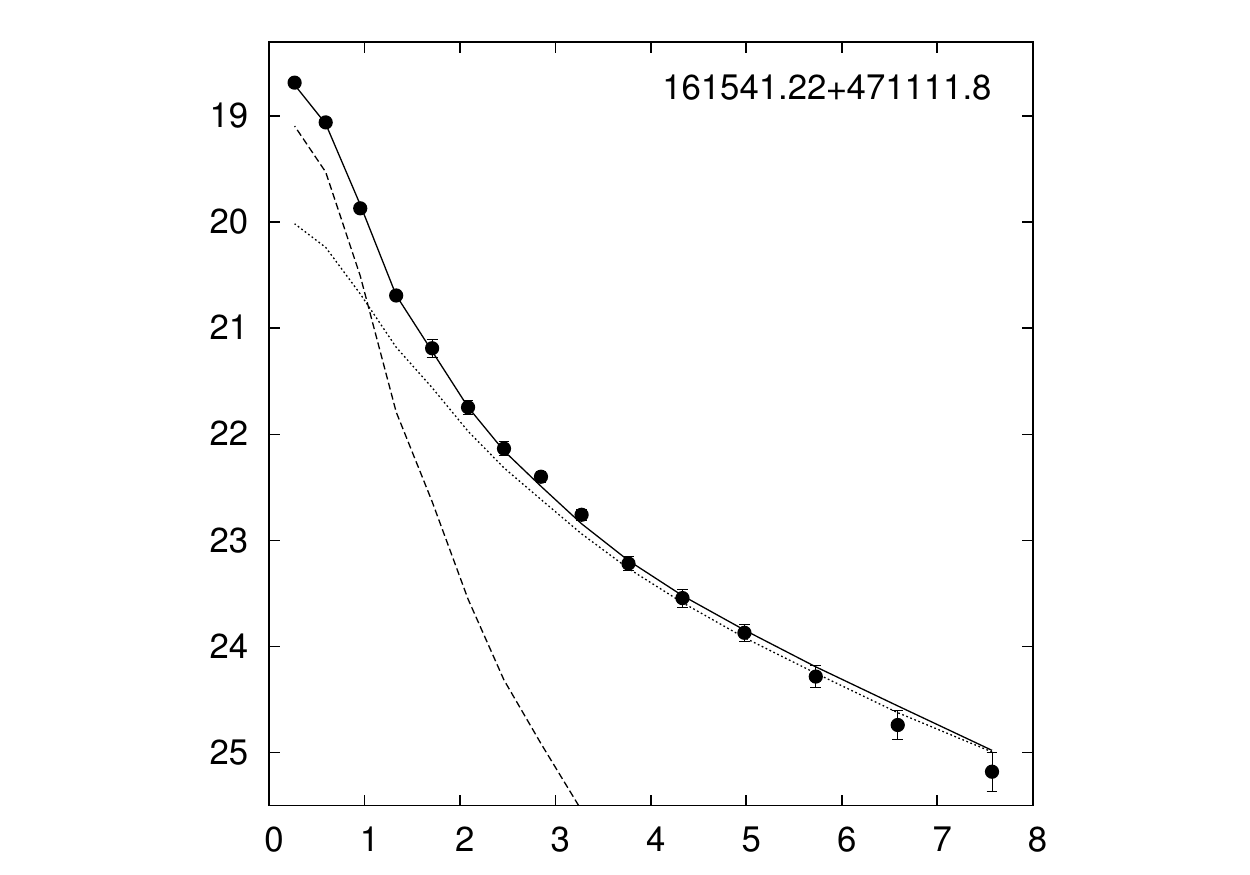}\\

\includegraphics[width=5.5cm]{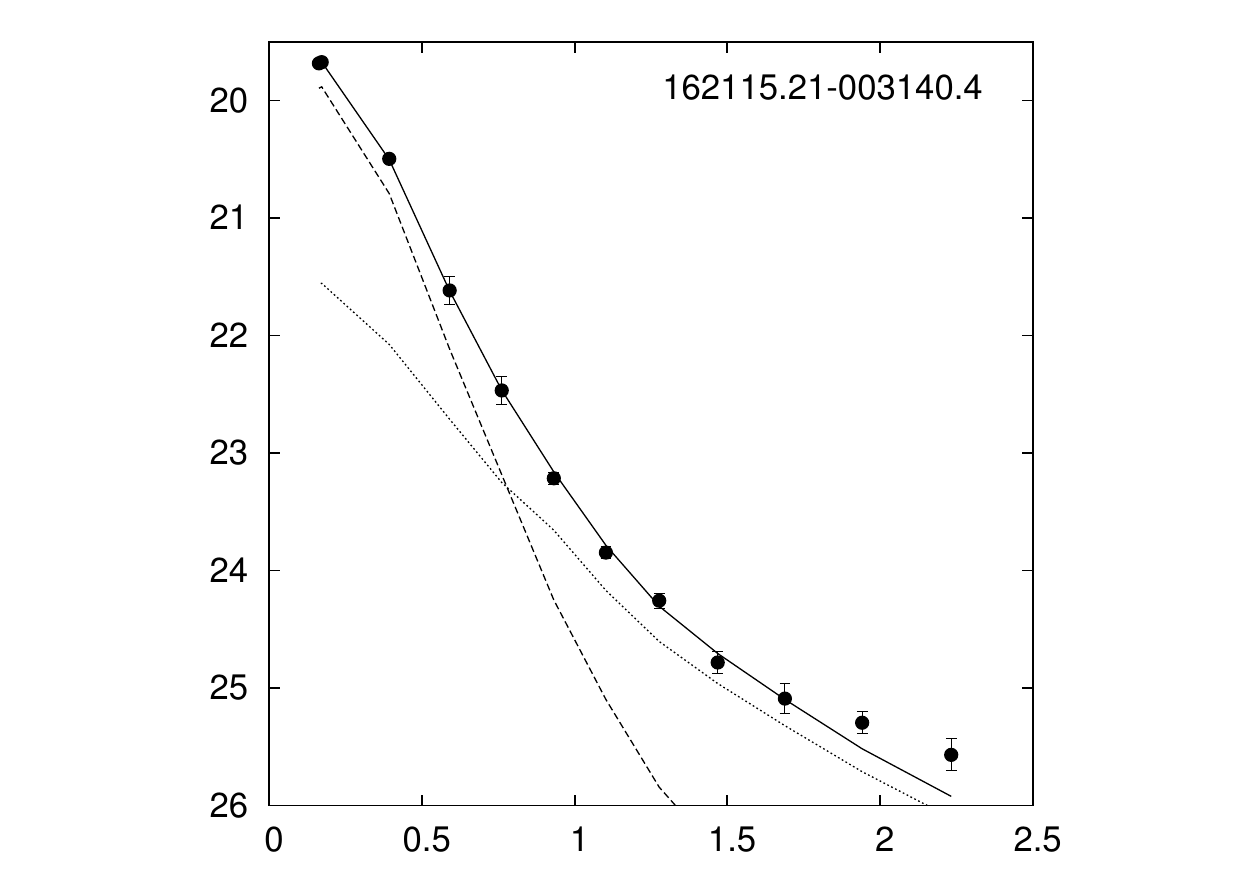}
\hspace*{-1.5cm}
\includegraphics[width=5.5cm]{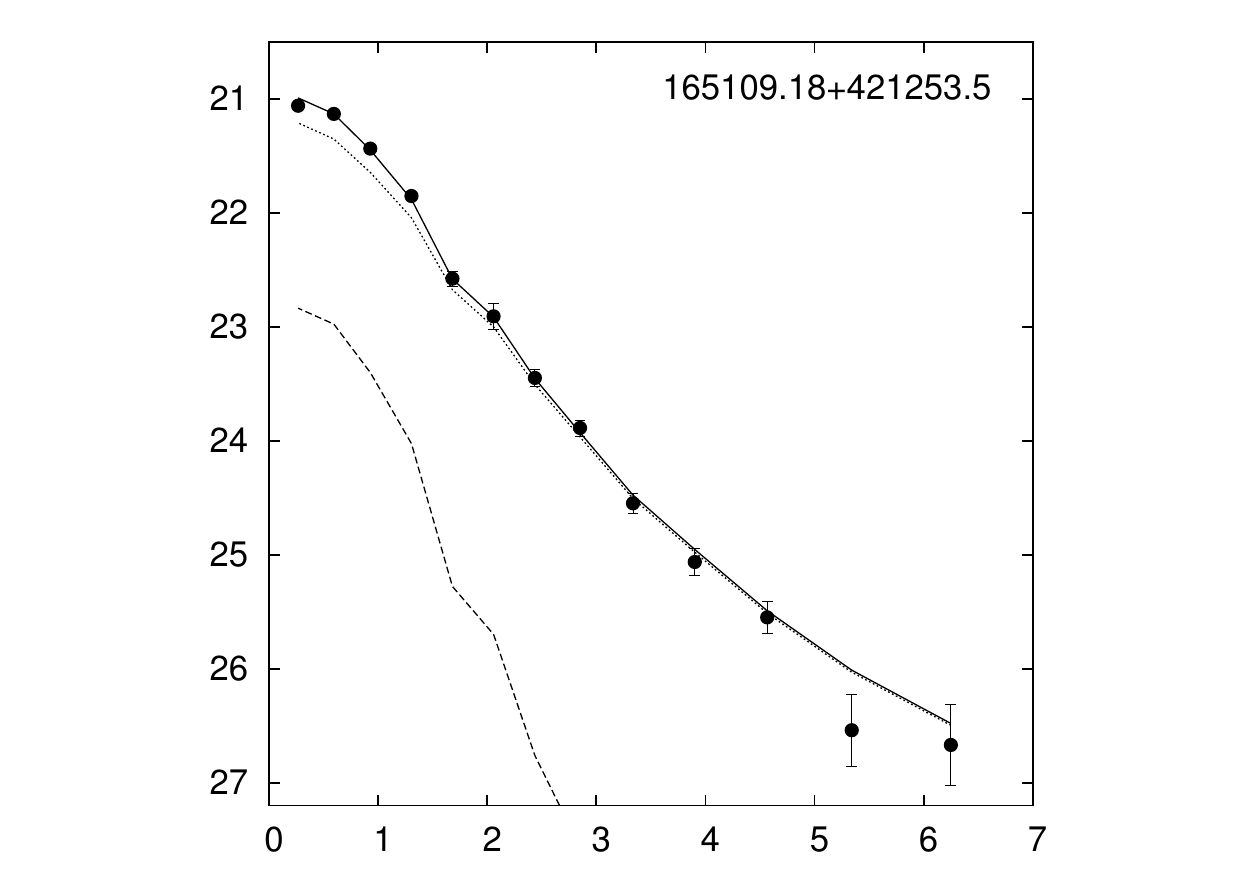}
\hspace*{-1.5cm}
\includegraphics[width=5.5cm]{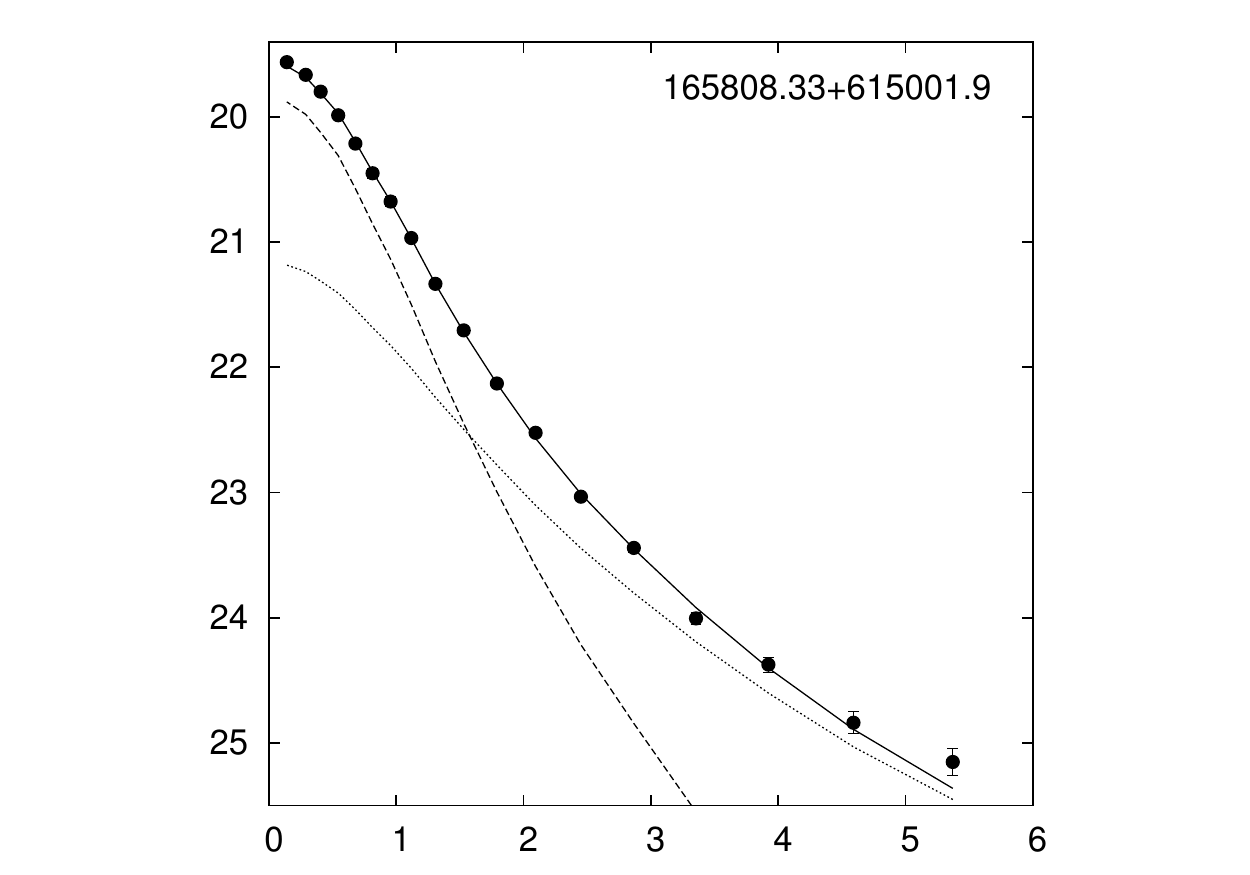}
\hspace*{-1.5cm}
\includegraphics[width=5.5cm]{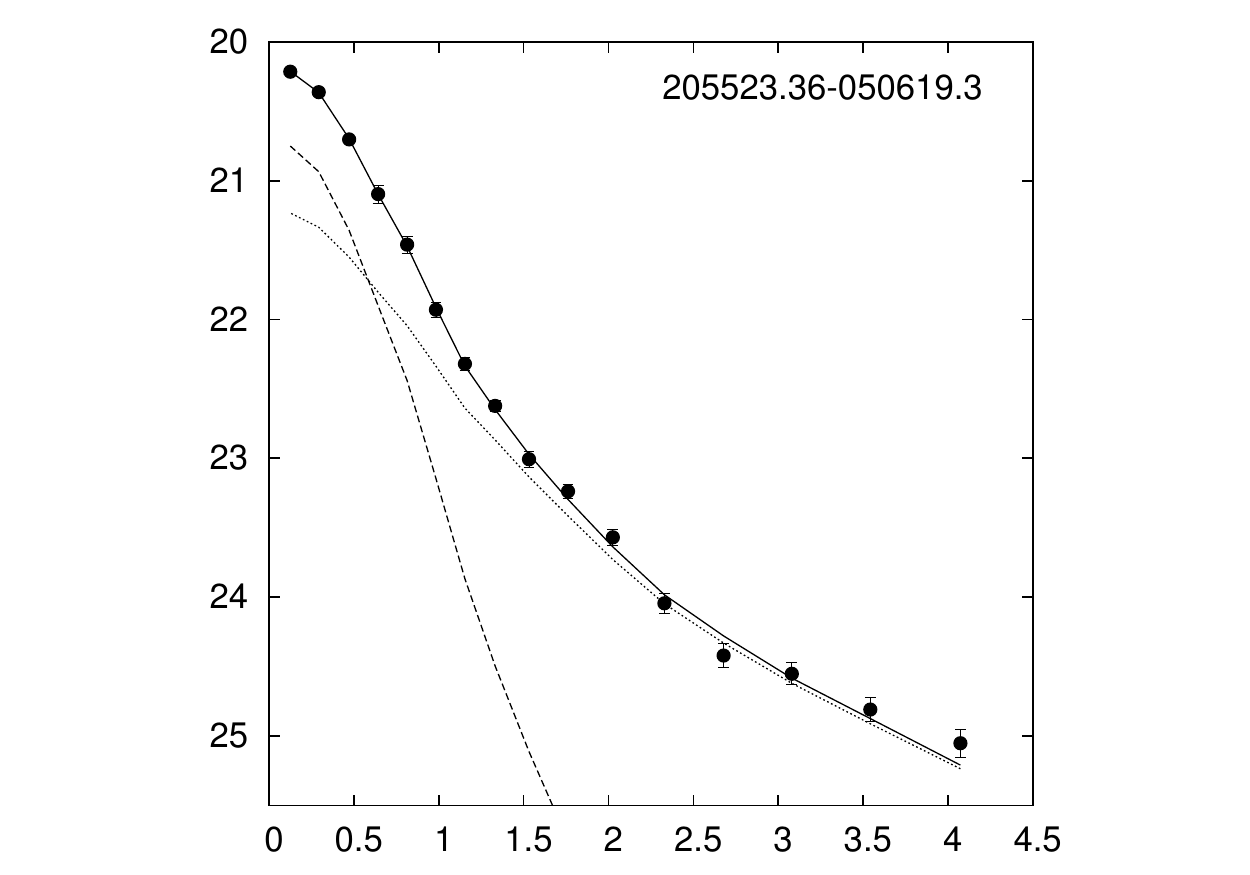}\\

\includegraphics[width=5.5cm]{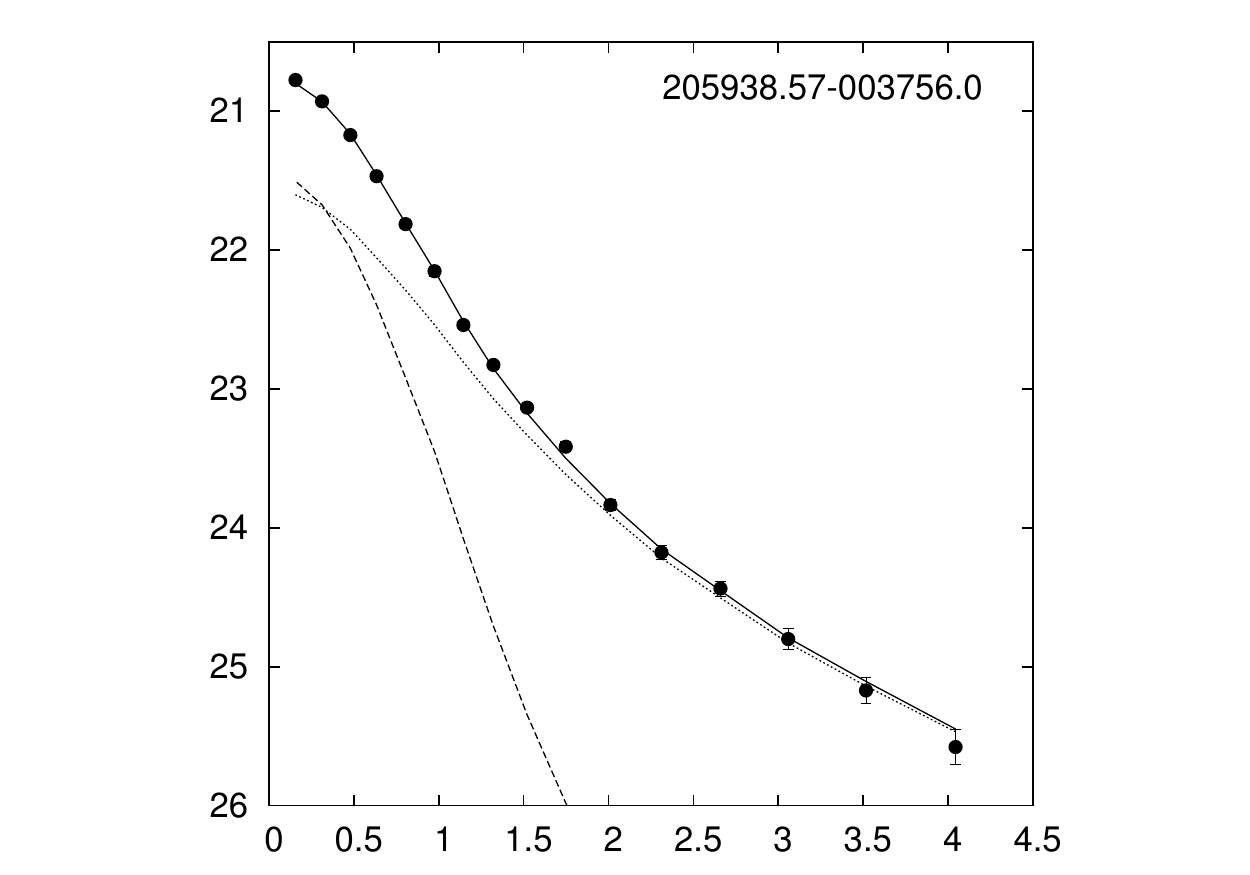}
\hspace*{-1.5cm}
\includegraphics[width=5.5cm]{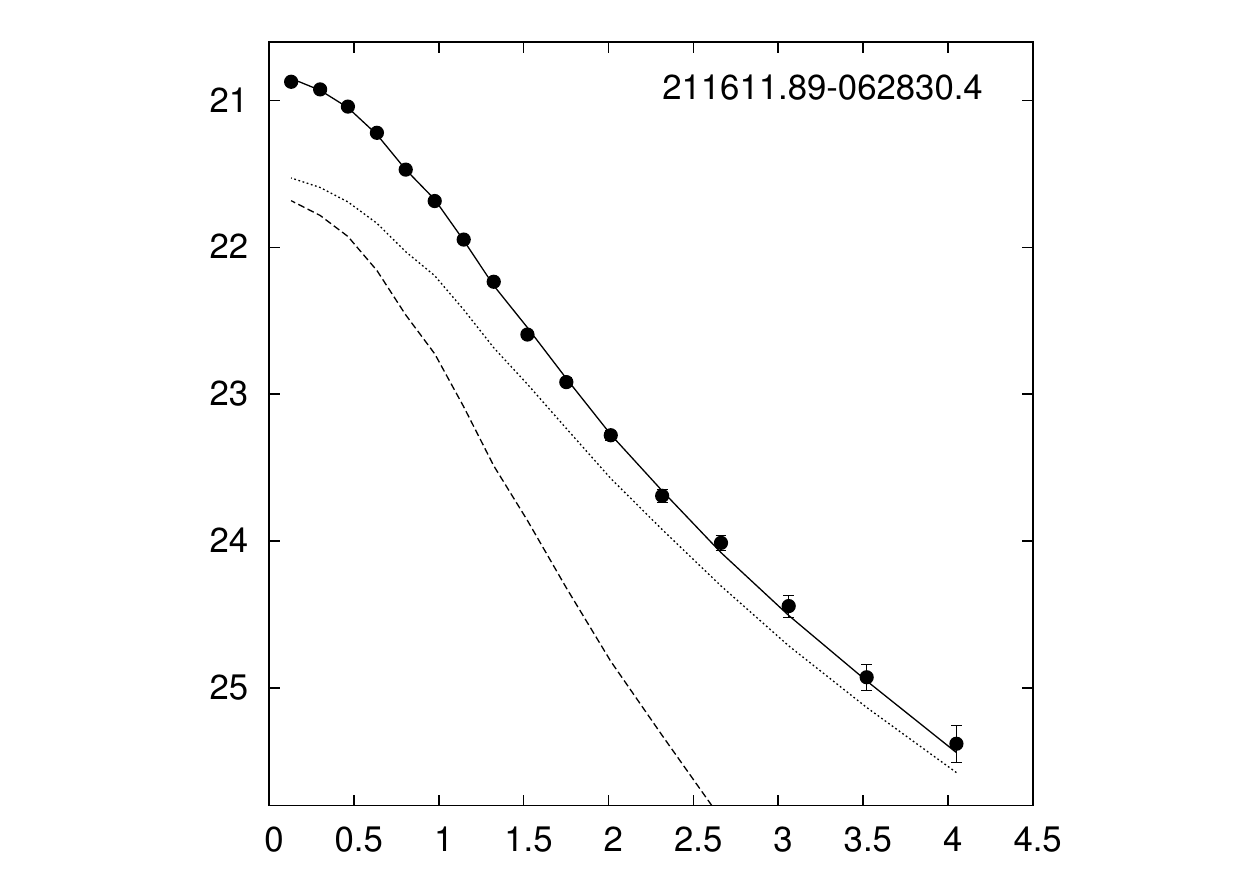}
\hspace*{-1.5cm}
\includegraphics[width=5.5cm]{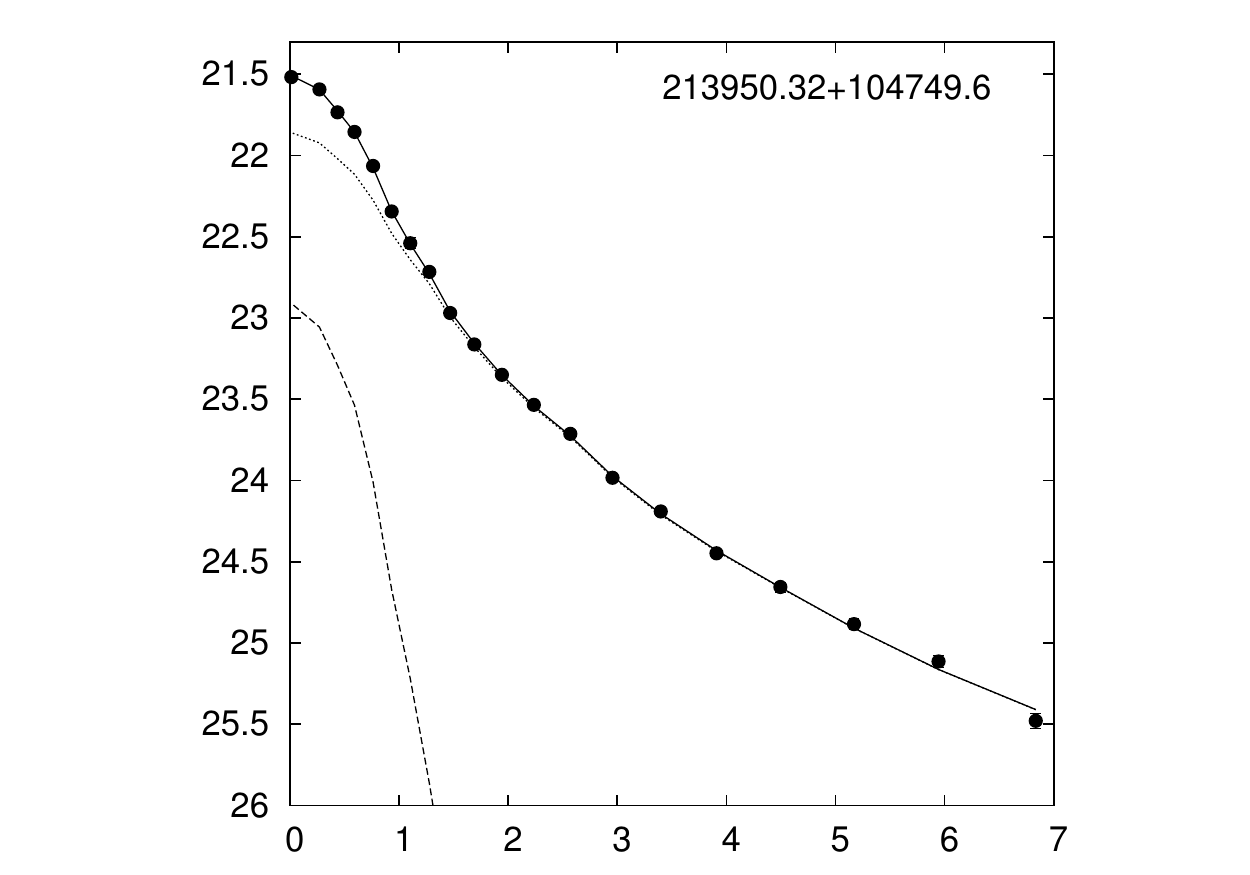}
\hspace*{-1.5cm}
\includegraphics[width=5.5cm]{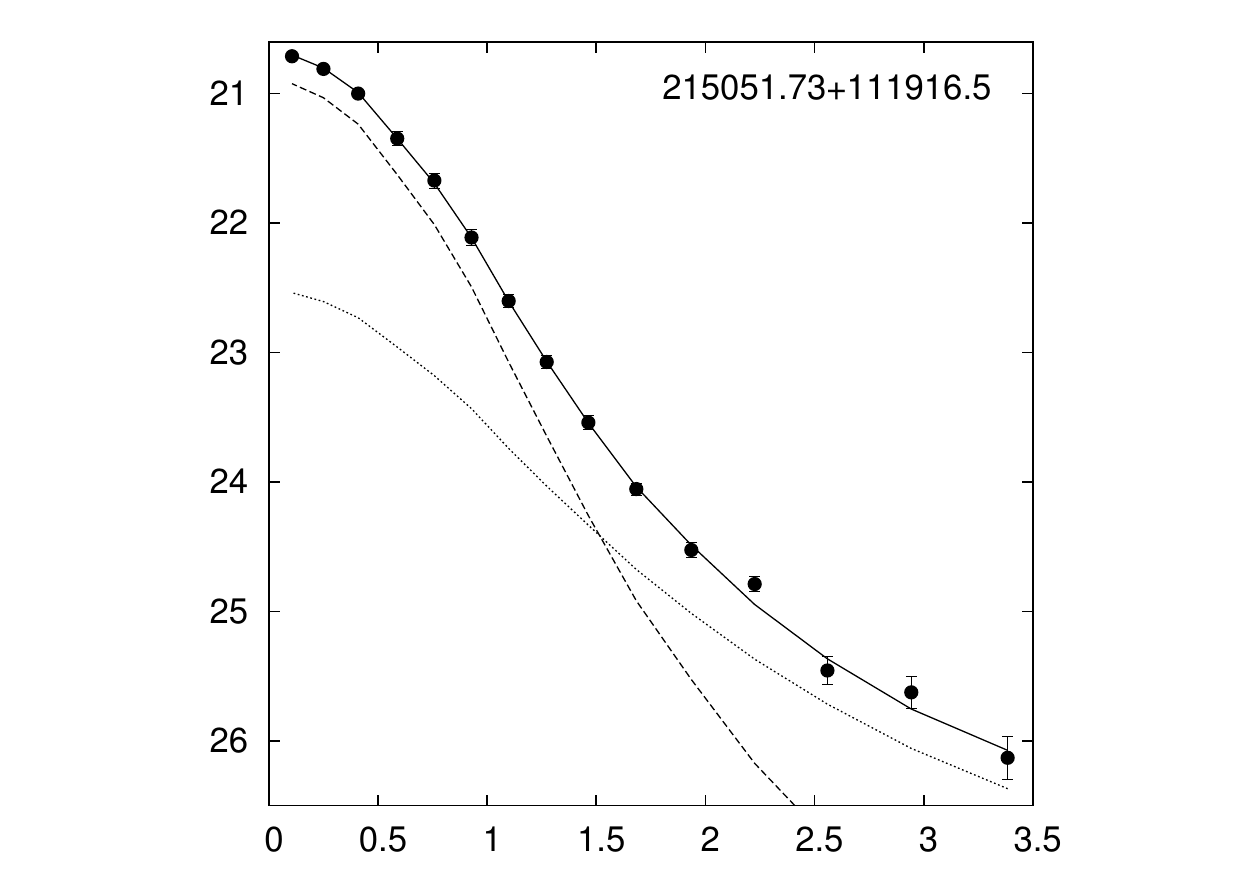}\\

\end{figure*}

\setcounter{figure}{0}

\begin{figure*}
\caption{--Continued.}

\includegraphics[width=5.5cm]{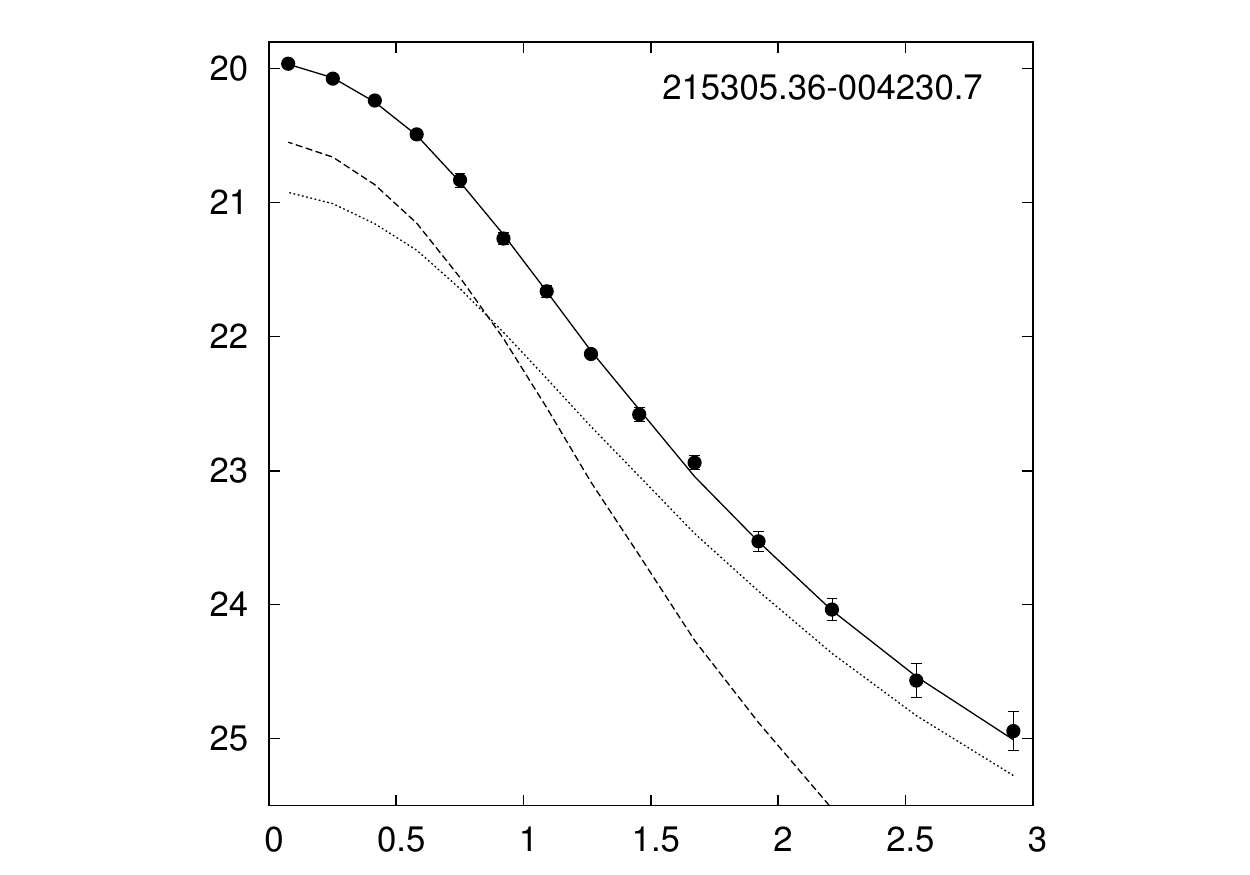}
\hspace*{-1.5cm}
\includegraphics[width=5.5cm]{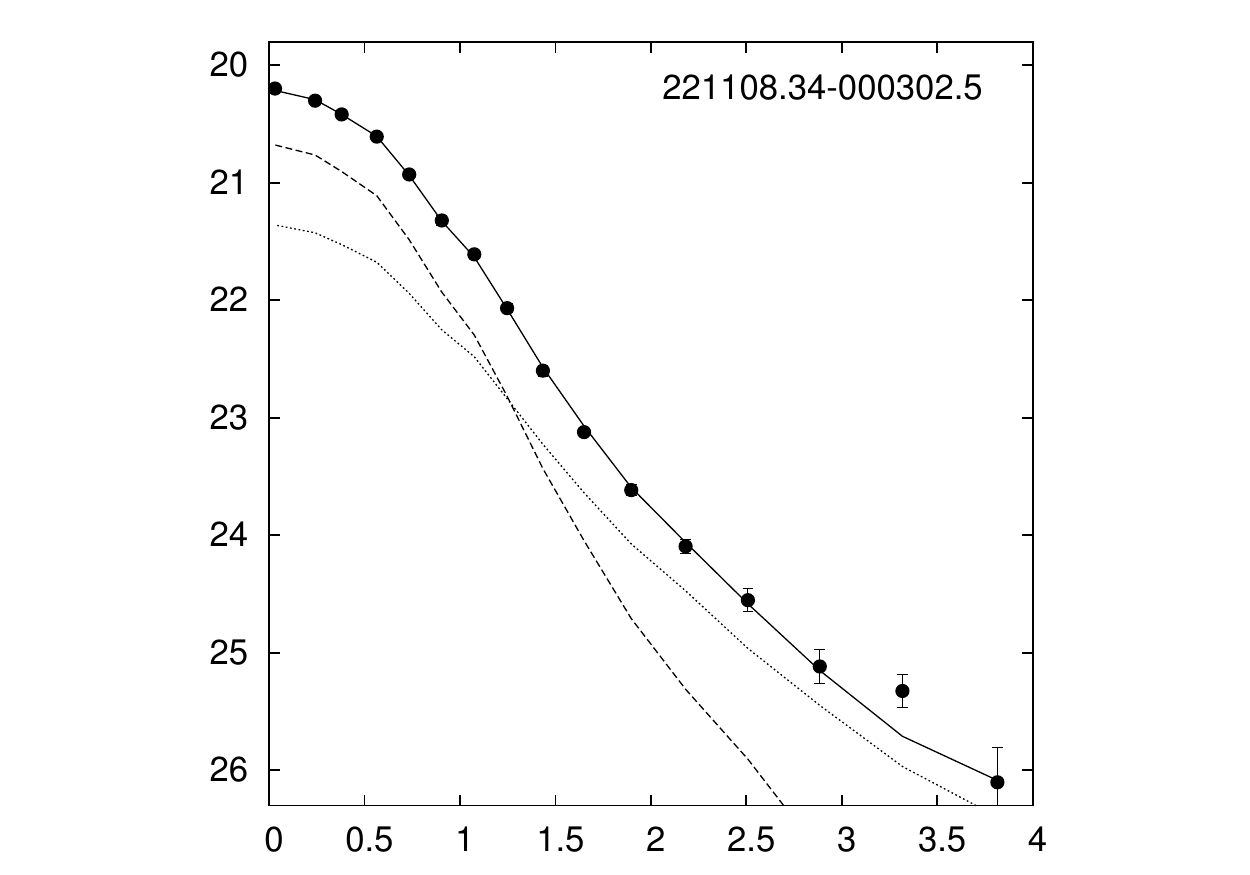}
\hspace*{-1.5cm}
\includegraphics[width=5.5cm]{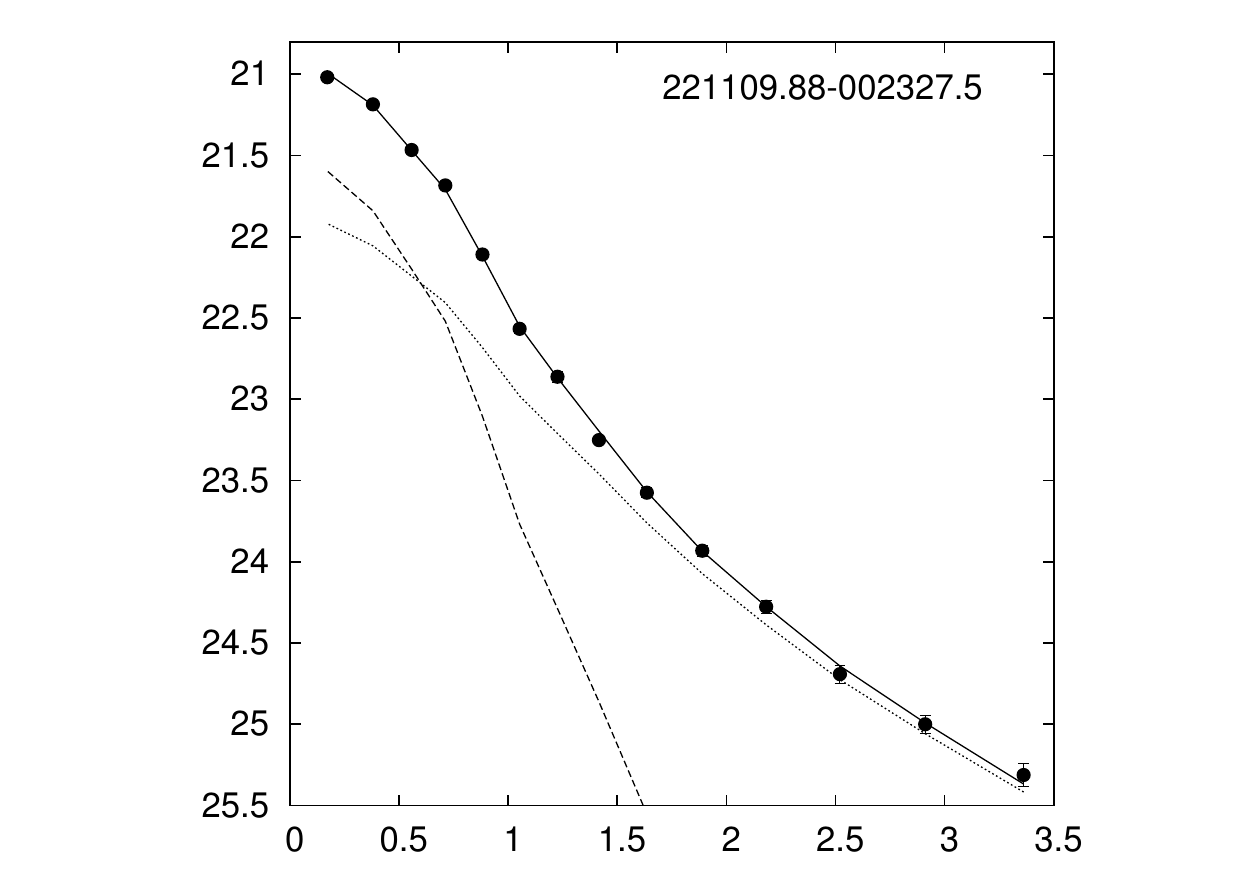}
\hspace*{-1.5cm}
\includegraphics[width=5.5cm]{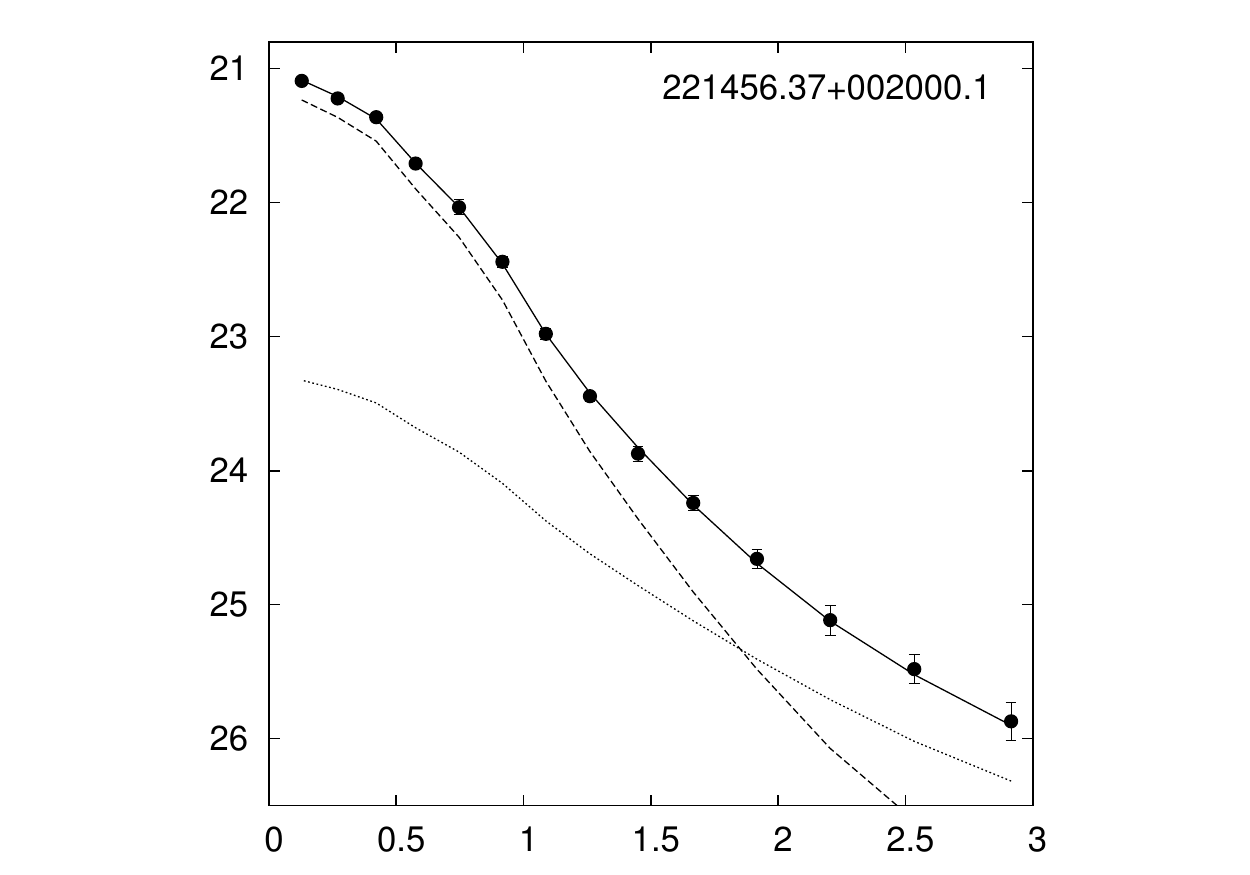}\\

\includegraphics[width=5.5cm]{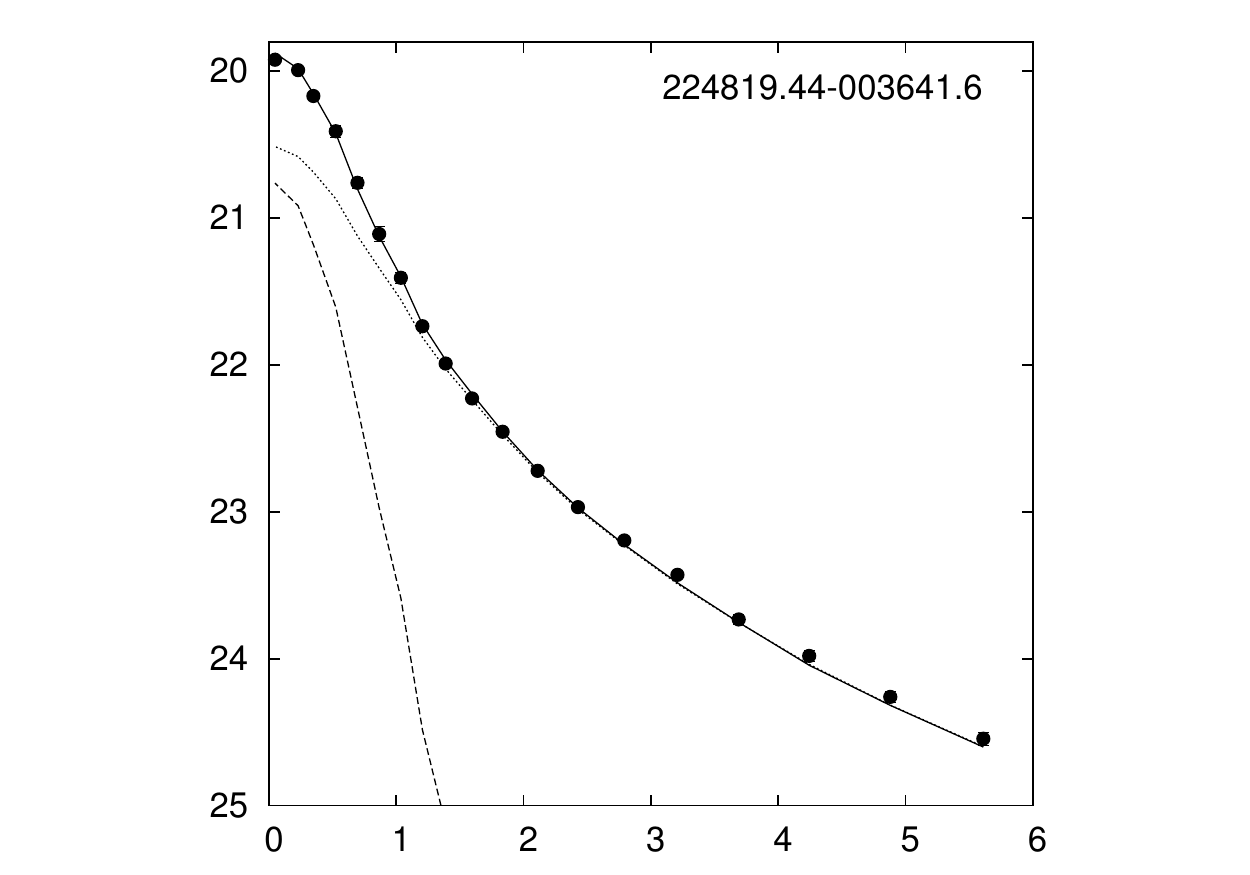}
\hspace*{-1.5cm}
\includegraphics[width=5.5cm]{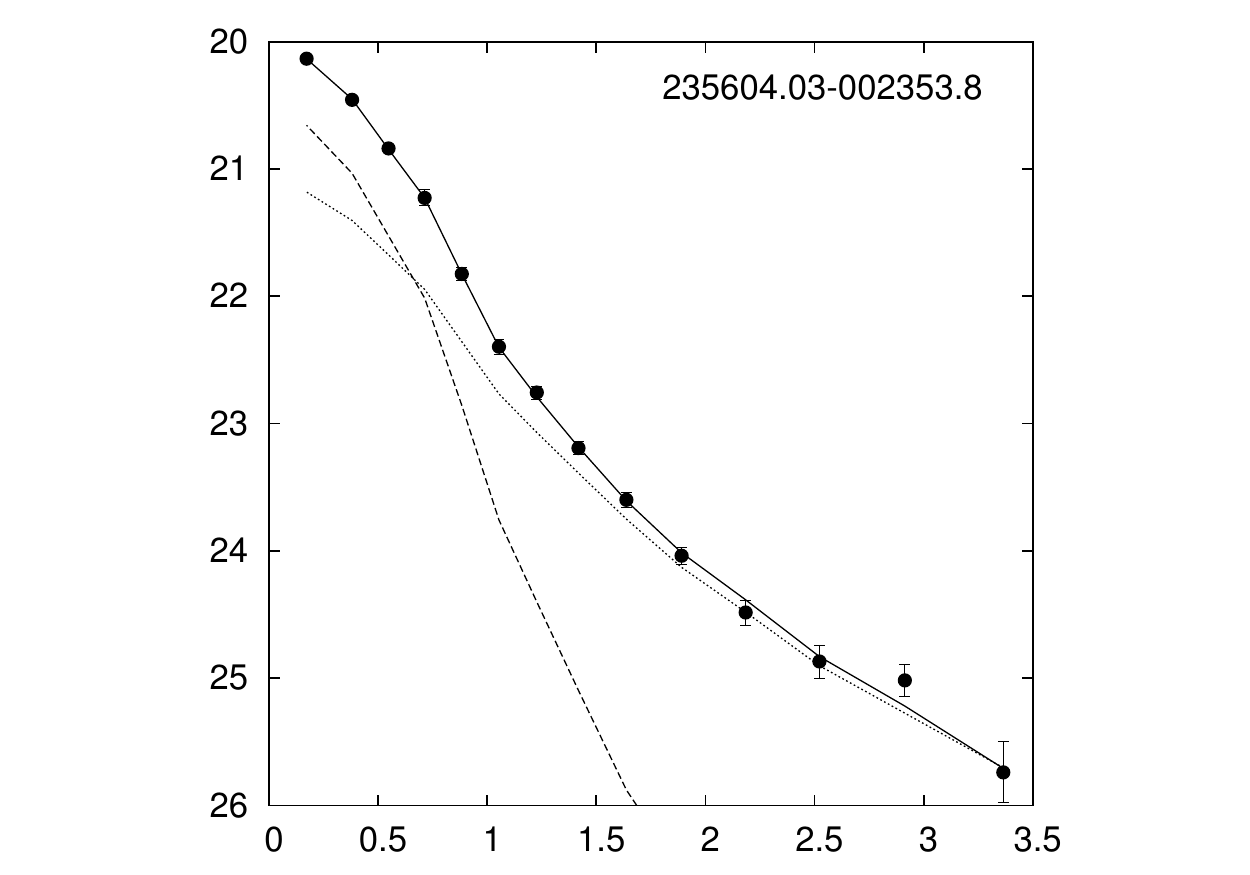}

\end{figure*}

\vspace*{2cm}

\begin{figure*}
\caption{Examples of unresolved (top row) and marginally resolved
(bottom row) targets. From left to right: NTT, CA, and NOT data.}
\includegraphics[width=5.5cm]{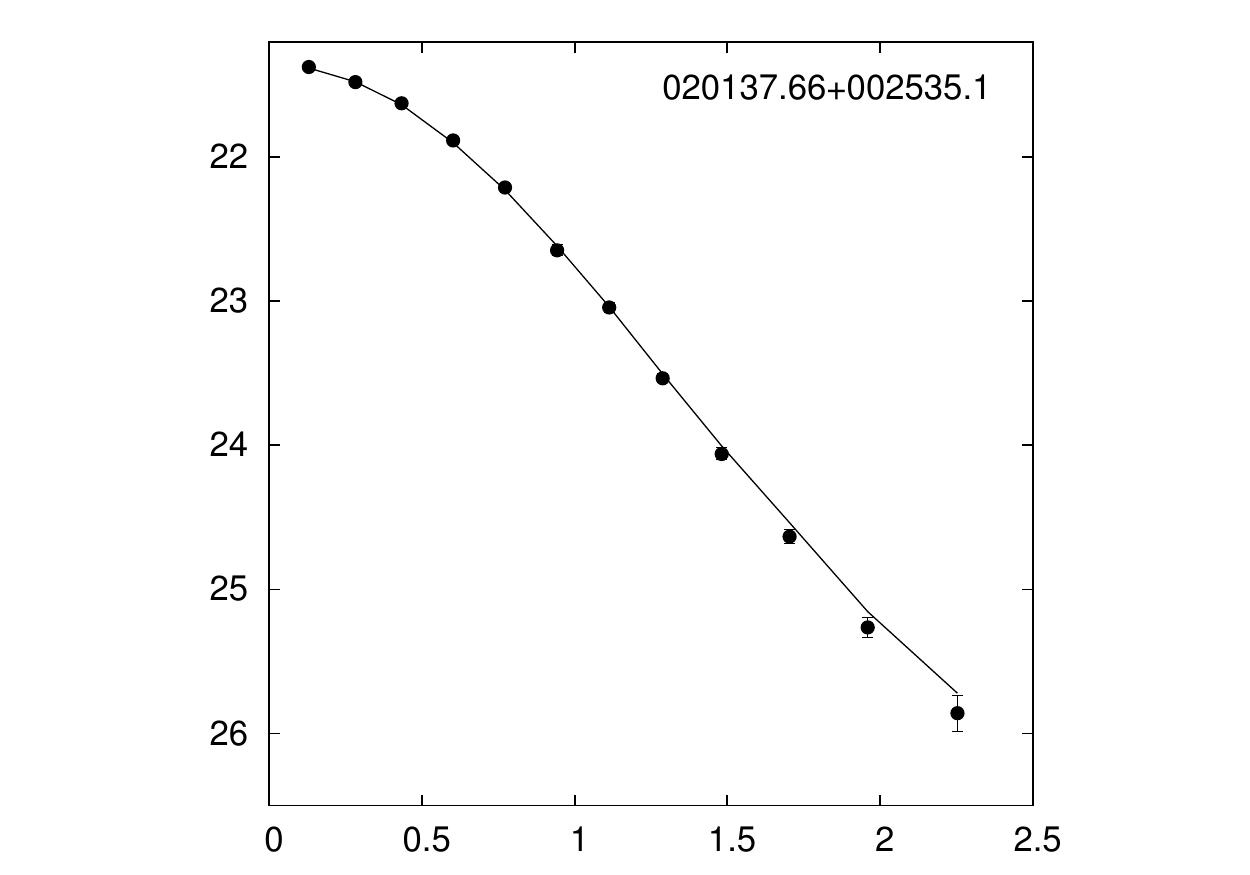}
\hspace*{-1.5cm}
\includegraphics[width=5.5cm]{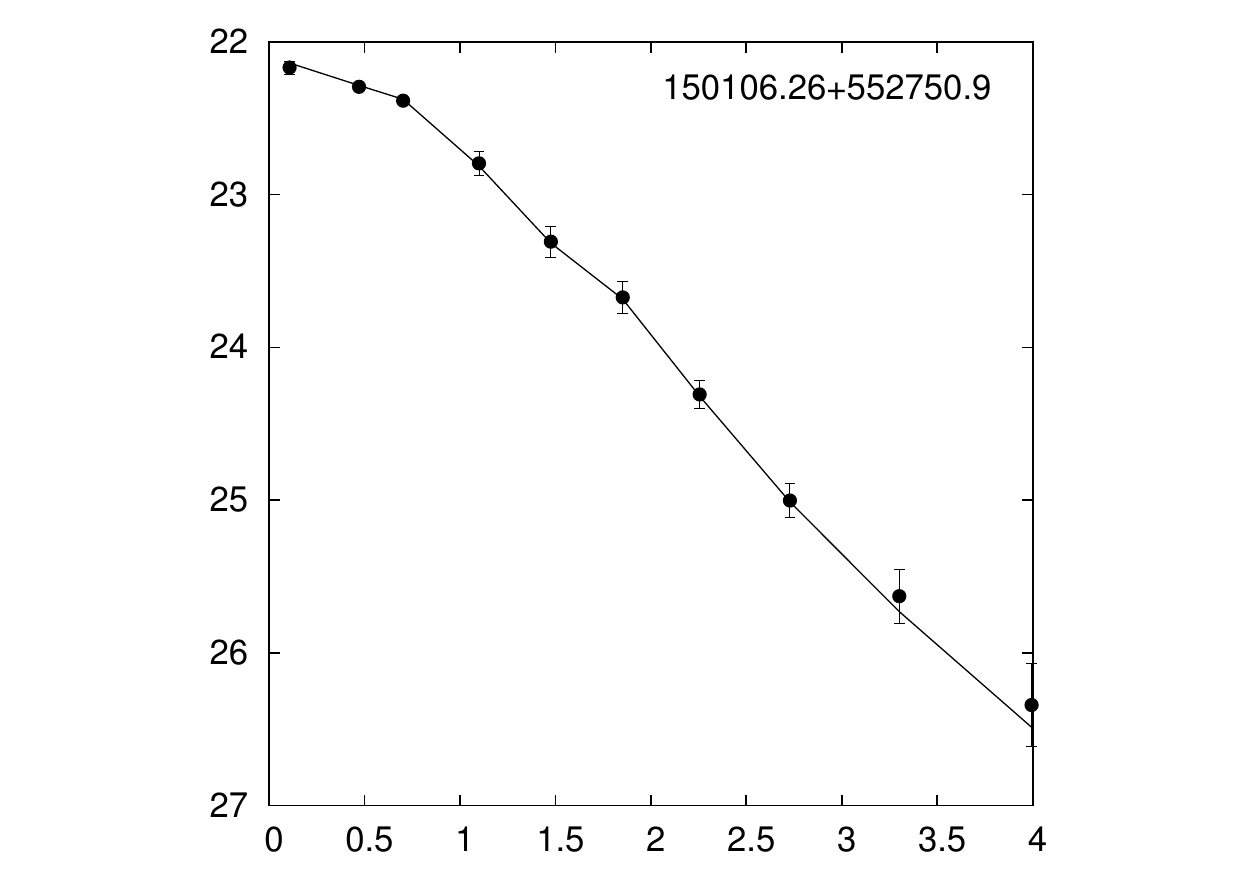}
\hspace*{-1.5cm}
\includegraphics[width=5.5cm]{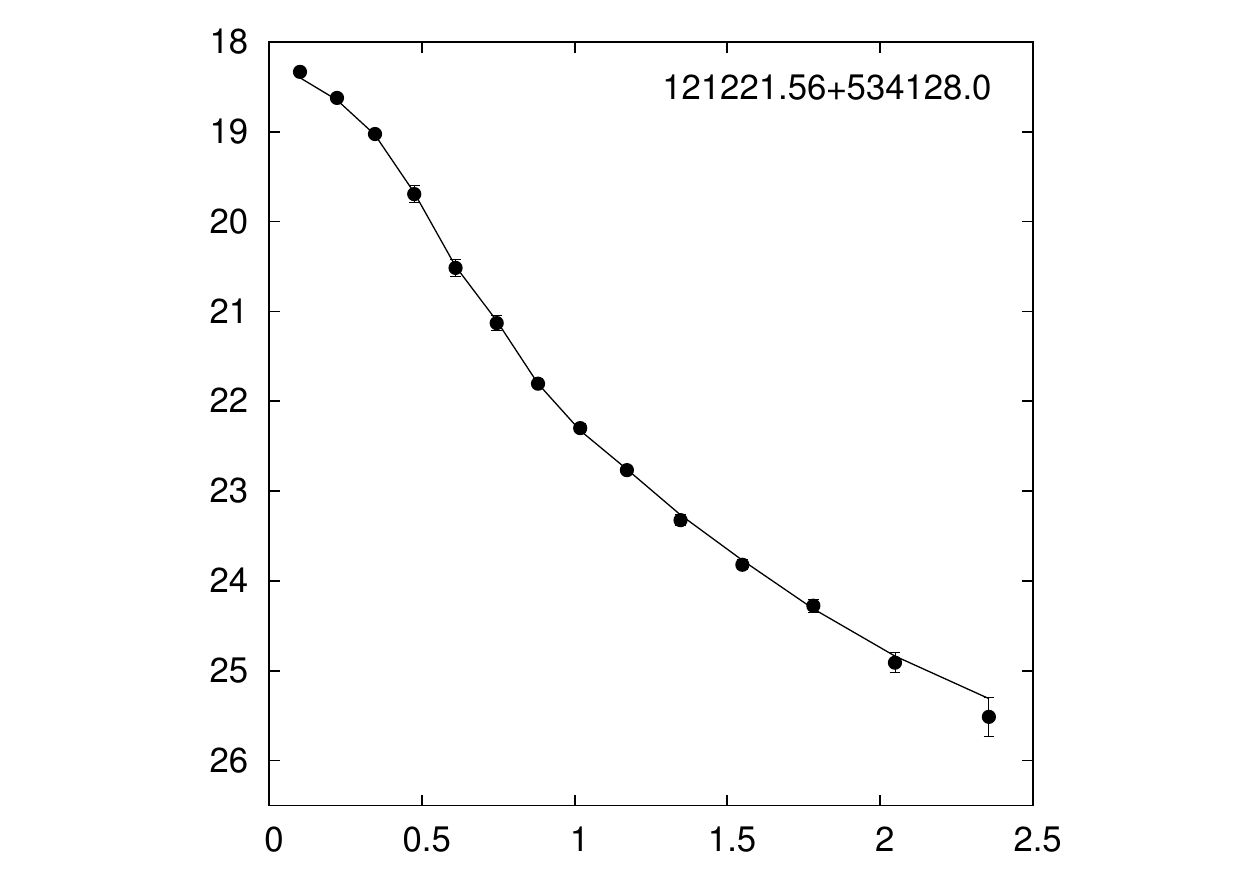}\\

\includegraphics[width=5.5cm]{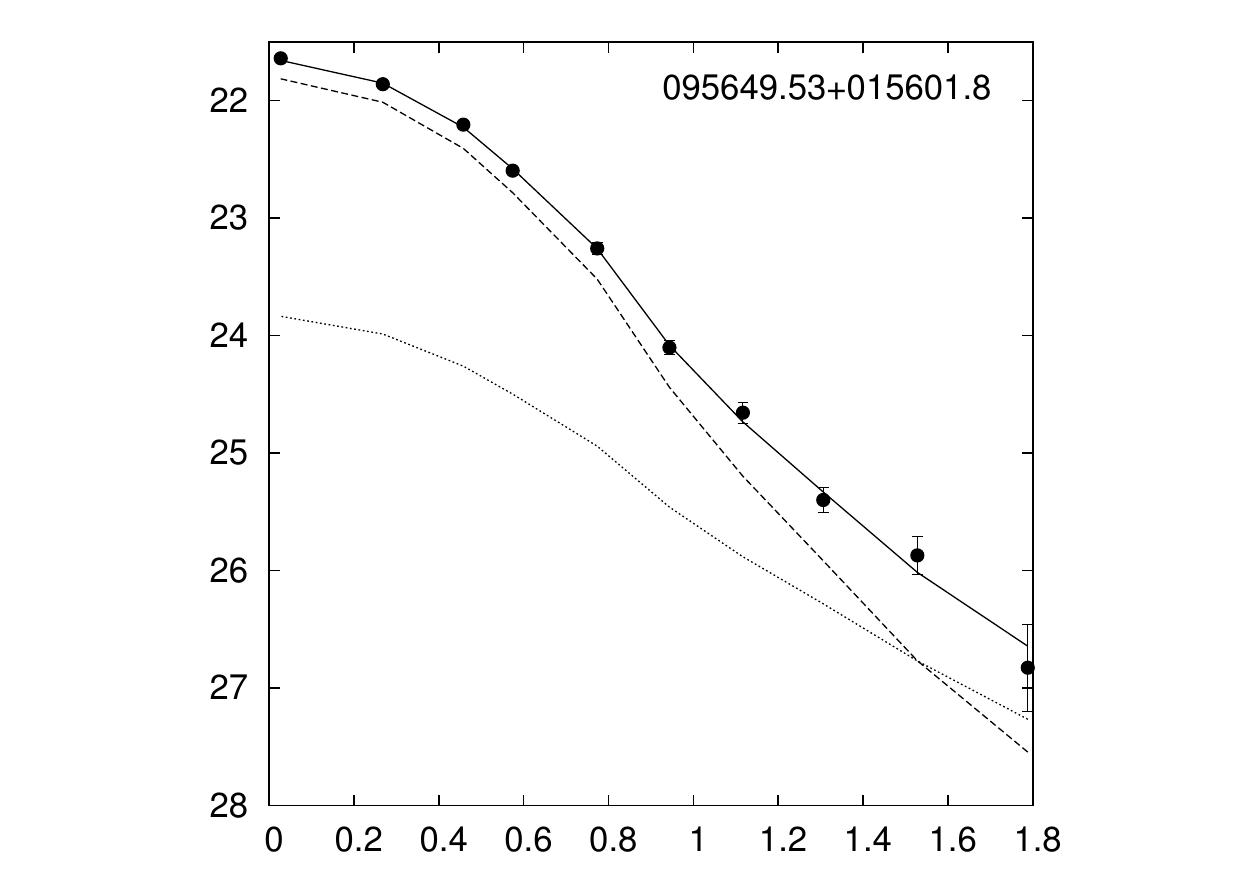}
\hspace*{-1.5cm}
\includegraphics[width=5.5cm]{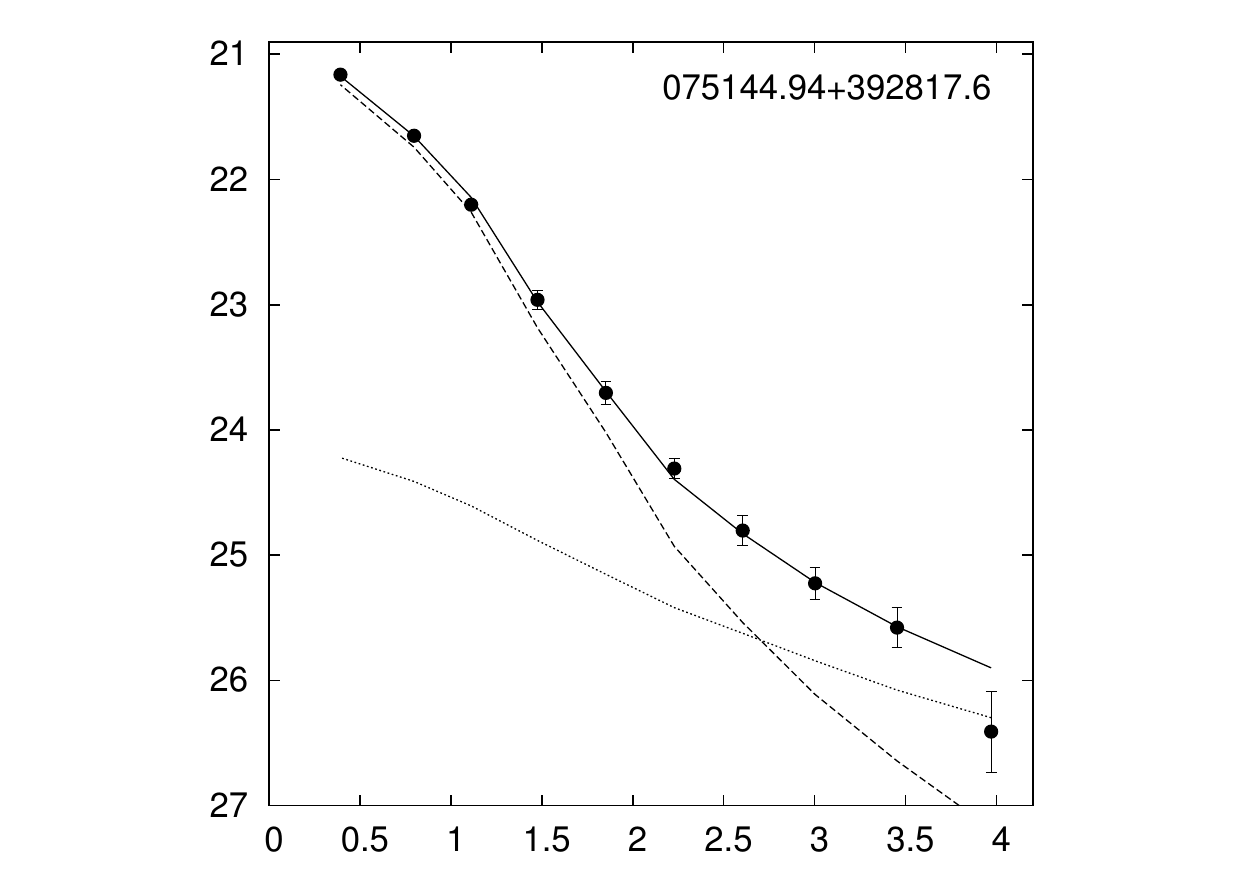}
\hspace*{-1.5cm}
\includegraphics[width=5.5cm]{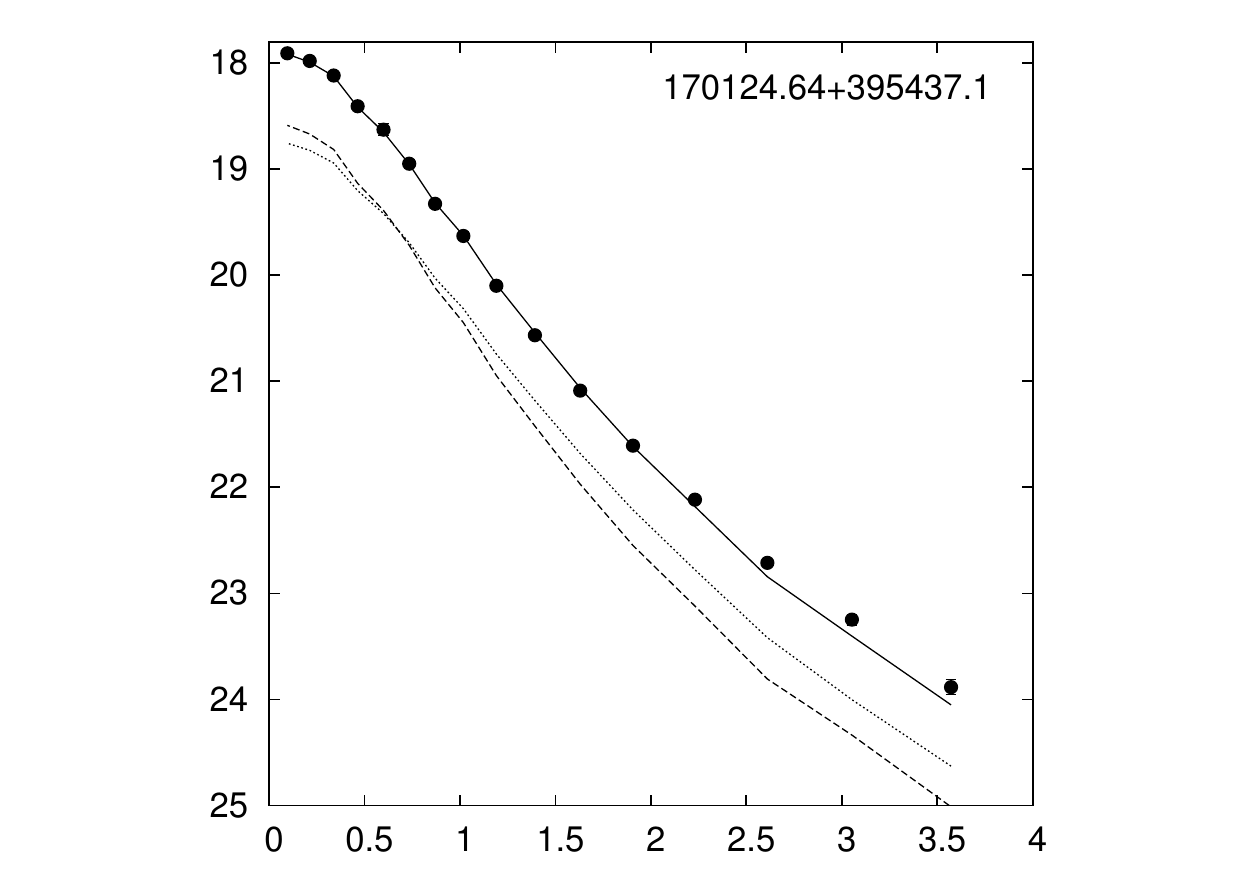}\\

\end{figure*}

\Online
\begin{longtable}{ccccccccccccc}
\caption {Main properties objects discussed in \cite{2011A&A...529A.162H} and the present work. 
Cols. 2 and 3 the spectroscopic redshift from SDSS and flags, Cols. 4 and 5
spectral indices $\alpha_{ox}$ and $\alpha_{ro}$. All these properties are from  \cite{2005AJ....129.2542C}.
Cols. 6 and 7 give the polarization and its error, Col. 8 the variability limit, Col. 9 the variability amplitude,
Cols. 10 and 11 the core (AGN) fraction over the total flux fraction as derived from the host galaxy fits and their
error, Col. 12 the peak frequency determined from SED fits, and Col. 13 a flag with respect to the SED fits.
The flags indicate whether the peak estimate is uncertain due to the shallowness of the radio
data (``radio'' if only one or no radio point available) or due to the presence of a strong host galaxy if the 
host galaxy fraction of the total flux exceeds 0.5 (flag is``galaxy'').
\label{summaryAllProps}}\\
Target name & z$_{\text{SDSS}}$ & z$_{\text{SDSS}}$ & aox & aro & pol & ePol & VL & Var & core & error & log$\left(\nu_\text{Peak}\right)$ & SED \tabularnewline 
 & & Flag & & &  & &  &  & frac & frac & & Flag \tabularnewline
$[$SDSS J$]$ &  &  &  &  & $[$\%$]$ & $[$\%$]$ & $[$mag$]$ & $[$mag$]$ &  &  & $[$log$\left(\text{Hz}\right)]$ & \tabularnewline
\hline
\hline
\endfirsthead
\caption* {\textbf{Table \ref{summaryAllProps}} $-$ Continued.}\\
Target name & z$_{\text{SDSS}}$ & z$_{\text{SDSS}}$ & aox & aro & pol & ePol & VL & Var & core & error & log$\left(\nu_\text{Peak}\right)$ & SED \tabularnewline 
 & & Flag & & &  & &  &  & frac & frac & & Flag \tabularnewline
$[$SDSS J$]$ &  &  &  &  & $[$\%$]$ & $[$\%$]$ & $[$mag$]$ & $[$mag$]$ &  &  & $[$log$\left(\text{Hz}\right)]$ & \tabularnewline
\hline
\hline
\endhead
\endfoot
\endlastfoot
000121.47$-$001140.3 & 0.4620 & - & $>$0.98 & 0.63 & 4.34 & 0.75 & 0.06 & -0.103 & 0.60 & 0.07 & 16.59 & - \tabularnewline
002142.26$-$090044.4 & 0.6481 & - & $>$1.07 & 0.51 & 12.80 & 0.83 & 0.065 & 0.104 & 1 & - & - & - \tabularnewline 
002200.95+000658.0 & 0.3057 & - & 0.94 & 0.28 & $<$2.56 & - & 0.064 & 0.259 & $<$0.16 & 0.07 & 15.21 & radio \tabularnewline 
002839.77+003542.2 & 0.6866 & u & $>$1.00 & 0.49 & 4.45 & 0.69 & 0.072 & 0.519 & 0.85 & 0.10 & 17.43 & - \tabularnewline 
003514.72+151504.1 & - & - & 1.16 & 0.25 & 6.90 & 0.96 & 0.087 & 1.057 & 1 & - & 17.29 & radio \tabularnewline 
004054.65$-$091526.8 & 5.0300 & - & $>$1.01 & $<$0.05 & $<$3.99 & - & 0.126 & 0.105 & 1 & - & - & - \tabularnewline 
005041.31$-$092905.1 & - & - & 1.28 & 0.55 & 19.01 & 0.35 & 0.05 & -0.641 & 1 & - & 15.08 & radio \tabularnewline 
010058.19$-$005547.8 & $>$0.6679 & u & $>$1.04 & 0.49 & 12.82 & 0.66 & 0.057 & -0.262 & 1 & - & 13.53 & - \tabularnewline 
010326.01+152624.8 & 0.2461 & - & $>$1.22 & 0.60 & 1.62 & 0.68 & 0.068 & 0.669 & 0.25 & 0.04 & 16.81 & radio \tabularnewline 
011012.66$-$004746.9 & 0.5477 & - & $>$0.94 & 0.54 & 5.52 & 0.90 & 0.074 & 0.122 & 0.69 & 0.05 & 14.34 & - \tabularnewline 
011452.77+132537.5 & - & - & $>$1.34 & 0.41 & 9.10 & 0.61 & 0.057 & 0.801 & 1 & - & 14.8 & radio \tabularnewline 
012155.87$-$102037.2 & 0.4695 & - & $>$1.02 & $<$0.23 & $<$0.90 & - & 0.06 & 0.005 & $<$0.20 & 0.08 & - & - \tabularnewline 
012227.38+151023.1 & - & - & $>$0.89 & 0.55 & 10.41 & 1.05 & 0.07 & 0.557 & 1 & - & 13.63 & - \tabularnewline 
012716.31$-$082128.9 & 0.3620 & u & $>$1.04 & 0.66 & 6.32 & 0.87 & 0.067 & 0.203 & 1 & - & - & - \tabularnewline 
012750.83$-$001346.6 & 0.4376 & - & $>$0.98 & 0.40 & 4.13 & 0.78 & 0.07 & -0.032 & 0.39 & 0.09 & 14.33 & radio \tabularnewline 
013408.95+003102.5 & - & - & $>$0.93 & $<$0.29 & $<$2.03 & - & 0.067 & 0.035 & 1 & - & - & - \tabularnewline 
014125.83$-$092843.7 & $>$0.5000 & u & 1.45 & 0.60 & 4.88 & 0.56 & 0.067 & 0.88 & 1 & - & 13.93 & - \tabularnewline 
020106.18+003400.2 & 0.2985 & - & 0.89 & 0.38 & 1.37 & 0.67 & 0.061 & 0.424 & 0.49 & 0.09 & 15.83 & radio \tabularnewline 
020137.66+002535.1 & - & - & $>$0.99 & $<$0.26 & $<$2.18 & - & 0.065 & 0.109 & 1 & - & - & - \tabularnewline 
022048.46$-$084250.4 & 0.5252 & u & 1.09 & 0.49 & 6.09 & 0.45 & 0.054 & -0.355 & 1 & - & 19.42 & - \tabularnewline 
023813.68$-$092431.4 & 0.4188 & - & 1.09 & 0.34 & 2.51 & 0.70 & 0.067 & 0.205 & $<$0.07 & 0.06 & 14.52 & galaxy \tabularnewline 
024156.38+004351.6 & 0.9900 & - & $>$0.90 & 0.11 & 2.60 & 0.72 & 0.06 & 0.026 & 1 & - & - & - \tabularnewline 
024157.37+000944.1 & 0.7896 & u & 0.83 & 0.32 & 4.03 & 1.63 & 0.114 & 0.303 & 1 & - & 14.14 & radio \tabularnewline 
024302.93+004627.3 & 0.4089 & - & $>$0.96 & 0.53 & $<$3.17 & - & 0.07 & 0.106 & 0.33 & 0.09 & - & - \tabularnewline 
024752.13+004106.3 & 0.3929 & - & $>$0.82 & 0.41 & $<$2.40 & - & 0.066 & -0.086 & 0.44 & 0.07 & 13.91 & radio \tabularnewline 
025046.48$-$005449.0 & - & - & $>$0.74 & $<$0.28 & $<$2.95 & - & 0.082 & 0.095 & 1 & - & - & - \tabularnewline 
025612.47$-$001057.8 & 0.6302 & - & $>$-0.11 & $<$0.28 & $<$2.42 & - & 0.076 & -0.284 & 0.62 & 0.07 & 14.11 & radio \tabularnewline 
030235.78$-$075027.0 & - & - & $>$0.71 & 0.45 & 8.42 & 0.71 & 0.064 & -0.497 & 1 & - & - & - \tabularnewline 
030240.30+003849.9 & - & - & $>$0.73 & 0.33 & $<$1.93 & - & 0.066 & 0.041 & 1 & - & 14.04 & - \tabularnewline 
030433.96$-$005404.7 & 0.5112 & - & 1.03 & 0.44 & 1.83 & 0.60 & 0.059 & -0.211 & 0.80 & 0.14 & 14.99 & radio \tabularnewline 
031712.23$-$075850.4 & 2.6993 & - & $>$1.03 & - & $<$3.20 & - & 0.076 & 0.075 & 1 & - & - & - \tabularnewline 
032343.62$-$011146.1 & - & - & 1.35 & 0.32 & 5.22 & 0.98 & 0.067 & 0.789 & 1 & - & 15.4 & radio \tabularnewline 
032356.64$-$010829.6 & 0.3923 & - & 0.89 & 0.41 & $<$2.13 & - & 0.071 & 0.052 & 0.50 & 0.07 & 13.38 & radio \tabularnewline 
040911.36$-$055529.4 & - & - & $>$0.93 & 0.53 & 9.76 & 1.33 & 0.079 & 0.629 & 1 & - & - & - \tabularnewline 
045128.96$-$002911.5 & - & - & $>$0.75 & 0.80 & $<$3.98 & - & 0.078 & 0.138 & 1 & - & - & - \tabularnewline 
074054.60+322601.0 & $>$0.9460 & u & $>$1.09 & 0.38 & 5.44 & 0.83 & 0.051 & 0.002 & 1 & - & - & - \tabularnewline 
075144.94+392817.6 & 0.4338 & u & $>$0.80 & 0.44 & 3.99 & 1.78 & 0.099 & -0.921 & 1 & - & - & - \tabularnewline 
075602.72+414039.8 & 0.5788 & - & $>$0.94 & 0.36 & $<$6.09 & - & 0.106 & -0.216 & 1 & - & - & - \tabularnewline 
081840.06+315348.2 & - & - & $>$1.01 & 0.57 & 14.48 & 1.17 & 0.077 & -0.381 & 1 & - & - & - \tabularnewline 
083413.90+511214.7 & - & - & $>$1.05 & 0.26 & $<$5.61 & - & 0.092 & -0.156 & 1 & - & - & - \tabularnewline 
083918.75+361856.1 & 0.3343 & - & $>$0.95 & 0.40 & $<$2.92 & - & 0.061 & 0.066 & 0.42 & 0.09 & 18.42 & galaxy \tabularnewline 
084225.52+025252.7 & 0.4251 & - & 1.15 & 0.45 & $<$1.57 & - & 0.033 & 0.155 & 0.45 & 0.04 & 17.53 & radio \tabularnewline 
084908.81+020622.5 & - & - & $>$1.06 & 0.55 & 10.94 & 0.71 & 0.033 & 0.206 & 1 & - & 13.79 & radio \tabularnewline 
085638.50+014000.7 & 0.4479 & - & $>$1.03 & 0.38 & $<$3.78 & - & 0.034 & 0.088 & 0.66 & 0.07 & 13.33 & - \tabularnewline 
085749.80+013530.3 & 0.2812 & - & 1.20 & 0.51 & 4.50 & 0.68 & 0.031 & 0.051 & 0.39 & 0.05 & 14.97 & radio \tabularnewline 
085920.56+004712.1 & - & - & 0.84 & 0.52 & 4.20 & 0.91 & 0.057 & 0.518 & 1 & - & 14.98 & radio \tabularnewline 
090133.43+031412.5 & 0.4591 & - & $>$1.04 & $<$0.21 & $<$2.08 & - & 0.033 & 0.03 & 1 & - & 20.39 & radio \tabularnewline 
090939.84+020005.3 & - & - & $>$1.00 & 0.73 & 18.99 & 0.49 & 0.035 & -1.101 & 1 & - & 13.33 & radio \tabularnewline 
091848.57+021321.8 & - & - & $>$1.13 & 0.32 & 4.29 & 1.16 & 0.058 & 0.896 & 1 & - & 14.49 & radio \tabularnewline 
092542.87+595816.3 & - & - & 1.03 & 0.40 & 8.65 & 1.10 & 0.078 & 0.171 & 1 & - & - & - \tabularnewline 
092638.88+541126.7 & 0.8500 & u & $>$1.09 & 0.51 & 7.02 & 0.93 & 0.077 & -0.55 & 1 & - & - & - \tabularnewline 
092912.25+030029.9 & - & - & $>$1.00 & 0.24 & 9.41 & 0.69 & 0.04 & -0.345 & 1 & - & 12.9 & radio \tabularnewline 
094245.30+541620.4 & - & - & $>$0.99 & 0.45 & 7.51 & 2.17 & 0.104 & 0.024 & 1 & - & 15.39 & - \tabularnewline 
094257.81$-$004705.2 & 1.3600 & - & $>$1.02 & 0.23 & $<$1.62 & - & 0.033 & -0.229 & 1 & - & 13.84 & radio \tabularnewline 
094432.33+573536.2 & - & - & 0.85 & 0.32 & 3.41 & 1.29 & 0.088 & -0.32 & 0.73 & 0.10 & - & - \tabularnewline 
094432.33+573536.2 & - & - & 0.85 & 0.32 & 8.06 & 0.61 & 0.088 & -0.32 & 0.73 & 0.10 & - & - \tabularnewline 
094441.48+555753.1 & - & - & $>$1.05 & 0.53 & $<$4.48 & - & 0.088 & -0.198 & 1 & - & - & - \tabularnewline 
094542.24+575747.7 & 0.2289 & - & 1.53 & 0.48 & 4.90 & 0.22 & 0.04 & -0.532 & 0.72 & 0.11 & 14.87 & - \tabularnewline 
094620.21+010452.1 & 0.5775 & - & 0.81 & 0.42 & 4.86 & 0.56 & 0.038 & -0.516 & 0.74 & 0.13 & 14.03 & radio \tabularnewline 
095127.82+010210.2 & - & - & 1.13 & 0.46 & 4.36 & 0.69 & 0.057 & -0.029 & 0.55 & 0.12 & 15.58 & - \tabularnewline 
095649.53+015601.8 & - & - & 0.80 & 0.43 & 7.10 & 1.60 & 0.043 & 0.403 & 1 & - & 16.01 & - \tabularnewline 
100050.22+574609.1 & 0.6392 & - & $>$1.16 & 0.58 & 5.14 & 1.22 & 0.088 & 0.311 & 1 & - & 13.63 & - \tabularnewline 
100326.63+020455.7 & - & - & $>$1.00 & 0.38 & 10.21 & 1.07 & 0.035 & 0.363 & 1 & - & 14.62 & - \tabularnewline 
100612.23+644011.6 & - & - & 1.07 & 0.53 & 8.81 & 1.01 & 0.074 & -0.49 & 1 & - & - & - \tabularnewline 
100959.63+014533.8 & $>$1.0900 & u & $>$0.99 & 0.55 & 14.98 & 1.35 & 0.038 & 1.034 & 1 & - & 13.36 & - \tabularnewline 
101115.64+010642.7 & 0.8615 & u & $>$0.95 & 0.70 & $<$1.64 & - & 0.058 & 0.203 & 1 & - & 14.31 & radio \tabularnewline 
101858.55+591127.8 & - & - & $>$1.40 & 0.50 & 19.08 & 0.46 & 0.046 & 0.102 & 1 & - & 19.47 & - \tabularnewline 
101950.87+632001.6 & - & - & $>$1.28 & 0.56 & 7.80 & 0.65 & 0.072 & -0.885 & 1 & - & 13.68 & - \tabularnewline 
102013.78+625010.1 & 0.2495 & - & 1.22 & 0.49 & $<$3.77 & - & 0.076 & 0.236 & 0.35 & 0.05 & - & - \tabularnewline 
102243.73$-$011302.5 & - & - & 0.93 & 0.39 & $<$2.56 & - & 0.031 & -0.122 & 1 & - & 20.08 & radio \tabularnewline 
102523.04+040229.0 & 0.2078 & - & 1.11 & 0.52 & 1.49 & 0.61 & 0.032 & 0.171 & 0.44 & 0.04 & 15.36 & radio \tabularnewline 
102724.97+631753.1 & $>$0.5816 & - & 1.14 & 0.41 & 11.51 & 1.06 & 0.077 & -0.181 & 1 & - & - & - \tabularnewline 
103208.36+040157.0 & - & - & $>$0.99 & 0.65 & 11.84 & 0.49 & 0.057 & -0.024 & 1 & - & - & - \tabularnewline 
103220.29+030949.2 & 0.3233 & - & $>$1.18 & 0.45 & 4.92 & 0.71 & 0.032 & 0.371 & 0.81 & 0.09 & 13.8 & radio \tabularnewline 
103239.07+662323.3 & - & - & $>$1.13 & 0.60 & 3.26 & 1.09 & 0.074 & -0.074 & 1 & - & - & - \tabularnewline 
103940.70+053609.3 & 0.5103 & - & $>$0.73 & 0.38 & $<$4.01 & - & 0.032 & -1.026 & 0.52 & 0.10 & 15.01 & radio \tabularnewline 
104523.86+015722.1 & - & - & $>$0.96 & 0.30 & $<$1.61 & - & 0.033 & -0.186 & 1 & - & - & - \tabularnewline 
104833.57+620305.0 & - & - & $>$1.07 & $<$0.27 & $<$1.72 & - & 0.055 & -0.296 & 1 & - & - & - \tabularnewline 
105151.84+010310.7 & 0.2654 & - & $>$1.06 & 0.42 & 1.92 & 0.35 & 0.034 & -0.74 & 0.84 & 0.07 & 14.21 & radio \tabularnewline 
105151.84+010310.7 & 0.2654 & - & $>$1.06 & 0.42 & 4.86 & 0.33 & 0.034 & -0.74 & 0.84 & 0.07 & 14.21 & radio \tabularnewline 
105606.62+025213.5 & 0.2360 & - & 0.72 & 0.32 & 1.16 & 0.39 & 0.056 & 0.089 & 0.36 & 0.06 & 15.77 & - \tabularnewline 
105752.79$-$005908.3 & 0.4678 & u & 0.94 & 0.29 & $<$3.90 & - & 0.033 & 0.254 & 0.72 & 0.13 & 15.37 & radio \tabularnewline 
105829.62+013358.8 & 0.8862 & u & 1.16 & 0.76 & 15.10 & 0.27 & 0.031 & -1.712 & 1 & - & 12.77 & radio \tabularnewline 
110356.15+002236.4 & 0.2747 & - & $>$1.12 & 0.52 & 4.38 & 0.36 & 0.056 & 0.192 & 0.40 & 0.03 & 14.89 & galaxy \tabularnewline 
110704.78+501037.9 & 0.7061 & - & $>$1.14 & 0.49 & 4.27 & 1.66 & 0.095 & 0.035 & 0.48 & 0.11 & - & - \tabularnewline 
110735.92+022224.5 & $>$1.0750 & u & $>$1.09 & 0.40 & 9.39 & 0.42 & 0.032 & -0.233 & 1 & - & 14.59 & radio \tabularnewline 
111717.55+000633.6 & 0.4511 & - & 0.87 & 0.44 & 1.46 & 0.52 & 0.036 & 0.803 & 0.67 & 0.08 & 15.46 & radio \tabularnewline 
111717.55+000633.6 & 0.4511 & - & 0.87 & 0.44 & 1.87 & 0.52 & 0.036 & 0.803 & 0.67 & 0.08 & 15.46 & radio \tabularnewline 
113115.50+023450.2 & $>$0.4538 & u & $>$1.16 & 0.38 & $<$2.79 & - & 0.033 & 0.811 & 1 & - & 14.22 & - \tabularnewline 
113234.38+023740.3 & - & - & $>$1.02 & 0.59 & $<$1.83 & - & 0.034 & 0.078 & 1 & - & 17.99 & radio \tabularnewline 
113245.61+003427.7 & - & - & 1.48 & 0.61 & 6.35 & 0.68 & 0.056 & 0.092 & 1 & - & 13.54 & radio \tabularnewline 
113523.70+660941.0 & - & - & $>$1.18 & 0.56 & 13.52 & 2.16 & 0.051 & 0.731 & 1 & - & - & - \tabularnewline 
113523.70+660941.0 & - & - & $>$1.18 & 0.56 & 12.43 & 0.83 & 0.051 & 0.731 & 1 & - & - & - \tabularnewline 
114153.35+021924.4 & 3.5979 & - & $>$1.19 & 0.11 & $<$1.26 & - & 0.056 & 0.223 & 1 & - & 13.91 & radio \tabularnewline 
114312.11+612210.8 & - & - & 1.46 & 0.48 & 5.97 & 0.97 & 0.072 & -0.701 & 1 & - & 13.63 & - \tabularnewline 
114926.13+624332.5 & 0.7620 & u & $>$1.16 & 0.41 & 4.88 & 1.81 & 0.046 & 0.656 & 1 & - & - & - \tabularnewline 
114926.13+624332.5 & 0.7620 & u & $>$1.16 & 0.41 & 7.94 & 0.60 & 0.046 & 0.656 & 1 & - & - & - \tabularnewline 
115404.54$-$001009.9 & 0.2535 & - & 0.95 & 0.37 & 2.19 & 0.69 & 0.031 & -0.086 & 0.55 & 0.07 & 14.6 & - \tabularnewline 
115548.41+613554.0 & - & - & $>$1.15 & 0.30 & $<$8.91 & - & 0.124 & 0.612 & 1 & - & - & - \tabularnewline 
115548.41+613554.0 & - & - & $>$1.15 & 0.30 & 3.73 & 1.18 & 0.124 & 0.612 & 1 & - & - & - \tabularnewline 
120303.50+603119.1 & 0.0653 & - & 1.44 & 0.49 & $<$1.57 & - & 0.073 & 0.627 & 0.46 & 0.03 & 14.62 & galaxy \tabularnewline 
120658.03+052952.2 & $>$0.7911 & - & $>$1.08 & 0.75 & 7.47 & 1.07 & 0.038 & 0.682 & 1 & - & 14.41 & radio \tabularnewline 
120938.33+021017.2 & - & - & 1.02 & 0.39 & 3.21 & 1.24 & 0.059 & 0.642 & 1 & - & 16.67 & radio \tabularnewline 
121221.56+534128.0 & 3.1900 & - & $>$1.22 & $<$0.03 & $<$0.69 & - & 0.042 & 0.154 & 1 & - & - & - \tabularnewline 
121300.80+512935.6 & 0.7957 & u & 1.08 & 0.45 & 13.38 & 1.17 & 0.075 & 0.16 & 1 & - & 14.57 & radio \tabularnewline 
121348.81+642520.2 & $>$0.4157 & - & $>$1.07 & 0.43 & 22.01 & 1.83 & 0.133 & 0.419 & 1 & - & - & - \tabularnewline 
121500.80+500215.6 & - & - & $>$1.38 & 0.44 & 8.00 & 1.57 & 0.04 & 0.254 & 1 & - & - & - \tabularnewline 
121500.80+500215.6 & - & - & $>$1.38 & 0.44 & 13.55 & 0.22 & 0.04 & 0.254 & 1 & - & - & - \tabularnewline 
121649.97+054136.7 & - & - & $>$1.03 & 0.59 & 16.66 & 0.88 & 0.035 & 0.135 & 1 & - & 13.96 & radio \tabularnewline 
121758.72$-$002946.2 & 0.4188 & - & $>$1.16 & 0.62 & 11.88 & 0.97 & 0.034 & 1.034 & 0.59 & 0.08 & 13.19 & radio \tabularnewline 
121834.93$-$011954.3 & 0.5545 & u & $>$0.95 & 0.56 & 11.24 & 0.36 & 0.055 & -0.607 & 1 & - & 13.75 & radio \tabularnewline 
121944.98+044622.4 & 0.4891 & - & $>$1.31 & 0.33 & 5.22 & 0.63 & 0.031 & 0.441 & 0.84 & 0.08 & 14.39 & - \tabularnewline 
121945.70$-$031424.0 & 0.2987 & - & 1.08 & 0.37 & 7.10 & 0.31 & 0.031 & -0.936 & 1 & - & 14.3 & - \tabularnewline 
122012.14$-$000306.8 & - & - & 1.07 & 0.42 & 6.65 & 0.82 & 0.034 & 0.369 & 1 & - & 16.74 & radio \tabularnewline 
122300.31+515313.9 & 0.3650 & - & $>$1.06 & 0.36 & 4.02 & 1.16 & 0.081 & 0.018 & 0.57 & 0.05 & - & - \tabularnewline 
122809.13$-$022136.1 & 0.3227 & - & 0.96 & 0.27 & $<$1.63 & - & 0.033 & 0.088 & 0.27 & 0.06 & 15.47 & radio \tabularnewline 
123132.38+013814.0 & 3.2300 & - & $>$1.17 & 0.23 & 1.48 & 0.67 & 0.059 & 0.173 & 1 & - & 20.99 & radio \tabularnewline 
123132.38+013814.0 & 3.2300 & - & $>$1.17 & 0.23 & $<$2.93 & - & 0.059 & 0.173 & 1 & - & 20.99 & radio \tabularnewline 
123341.33$-$014423.7 & - & - & $>$1.17 & 0.45 & 10.64 & 0.64 & 0.032 & 0.406 & 1 & - & 14.35 & radio \tabularnewline 
123743.09+630144.9 & 3.5347 & - & $>$1.18 & $<$0.04 & $<$1.38 & - & 0.046 & 0.146 & 1 & - & - & - \tabularnewline 
124225.39+642919.1 & 0.0424 & - & $>$1.40 & - & $<$1.49 & - & 0.074 & 0.342 & $<$0.05 & 0.05 & - & - \tabularnewline 
124425.30+044459.7 & 0.3999 & - & $>$1.01 & 0.41 & 2.33 & 0.78 & 0.035 & 0.213 & $<$0.16 & 0.07 & 13.57 & galaxy \tabularnewline 
124533.79+022825.2 & $>$1.0900 & u & $>$1.04 & 0.45 & 5.95 & 0.50 & 0.033 & -0.14 & 1 & - & - & - \tabularnewline 
124602.52+011318.8 & 0.3864 & - & $>$1.36 & 0.42 & 3.37 & 0.78 & 0.059 & 1.838 & 1 & - & 14.06 & radio \tabularnewline 
124834.30+512807.8 & 0.3508 & - & 1.41 & 0.50 & 8.13 & 1.36 & 0.078 & 0.188 & 0.71 & 0.11 & 14.0 & - \tabularnewline 
125032.59+021632.2 & - & - & $>$0.97 & 0.75 & 9.70 & 0.98 & 0.035 & 0.929 & 1 & - & 13.65 & radio \tabularnewline 
125359.32+624257.5 & $>$0.8680 & - & 1.25 & 0.37 & 17.28 & 1.18 & 0.089 & -0.215 & 1 & - & - & - \tabularnewline 
125820.79+612045.6 & 0.2235 & - & $>$1.19 & 0.39 & $<$1.62 & - & 0.08 & -0.346 & $<$0.00 & 0.00 & - & - \tabularnewline 
131106.48+003510.0 & - & - & 1.19 & 0.38 & 13.80 & 0.53 & 0.031 & 0.095 & 1 & - & 14.86 & radio \tabularnewline 
131330.15+020105.9 & 0.3558 & - & 1.22 & 0.52 & 5.58 & 0.53 & 0.032 & 0.042 & 0.73 & 0.04 & 14.67 & radio \tabularnewline 
132301.01+043951.4 & 0.2244 & - & 0.96 & 0.45 & 1.39 & 0.53 & 0.031 & -0.093 & 0.58 & 0.05 & 14.09 & radio \tabularnewline 
132541.91$-$022810.1 & 0.8073 & u & 0.82 & 0.43 & 5.21 & 1.04 & 0.067 & 0.318 & 1 & - & 14.08 & radio \tabularnewline 
132759.76+645811.3 & 0.4468 & - & $>$1.11 & 0.52 & $<$2.99 & - & 0.092 & -0.069 & $<$0.04 & 0.06 & - & - \tabularnewline 
133105.71$-$002221.2 & 0.2426 & - & 1.23 & 0.40 & 1.83 & 0.55 & 0.032 & 0.04 & 0.14 & 0.04 & 14.52 & radio \tabularnewline 
133219.65+622715.9 & 3.1500 & - & $>$1.10 & 0.19 & $<$1.68 & - & 0.089 & -0.045 & 1 & - & - & - \tabularnewline 
134037.59$-$014847.6 & 0.5130 & - & $>$0.98 & 0.43 & 4.10 & 0.70 & 0.038 & -0.355 & 0.69 & 0.09 & 14.08 & radio \tabularnewline 
135738.70+012813.6 & 0.5640 & u & 1.11 & 0.46 & 9.08 & 0.33 & 0.041 & 0.552 & 1 & - & 15.26 & radio \tabularnewline 
135738.70+012813.6 & 0.5640 & u & 1.11 & 0.46 & 9.18 & 3.01 & 0.041 & 0.552 & 1 & - & 15.26 & radio \tabularnewline 
140450.91+040202.2 & - & - & 1.26 & 0.32 & 7.75 & 0.77 & 0.032 & 0.744 & 1 & - & - & - \tabularnewline 
141003.92+051557.7 & 0.5440 & - & $>$0.49 & 0.40 & 1.31 & 0.59 & 0.035 & -0.044 & 0.55 & 0.04 & 14.3 & radio \tabularnewline 
141004.65+020306.9 & $>$1.1150 & u & $>$1.26 & 0.56 & 3.16 & 0.90 & 0.055 & 1.491 & 1 & - & 13.11 & radio \tabularnewline 
141004.65+020306.9 & $>$1.1150 & u & $>$1.26 & 0.56 & $<$6.76 & - & 0.055 & 1.491 & 1 & - & 13.11 & radio \tabularnewline 
141030.84+610012.8 & 0.3833 & - & 0.89 & 0.36 & $<$3.10 & - & 0.08 & 0.175 & 0.36 & 0.06 & 15.67 & radio \tabularnewline 
141826.33$-$023334.1 & - & - & $>$1.41 & 0.34 & 2.84 & 0.44 & 0.031 & 0.233 & 1 & - & 14.18 & - \tabularnewline 
141927.50+044513.8 & $>$1.6850 & - & $>$1.24 & 0.34 & 8.46 & 0.70 & 0.032 & 0.425 & 1 & - & 17.38 & - \tabularnewline 
142409.49+043452.1 & 0.6654 & u & $>$1.32 & 0.51 & 7.23 & 0.45 & 0.031 & -0.167 & 1 & - & 13.56 & radio \tabularnewline 
142505.61+035336.2 & 2.2476 & u & $>$1.13 & $<$0.09 & $<$0.57 & - & 0.032 & -0.247 & 1 & - & - & - \tabularnewline 
142526.20$-$011825.8 & - & - & $>$1.05 & 0.41 & $<$4.28 & - & 0.037 & 0.638 & 1 & - & 15.77 & radio \tabularnewline 
143657.71+563924.8 & - & - & 0.99 & 0.41 & 6.06 & 0.95 & 0.063 & -0.166 & 1 & - & 21.6 & radio \tabularnewline 
145111.69+580003.0 & 0.4053 & - & $>$1.13 & 0.39 & $<$3.86 & - & 0.085 & -0.084 & 0.33 & 0.08 & - & - \tabularnewline 
145507.44+025040.3 & - & - & $>$1.01 & 0.62 & 14.42 & 0.80 & 0.035 & 0.351 & 1 & - & 14.01 & radio \tabularnewline 
150006.49+012956.0 & 0.7083 & - & $>$0.92 & 0.34 & $<$2.53 & - & 0.038 & -0.013 & 0.83 & 0.11 & 14.16 & radio \tabularnewline 
150106.26+552750.9 & - & - & $>$1.01 & 0.50 & $<$4.10 & - & 0.052 & 0.069 & 1 & - & - & - \tabularnewline 
150106.26+552750.9 & - & - & $>$1.01 & 0.50 & 5.37 & 0.72 & 0.052 & 0.069 & 1 & - & - & - \tabularnewline 
150818.97+563611.2 & 2.0521 & u & $>$0.95 & $<$0.21 & $<$3.73 & - & 0.083 & 0.229 & 1 & - & - & - \tabularnewline 
151115.49+563715.4 & - & - & $>$1.06 & $<$0.25 & $<$5.05 & - & 0.115 & -0.223 & 1 & - & - & - \tabularnewline 
153058.17+573625.2 & 1.0998 & u & $>$1.08 & 0.50 & 8.07 & 1.11 & 0.085 & -0.437 & 1 & - & 14.93 & - \tabularnewline 
154515.78+003235.2 & 1.0114 & u & $>$1.00 & $<$0.23 & $<$1.79 & - & 0.05 & 0.543 & 1 & - & 24.04 & radio \tabularnewline 
154515.78+003235.2 & 1.0114 & u & $>$1.00 & $<$0.23 & $<$5.56 & - & 0.05 & 0.543 & 1 & - & 24.04 & radio \tabularnewline 
155848.38+022818.6 & - & - & $>$0.96 & 0.48 & 6.56 & 0.57 & 0.034 & -0.28 & 1 & - & - & - \tabularnewline 
160339.49+500955.5 & 0.6209 & u & $>$1.11 & 0.50 & 7.07 & 1.02 & 0.077 & -0.205 & 1 & - & - & - \tabularnewline 
160519.05+542059.9 & 0.2117 & - & 0.81 & 0.36 & 1.68 & 0.70 & 0.056 & 0.297 & 1 & - & 16.19 & radio \tabularnewline 
161541.22+471111.8 & 0.1986 & - & $>$1.35 & 0.52 & 3.55 & 1.16 & 0.081 & 0.225 & 0.53 & 0.05 & - & - \tabularnewline 
162115.21$-$003140.4 & 0.4132 & u & 0.95 & 0.41 & 4.31 & 0.74 & 0.033 & 0.328 & 0.68 & 0.06 & - & - \tabularnewline 
162259.24+440142.9 & - & - & 1.13 & 0.37 & $<$4.93 & - & 0.104 & -0.08 & 1 & - & 17.12 & radio \tabularnewline 
165109.18+421253.5 & 0.2686 & - & $>$1.07 & 0.49 & 3.38 & 1.10 & 0.094 & -0.226 & $<$0.15 & 0.11 & - & - \tabularnewline 
165248.44+363212.6 & 0.6470 & u & $>$1.13 & 0.54 & 8.01 & 0.89 & 0.076 & -0.258 & 1 & - & - & - \tabularnewline 
165806.77+611858.9 & $>$1.4100 & u & $>$1.09 & $<$0.14 & $<$5.52 & - & 0.121 & -0.371 & 1 & - & - & - \tabularnewline 
165806.77+611858.9 & $>$1.4100 & u & $>$1.09 & $<$0.14 & $<$5.59 & - & 0.121 & -0.371 & 1 & - & - & - \tabularnewline 
165808.33+615001.9 & 0.3742 & - & $>$1.19 & 0.47 & 0.86 & 0.27 & 0.042 & 0.027 & 0.64 & 0.10 & - & - \tabularnewline 
170108.90+395443.1 & 1.8900 & u & $>$1.07 & 0.26 & $<$1.20 & - & 0.105 & 0.143 & 1 & - & 23.79 & - \tabularnewline 
170124.64+395437.1 & 0.5071 & u & $>$1.46 & 0.50 & 6.75 & 0.21 & 0.04 & 0.132 & 1 & - & - & - \tabularnewline 
171445.55+303628.0 & 0.8500 & u & 1.14 & 0.25 & $<$0.54 & - & 0.044 & 0.025 & 1 & - & 15.63 & radio \tabularnewline 
171501.36+292912.3 & - & - & $>$0.97 & 0.54 & $<$4.33 & - & 0.116 & -0.069 & 1 & - & - & - \tabularnewline 
172640.50+595550.2 & 0.3471 & u & 1.01 & 0.50 & 2.82 & 1.12 & 0.09 & -0.857 & 1 & - & - & - \tabularnewline 
173719.12+570216.5 & - & - & $>$0.94 & 0.49 & 6.85 & 2.06 & 0.143 & -0.197 & 1 & - & - & - \tabularnewline 
205523.36$-$050619.3 & 0.3426 & - & $>$1.06 & 0.45 & 3.28 & 0.96 & 0.071 & 0.318 & 0.38 & 0.05 & - & - \tabularnewline 
205938.57$-$003756.0 & 0.3354 & - & 0.99 & 0.45 & $<$3.46 & - & 0.071 & 0.257 & 0.37 & 0.06 & - & - \tabularnewline 
211552.88+000115.5 & - & - & $>$0.95 & 0.10 & $<$2.22 & - & 0.073 & 0.102 & 1 & - & - & - \tabularnewline 
211611.89$-$062830.4 & 0.2916 & - & $>$1.08 & 0.41 & 3.56 & 0.85 & 0.071 & -0.086 & 0.37 & 0.11 & - & - \tabularnewline 
212019.13$-$075638.4 & - & - & $>$0.96 & $<$0.23 & $<$3.16 & - & 0.082 & 0.079 & 1 & - & - & - \tabularnewline 
213950.32+104749.6 & 0.2960 & - & 1.02 & - & $<$2.37 & - & 0.073 & 0.646 & 0.11 & 0.04 & - & - \tabularnewline 
215051.73+111916.5 & 0.4946 & u & $>$1.03 & 0.38 & 4.88 & 0.89 & 0.063 & 0.172 & 0.70 & 0.11 & - & - \tabularnewline 
215305.36$-$004230.7 & 0.3416 & - & 0.91 & 0.39 & 5.86 & 0.75 & 0.066 & -0.069 & 0.50 & 0.14 & - & - \tabularnewline 
215650.34$-$085535.4 & $>$1.0179 & u & $>$1.11 & 0.40 & 3.24 & 1.21 & 0.085 & 0.424 & 1 & - & - & - \tabularnewline 
221108.34$-$000302.5 & 0.3619 & - & 0.94 & 0.43 & 6.03 & 0.70 & 0.064 & -0.323 & 0.58 & 0.11 & - & - \tabularnewline 
221109.88$-$002327.5 & 0.4476 & - & $>$0.98 & 0.54 & 5.03 & 0.73 & 0.066 & -0.121 & 0.43 & 0.07 & - & - \tabularnewline 
221456.37+002000.1 & - & - & $>$0.98 & 0.55 & 10.70 & 1.18 & 0.071 & 0.239 & 0.77 & 0.10 & - & - \tabularnewline 
224448.11$-$000619.3 & - & - & $>$0.68 & 0.35 & 4.24 & 0.60 & 0.056 & -0.262 & 1 & - & - & - \tabularnewline 
224730.19+000006.5 & - & - & $>$1.01 & 0.63 & 4.24 & 0.52 & 0.054 & -0.194 & 1 & - & - & - \tabularnewline 
224819.44$-$003641.6 & 0.2123 & - & $>$1.03 & 0.52 & 1.27 & 0.57 & 0.064 & 1.43 & 0.19 & 0.04 & - & - \tabularnewline 
225624.27+130541.7 & - & - & $>$1.10 & 0.49 & 15.56 & 0.80 & 0.058 & -0.038 & 1 & - & - & - \tabularnewline 
231000.81$-$000516.3 & $>$1.6800 & u & $>$1.04 & $<$0.13 & $<$3.21 & - & 0.066 & 0.036 & 1 & - & - & - \tabularnewline 
233445.56+154711.1 & - & - & $>$0.72 & 0.44 & 11.73 & 0.82 & 0.064 & 0.093 & 1 & - & - & - \tabularnewline
235604.03$-$002353.8 & 0.2830 & - & 1.28 & 0.33 & $<$3.48 & - & 0.066 & 0.337 & 0.46 & 0.08 & - & - \tabularnewline
\end{longtable}

\end{appendix}
\end{document}